\pdfoutput=1
\documentclass[PAPER, atlasdraft=false, UKenglish, cernpreprint, texmf, orcidlogo]{atlasdoc}
\usepackage[subcaption, rotating, longtable, multirow]{atlaspackage}
\usepackage{atlasbiblatex}

\usepackage{atlasphysics}

\usepackage{adjustbox}
\usepackage{pifont}

\usepackage{cleveref}

\graphicspath{{logos/}{figures/}}

\usepackage{ANA-TOPQ-2023-34-PAPER-defs}

\AtlasTitle{Multi-approach Standard Model Effective Field Theory analyses of top-quark associated production in leptonic final states at $\sqrt{s}=13$ TeV with the ATLAS detector}
\AtlasAbstract{%
Standard Model Effective Field Theory (SMEFT) interpretations of measurements of
top-quark associated production processes in leptonic final states are
presented using the full Run 2 proton--proton collision dataset
recorded by the ATLAS detector at the Large Hadron Collider at
$\sqrt{s}=13$~TeV, corresponding to an integrated luminosity of 140
fb$^{-1}$. A simultaneous likelihood fit is performed across final
states sensitive to \tty, \ttZ, \ttW, \ttH, \tttt, \tZq, \tWZ, \tHq,
and \tHW production, with harmonised reconstructed final-state objects
and a coherent treatment of detector-level SMEFT effects across all
contributing processes. Two complementary analysis strategies are
investigated: an approach employing a more generic event
categorisation based on counting reconstructed objects, and a more
complex approach combining orthogonal regions optimised to
specifically identify individual processes, as employed by previous
dedicated measurements. Constraints are obtained on the Wilson coefficients of
dimension-six SMEFT operators affecting top-quark interactions,
including top-boson, four-quark and top-lepton operators,
together with
complex operator contributions.
The results do not show a significant deviation from the Standard
Model expectation and 95\% confidence interval limits are placed in a
simultaneous fit to 33 SMEFT Wilson coefficients.  The comparison of the two
analysis strategies provides new insight into the design of
increasingly complex SMEFT interpretations. The second, more complex, approach targeting individual processes results in the best
sensitivity to the majority of Wilson coefficients.
}

\AtlasRefCode{TOPQ-2023-34}

\PreprintIdNumber{CERN-EP-2026-270}

\AtlasJournal{JHEP}
\AtlasCoverEgroupAnalysisTeam{atlas-topq-2023-34-analysis-team@cern.ch}

\hypersetup{pdftitle={ATLAS document},pdfauthor={The ATLAS Collaboration}}

\begin{document}

\maketitle

\tableofcontents

\section{Introduction}
\label{sec:intro}

% The next lines are included from the .//Sections/Intro.tex input file

%
%
%
%
%
%
%
%
%
%
%
%
%
%

The Large Hadron Collider (LHC)~\cite{Evans:2008zzb} enables an
extensive programme of direct searches for physics beyond the Standard
Model (BSM), exploiting its unprecedented energy reach. At the same
time, increasingly precise measurements of Standard Model (SM) processes
provide a complementary approach to probing BSM physics, with
deviations from SM predictions potentially providing
indirect evidence for new phenomena. As the heaviest known elementary
particle, the top quark plays a particularly important role in tests
of the SM and the mechanism of electroweak (EW) symmetry
breaking (EWSB). Rare processes involving the associated production of
top quarks with electroweak bosons, collectively denoted as \topx, as
well as being interesting in their own right also provide sensitive
probes of EW couplings to the top-quark. In particular, they provide access
to interactions involving top quarks and electroweak bosons that are
not directly accessible in other particle-collision
processes~\cite{Maltoni:2019aot,Brivio:2019ius}. The search for
deviations from SM predictions in \topx processes within
an Effective Field Theory (EFT) framework therefore provides a
complementary probe of BSM signatures, testing the structure of the
top-quark interactions and their relation to EWSB.

The SM Effective Field Theory
(SMEFT)~\cite{Buchmuller:1985jz,Grzadkowski:2010es,Brivio:2017vri,Falkowski:2019tft}
offers a self-consistent framework to study the combined effect of
possible BSM signatures across different measurements, providing a
coherent description of the impact of heavy new-physics states at a
mass scale $\Lambda$ significantly above the EW scale.
Higher-dimensional operators can be added to the SM Lagrangian to approximate deviations from the SM due to the low-energy effects of heavy new physics.
The shorthand \EFT will be used in place of SMEFT in all of the following.
The leading order BSM effects on the Lagrangian considered here are parameterised by dimension-six operators.
Measurements of
observables sensitive to the effect of \EFT operators can be used to
constrain Wilson coefficients (WCs) associated with these
operators.

ATLAS and CMS have published multiple inclusive and
differential SM cross-section measurements of \topx
processes including those using the \EFT framework to probe for BSM physics~\cite{TOPQ-2021-08,TOPQ-2020-20,TOPQ-2018-24,TOPQ-2019-30,TOPQ-2023-02,TOPQ-2019-18-fix,HIGP-2024-08,TOPQ-2021-21,HIGP-2025-16,CMS-TOP-17-019,CMS-TOP-18-009,CMS-TOP-19-001,CMS-TOP-21-001,CMS-TOP-18-010,CMS-TOP-21-004,CMS-TOP-21-003,CMS-TOP-22-013,CMS-TOP-22-006,CMS-TOP-24-012}.
Other measurements from ATLAS and CMS in \ttbar and single-top-quark final states also have sensitivity to relevant \EFT operators~\cite{TOPQ-2019-23,TOPQ-2018-10,TOPQ-2019-28,TOPQ-2018-11,TOPQ-2020-06,TOPQ-2020-17,ATL-PHYS-PUB-2025-037,CMS-TOP-17-014,CMS-TOP-18-006,CMS-TOP-17-020,CMS-TOP-19-004}.

Several operators affect multiple top-quark processes simultaneously,
and these should be treated coherently for a robust interpretation and to extract the highest
possible sensitivity~\cite{Hartland:2019bjb}.  The CMS Collaboration have published a
simultaneous \EFT interpretation of multiple \topx processes in leptonic
final states~\cite{CMS-TOP-22-006}. The result presented here is the first of its
kind for ATLAS and significantly extends the scope of
Ref.~\cite{CMS-TOP-22-006}.

``Global'' \EFT analyses have been carried out within
the LHC collaborations that provide constraints on dimension-six WCs
from measurements of Higgs boson, electroweak, top-quark, and diboson
processes~\cite{HIGG-2022-17,CMS-SMP-24-003-fix,STDM-2023-32}. Outside the
collaborations, similar global analyses have been
performed~\cite{Falkowski:2015ModelIndependent,Ellis:2018UpdatedGlobalSMEFT,Falkowski:2020FlavourfulLikelihood,Dawson:2020PuttingSMEFTFitsToWork,Ethier:2021CombinedSMEFT,Giani:2023smefit,Grunwald:2023nli}.
While the work outside the collaborations achieves impressive sensitivity, it can be limited by incomplete information on uncertainty correlations and overlapping phase spaces. The inclusion of full uncertainty correlations and the correct statistical treatment of overlapping measurements are important both to improve sensitivity and to ensure correctness. This can be handled more efficiently inside the experimental collaborations that performed the measurements. However, once incorporated, the information can be made public and shared more widely for the benefit of the whole community.

This paper focuses on \EFT interpretations of measurements of \topx processes, including:
\tty, \ttZ, \ttW, \ttH, \tttt, \tZq, \tWZ, \tHq, and \tHW.
A simultaneous measurement of multiple \topx processes is
made which allows for the inclusion of modifications to the \topx
processes via the dependence on the dimension-six WCs implemented at detector-level
coherently across all relevant processes.
The measurement targets final states sensitive to leptonic top quark decays that are enriched in the \topx processes of interest.

There are several motivations for such a measurement of multiple processes to simultaneously probe multiple \EFT operators.
Many processes are affected by overlapping sets of operators, so the combination can improve sensitivity and resolve degeneracies.
In this way the \EFT framework can be thought of as a multi-messenger~\cite{Meszaros:2019xej} approach; measurements of single processes may not be able to separate effects
from multiple \EFT operators, while the combination of multiple processes exploiting complementary
information from multiple sources can enhance sensitivity.
One of the challenges of this simultaneous measurement approach is that
there can be overlapping signals, overlapping event selections and
strong correlations in systematic uncertainties.  These challenges are addressed here
by harmonisation of object definitions, orthogonal event categories
and correlation of systematic uncertainties including theory
modelling. Additionally, consistent parameterisations of the dependence on the dimension-six WCs across the
relevant processes are applied including for processes which act as
backgrounds to one another. The \EFT samples are propagated through
detector simulation and reconstruction, so \EFT-induced changes to
acceptance, selection efficiencies and migrations between regions and observable bins
are retained.

Previous approaches for \topx combinations, such as
Ref.~\cite{CMS-TOP-22-006}, were based on a generic, relatively
process-agnostic definition of analysis regions through the counting
of reconstructed objects. For the first time, this kind of strategy,
referred to here as the Object Based (OB) approach, is directly
compared with an alternative strategy that places greater emphasis on
defining regions to isolate individual \topx processes, referred to as
the Process Based (PB) approach. This comparison aims to assess
whether the PB approach can better exploit the complementary
sensitivities of the different \topx processes to a broad range
of \EFT operators.

The PB approach implementation
benefits from the significant work already performed by many analysis
teams to optimise sensitivity to dedicated cross-section measurements
of individual \topx processes. The source analyses are summarised in
Table~\ref{tab:analyses} with ATL-$XYZ$ analysis labels defined that are referred to
later in the paper. In the PB approach these analyses are coherently combined
using slight modifications to their existing region
definitions, and generally rely on dedicated machine learning
algorithms (ML) to separate between processes.
This improved process separation can increase the sensitivity to
dimension-six operators whose effects differ between the contributing
processes.  Potential downsides include reduced data statistics due to
more restrictive event selections and high dependence on the modelling
of signal and background processes. The latter can enter in two ways:
due to the reliance of ML discriminants on complex event topologies
that can lead to increased systematic uncertainties, and due to the
assumption of SM kinematics in the inputs used for ML classifier
training. It is therefore unclear a priori which approach provides
the best sensitivity.
In general, these analyses are the most sensitive to the relevant
operators, but there are cases where more precise measurements exist
but with a different final state. For example, the latest measurements
of Higgs boson production in other decay modes, such as
$H\to\gamma\gamma$~ and
$H\to b\bar{b}$, may have stronger sensitivity to relevant operators~\cite{HIGG-2020-16,HIGG-2020-24,HIGP-2024-24,HIGP-2025-16}.

\begin{table}[!htb]
\footnotesize
\caption{A summary of the dedicated per-process source analyses used to inform reconstructed object definitions, and the event selection region definitions (and ML models) used for the PB approach. The shorthand labels used in this paper are given as well as the references.}
\label{tab:analyses}
\def\arraystretch{1.3}
\begin{center}
\begin{tabular}{ccc}
\toprule
Process & Label & Publication \\
\midrule
\tttt & \aftops & Eur. Phys. J. C 83 (2023) 496\\
\ttZ & \attz & JHEP 07 (2024) 163\\
\ttW & \attw & JHEP 05 (2024) 131\\
\tty & \atty & JHEP 10 (2024) 191\\
\ttll & \attll & Eur. Phys. J. C 85 (2025) 1434\\
\ttWEW & \attwj & Phys. Rev. D 114 (2026) 012001\\
\ttH & \atth & JHEP 05 (2026) 183\\
\tZq & \atzq & Submitted to Eur. Phys. J. C \\
\bottomrule
\end{tabular}
\end{center}
\end{table}

The results of this comparison between OB and PB approaches can inform
future strategies for such measurements. This paper also considers further extensions to the
approach taken in Ref.~\cite{CMS-TOP-22-006} by including additional
regions sensitive to \tty, forward jets to better isolate
$t$-channel single-top quark topologies, and imaginary components of complex WCs.

The paper is organised as follows: the ATLAS detector is described in
Section~\ref{sec:detector}, the analysis strategy is described in
Section~\ref{sec:analysis}, and an overview of the dataset and
simulated data used is provided in Section~\ref{sec:datamc}. The
reconstructed objects are defined in
Section~\ref{sec:objectdefintions} and the event selection criteria
used to define the analysis regions are provided in
Section~\ref{sec:regions}. The estimation of the major backgrounds is
described in Section~\ref{sec:bkgEst}, the systematic uncertainties
are outlined in Section~\ref{sec:systematics} and the results and conclusion are
provided in Sections~\ref{sec:result} and~\ref{sec:conclusion},
respectively.

\FloatBarrier

% End of text imported from the .//Sections/Intro.tex input file

%

%

%
\section{ATLAS detector}
\label{sec:detector}

% The next lines are included from the .//Sections/Detector.tex input file

%
\newcommand{\AtlasCoordFootnote}{%
ATLAS uses a right-handed coordinate system with its origin at the nominal interaction point (IP)
in the centre of the detector and the \(z\)-axis along the beam pipe.
The \(x\)-axis points from the IP to the centre of the LHC ring,
and the \(y\)-axis points upwards.
Polar coordinates \((r,\phi)\) are used in the transverse plane,
\(\phi\) being the azimuthal angle around the \(z\)-axis.
The pseudorapidity is defined in terms of the polar angle \(\theta\) as \(\eta = -\ln \tan(\theta/2)\) and is equal to the rapidity
$ y = \frac{1}{2} \ln \left( \frac{E + p_z}{E - p_z} \right) $ in the relativistic limit.
Angular distance is measured in units of \(\Delta R \equiv \sqrt{(\Delta y)^{2} + (\Delta\phi)^{2}}\).}

The ATLAS detector~\cite{PERF-2007-01} at the LHC covers nearly the entire solid angle around the collision point.\footnote{\AtlasCoordFootnote}
It consists of an inner tracking detector surrounded by a thin superconducting solenoid, electromagnetic and hadronic calorimeters,
and a muon spectrometer incorporating three large superconducting air-core toroidal magnets.

The inner-detector system (ID) is immersed in a \qty{2}{\tesla} axial magnetic field
and provides charged-particle tracking in the range \(|\eta| < 2.5\).
The high-granularity silicon pixel detector covers the vertex region and typically provides four measurements per track,
the first hit generally being in the insertable B-layer (IBL) installed before Run~2~\cite{ATLAS-TDR-19,PIX-2018-001}.
It is followed by the SemiConductor Tracker (SCT), which usually provides eight measurements per track.
These silicon detectors are complemented by the transition radiation tracker (TRT),
which enables radially extended track reconstruction up to \(|\eta| = 2.0\).
The TRT also provides electron identification information
based on the fraction of hits (typically 30 in total) above a higher energy-deposit threshold corresponding to transition radiation.

The calorimeter system covers the pseudorapidity range \(|\eta| < 4.9\).
Within the region \(|\eta|< 3.2\), electromagnetic calorimetry is provided by barrel and
endcap high-granularity lead/liquid-argon (LAr) calorimeters,
with an additional thin LAr presampler covering \(|\eta| < 1.8\)
to correct for energy loss in material upstream of the calorimeters.
Hadronic calorimetry is provided by the steel/scintillator-tile calorimeter,
segmented into three barrel structures within \(|\eta| < 1.7\), and two copper/LAr hadronic endcap calorimeters.
The solid angle coverage is completed with forward copper/LAr and tungsten/LAr calorimeter modules
optimised for electromagnetic and hadronic energy measurements, respectively.

The muon spectrometer (MS) comprises separate trigger and
high-precision tracking chambers measuring the deflection of muons in a magnetic field generated by the superconducting air-core toroidal magnets.
The field integral of the toroids ranges between \num{2.0} and \qty{6.0}{\tesla\metre}
across most of the detector.
Three layers of precision chambers, each consisting of layers of monitored drift tubes, cover the region \(|\eta| < 2.7\),
complemented by cathode-strip chambers in the forward region, where the background is highest.
The muon trigger system covers the range \(|\eta| < 2.4\) with resistive-plate chambers in the barrel, and thin-gap chambers in the endcap regions.

The luminosity is measured mainly by the LUCID--2~\cite{LUCID2} detector that records Cherenkov light produced in the quartz windows of photomultipliers located close to the beampipe.

Events were selected by the first-level trigger system implemented in custom hardware,
followed by selections made by algorithms implemented in software in the high-level trigger~\cite{TRIG-2016-01}.
The first-level trigger accepted events from the \qty{40}{\MHz} bunch crossings at a rate close to \qty{100}{\kHz},
which the high-level trigger further reduced in order to record complete events to disk at about \qty{1.25}{\kHz}.

A software suite~\cite{SOFT-2022-02} is used in data simulation, in the reconstruction
and analysis of real and simulated data, in detector operations, and in the trigger and data acquisition
systems of the experiment.

% End of text imported from the .//Sections/Detector.tex input file

%

%

%
\section{Analysis strategy}
\label{sec:analysis}

% The next lines are included from the .//Sections/AnalysisStrategy.tex input file

The primary focus of this paper is an EFT interpretation performed at
the detector-level. Each bin of the differential detector-level
observables is parameterised to account for the modifications to the
SM prediction from the WCs. The parameterisation is calculated for
every \topx process on which the effect of \EFT operators is
considered: \tty, \ttZ, \ttW, \ttH, \tttt, \tZq, \tWZ, \tHq,
and \tHW. Effects from the dimension-six \EFT operators are only
considered in the production mechanism of the processes, not in their
decay. While there can be sensitivity to \EFT operators in top-quark
decays~\cite{Boughezal:2019xpp}, the sensitivity to such operators is expected to be higher in \ttbar and single-top measurements with cross-sections that are significantly larger than those targeted here.

Following the convention used in \atty, the \tty simulation is split into \ttyprod
and \ttydecay components at generator level.
The \ttZ and \tZq processes include both contributions from on-shell
$Z$ bosons and off-shell $\gamma^{*}$ (including interference) leading
to the \ttll and \tllq final states.  The \ttW process is also
separated, with the bulk \ttWQCD contribution isolated from the \ttWEW contribution
(which arises in NLO corrections at order $\alpha^3\alpha_s$) due to the complementary
sensitivity to different \EFT operators from each, as done
in \attwj. The details of how these processes are defined and
simulated are described in Appendix~\ref{app:MC}. The details of
the \EFT implementation are described in Section~\ref{sec:ana:smeft}.

The OB and PB approaches used in this analysis are described in
Sections~\ref{sec:ana:ob} and~\ref{sec:ana:pb}, respectively.
In advance of the \EFT interpretation,
normalisation factors of the different processes are measured to compare the sensitivity to each process between the OB and PB approaches, as well as to validate them against the dedicated measurements.

In general, the choice of the physics object definitions, described in Section~\ref{sec:objectdefintions}, is carefully made to
ensure orthogonality between selections while maintaining the best
sensitivity to the \topx processes. This harmonisation is described in
more detail in Section~\ref{sec:ana:pb}.

Since both approaches target events containing processes with at least
one leptonically decaying top quark, typical requirements are for one
or more charged leptons (electron or muon) and one or more jets
identified as originating from a $b$-quark ($b$-tagged jets). No requirements are applied to select events with hadronically-decaying $\tau$-leptons. Processes
with additional top quark decays are selected through the
requirement of additional $b$-tagged jets and potentially light jets
originating from hadronically-decaying top quarks. Processes involving
photons, \tty, are identified through the requirement of a
reconstructed photon. Processes with additional leptonically decaying
particles, \ttZ, \ttW, \ttH and \tttt, are identified with 2-, 3-
and 4-lepton selections.  Events containing $t$-channel single-top-quark
processes, \tZq and \tHq, are enriched with selections targeting
the so-called spectator quark which has a topology with, on average, a
higher rapidity (forward) light jet. More complex final states, such
as \ttH and \tttt, can be further isolated by requiring even higher
jet and $b$-tagged jet multiplicities.

The OB and PB approaches described below differ primarily in the
definitions of the signal regions. Both methodologies utilise control
regions, most of which are common to both, for estimating
normalisations and controlling both irreducible and reducible
background processes and their corresponding systematic
uncertainties. The major backgrounds and their common control regions are introduced in Section~\ref{sec:ana:background}.

\subsection{EFT parameterisation}
\label{sec:ana:smeft}

The SMEFT provides a complete
phenomenological extension of the SM beyond dimension-four terms in the
Lagrangian.
New physics effects from BSM theories characterised by a higher mass scale than the
energies probed in a typical LHC collision can therefore be related to
a set of operators $\left\{\mathcal{Q}\right\}$, together with their
WCs $\left\{C\right\}$.  Observed deviations from the
SM can then be expressed in terms of $\left\{C\right\}$, without
specifying a particular BSM model.  Similarly, LHC data found to be in
agreement with the SM lead to constraints in the phenomenological
space spanned by the WCs~$\left\{C\right\}$, which in
turn can be translated into exclusions of various BSM scenarios.  The
Lagrangian of the \EFT thus reads:
\begin{equation}%
\mathcal{L}_\text{EFT} = \mathcal{L}_\text{SM} + \sum_{d>4}\mathcal{L}^{(d)},\quad \mathcal{L}^{(d)}=\sum_{i=1}^{n_d}\dfrac{C_i^{(d)}}{\Lambda^{d-4}}\mathcal{Q}_i^{(d)},
\end{equation}
where $\Lambda$ is a suitable cutoff scale and $d$ is the number of dimensions considered.
Operators contributing to $\mathcal{L}^{(5)}$ include
lepton-number violating terms, and are therefore ignored
in this analysis.
The EFT parameterisation is restricted to dimension-six operators, retaining both their interference with the SM and their quadratic contributions. Dimension-eight operators, which contribute at the same order in \(1/\Lambda\) as the dimension-six squared terms, are not considered.

The \EFT Monte Carlo simulated samples used are produced with
the \texttt{SMEFTsim 3.0} UFO model \cite{Brivio:2020onw} with
the \texttt{top} flavour scheme. The event generation is conducted considering only single operator insertions at leading order in
QCD.  Further details are provided in Section~\ref{sec:datamc}.
The \texttt{top} flavour scheme assumes a $U(2)^3$ flavour symmetry in
the quark sector while allowing independent lepton-flavour
coefficients for the three generations. However, the lepton flavour
generations are recombined in this analysis to have a single lepton
definition inclusive in flavour, assuming lepton-flavour universality
for the top-lepton operators. This choice is largely motivated by the
fact that Ref.~\cite{TOPQ-2023-02} already demonstrated that the
sensitivity to flavour-separated WCs is still limited with the current
dataset.

The dependence of a generic observable $\mathcal{O}$ (e.g. an
inclusive or differential cross section) on a set of WCs~$\left\{C\right\}$ can be parameterised as:
\begin{equation}\label{eq:smeft-parameterisation}
\mathcal{O} = \mathcal{O}_\text{SM} + \sum_i C_i A_i + \sum_{i,j} C_i C_j B_{ij}.
\end{equation}
A second-degree multivariate polynomial fit to
any observable of interest therefore yields the coefficients for the ``linear terms'', $A_i$,
which represents the interference between the SM and the dimension-six operators, and
the ``quadratic terms'', $B_{ij}$, which represents dimension-six-squared contributions and cross-terms.
This parameterisation is calculated for each measurement bin and for each process.
In this analysis, a scaling approach is adopted to apply the dependence on the WCs as follows:
\begin{equation}\label{eq:smeft-lo-nlo-correction}
\mathcal{O} \mapsto \mathcal{O}^\prime = \dfrac{\mathcal{O}_\text{SM}^\text{NLO}}{\mathcal{O}_\text{SM}^\text{LO}} \times \left( \mathcal{O}_\text{SM}^\text{LO} + \sum_i C_i A_i + \sum_{i,j} C_i C_j B_{ij} \right),
\end{equation}
where $\mathcal{O}_\text{SM}^\text{NLO}$ relies on the modelling of
the nominal \topx samples (detailed in~\cref{tab:MCSamples}) and $\mathcal{O}_\text{SM}^\text{LO}$ is the
SM point of the EFT samples (with all WCs set to zero).  Once the
dedicated signal regions are defined, the events within them are binned
in variables sensitive to dimension-six operator variations, as
described in Section~\ref{sec:regions}. The effect of EFT operators in
the analysis is defined by the parameterisation in
Eq.~(\ref{eq:smeft-lo-nlo-correction}) and is extracted for each
process in each bin.  For some operators and
\topx processes the higher order corrections for the EFT operator
dependence are different than those for the
SM~\cite{Maltoni:2016yxb,Degrande:2018fog,Degrande:2020evl}. For
example, this is true for the effect of the four-heavy-quark operators
on the \tttt process where the EFT correction is smaller than the SM
correction \cite{Degrande:2020evl}.
In the context of working with the LO \texttt{SMEFTsim 3.0} model without dedicated calculations for all cases where this could be relevant, and for consistency in the treatment across processes, no correction is applied
to the parameterisations beyond those discussed in Eqn.~\ref{eq:smeft-lo-nlo-correction}. However, this effect was
studied specifically for the case of \tttt, and the impact was 20-30\%
worse limits on the expected 68\% confidence intervals for the four-heavy-quark operators.%

An overview of all the EFT operators considered and to which processes
they apply is given in Table~\ref{tab:smeft-operators-lagranges}. They
broadly fall into three classes: top-boson operators, four-quark
operators and top-lepton operators.  The top-boson operator WCs \ctHRe, \ctHIm, \ctWRe, \ctWIm, \ctBRe, \ctBIm, \ctGRe and \ctGIm are the only Wilson coefficients
that are split into imaginary and real components in \texttt{SMEFTsim 3.0}, since they have a meaningful phase.
The imaginary components are CP-violating and although they can interfere with the SM this contribution doesn't affect the overall normalisation. Therefore, the dimension-six-squared contributions of these CP-odd components are dominant in this analysis.

Out of the top-lepton operators, the operators \cQl1 and \cQl3 cannot be constrained in the fit simultaneously as
there is a blind direction.  Instead, linear combinations of the two
are considered: \(c_{lQ}^{(\pm)} = c_{lQ}^{(1)} \pm c_{lQ}^{(3)}\), where
\cQl+ impacts vertices with two top quarks and two
neutrinos and \cQl- impacts vertices with two top quarks and
two charged-leptons.  As there is no sensitivity to \cQl+ from either the OB or PB measurements, only \cQl- is considered.
For \cHQ the situation is similar, the linear combinations $\cHQ\pm = \cHQ1 \pm \cHQ3$ correspond to vertices with a bottom quark and a
$Z$ boson (\cHQ+) and vertices with a top quark and a
$Z$ boson (\cHQ-). As \cHQ+ is far more highly constrained by \( Z \rightarrow b\bar{b}\) measurements than would be possible by this analysis~\cite{Grunwald:2023nli}, only \cHQ- is included.
Of the four-heavy-quark operators, \cQQ1 and \cQQ8 have a blind
direction in \cQQ8 \(= -3\) \cQQ1, and therefore cannot be constrained
in the fit at the same time.  Instead, a linear combination of the
two, $\cQQ+ = \frac{1}{2} \cQQ1 + \frac{1}{6} \cQQ8$
is included, as described in Ref.~\cite{DHondt:2018cww}.

The complexity of the fits described in Section~\ref{sec:result} becomes very large with $\mathcal{O}(100)$
regions, $\mathcal{O}(200)$ SR bins, $\mathcal{O}(10)$ SM processes and $\mathcal{O}(30)$ EFT operators.
This corresponds to up to $\mathcal{O}(100\textrm{k})$ EFT parameterisation values (i.e. linear or quadratic terms for a given operator, process, and bin) and $\mathcal{O}(1\textrm{M})$ input parameters (e.g. EFT parameterisation or nuisance parameter values for a given process and bin) to build the fit model.

\subsection{Object-based approach}
\label{sec:ana:ob}
A relatively straightforward strategy for building a coherent EFT
interpretation in the \topx sector is the OB approach. This is based
on building selections from the counting of reconstructed
objects and their properties that roughly separate the data into signal regions with different final states.
These features include
lepton, jet, $b$-tagged jet and photon multiplicities,
lepton charges and invariant masses of the charged leptons.
Although the regions are defined by simple features, they do provide
separation between physics processes despite in this approach not
strictly being optimised to do so. Nevertheless, the final binning of
the regions in EFT-optimised observables still enables sensitivity.

\clearpage
% The next lines are included from the .//Sections/SMEFTTable.tex input file
\begin{table}[!htb]
\scriptsize
\caption{Definitions of the relevant dimension-6 SMEFT operators and their impact on \topx processes.
For the four top-boson operators, indicated with a $\left(\star\right)$, the real and imaginary parts of the corresponding Wilson coefficients are distinguished.
The ``\checkmark'' indicates that a given operator (row) is applied to a given process (column).
Operators presented in grey are used in the definitions of linear combinations of operators.
}
\label{tab:smeft-operators-lagranges}
\def\arraystretch{1.67}
\begin{center}
\setlength{\tabcolsep}{4pt}
\begin{tabular}{llccccccccccc}%
\toprule
& Operator & Definition & & \ttZ & \tZq & \ttH & \ttW & \ttWEW & \ttyprod & \tttt & \tWZ & \tHq/\tHW \\
\midrule
\multirow{6}{*}{\rotatebox[origin=c]{90}{\normalsize top-boson}}
& $\mathscr{O}_{tW}$         & $(\bar Q \sigma^{\mu\nu} t) \sigma^i \tilde H W_{\mu\nu}^i$ & $\left(\star\right)$ & \checkmark & \checkmark &  & \checkmark & \checkmark & \checkmark & \checkmark & \checkmark & \checkmark \\
& $\mathscr{O}_{tB}$         & $(\bar Q \sigma^{\mu\nu} t)  \tilde H B_{\mu\nu}$           & $\left(\star\right)$ & \checkmark & \checkmark &  &  & \checkmark & \checkmark & \checkmark & \checkmark & \\
& $\mathscr{O}_{tG}$         & $(\bar Q \sigma^{\mu\nu} T^a t) \tilde H G_{\mu\nu}^a$      & $\left(\star\right)$ & \checkmark & & \checkmark & \checkmark & \checkmark & \checkmark & \checkmark & \checkmark & \\
& $\mathscr{O}_{HQ}^{(-)}$   & $\mathscr{O}_{HQ}^{(1)} - \mathscr{O}_{HQ}^{(3)}$ & & \checkmark & \checkmark &  &  & \checkmark & \checkmark & \checkmark & \checkmark & \\
& \textcolor{gray}{$\mathscr{O}_{HQ}^{(1)}$}   & \textcolor{gray}{$(H^\dag i\overleftrightarrow{D}_\mu H) (\bar Q \gamma^\mu Q)$} \\
& \textcolor{gray}{$\mathscr{O}_{HQ}^{(3)}$}   & \textcolor{gray}{$(H^\dag i\overleftrightarrow{D}^i_\mu H) (\bar Q \sigma^i\gamma^\mu Q)$} \\
& $\mathscr{O}_{Ht}$         & $(H^\dag i\overleftrightarrow{D}_\mu H) (\bar t \gamma^\mu t)$ & & \checkmark & \checkmark &  &  & \checkmark & \checkmark & \checkmark & \checkmark & \\
& $\mathscr{O}_{tH}$         & $(H^\dag H) (\bar{Q} \tilde{H}t )$ & $\left(\star\right)$ & & & \checkmark & & \checkmark &  & \checkmark & & \checkmark \\

\midrule
\multirow{18}{*}{\rotatebox[origin=c]{90}{\normalsize four-quark}}
& $\mathscr{O}_{tu}^{(1)}$   & $(\bar t \gamma_\mu t)(\bar u \gamma^\mu u)$ & & \checkmark &  & \checkmark &  &  & \checkmark & \checkmark & \checkmark & \\
& $\mathscr{O}_{tu}^{(8)}$   & $(\bar t T^a\gamma_\mu t)(\bar u T^a\gamma^\mu u)$ & & \checkmark &  & \checkmark &  &  & \checkmark & \checkmark & \checkmark &  \\
& $\mathscr{O}_{td}^{(1)}$   & $(\bar t \gamma_\mu t)(\bar d \gamma^\mu d)$ & & \checkmark &  & \checkmark &  &  & \checkmark & \checkmark & \checkmark &  \\
& $\mathscr{O}_{td}^{(8)}$   & $(\bar t T^a\gamma_\mu t)(\bar d T^a\gamma^\mu d)$ & & \checkmark &  & \checkmark &  &  & \checkmark & \checkmark & \checkmark &  \\
& $\mathscr{O}_{tj}^{(1)}$   & $(\bar q \gamma_\mu q)(\bar t \gamma^\mu t)$ & & \checkmark &  & \checkmark & \checkmark &  & \checkmark & \checkmark & \checkmark &  \\
& $\mathscr{O}_{tj}^{(8)}$   & $(\bar q T^a\gamma_\mu q)(\bar t T^a\gamma^\mu t)$ & & \checkmark &  & \checkmark & \checkmark &  & \checkmark & \checkmark & \checkmark &  \\
& $\mathscr{O}_{Qt}^{(1)}$   & $(\bar Q \gamma_\mu Q)(\bar t \gamma^\mu t)$ & & \checkmark &  & \checkmark &  &  & \checkmark & \checkmark & \checkmark &  \\
& $\mathscr{O}_{Qt}^{(8)}$   & $(\bar Q T^a\gamma_\mu Q)(\bar t T^a\gamma^\mu t)$ & & \checkmark &  & \checkmark &  &  & \checkmark & \checkmark & \checkmark &  \\
& $\mathscr{O}_{Qu}^{(1)}$   & $(\bar Q \gamma_\mu Q)(\bar u \gamma^\mu u)$ & & \checkmark &  & \checkmark &  &  & \checkmark & \checkmark & \checkmark &  \\
& $\mathscr{O}_{Qu}^{(8)}$   & $(\bar Q T^a\gamma_\mu Q)(\bar u T^a\gamma^\mu u)$ & & \checkmark &  & \checkmark &  &  & \checkmark & \checkmark & \checkmark &  \\
& $\mathscr{O}_{Qd}^{(1)}$   & $(\bar Q \gamma_\mu Q)(\bar d \gamma^\mu d)$ & & \checkmark &  & \checkmark &  &  & \checkmark & \checkmark & \checkmark &  \\
& $\mathscr{O}_{Qd}^{(8)}$   & $(\bar Q T^a\gamma_\mu Q)(\bar d T^a\gamma^\mu d)$ & & \checkmark &  & \checkmark &  &  & \checkmark & \checkmark & \checkmark &  \\
& $\mathscr{O}_{QQ}^{(+)}$   & $\frac{1}{2} \mathscr{O}_{QQ}^{1} + \frac{1}{6} \mathscr{O}_{QQ}^{8}$ & & \checkmark &  & \checkmark &  &  & \checkmark & \checkmark & \checkmark &  \\
& \textcolor{gray}{$\mathscr{O}_{QQ}^{(1)}$}   & \textcolor{gray}{$(\bar Q \gamma_\mu Q)(\bar Q \gamma^\mu Q)$} \\
& \textcolor{gray}{$\mathscr{O}_{QQ}^{(8)}$}   & \textcolor{gray}{$(\bar Q T^a\gamma_\mu Q)(\bar Q T^a\gamma^\mu Q)$} \\
& $\mathscr{O}_{Qj}^{(1,1)}$ & $(\bar Q \gamma_\mu Q)(\bar q \gamma^\mu q)$ & & \checkmark &  & \checkmark & \checkmark &  & \checkmark & \checkmark & & \\
& $\mathscr{O}_{Qj}^{(3,1)}$ & $(\bar Q \sigma^i\gamma_\mu Q)(\bar q \sigma^i\gamma^\mu q)$ & & \checkmark & \checkmark & \checkmark & \checkmark &  & \checkmark & \checkmark & & \\
& $\mathscr{O}_{Qj}^{(1,8)}$ & $(\bar Q T^a\gamma_\mu Q)(\bar q T^a\gamma^\mu q)$ & & \checkmark &  & \checkmark & \checkmark &  & \checkmark & \checkmark & & \\
& $\mathscr{O}_{Qj}^{(3,8)}$ & $(\bar Q \sigma^iT^a\gamma_\mu Q)(\bar q \sigma^iT^a\gamma^\mu q)$ & & \checkmark & \checkmark & \checkmark & \checkmark &  & \checkmark & \checkmark & & \\

\midrule
\multirow{5}{*}{\rotatebox[origin=c]{90}{\normalsize top-lepton}}
& $\mathscr{O}_{et}$         & $(\bar e_p \gamma_\mu e_r)(\bar t \gamma^\mu t)$ & & \checkmark &  &  &  &  &  & & \checkmark &  \\
& $\mathscr{O}_{Qe}$         & $(\bar Q \gamma_\mu Q)(\bar e_p \gamma^\mu e_r)$ & & \checkmark &  &  &  &  &  & & \checkmark &  \\
& $\mathscr{O}_{lt}$         & $(\bar l_p \gamma_\mu l_r)(\bar t \gamma^\mu t)$ & & \checkmark &  &  &  &  &  & & \checkmark &  \\
& $\mathscr{O}_{lQ}^{(-)}$   & $\mathscr{O}_{lQ}^{(1)} - \mathscr{O}_{lQ}^{(3)}$ & & \checkmark & \checkmark &  &  &  &  & & \checkmark & \\
& \textcolor{gray}{$\mathscr{O}_{lQ}^{(1)}$}   & \textcolor{gray}{$(\bar l_p \gamma_\mu l_r)(\bar Q \gamma^\mu Q)$} \\
& \textcolor{gray}{$\mathscr{O}_{lQ}^{(3)}$}   & \textcolor{gray}{$(\bar l_p \sigma^i\gamma_\mu l_r)(\bar Q\sigma^i\gamma^\mu Q)$} \\
& $\mathscr{O}_{leQt}^{(1)}$ & $(\bar l_p^j e_r)\epsilon_{jk}(\bar{Q}^{k} t)$ & & \checkmark & \checkmark &  &  &  &  & & \checkmark &  \\
& $\mathscr{O}_{leQt}^{(3)}$ & $(\bar l_p^j \sigma_{\mu\nu} e_r)\epsilon_{jk}(\bar{Q}^{k} \sigma^{\mu\nu} t)$ & & \checkmark & \checkmark &  &  &  &  & & \checkmark & \\

\bottomrule
\end{tabular}
\end{center}
\end{table}

% End of text imported from the .//Sections/SMEFTTable.tex input file
\clearpage

\subsection{Process-based approach}
\label{sec:ana:pb}

Some processes, such as \ttW, \ttH and \tttt, are generally not easily
separable through object multiplicities alone. However, the full Run 2
dataset and wealth of ATLAS measurements optimised for sensitivity to
individual processes can be harnessed to give additional sensitivity
to each process in the context of this EFT interpretation.  The PB
approach utilises these process-specific analyses by repurposing their
event selections, including ML classifiers used to single out signal
and background processes, to define a set of orthogonal regions that
target individual processes with high purity.

The challenge of managing overlapping event selections and the resulting statistical correlations
between them is dealt with by careful treatment of the object definitions and modifications to overlapping
region definitions. In the result presented here, two general changes to
the regions established in previous works have are made:

\begin{itemize}
\item Harmonising all the lepton, photon, jet and $b$-tagging
object definitions. In each case, checks regarding the change
in analysis performance and sensitivity compared to the
original reference analysis are made. This harmonisation is
discussed in more detail in Section~\ref{sec:objectdefintions} and summarised in
Table~\ref{tab:harmonisation}.
\item Invoking orthogonality by assigning events passing
multiple selections to the regions with the priority given to reducing
the statistical uncertainty that is dominant for many WCs. This is done by giving
preference to assigning events passing multiple selections to
the signal regions with the fewest expected signal events for
the \topx process that respective region is optimised for.

\end{itemize}

\subsection{Fitted observables and binning}

The observable and binning choices within
each region are fine-tuned for sensitivity to EFT effects following
the event selection.
All WCs were considered in the optimisation procedure, the one with the strongest energy
growth in a given region was then selected for optimisation of the binning.
Many of the EFT effects considered in this measurement increase with the
characteristic energy scale of the event. In most lepton regions, the
fitted observable is
\ljzero, the transverse momentum of the hardest two-body system in the event,
evaluated over all possible combinations of two leptons, two jets, or a lepton and a jet,
which serves as a proxy for the hard scale of
the event, while the photon transverse momentum is used in the photon
regions. In the high-mass off-shell three-lepton regions,
the same-flavour dilepton invariant mass, \mllSF, is used
instead, providing sensitivity to the energy scale relevant for the
two-lepton-two-quark operators. The binning is optimised for
sensitivity to operators with strong energy growth while requiring the
MC statistical uncertainty in each bin to remain below 10\%.

\subsection{Background modelling}
\label{sec:ana:background}
Background contamination in the signal regions is due to contributions
from several SM processes which can feature the same final state
particles as the signal \topx processes. These are broadly split into two
categories based on the source of the reconstructed objects:
irreducible backgrounds, which have the same final state when
reconstructed objects are correctly identified; and reducible
backgrounds, where the objects have been incorrectly reconstructed, or
misidentified. In this analysis irreducible backgrounds have prompt
leptons, i.e. leptons produced in $W/Z$ boson decays or
leptonic $\tau$-lepton decays, and are generally dominated by diboson
production ($VV$) in association with heavy quark initiated jets. There
are several important sources of reducible backgrounds. Misidentified
photons (\NPP) can arise through electrons or jets (hadrons) being
misidentified as photons (\NPPX{E} and \NPPX{H}).  A significant
source of background is due to the selection of leptons from
non-prompt sources (\NPL), primarily from light (\NPLX{LF}) and heavy (\NPLX{HF}) hadron decays, that pass
the isolation requirements.
Sources of leptons with misidentified charge, primarily electron
charge misreconstruction (\NPQ) are important as they contribute
especially to final states with two same-sign leptons.  Additional
backgrounds from lepton-from-photon conversions (\NPC{}s) also
contribute to these final states. Specifically, these background arise either when a virtual
photon produces a lepton-antilepton pair (internal
conversions, \NPCX{Is}) or where a photon undergoes a conversion in
the detector material (material conversions, \NPCX{Ms}), with only
one of the resulting leptons being reconstructed.

The OB and PB approaches share control regions for background
contributions related to the reducible sources. %
\NPL{}s stemming from hadronic decays are separated by light-flavour and heavy-flavour sources, as
well as by lepton type (electron or muon), into four categories,
though the light-flavour muon component is extremely small and not
directly constrained by data.
The definitions of the control regions for these backgrounds are provided
in \cref{sec:nplCRs}. Other background regions are defined in the OB
and PB approaches, including those targeting $VV$ processes,
that in all cases are designed to be orthogonal to
the common CRs and SR selections.
In both approaches the control regions are used
to normalise the $VV$, \NPL{}, and \NPC{} processes to data.

Several backgrounds are estimated using data-driven
methods described in Section~\ref{sec:dataDrivenBkgEst}. These are applied
identically in both analysis approaches. The \NPPX E background is determined using electron--photon misidentification rate
scale factors (SFs). The \NPPX H background is determined using an `ABCD' method~\cite{ATL-PHYS-PUB-2026-014}. The \NPQ background
is determined by propagating charge-misidentification rates, derived in data from same-sign
di-electron events around the $Z$ boson mass, through the analysis.

%

% End of text imported from the .//Sections/AnalysisStrategy.tex input file

%

%

%
\section{Data and simulated event samples}
\label{sec:datamc}

% The next lines are included from the .//Sections/DataMC.tex input file

\subsection{Collision dataset}

The full \RunTwo data sample, collected in \pp collisions during the
years 2015--2018 and corresponding to a luminosity of \lumi~\ifb\ is
used.  Only events from time periods in which LHC beams were stable
and all ATLAS detectors were operational are
selected~\cite{DAPR-2018-01}.  The uncertainty in the combined
2015--2018 integrated luminosity is 0.83\% \cite{DAPR-2021-01},
obtained using the LUCID-2 detector~\cite{LUCID2} for the primary
luminosity measurements, complemented by measurements using the inner
detector and calorimeters.
The average number of interactions per bunch crossing has a mean of \num{33.7}.

The data were collected using a combination of single- and di-lepton triggers.
The single-electron and single-muon triggers have requirements on the lepton candidates to
maintain high efficiency across the full momentum range while
controlling the trigger rates~\cite{TRIG-2016-01,TRIG-2018-05,TRIG-2018-01,TRIG-2019-03}.
The use of a given trigger was dependent on the flavour and the transverse momenta of the leptons in the event, and on the
data-taking period.
For electrons and muons the lowest available trigger transverse momentum (\pT) thresholds
were from $\pT = 20$~\GeV to $\pT = 26$~\GeV.
Isolation requirements were applied to the
triggers with the lowest \pT thresholds of 26~\GeV for electrons
and muons.
The di-electron trigger required two electrons that satisfied loose
identification criteria with \pT thresholds that were raised as
the instantaneous luminosity increased during Run 2: 12 GeV in 2015, 17 GeV in 2016,
and 24 GeV in 2017–2018. Di-muon triggers utilised asymmetric \pT
thresholds for leading (sub-leading), in \pT, muons: 18 (8) GeV in 2015 and 22
(8) GeV in 2016–2018. An electron+muon trigger required events to have
an electron candidate satisfying loose identification with a 17 GeV
threshold and a muon candidate passing a 14 GeV threshold for all
data-taking periods.

\subsection{Simulated events}

Samples of simulated events were produced using MC
event generators to model the different SM processes and relevant EFT effects.
After generating events for each process of interest, the
detector response was modelled by either a full simulation of the
ATLAS detector based on \GEANT~\cite{SOFT-2010-01,Agostinelli:2002hh}
or a fast simulation~(\textsc{AtlFast~2}) relying on parameterised
showers in the calorimeter~\cite{ATL-SOFT-PUB-2018-002}.
The latter was used for the EFT signal samples described below, for the modelling of some rare background processes
and the modelling of processes with alternative generators (see Section~\ref{sec:systematics}).

\begin{table}[htb!]
\begin{center}
\caption{
The configurations used for event generation of signal and background processes.
The samples used to estimate the systematic uncertainties are indicated in parentheses and grey.
\MGNLO~\cite{Alwall:2014hca} is referred to here as MG5\_aMC.
}
\label{tab:MCSamples}
\vspace{0.25cm}
\small
\setlength\tabcolsep{1.5pt}
\begin{tabular}{~l^l^l^l^l^l}
\toprule
Process & Generator & ME order & Parton shower & ME PDF & Tune \\
\midrule
\ttyprod               & MG5\_aMC & NLO & \PYTHIA[8] & \NNPDF[3.0nlo] & A14 \\
\rowstyle{\color{gray}}
& (MG5\_aMC) & (NLO) &  (\HERWIG[7])  & (\NNPDF[3.0nlo]) &   (H7-UE-MMHT)\\
\rowstyle{\color{gray}}
& (MG5\_aMC) & (NLO) &  (\PYTHIA[8])  & (\NNPDF[3.0nlo]) &   (\textsc{A14 Var3c})\\
\ttydecay              & MG5\_aMC & LO & \PYTHIA[8] & \NNPDF[3.0nlo] & A14 \\
\rowstyle{\color{gray}}
& (MG5\_aMC) & (LO) &  (\HERWIG[7])  & (\NNPDF[3.0nlo]) &   (H7-UE-MMHT)\\
\rowstyle{\color{gray}}
& (MG5\_aMC) & (LO) &  (\PYTHIA[8])  & (\NNPDF[3.0nlo]) &   (\textsc{A14 Var3c})\\
\ttZ                  & MG5\_aMC & \textsc{NLO} & \PYTHIA[8]  &  \NNPDF[3.0nlo] &  A14 \\
\rowstyle{\color{gray}}
& (MG5\_aMC) & (NLO) &  (\HERWIG[7])  & (\NNPDF[3.0nlo]) &   (H7-UE-MMHT)\\
\rowstyle{\color{gray}}
& (MG5\_aMC) & (NLO) &  (\PYTHIA[8])  & (\NNPDF[3.0nlo]) &   (\textsc{A14 Var3c})\\
$\ttbar W$             & \textsc{Sherpa 2.2.10} & \textsc{MePs@Nlo} & \textsc{Sherpa} &  \NNPDF[3.0nnlo] & \textsc{Sherpa} default \\
\rowstyle{\color{gray}}
& (MG5\_aMC)  & (FxFx NLO) & (\PYTHIA[8]) & (\NNPDF[3.0nlo]) & (A14)   \\
\rowstyle{\color{gray}}
& (\POWHEG)  & (NLO) & (\PYTHIA[8]) & (\NNPDF[3.0nlo]) & (A14)   \\
\rowstyle{\color{gray}}
& (\POWHEG)  & (NLO) & (\HERWIG[7]) & (\NNPDF[3.0nlo]) & (H7-UE-MMHT)   \\
\ttWEW       & \textsc{Sherpa 2.2.10} & LO & \textsc{Sherpa} &  \NNPDF[3.0nnlo] & \textsc{Sherpa} default \\ %
\rowstyle{\color{gray}}
& (MG5\_aMC)  & (LO) & (\PYTHIA[8]) & (\NNPDF[3.0nlo]) & (A14)   \\
\ttH                   & \textsc{Powheg-BOX}  & NLO & \PYTHIA[8]\ & \NNPDF[3.0nlo] & A14 \\
\rowstyle{\color{gray}}
& (\textsc{Powheg-BOX}) & (NLO) & (\HERWIG[7]) &  (\NNPDF[3.0nlo]) &  (H7-UE-MMHT) \\
\rowstyle{\color{gray}}
& (MG5\_aMC)  & (NLO) & (\PYTHIA[8]) & (\NNPDF[3.0nlo]) & (A14)   \\
$t\bar t t\bar t$      & MG5\_aMC & NLO & \PYTHIA[8] & \NNPDF[3.1nlo] & A14 \\
\rowstyle{\color{gray}}
& (\textsc{Sherpa 2.2.11}) & (\textsc{MePs@Nlo}) & (\textsc{Sherpa}) &  (\NNPDF[3.0nnlo]) & (\textsc{Sherpa} default) \\
\rowstyle{\color{gray}}
& (\textsc{Powheg-BOX}) & (NLO) &  (\HERWIG[7])  & (\NNPDF[3.0nlo]) &   (H7-UE-MMHT)\\
\tZq                   & MG5\_aMC & \textsc{NLO} & \PYTHIA[8] & \NNPDF[3.0nlo] & A14 \\
\rowstyle{\color{gray}}
& (MG5\_aMC) & (NLO) &  (\HERWIG[7])  & (\NNPDF[3.0nlo]) &   (H7-UE-MMHT)\\
\tWZ                   & MG5\_aMC & \textsc{NLO} & \PYTHIA[8] & \NNPDF[2.3lo] & A14 \\

\tHq, \tHW             & MG5\_aMC & NLO & \PYTHIA[8] & \NNPDF[2.3lo] & A14 \\
$\ttbar$               & \textsc{Powheg-BOX} & NLO & \PYTHIA[8] & \NNPDF[3.0nlo] & A14  \\
\rowstyle{\color{gray}}
& (\textsc{Powheg-BOX}) & NLO & (\textsc{\HERWIG[7]}) & (\NNPDF[3.0nlo]) &  (H7-UE-MMHT) \\
$t\bar t t$            & MG5\_aMC & LO & \PYTHIA[8] & \NNPDF[2.3lo] & A14 \\
Single top             & \textsc{Powheg-Box} & NLO & \PYTHIA[8] & \NNPDF[3.0nlo] & A14   \\
($t$-, $Wt$-, $s$-channel) &  &  &                                     &  &  \\
$V$, $VV$, $qqVV$, $VVV$    & \textsc{Sherpa} 2.2.2(1)  & \textsc{MePs@Nlo} & \textsc{Sherpa} & \NNPDF[3.0nnlo] & \textsc{Sherpa} default  \\
$Z\gamma$,$W\gamma$  & Sherpa 2.2.4 & \textsc{MePs@Nlo} & \textsc{Sherpa} & \NNPDF[3.0nnlo] & \textsc{Sherpa} default \\
$tW\gamma$ & MG5\_aMC  &  LO & \PYTHIA[8] & \NNPDF[2.3lo] & A14 \\
$tq\gamma$ & MG5\_aMC  &  NLO & \PYTHIA[8] & \NNPDF[2.3lo] & A14 \\
$VH$ & \PYTHIA[8] & LO & \PYTHIA[8] & \NNPDF[2.3lo] & A14 \\
$t\bar{t}WW$ & MG5\_aMC & LO & \PYTHIA[8] & \NNPDF[2.3lo] & A14 \\
\bottomrule
\end{tabular}
\end{center}
\end{table}

An overview of the processes and MC generators used is provided in
Table~\ref{tab:MCSamples}. The choice of MC generator used for a given
process follows the relevant dedicated analysis.
A detailed description of all the MC generator configurations used
is given in Appendix~\ref{app:MC}.

\begin{table}[h!]
\begin{center}
\caption{A summary of the cross-section values and their systematic uncertainties from the referenced theory predictions. Where
no dedicated theory prediction is utilised the generator and order in the QCD coupling is provided.
The \ttZ cross-section refers to the one calculated for an on-shell $Z$ boson,
but is consistently applied to the off-shell \ttll contribution as described in Ref.~\cite{ATL-PHYS-PUB-2016-005}.
For \tZq the `5-fl' label refers to the flavour scheme used. The quoted cross-sections are inclusive, before any branching fractions are applied. }
\label{tab:TopXSMUncertainties}
{\renewcommand{\arraystretch}{1.2}
\small
\begin{tabular}{ lccc }
\toprule
Process & $\sigma$ [fb] & $\Delta \sigma$ [\%] & Reference(s) \\
\midrule
\ttyprod & 2800 & $\pm$14 & \MGNLO (NLO) \\
\ttZ    & 840 & +10/-12 & Refs.~\cite{deFlorian:2016spz,ATL-PHYS-PUB-2016-005} \\
\ttWQCD & 710 & $\pm$7 & Ref.~\cite{PhysRevLett.131.231901} \\
\ttWEW  & 48 & $\pm$7 & Ref.~\cite{Frederix:2021agh} \\
\ttH    & 510 & +7/-10 & Refs.~\cite{deFlorian:2016spz,Beenakker:2002nc,Dawson:2003zu,Yu:2014cka,Frixione:2014qaa} \\
\tttt   & 13.4 & +25/-22 & Ref.~\cite{Frederix:2017wme} \\
\tZq    & 850 & +1.4/-2.0 & \MGNLO (NLO) (5-fl) \\
\tWZ    & 16 & $\pm$30 & \MGNLO (NLO) \& Ref.~\cite{CMS-TOP-24-009-fix} \\
\tHq    & 74 & +6.6/-13.3 & Ref.~\cite{deFlorian:2016spz} \\
\tHW    & 15 & +6.6/-13.3 & Ref.~\cite{deFlorian:2016spz} \\
\bottomrule
\end{tabular}
}
\end{center}
\end{table}

Table~\ref{tab:TopXSMUncertainties} details the cross-section values,
their uncertainties and where appropriate the references for the
theoretical calculations used for each of the \topx processes. Where
no dedicated theory prediction is utilised the generator and order in
the QCD coupling is provided.

Dedicated EFT samples, used to determine the WC dependence of reconstructed yields and distributions, are produced at LO in QCD for the \ttZ and \tZq
processes, using the \MADGRAPH[2.9.3] generator and \PYTHIA[8.245]
parton shower and for \tWZ, \tHq, \tHW, \tty, \ttH,
\tttt, \ttW and \ttWEW using the \MADGRAPH[2.9.9] generator and
\PYTHIA[8.307] parton shower. All EFT MC samples use the \NNPDF[3.0lo] PDF set for the matrix element (ME)
calculation. They include the \texttt{SMEFTsim 3.0} model
\cite{Brivio:2020onw} in the ``$G_\mu$,$m_Z$,$m_W$'' EW input scheme
\cite{Brivio:2021yjb} with the \texttt{top} flavour restrictions.
The events are initially generated according to the SM, and the \MADGRAPH
reweighting module is used to compute a large number of alternative
event weights corresponding to the inclusion of dimension-six EFT
vertices and propagators in the production Feynman diagrams.  These
weights are used to extract the dependence of various
observables on a large number of
different EFT operators related to \ttbar production, top-boson
vertices, four-quark and top-lepton interactions as described in
Section~\ref{sec:ana:smeft}.
In general, the reweighting is performed for WC values close to and
below the current experimental limits such that only a minimal
extrapolation is needed away from the points generated.
This is only not the case for \ctHRe and \ctHIm, for which WC values of $\pm 1$ were used in the generation.
As the dimension-six variations are based
on a sample using reweighting, the uncertainty in the parameterisation
due to the limited MC statistics is assumed to be negligible.  Since
the renormalisation-group evolution of the WCs is not included in the
\MADGRAPH setup, the renormalisation and factorisation scales are kept
fixed at $\mu=\sum_i m_i$ (where $i$ runs over the massive final state
resonances).
%

%

% End of text imported from the .//Sections/DataMC.tex input file

%

%

%
\section{Reconstructed objects}
\label{sec:objectdefintions}

% The next lines are included from the .//Sections/Objects.tex input file

The definition of the reconstructed objects follows closely those
used in the dedicated \topx analyses already described. Similarly to
the simulated MC samples, the object definitions are generally taken
from the most relevant dedicated analysis.
However, there are some
modifications described in the following. A summary of the object
definitions used in this analysis is given in
Tables~\ref{tab:lepdef} and~\ref{tab:jetdef}. More details of each object and the procedure
to remove overlapping objects are provided in
Appendix~\ref{app:objectdefinitions}.

As discussed in Section~\ref{sec:analysis}, in order to facilitate the
combination of the different process-based analyses it is necessary to
harmonise the object definitions as defined above. Summaries of the
object definitions for the original analyses which provide inputs to
the PB analysis are given in Table~\ref{tab:harmonisation}.
The final choice object definitions represents a compromise between the choices made in the dedicated analyses.
The guiding principle is to keep the
background rejection high enough for the processes that need it most, such
as \ttW and \tttt, especially in the $2\ell$SS and $3\ell$ regions,
but not to use the definitions with the very highest background rejection as
this significantly decreases the signal efficiency for the cleaner processes like \tty and (on-shell) \ttZ.
It is found
that the impact of these choices is not significant but are a necessary step in
ensuring orthogonality.

One particularly important difference is the additional isolation
requirements applied to electrons and muons from the non-prompt lepton BDT discriminants
(\nplbdt)~\cite{HIGG-2017-02}. These are based on isolation and
lifetime information about a track-based jet that matches the selected
light lepton. Two versions of the discriminant exist; \plv which has
two working points (WP) `Loose' and `Tight', and \pliv which has
`Tight' and `VeryTight'. The \plv{Tight} WP is defined with a prompt
electron (muon) efficiency of 65\% (80\%) for a candidate lepton
$\pt=20$~\GeV, increasing to 90\% (95\%) at $\pt=45$~\GeV.  The
efficiency for the tightest WP, \pliv{VeryTight}, for muons
(barrel/endcap electrons)
is about 55\% (55\%/60\%) for
$\pt \sim 20$~\GeV and reaches a plateau of 90\% (85\%/83\%) for
$\pt \sim 40 (40/65)$~\GeV. See Appendix~\ref{app:objectdefinitions}
for the rejection factors of these WPs.
For the harmonised object definitions in this work, the \plv{Tight} WP
is chosen as it offers a compromise between efficiency and background
rejection.

\begin{table}[!htb]
\footnotesize
\caption{Summary of the electron, muon and photon object definitions. The selection requirements for signal electrons and muons are applied in addition to the pre-selected objects used for the removal of overlapping objects.}
\label{tab:lepdef}
\def\arraystretch{1.3}
\begin{center}
\begin{tabular}{ lcc|c }
\toprule
& \textbf{Pre-selected Electron} & \textbf{Pre-selected Muon} & \textbf{Selected Photon} \\
\midrule
Acceptance     & $\pT > 7$~\GeV, $|\eta^\mathrm{clust}| < 2.47$ & $\pT > 7$~\GeV, $\abseta < 2.5$   & $\pT > 20$~\GeV, $|\eta| < 2.37$\\
&  except $1.37<|\eta^\mathrm{clust}|<1.52$ &       & except $1.37<|\eta|<1.52$ \\
\midrule
Quality & \texttt{LooseAndBLayerLH} & \texttt{Medium} & \texttt{Tight} \\
Isolation & \plv{Loose} & \plv{Loose} & $\ET^{\textrm{iso}}\big|_{\Delta R=0.4}<0.022\cdot\ET(\gamma)+2.45$~\GeV \\
&               &             & and $\pT^{\textrm{iso}}\big|_{\Delta R=0.2}<0.05\cdot\ET(\gamma)$ \\
\midrule
Impact parameter & $|d_{0}/\sigma(d_{0})|<5.0$                   & $|d_{0}/\sigma(d_{0})|<3.0$ & -- \\
& $|z_{0} \cdot \mathrm{sin}(\theta)|<0.5$~mm   & $|z_{0} \cdot \mathrm{sin}(\theta)|<0.5$~mm & -- \\
\bottomrule
\addlinespace[0.5em]
& \textbf{Signal Electron} & \textbf{Signal Muon} & -- \\
\midrule
Quality     & \texttt{TightLH} &  \texttt{Medium} & -- \\
Isolation   & \plv{Tight}  & \plv{Tight} & -- \\
\bottomrule
\end{tabular}
\end{center}
\end{table}

\begin{table}[!htb]
\footnotesize
\caption{Summary of the jet object definitions. The selection requirements for jets and the additional requirements to determine them as originating from $b$-quarks are provided.}
\label{tab:jetdef}
\def\arraystretch{1.3}
\begin{center}
\begin{tabular}{ lc|c }
\toprule
& \textbf{Selected Jet} & \textbf{Selected Forward Jet} \\
\midrule
Acceptance     & $\pT > 25\,\GeV, \abseta < 2.5$ & $\pT > 25\,\GeV, 2.5 \leq \abseta < 4.5$ \\
\midrule
(Forward) Jet  & reject jets with $\pT < 60$~\GeV, $\abseta < 2.4$ & reject jets with $\pT < 60$~\GeV, $2.5 < \abseta < 4.5$,\\
Vertex Tagger  & and $\mathrm{JVT}<0.5$ & $\mathrm{fJVT}>0.4$ and $|\mathrm{timing}|>10$~ns\\
\bottomrule
\addlinespace[0.5em]
& \textbf{Selected $b$-tagged jet} & \\
\midrule
Acceptance    & $\pT > 25\,\GeV, \abseta < 2.5$ & \\
$b$-tagging efficiencies  & $60$, $70$, $77$ and $85\%$ & \\
\bottomrule
\end{tabular}
\end{center}
\end{table}

\begin{table}[!htb]
\footnotesize
\caption{Object definitions for the dedicated source analyses are compared against the definitions used in this analysis for both the OB and PB approaches.}
\begin{adjustbox}{width=\columnwidth,center}
\label{tab:harmonisation}
\def\arraystretch{1.3}
\begin{tabular}{lcccccccc}
\toprule
&                        This Analysis &                   \attz &          \attll &         \aftops &         \attw/\atth &     \attwj &         \atzq &          \atty\\
\midrule
Electron Quality &        TightLH &               MediumLH &      TightLH &       TightLH &       TightLH &       TightLH &       MediumLH &      MediumLH\\
Electron Isolation &      \plv{Tight} &           \plv{Loose} &    \plv{Tight} &   \pliv{Tight} &   \pliv{VeryTight} & \plv{Tight} &  \plv{Loose} &  \plv{Loose}\\
\nplqbdt &                   \cmark &                \xmark &        \cmark &        \cmark &        \cmark &        \cmark &        \xmark &        \xmark\\
\midrule
Muon Quality &            Medium &                Medium &        Medium &        Medium &        Medium &        Medium &        Medium &        Medium\\
Muon Isolation &          \plv{Tight} &            \plv{Loose} &   \plv{Tight} &   \pliv{Tight} &  \pliv{VeryTight} & \plv{Tight} & \plv{Loose} &  \plv{Loose}\\
\midrule
Photon Quality &          Tight &                 --- &           --- &           --- &           --- &           --- &           --- &           Tight\\
Photon Isolation &        FixedCutTight &         --- &           --- &           --- &           --- &           --- &           --- &           FixedCutTight\\
\midrule
Jet \pt[\GeV{}] &      $\geq 25$ &             $\geq 25$ &     $\geq 25$ &     $\geq 20$ &     $\geq 25$ &     $\geq 25$ &     $\geq 25$ &     $\geq 25$\\
Jet $|\eta|$ &            $<4.5$ &                $<2.5$ &        $<2.5$ &        $<2.5$ &        $<2.5$ &        $<4.5$ &        $<4.5$ &        $<2.5$\\
\bottomrule
\end{tabular}
\end{adjustbox}
\end{table}

% End of text imported from the .//Sections/Objects.tex input file

%

%

%
\section{Analysis regions}
\label{sec:regions}

% The next lines are included from the .//Sections/Regions.tex input file

A set of event selection requirements is applied to enrich
the different analysis signal regions in the corresponding \topx processes of
interest and to define control regions to fit the background normalisations to data. A
detailed discussion of the backgrounds and corresponding control
regions is left to Section~\ref{sec:bkgEst}. The common
preselection requirements are applied, followed by more
targeted region selections of the separate analysis approaches.
Initial requirements on the number of
photons, \nphotons, and charged light leptons, \nleps, including for
the latter the flavour and charge of lepton pairs define five broad categorisations:
\PreselOnelY, \PreselTwolOSY, \PreselTwolSS, \PreselThreel and \PreselFourl.
Some of these regions also include requirements of opposite-sign same-flavour lepton pairs, \nossf, and total
lepton charge, \modsumq.
Then additional requirements are imposed on the
number of jets with $|\eta|<4.5$, \ntjets, and $b$-tagged jets, \nbjets, as well as
requirements on the lepton transverse momentum, \ptlep{}.
Two categories of jets are considered in the analysis, central jets (those with $|\eta_{j}|<2.5$)
and forward jets (those with $2.5<|\eta_{j}|<4.5$).
The details are given in Table~\ref{tab:PBPreselection}.

These preselection requirements use the harmonised object definitions
previously described in Section~\ref{sec:objectdefintions}. Events
passing each preselection are then subdivided independently according
to the OB and PB region definitions, described in the following.

\begin{table}[h]
\begin{center}
\caption{A summary of the pre-selection requirements for each analysis category. The definition of the symbols is given in the text. The pre-selections are the same for both OB and PB approaches.}
\label{tab:PBPreselection}
{\renewcommand{\arraystretch}{1.2}
\begin{tabular}{ cccccc }
\toprule
& \PreselOnelY & \PreselTwolOSY & \PreselTwolSS & \PreselThreel & \PreselFourl\\
\midrule
\nleps & 1 & 2 & 2 & 3 & 4\\
\nossf & --- & $\geq1$ & 0 & $\geq1$ & $\geq1$ \\
\modsumq & --- & 0 & 2 & 1 & 0 \\
\nphotons & 1 & 1 & --- & --- & --- \\
\ntjets & $\geq4$ & $\geq2$ & $\geq3$ & $\geq2$ & $\geq2$\\
\nbjets & $\geq1@85\%$ & $\geq1@85\%$ & $\geq1@85\%$ & --- & --- \\
\ptlep{1} & \multicolumn{5}{c}{$\geq 27$ \GeV} \\
\ptlep{2} & --- & \multicolumn{3}{|c|}{$\geq15$ \GeV} & $\geq 7$ \GeV\\
\ptlep{3} & --- & --- & --- & $\geq10$ \GeV & $\geq 7$ \GeV \\
\ptlep{4} & --- & --- & --- & --- & $\geq 7$ \GeV \\
\bottomrule
\end{tabular}
}
\end{center}
\end{table}

\subsection{Object-based regions}

For the OB approach, regions are grouped based on the
preselection requirements and the signal regions are defined by
multiplicities of photons, leptons, central jets and forward jets, as well as lepton
charges.
Regions with explicit requirements on photons are referred to as the photon regions while those without such requirements are referred to as lepton regions.
A total of 47 regions are defined in the OB approach.
A summary of the OB regions before the fit to data is shown in
Figure~\ref{fig:Regions:OBRegionsSummary}.

The photon regions, \OBonely and \OBtwolOSy, target \tty by selecting
events consistent with a semi- and di-leptonically decaying top-quark pair,
respectively. A detailed description of the selection is given in Appendix~\ref{sec:app:OBregions},
Table~\ref{tab:OBPhotonRegions}. The lepton signal regions are broadly
separated into two categories, those that have a same-flavour
opposite-charge lepton pair close to the $Z$ boson mass, \threelZ, which
enhance \ttZ and \tZq contributions and those that invert this
requirement to enhance contributions without a resonant $Z$ boson/dilepton decay, \threelnoZ
and \twolSS, such as \ttW, \ttH and \ttll. The \threelZ regions are further
subdivided into those containing a forward jet, with suffix ``-FJ'',
and those without. These regions are described in more detail in
Tables~\ref{tab:OBLeptonRegions_reduced} and ~\ref{tab:OBLeptonRegions_forward} of Appendix~\ref{sec:app:OBregions}.

\begin{figure}[!t]
\centering
\subfloat[]{\includegraphics[width=0.5\linewidth]{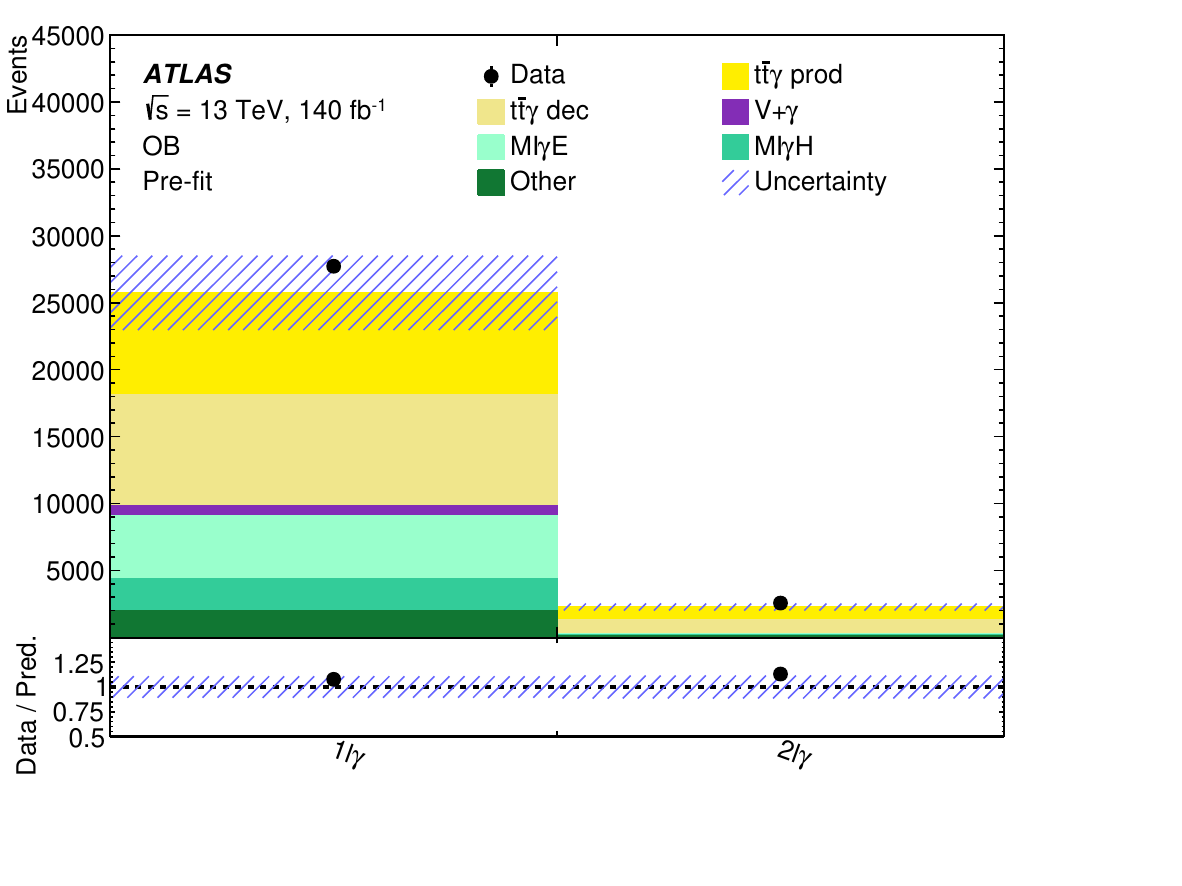}}
\subfloat[]{\includegraphics[width=0.5\linewidth]{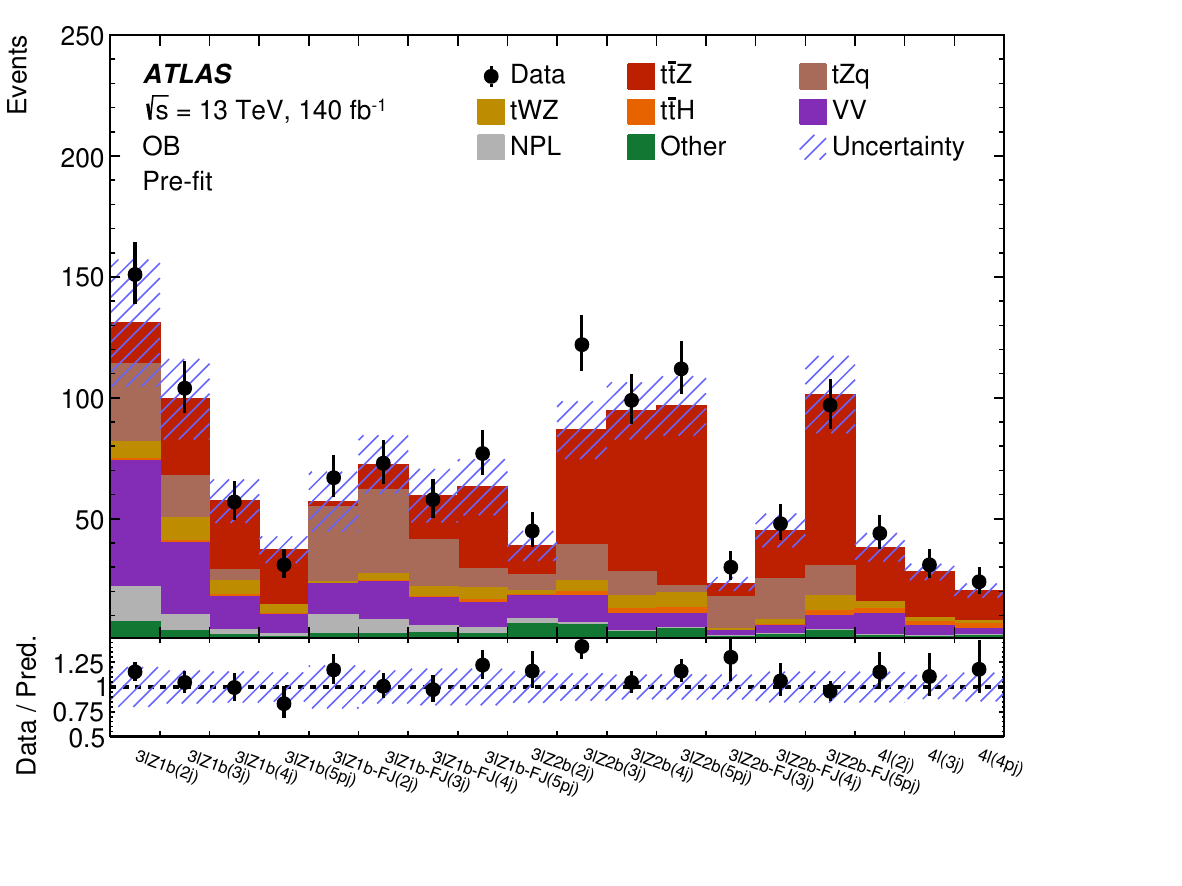}}\\
\subfloat[]{\includegraphics[width=0.5\linewidth]{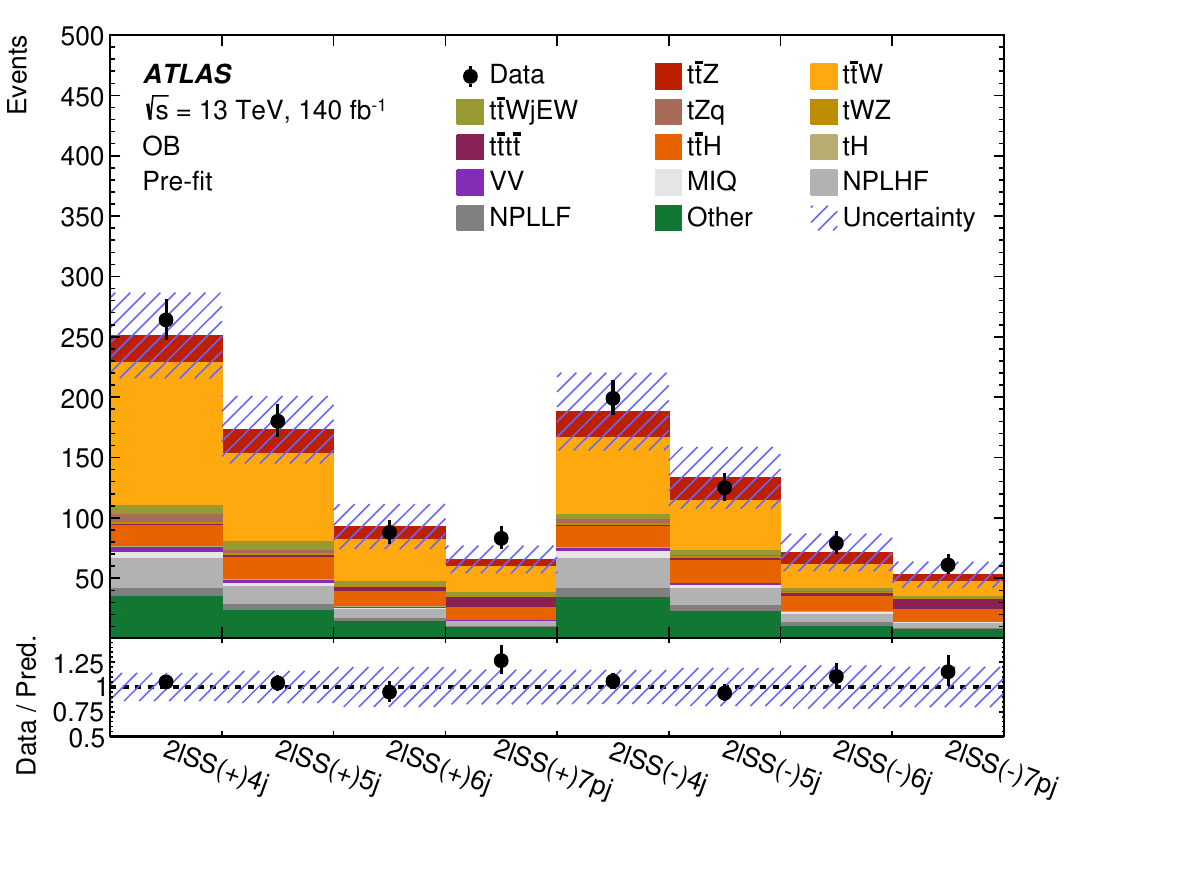}}
\subfloat[]{\includegraphics[width=0.5\linewidth]{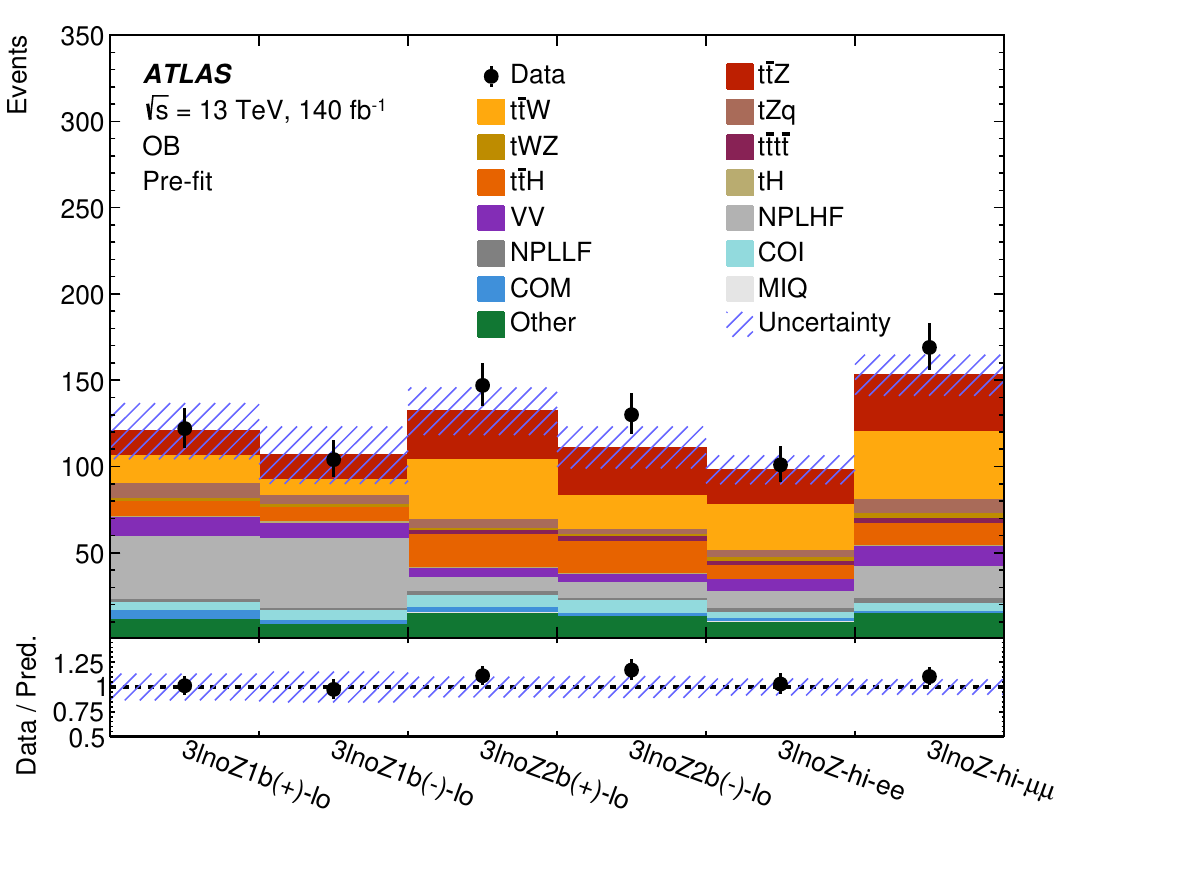}}\\
\subfloat[]{\includegraphics[width=0.5\linewidth]{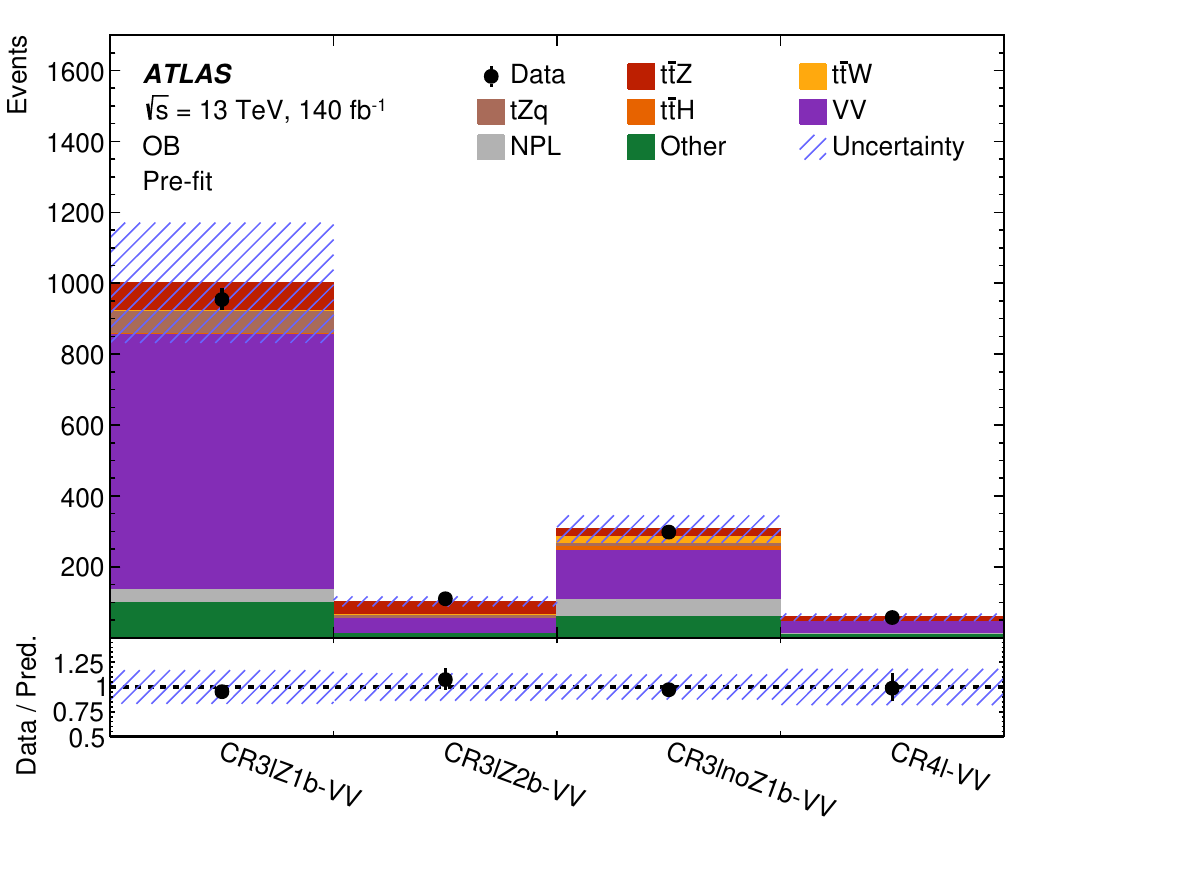}}
\caption{Event yields for the OB approach regions before the fit to data grouped into five categories: (a) photon regions, (b) three- and four-lepton on-shell $Z$ boson regions, (c) two-lepton same-sign regions, (d) three-lepton off-shell regions and (e) the diboson control regions. \UncertaintyBandSummary \OtherCategoryDescription}
\label{fig:Regions:OBRegionsSummary}
\end{figure}

\subsection{Process-based regions}

The PB regions are split into three
orthogonal categories described in the following. A total of 56 regions are defined in the PB approach.
A summary of the PB regions before the fit to data is
shown in Figure~\ref{fig:Regions:PBRegionsSummary}.

The photon regions are defined similarly to the dedicated \atty
analysis. They are orthogonal to all other regions (and each other) by
default due to lepton multiplicity requirements; exactly one lepton
and one photon regions (referred to as \PBonelGamma regions),
and \PBtwolOppSignGamma regions which require exactly one photon and
two opposite-sign leptons. In the \PBonelGamma regions, a multi-class
neural network (NN) is used to separate \ttyprod from three categories of background
and in the \PBtwolOppSignGamma region a binary classification NN is used
to separate \ttyprod from all background sources.

The next set of regions are designed to select events with three and
four leptons that also contain a lepton-pair consistent with a
$Z$ boson. These are referred to as the \PBthreelOnZnd and \PBfourlOnZ
regions and are taken from \attz and \atzq.  The \PBthreelOnZnd regions
from the \attz and \atzq analyses overlap. In the dedicated analyses
the binary NN from the \atzq analysis is used to separate out a \tZq
signal region \PBthreelOnZtZqSR and a control region \PBthreelOnZtZqCR
while the three-class NN from \attz defines three orthogonal regions:
a \ttZ signal region (\PBthreelOnZttZttZ), a \tZq region (\PBthreelOnZttZtZq)
and a \WZ control region (\PBthreelOnZttZWZ).  As the three regions from \attz overlap with the two
from \atzq, a prioritisation is applied to ensure orthogonality. Overlapping events are assigned first to
the \PBthreelOnZtZqSR region since there are fewer \tZq events than
there are \ttZ events in \PBthreelOnZttZttZ. Events are then assigned
to the three \ttZ regions and finally to \PBthreelOnZtZqCR. The detailed selection criteria are given in Appendix~\ref{app:regions}.

Again without intervention overlapping phase-space exists between the \PBtwolSSttH, \PBtwolSSttW, \PBtwolSSttWj, \PBthreelOffZnd and \PBtttt regions, and as such to ensure orthogonality a prioritisation procedure is applied. First overlapping events are assigned to the \PBtttt region, if they are in the high score region GNN$>0.6$, since \tttt has lower yields than other \topx processes.
Next overlapping events from the low score region (GNN$\leq0.6$) are assigned with priority to the \PBtwolSSttWj region, followed finally by the \PBtwolSSttH, \PBtwolSSttW and \PBthreelOffZnd regions.

\begin{figure}[!t]
\centering
\subfloat[]{\includegraphics[width=0.5\linewidth]{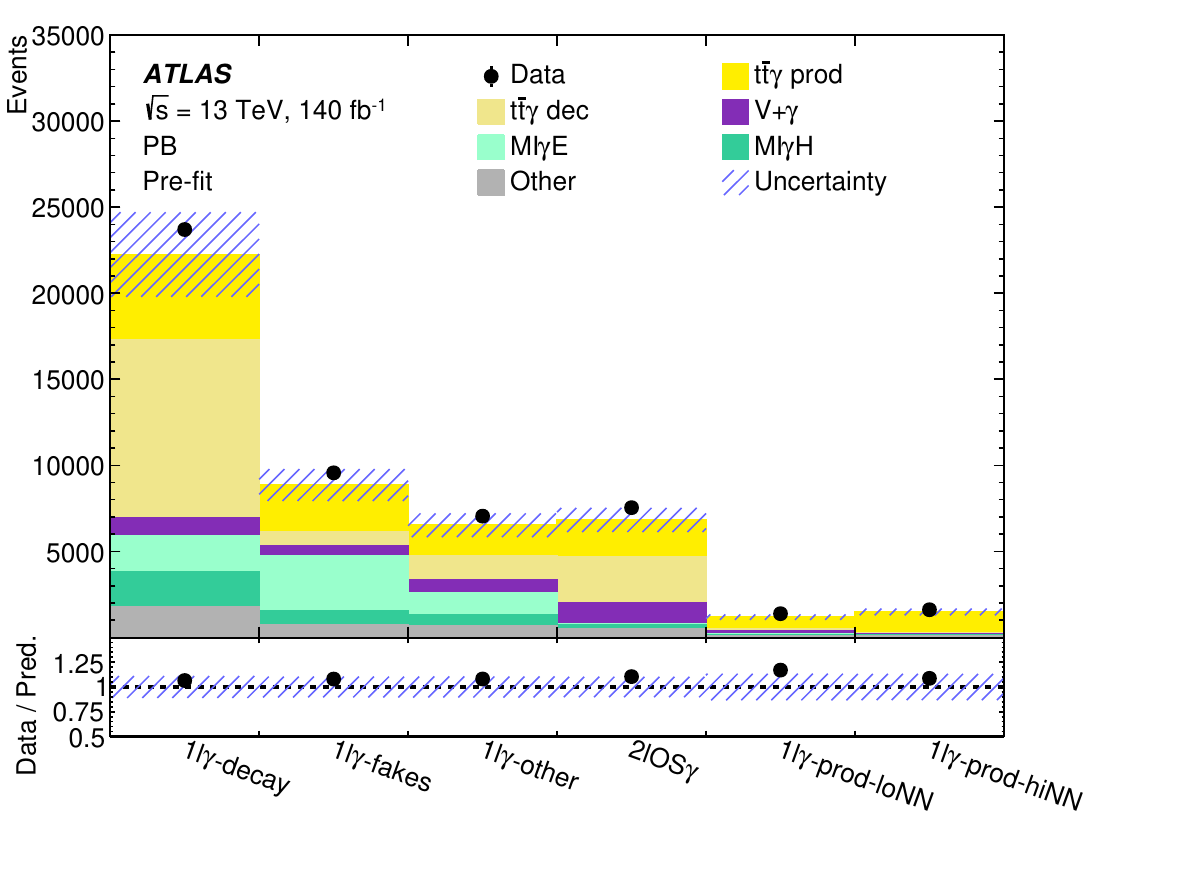}}
\subfloat[]{\includegraphics[width=0.5\linewidth]{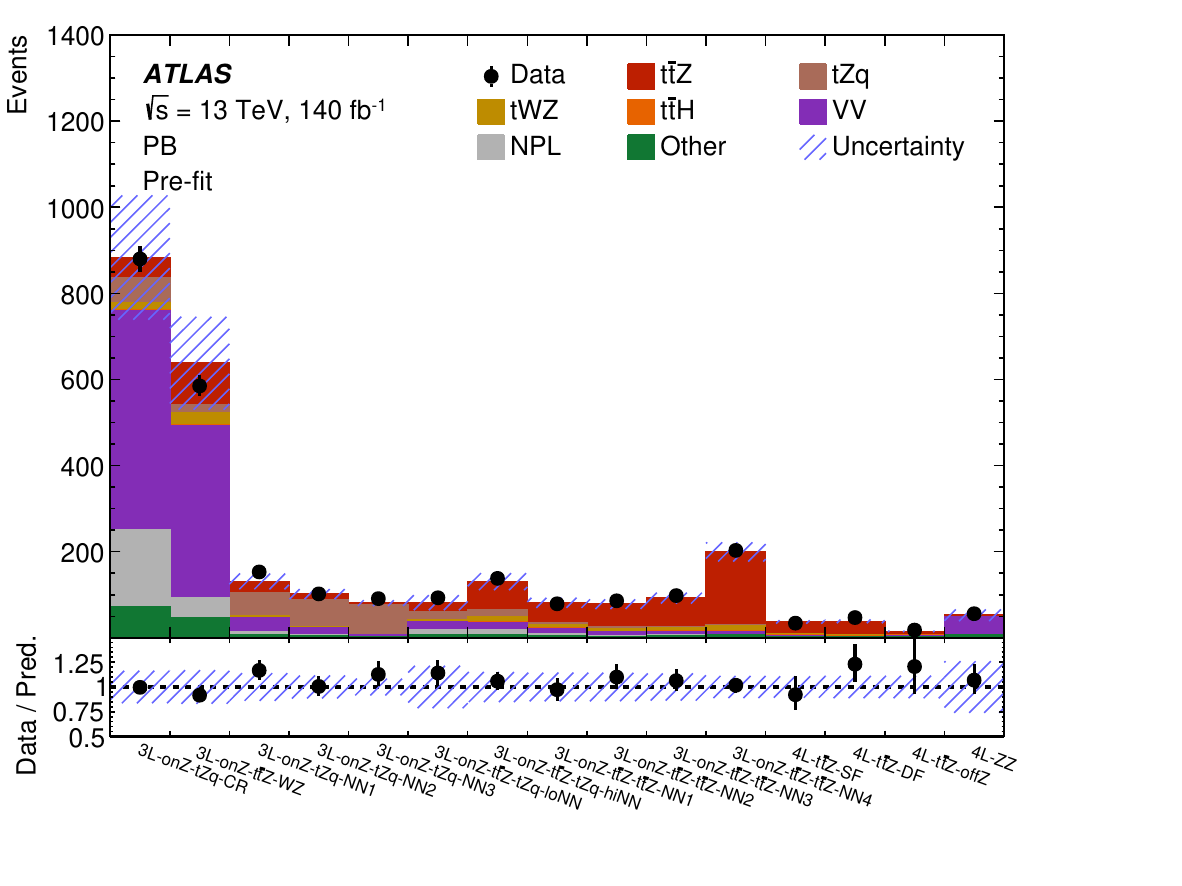}}\\
\subfloat[]{\includegraphics[width=0.5\linewidth]{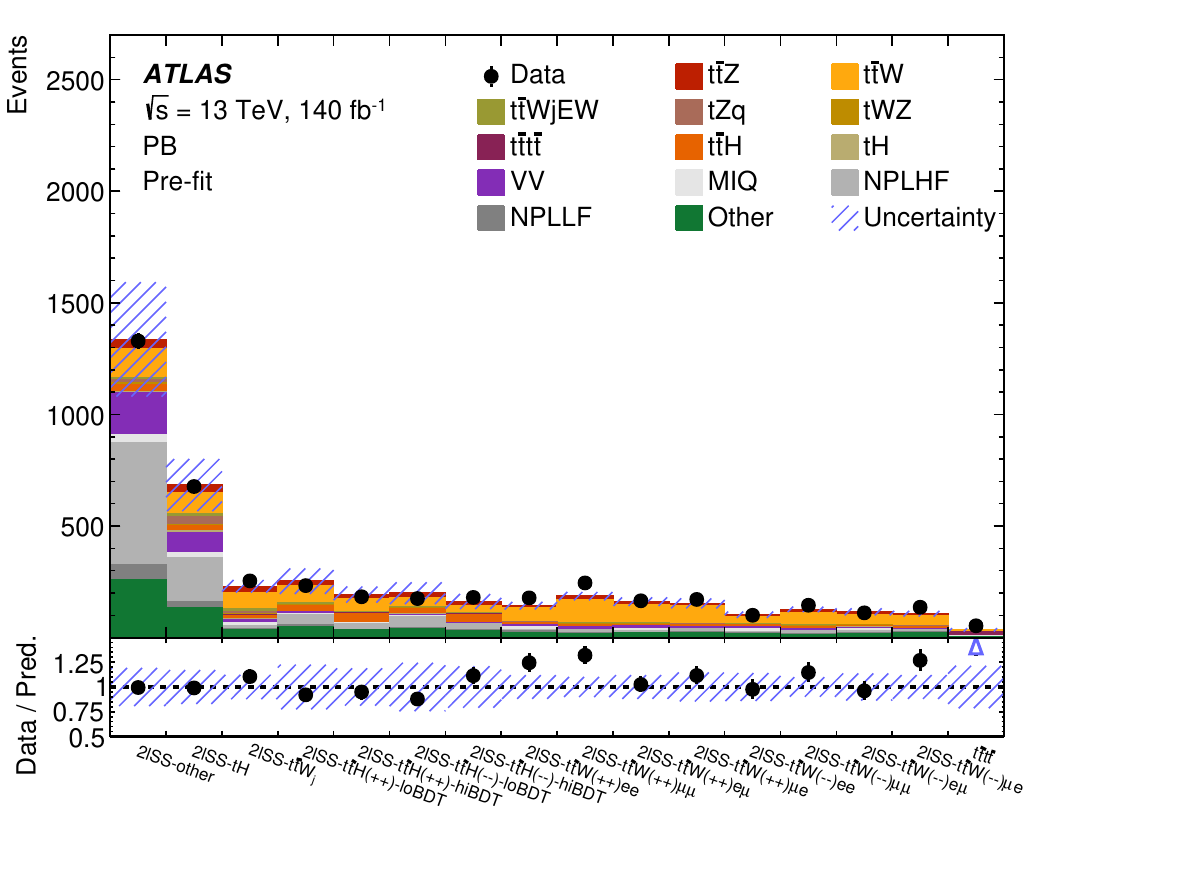}}
\subfloat[]{\includegraphics[width=0.5\linewidth]{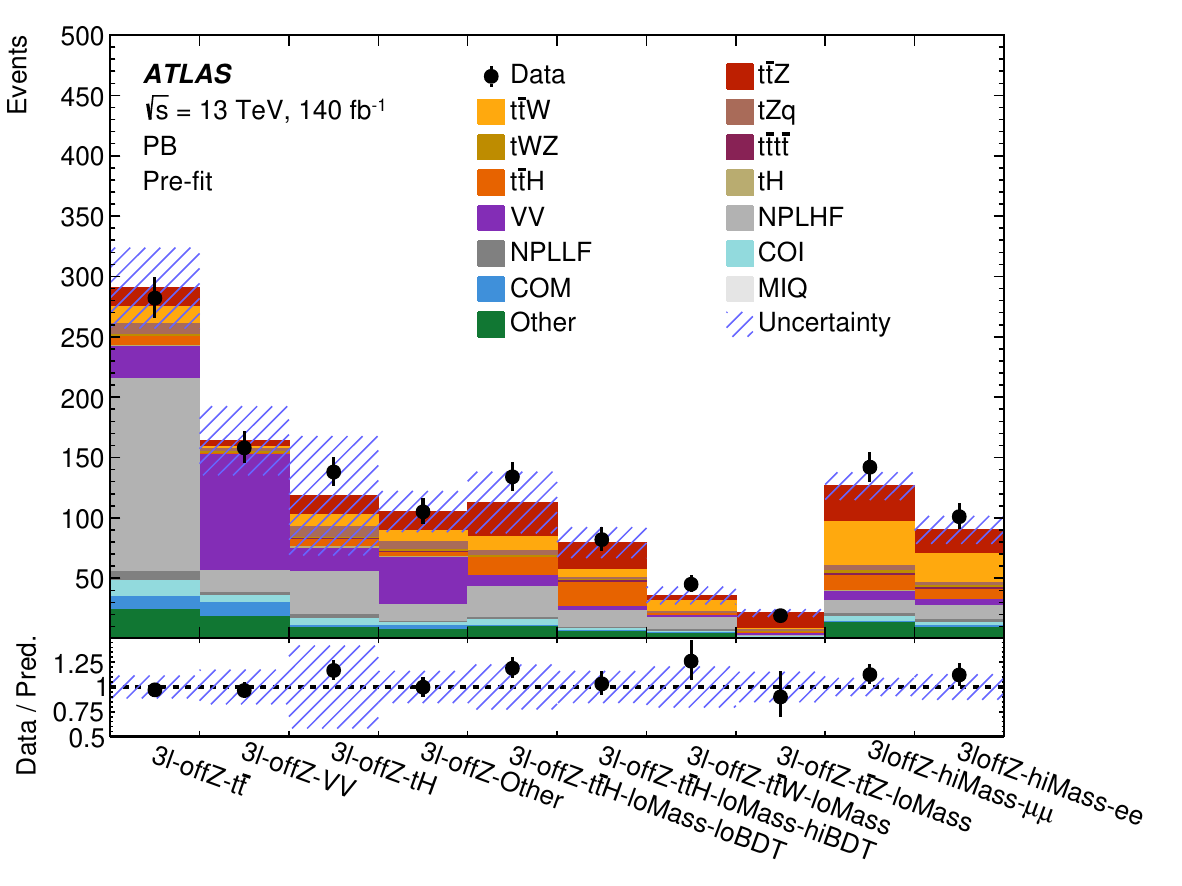}}
\caption{Event yields for the PB approach analysis regions before the fit to data, grouped into four categories: (a) photon regions, (b) three- and four-lepton on-shell $Z$ boson regions, (c) two-lepton same-sign regions and (d) three-lepton off-shell $Z$ boson regions. \UncertaintyBandSummary \OtherCategoryDescription}
\label{fig:Regions:PBRegionsSummary}
\end{figure}

\FloatBarrier

% End of text imported from the .//Sections/Regions.tex input file

%

%

%
\section{Background estimation}
\label{sec:bkgEst}

% The next lines are included from the .//Sections/BackgroundEstimation.tex input file

%
%

\subsection{Non-prompt and conversion lepton control regions}
\label{sec:nplCRs}

The different \NPL and \NPC background contributions, described in Section~\ref{sec:ana:background}, are isolated using MC event generator-level information. Unless otherwise stated all simulated samples are considered for each category. Estimation of the \NPL and \NPC contributions is made using the dedicated
control regions enriched in each of these backgrounds and targeting
the signal region phase spaces in which they appear. The same control
region definitions are used in both the OB and PB
approaches. Dedicated normalisation factors, $\lambda$, are assigned to each
background component which can be modified in the fit to data.

The detailed selection applied to define the control regions is given
in Appendix~\ref{app:backgroundest}. The main differences between control regions
and their respective signal regions are described in the
following. The \FakeCRtwoSSHF
regions require exactly one of the
leptons to fail the `Tight' isolation requirement but pass a looser
requirement (fail-tight), with this lepton defining the assignment between
electron (e) and muon (m) regions.
The \FakeCRthreelHFe, \FakeCRthreelHFm and \FakeCRthreelLFe
regions also require exactly one fail-tight lepton. Events with one or more $b$-jets
are assigned to the heavy flavour (HF) regions while those with none
are assigned to the light flavour (LF) region. The \NPC regions target
events with a $Z$ boson decaying to muons produced in association with a
conversion candidate. The \FakeCRthreelExtConv region requires events
to have exactly one external conversion electron, two opposite charge
muons and no $b$-tagged jets. The \FakeCRthreelIntConv requires exactly one
internal conversion electron, two opposite charge muons and no
$b$-tagged jets. Events in \FakeCRthreelPromptConv also require no $b$-tagged jets and
two opposite charge muons but must have an electron that has no
identified conversion vertex. Since these are the most
signal-region-like events, this region acts as a validation of
the extrapolation of the conversion background corrections to the SRs. An additional region \FakeCRfourlZeroB is defined
requiring 4 tight leptons, no $b$-tagged jets and at least one opposite-sign
same-flavour lepton pair, in order to help constrain the $ZZ$+light
background component in the \FakeCRthreelPromptConv region.

In general, this approach is similar to those used for \atth and \attz.
It assumes that these background processes are well modelled within the
uncertainties in the simulated samples in terms of the relative efficiencies
of the isolation requirements and the event topologies, while the normalisation
factors account for any mismodelling that affects the overall background normalisations.

The common control regions are shown in
Figure~\ref{fig:BkgEst:CommonCRSummary} before the fit to data.

\begin{figure}[t]
\centering
\includegraphics[width=0.8\textwidth]{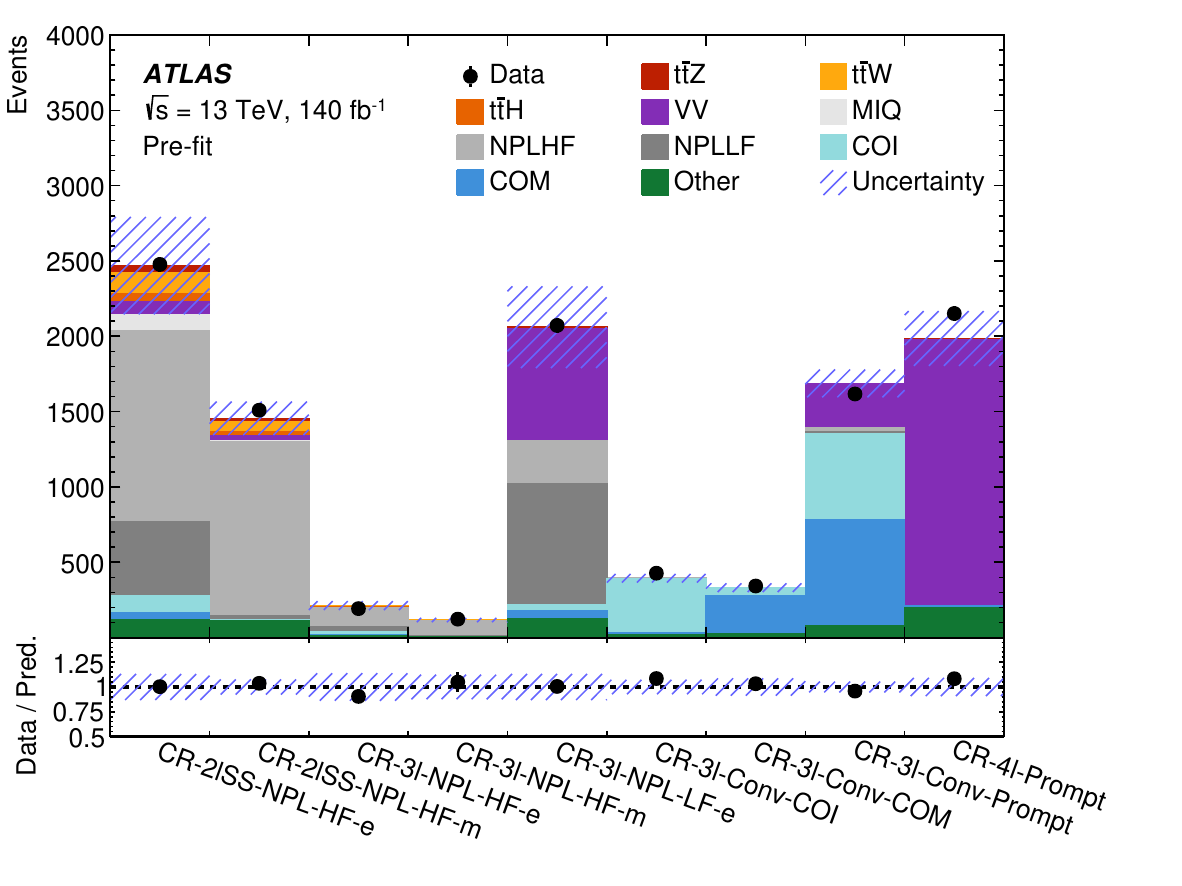}
\caption{Event yields for the common control regions before the fit to data. The uncertainty band includes the statistical as well as the systematical components. \OtherCategoryDescription}
\label{fig:BkgEst:CommonCRSummary}
\end{figure}

\subsection{Data-driven background estimates}
\label{sec:dataDrivenBkgEst}

\subsubsection*{Misidentified photon background estimation}

The contribution from processes where an electron mimicks the
signature of a photon in the detector, referred to in the following as
\NPPX{Es}, originates mainly from dileptonic \ttbar events where an
electron from the top quark decay chain is misidentified as a
photon. It is estimated by correcting the \NPPX{E} fake rate
determined in data to the one determined from simulation following the
method in Ref.~\cite{TOPQ-2018-24}. The \NPPX{E} fake rates are
derived from the numbers of electron–positron candidates from $Z \to
e^{+}e^{-}$ decays.  Two regions are defined: one targetting $e\gamma$
events that are enriched in \NPPX{E} candidates, with an electron and
a photon satisfying the signal object definitions; and one targetting
$e^{+}e^{-}$ events that are used as reference. The \NPPX{E} rate is
determined separately for converted and unconverted photons as a
function of photon $\pt$ and $\eta$ by comparing the observed
$e\gamma$ and $ee$ event yields around the $Z$ boson peak in
these two regions, after subtracting the respective background
contributions, estimated from the sidebands around the peak
region. The correction of fake rates is then applied to the \NPPX{E}
estimation from simulation.  Systematic uncertainties associated with
the estimation arise from the choice of function describing the $Z$
boson mass peak and the mass range considered, as well as from the
assumed function for the background description.

To estimate the background contributions from events where the photon
signature arises from hadron decays or hadronic energy depositions in
the electromagnetic calorimeter, referred to as \NPPX{H}, an `ABCD'
data-driven method~\cite{ATL-PHYS-PUB-2026-014} is used. In this method three regions are defined that are
orthogonal to the signal regions, and each other, and are enriched in \NPPX{H} photon events are defined by either
inverting the photon isolation or by loosening some of the photon
identification criteria.
Since the majority of the
\NPPX{H} contribution originates from \ttbar production, modelling
uncertainties considered on the \ttbar process, described in the
following, are propagated to the estimation, constituting the
dominant source of the uncertainty. The effect of the QCD scale
uncertainty is obtained by separately varying the renormalisation and
factorisation scales relative to their nominal values by a factor of
two up and down. The uncertainty in the choice of parton shower and
hadronisation model is estimated by comparing the nominal simulated samples
interfaced to \Pythia 8 with samples interfaced to
\HERWIG[7.2.1]\cite{Bahr:2008pv,Bellm:2015jjp}. The uncertainty in the
modelling of the initial-state radiation is estimated by varying the
amount of radiation corresponding to the Var3c eigentune
uncertainty~\cite{ATL-PHYS-PUB-2014-021} in the \tune{A14} tune. An
additional uncertainty is applied by comparing the nominal sample to
an alternative one with the $h_{\textrm{damp}}$ parameter in \POWHEG
increased by a factor of two~\cite{ATL-PHYS-PUB-2018-009}. Additional
contributions arise from the normalisation uncertainties in the prompt
photon backgrounds and the uncertainties in the data-driven \NPPX{E}
background entering the \NPPX{H} enriched regions.

\subsubsection*{Misidentified lepton charge background estimation}

Finally, a data-driven estimate of electron charge misidentification, \NPQ, is used in the $2\ell$SS
regions. Charge-misidentification rates are derived for electrons in $\pT$ and
$\eta$ bins using two $Z\rightarrow ee$ regions, one with
opposite-sign (OS) electrons and one with same-sign (SS) electrons. Events are selected with exactly two signal electrons, at most one jet and
no $b$-tagged jet. Additionally a $Z$ peak region is defined with
$\SI{71}{\giga\eV} < m(ee) < \SI{101}{\giga\eV}$, where a
background subtraction is performed assuming linear background
behaviour by subtracting the event yields from two side-band regions
($\SI{61}{\giga\eV} < m(ee) < \SI{71}{\giga\eV}$ and
$\SI{101}{\giga\eV} < m(ee) < \SI{111}{\giga\eV}$). The
charge-misidentification rates are then obtained by maximising a likelihood
function where the charge-misidentification rates in each lepton $\pT$ and $\eta$
bin are the parameters of interest. The charge-misidentification estimate in the
$2\ell$SS regions is obtained by applying these rates to a region with
identical selection except with the SS selection switched to an OS one.

% End of text imported from the .//Sections/BackgroundEstimation.tex input file

%

%

%
\section{Systematic uncertainties}
\label{sec:systematics}

% The next lines are included from the .//Sections/Systematics.tex input file
The systematic uncertainties described in the following are separated
into detector systematic uncertainties, which are related to the
reconstruction and identification of physics objects in the detector, and
theory uncertainties related to the modelling of the different
processes.  The systematic uncertainties associated with the
estimation of the \NPL{} and \NPQ{} backgrounds were already
outlined in Section~\ref{sec:bkgEst}.

\subsection{Detector-related uncertainties}

The luminosity estimate has a relative uncertainty of 0.83\%, derived
following the prescriptions described in Ref.~\cite{DAPR-2021-01} and
is applied to all processes determined from MC simulations.
Additionally, the distribution of the average number of interactions per bunch crossing in simulation is rescaled to match
that in data. The uncertainty in this rescaling process is propagated by varying the corresponding reweighting
factors.

The uncertainty in the jet energy scale (JES) is derived combining
information from test-beam data, LHC collision data and
simulation~\cite{JETM-2018-05}.  The jet energy resolution (JER) is
measured separately for data and MC using two in situ
techniques~\cite{JETM-2018-05}.  A systematic uncertainty is defined
as the quadratic difference between the jet energy resolutions for
data and simulation.
Uncertainties associated with the JVT are applied and account for the choice of MC
generator and residual contamination from pile-up jets after pile-up
suppression.

Uncertainties are assigned to the scale and resolution of the soft
track component of the \met. They are derived from differences between
data and MC simulation in the measurement of the \pT balance between
the hard and soft \met components~\cite{PERF-2016-07}.

The effects of uncertainties in efficiencies for the heavy-flavour
identification of jets by the $b$-tagging algorithm are evaluated.
These efficiencies are measured from data and depend on the jet
flavour and kinematic quantities of the jets.  The SFs and their
uncertainties are applied to each jet in the simulation depending on
its flavour, \pt and $\eta$~\cite{FTAG-2019-07,FTAG-2018-01,FTAG-2020-08,FTAG-2019-02}.  Additionally,
flavour tagging extrapolation uncertainties for jets with $\pT$ above
the calibration range are considered.

Uncertainties in the SFs applied to correct for the difference between data and
simulation are derived separately for electron identification, isolation and trigger efficiencies.
The accuracy of the momentum scale and resolution of
electrons in simulation is checked using reconstructed distributions
of the $Z \rightarrow e^{+}e^{-}$ and $J/\psi \rightarrow e^{+}e^{-}$
masses.  Small discrepancies are observed between data and simulation,
and corrections for the energy scale and resolution are applied. A
detailed description can be found in Ref.~\cite{EGAM-2018-01}.

The correction factors applied to the simulated samples to improve the
description of the photon energy scale, identification and isolation efficiency~\cite{EGAM-2018-01} are varied within their
uncertainties to estimate the corresponding systematic uncertainties.

SFs are also applied to correct for the difference between data and
simulation in the muon identification, isolation and trigger
efficiency, as well as track-to-vertex association~\cite{MUON-2018-03}
and associated uncertainties in these SFs are considered.  Momentum
scale and resolution corrections are applied for muons in the
simulation.  Uncertainties in both the momentum scale and resolutions
in the MS and the ID are considered and varied separately.  Additional
uncertainties are considered to account for the charge-dependent scale
correction applied on data.  A more detailed description can be found
in Refs.~\cite{MUON-2018-03,MUON-2022-01}.

The aforementioned detector-related uncertainties are correlated
across all processes.

\subsection{Modelling uncertainties}
\label{sec:syst:mc}

For all \topx and diboson processes, uncertainties related to missing higher orders in the calculations
are determined by varying the renormalisation and factorisation scales simultaneously up and down by a factor of two and
compared to the nominal predictions. Uncertainties in the PDF are
evaluated following the recommended \tool{PDF4LHC}
prescription \cite{Butterworth:2015oua}.

In addition to the scale and PDF variations, all \topx processes have
systematic variations relating to the parton shower,
hadronisation and underlying event modelling. Either the parton shower
parameters within a generator are varied (for example A14 eigentune
variations), or the generator used for the interface of parton
showering and underlying event simulation is changed (for
example, \PYTHIA changed to \HERWIG). For \tHq and \tHW only the scale and PDF variations are included.
None of the modelling uncertainties are treated as correlated across processes.
A detailed description of the generator variations used is given in
Appendix~\ref{app:MC}.

Some minor backgrounds do not have systematic uncertainties from
dedicated MC generator variations, but are instead assigned a single
normalisation uncertainty. Unless otherwise stated these uncertainties
are motivated by the individual analyses where the processes are most
important. Production of $V\gamma$ is assigned a
$20\%$ uncertainty, $VV$ with less than two leptons in the decay,
$VH$, $VVV$, $\ttWW$ and three-top-quark production are all assigned an
uncertainty of $50\%$. These minor backgrounds contribute
approximately $2.5\%$ of the total event yields in the signal regions.

Uncertainties related to the modelling of events with $VV$+$b$, $VV$+$c$ and $VV$+light-hadron
content are treated as uncorrelated between the contributions.
As the $VV$+$c$ and $VV$+$b$ components are affected differently by additional heavy-flavour radiation
compared to the $VV$+light component, which dominates the regions that primarily constrain these processes,
additional uncertainties of $40\%$ and $50\%$ are applied to the $VV$+$c$ and $VV$+$b$ components, respectively.
Uncertainties in the modelling of the $VV$ processes related to the
ME matching scale, resummation scale and recoil scheme are
estimated by using alternative generator-level samples.

Modelling uncertainties are applied to the \ttbar process which
constitutes a major fraction of the contributions to the \NPL
backgrounds in this analysis. However, different sources of
mismodelling are observed in the \NPL control regions compared to the
signal regions, these uncertainties are therefore not correlated
between the different subcomponents (\NPLX{LF} and \NPLX{HF}).

The SM \topx processes are assigned cross-section uncertainties,
as discussed in Table~\ref{tab:TopXSMUncertainties}. To avoid double-counting
the normalisation uncertainties, the total normalisation component is
removed from the other modelling variations such that only their effects on the
acceptance and shapes of the distributions remain.
In the case of off-shell \ttll and \tllq contributions, no dedicated
modelling uncertainty is applied to extrapolate from the on-shell to off-shell regions. With the high \mllSF region being largely dominated by statistical uncertainty, such an uncertainty would have minimal impact on the sensitivity.

The sensitivity to the majority of the EFT WCs is limited by the
available data statistics rather than systematic uncertainties. The
WCs where this is not the case are \cHQ-, \cHt, \ctHIm and \cQj18.
For \cHQ- and \cHt
the main contributions are related to the MC modelling of \tZq, \ttZ
and \ttWEW. For \ctHIm the modelling of \tttt is most relevant while
for \cQj18 the modelling of \tty and \ttWQCD dominates.

% End of text imported from the .//Sections/Systematics.tex input file

%

%

%
%
%
%
%

%

%
\section{Statistical analysis and results}
\label{sec:result}

% The next lines are included from the .//Sections/Results.tex input file

\subsection{Likelihood fit}

Results are extracted using maximum-likelihood fits of the predictions
in the analysis regions to the event yields in data implemented in
\tool{TRExFitter}~\cite{TRExFitter}.  The likelihood
function ${\cal L}(\vec{s},\vec{\lambda},\vec{\theta})$ is constructed
as a product of Poisson probability terms over all bins considered in
the measurement, and depends on a set of signal parameters, $\vec{s}$, the
normalisation factors for several backgrounds (see
Section~\ref{sec:analysis}), $\vec{\lambda}$, and a set of
nuisance parameters (NP), $\vec{\theta}$, encoding systematic
uncertainties in the signal and background
expectations~\cite{Cowan:2010js,Cranmer:2012sba}. Systematic
uncertainties can impact the estimated signal and background rates,
the migration of events between regions, and the shape of the
fitted distributions. Both $\vec{s}$ and $\vec{\lambda}$ are treated
as free parameters in the likelihood fit. The NPs $\vec{\theta}$ allow
variations of the expectations for signal and background according to
the systematic uncertainties, subject to Gaussian constraints in
the likelihood fit.  Their fitted values represent the deviations from
the nominal expectations that are needed to provide the best fit to
the data.  Statistical uncertainties in each bin due to the limited
size of the simulated event samples are taken into account with
dedicated parameters, using the Beeston--Barlow ``lite''
technique~\cite{Barlow:1993dm}.  This methodology follows the
conventions of Ref.~\cite{Cowan:2010js}.
A pruning procedure is implemented to remove uncertainties that are too small
to meaningfully impact the results, improving the fit’s convergence and reducing
computational time. In addition, a smoothing procedure is applied, where necessary,
to reduce the impact of statistical fluctuations in the systematic templates.

A validation of the analysis is performed using a fit to data in
which the normalisations of the SM top+X predictions are allowed to float, while
their SM kinematic distributions are retained.
In this case, $\vec{s}$ is composed of the normalisation factors
for each \topx process, defined as the ratio of the observed
production cross section ($\sigma_{\textrm{obs}}$) to the theoretical
cross section ($\sigma_{\text{SM}}$) and denoted by $\zeta =
\sigma_{\textrm{obs}}/\sigma_{\text{SM}}$. These are free parameters
in the likelihood fit.

The primary interpretation is a fit to data under
the \EFT hypothesis where \EFT operators are allowed to affect the SM \topx
processes by an amount defined by parameterisations bin-by-bin of the
per-process cross-section as a function of the dimension-six WCs. In this mode
of fitting the floating normalisation factor for that process is
replaced with a dedicated nuisance parameter constrained to the corresponding
theoretical cross-section uncertainty.

Both fits utilise the same methodology for likelihood calculation.
The only change is the inclusion of the WC
dependence and the removal of the SM normalisation factors and their
replacement with cross section uncertainties as described in
Section~\ref{sec:systematics}.

\subsection{Standard Model normalisation fits}
\label{sec:SMfit}

Simultaneous fits of \topx normalisation factors relative to the SM
predictions are performed first. These fits allow a comparison between the
central values, dominant uncertainties, and the sensitivity to each
SM \topx process between the OB and PB approaches themselves and
to dedicated measurements.
This gives an indication of the impact of the modified
object and region definitions in the PB fit as well as the overall impact
of the difference between OB and PB selections.

The results of the OB and PB fits to the data and comparisons to the
dedicated analyses are shown in Figure~\ref{fig:SMFit:SMNFComparison}
and Table~\ref{tab:smfit:SMNFcomparison}.

In general there is good agreement between the approaches introduced
here and the dedicated analyses. In the PB results, only small
differences are observed, which are expected due to the slightly
modified object and region definitions.  In the OB fit a slight
difference is observed with respect to the other approaches for the
\ttW and \ttH normalisation factors. There are a number of reasons
contributing to this, including the relatively large anticorrelation
between these normalisation factors and the fact that the OB selection
is significantly more restrictive for the relevant regions than the PB selection, meaning that
there is a significant non-overlap in events for the two
approaches. The different normalisation factors reflect the difference
in the agreement between the data and predictions before the fit to
data. The residual difference in, for example, the \ttW and \ttH
normalisation factors between the PB and dedicated analyses is likely
due to the difference in object definitions between the two. A notable
difference, highlighted in Table~\ref{tab:harmonisation}, is that the
lepton isolation used in the PB approach is slightly more efficient
but has reduced background rejection than the one used in the dedicated
analyses.

\begin{table}[t]
\centering
\caption{Tabulated values of the SM normalisation factors and their 1$\sigma$ uncertainties after the fit to data for the OB and PB approaches. These are compared to the values observed in the dedicated analyses.}   \label{tab:smfit:SMNFcomparison}
\renewcommand{\arraystretch}{1.4}

\begin{tabular}{ cccc }
\toprule
& OB & PB & Dedicated analysis \\
\midrule
\tZq
& $1.25^{+0.23}_{-0.21}$
& $1.13^{+0.14}_{-0.12}$
& $0.98^{+0.11}_{-0.10}$ \\
\ttH
& $1.34^{+0.55}_{-0.54}$
& $0.58^{+0.34}_{-0.34}$
& $0.63^{+0.22}_{-0.20}$ \\
\ttW
& $1.12^{+0.19}_{-0.17}$
& $1.35^{+0.12}_{-0.12}$
& $1.43^{+0.14}_{-0.14}$ \\
\ttZ
& $1.15^{+0.12}_{-0.12}$
& $1.07^{+0.09}_{-0.08}$
& $1.03^{+0.06}_{-0.06}$ \\
\tttt
& $2.56^{+1.13}_{-1.05}$
& $2.11^{+0.92}_{-0.60}$
& $1.89^{+0.77}_{-0.54}$ \\
\ttyprod
& $1.07^{+0.08}_{-0.07}$
& $1.12^{+0.05}_{-0.05}$
& $1.08^{+0.05}_{-0.05}$ \\
\ttydecay
& $1.21^{+0.13}_{-0.12}$
& $0.98^{+0.08}_{-0.07}$
& $0.98^{+0.06}_{-0.06}$ \\
\bottomrule
\end{tabular}
\end{table}

\begin{figure}[t]
\centering
\includegraphics[width=0.5\linewidth]{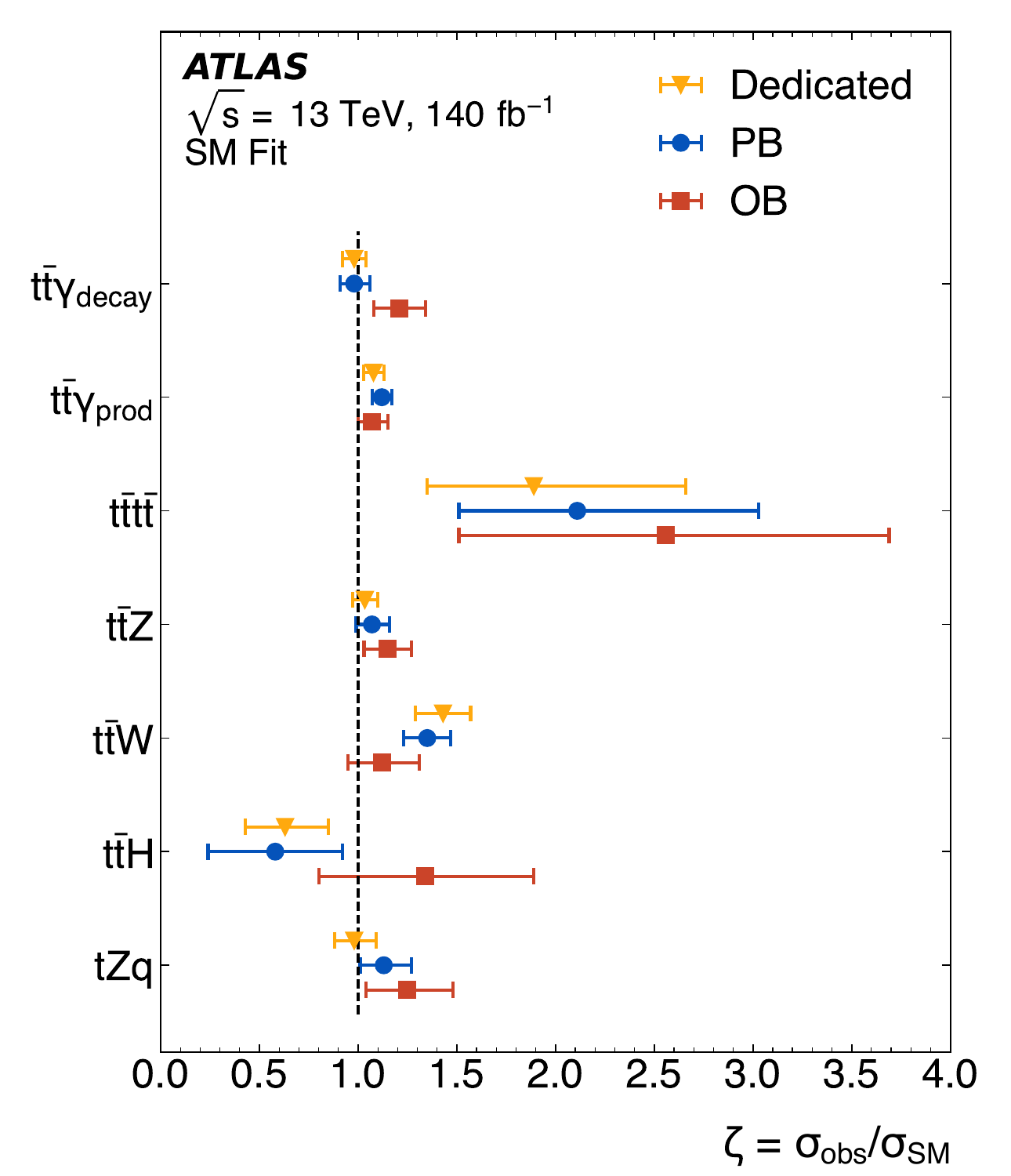}
\caption{Comparison between the SM normalisation factors, $\zeta = \sigma_{\textrm{obs}}/\sigma_{\textrm{SM}}$, and 1$\sigma$ uncertainties after the fit to data in the
OB (square marker) and PB (circle marker) approaches. Also shown are the normalisation factors for the dedicated analyses (triangle marker).}
\label{fig:SMFit:SMNFComparison}
\end{figure}

\subsection{Distributions after the WC fits to data}

The results of the OB approach fit to data are shown in
Figures~\ref{fig:Results:OB_Giga_2LSS_3LnoZ}-\ref{fig:Results:OB_Giga_tty_4L_CR} and in Figures~\ref{fig:Results:PB_Giga_2LSS}-\ref{fig:Results:PB_Giga_gamma} for the PB approach. In both cases all 33 WCs are free parameters in the fit.

In general the agreement between data and predictions after the fit to
data is very good.  The goodness-of-fit of each fit can also be
assessed relative to a saturated model~\cite{Cousins:2018tiz},
constructed using a likelihood-ratio statistic that compares the
fitted model to a saturated (most flexible possible) model.  For the
OB approach, a $p$-value of 0.74 is obtained, which is lower than in
the OB fit to SM normalisation factors, where it is 0.97. The same is
observed for the PB approach; the goodness-of-fit $p$-value is 0.17,
compared to 0.59 for the fit to SM normalisation factors. This is in
part due to the addition of the cross-section uncertainty nuisance
parameters instead of the SM normalisation factors which are then
modified by the fit. The pulls of the corresponding nuisance
parameters follow the general trends of the SM normalisation factors
described in Section~\ref{sec:SMfit}.  For example, in the PB fit the
\ttW cross-section uncertainty nuisance parameter value is
approximately 60\% higher than the nominal value, \tttt is
approximately 20\% higher and \ttH is approximately 10\% lower.
However, the size of the effect from the nuisance parameter pulls is smaller than the change in normalisation factors observed in the SM fit.
This implies that the effect of the EFT operators generally does not
produce single process normalisation changes across the relevant observables in
the same way that can be achieved by a non-unity SM normalisation
factor.

The compatibility with the SM is assessed using the profile-likelihood ratio (PLR), built using the negative log-likelihood (NLL) values
$2\Delta\mathrm{NLL}=2(\mathrm{NLL}_{\mathrm{WCs=0}}-\mathrm{NLL}_{\mathrm{best-fit}})$,
and assuming the asymptotic $\chi^2$ distribution with 33 degrees of
freedom. The OB approach fit yields a $2\Delta\mathrm{NLL}=9.18$ relative to
the SM hypothesis, corresponding to a $p$-value of approximately
$1.00$. The PB fit yields a $2\Delta\mathrm{NLL}=14.94$ relative to
the SM hypothesis, corresponding to a $p$-value of approximately
$1.00$. Neither fit shows a significant preference for non-zero dimension-six WCs over the SM point.

In the OB approach there are very few significant shifts or constraints in the fitted
nuisance parameters. Some exceptions are the cross-section
uncertainties for some \topx processes which, as discussed,
attempt to account for overall shifts in process normalisations not covered by
the dimension-six WC variations. Some pulls are also observed in the
\ttbar modelling uncertainties, but these are due to the agreement
between data and MC in the common control regions and not the OB signal regions. The background normalisation factors
associated with the $VV$, \NPL{}, and \NPC{} backgrounds,
are in good agreement with unity:
$\lambda_{ZZ}=1.09^{+0.10}_{-0.09}$,
$\lambda_{WZ}=1.24^{+0.20}_{-0.17}$,
$\lambda_{\textrm{\NPL{LFe}}}=1.06^{+0.22}_{-0.19}$,
$\lambda_{\textrm{\NPL{HF$\mu$}}}=1.07^{+0.09}_{-0.08}$,
$\lambda_{\textrm{\NPL{HFe}}}=0.89^{+0.18}_{-0.19}$,
$\lambda_{\textrm{\NPC{I}}}=1.04^{+0.09}_{-0.08}$ and
$\lambda_{\textrm{\NPC{M}}}=0.90^{+0.09}_{-0.09}$.

\begin{figure}[ht]
\centering
\subfloat[]{\includegraphics[width=0.75\linewidth]{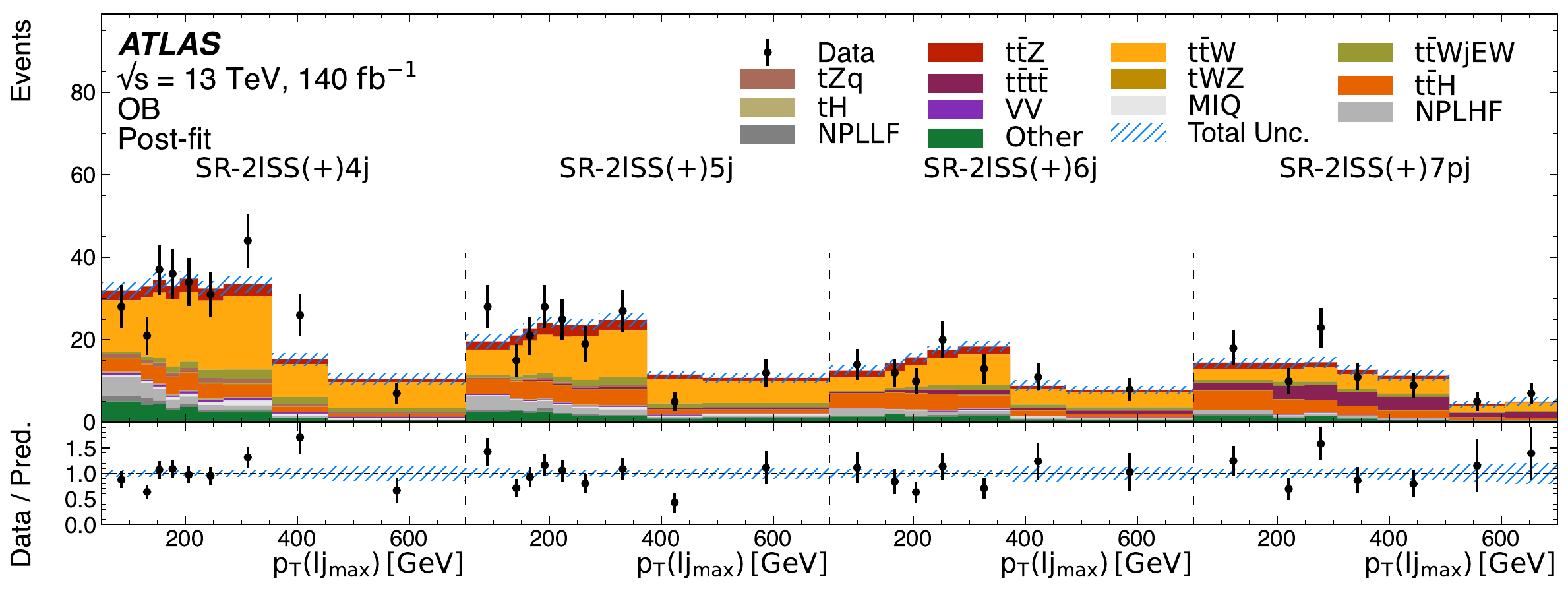}}\\
\subfloat[]{\includegraphics[width=0.75\linewidth]{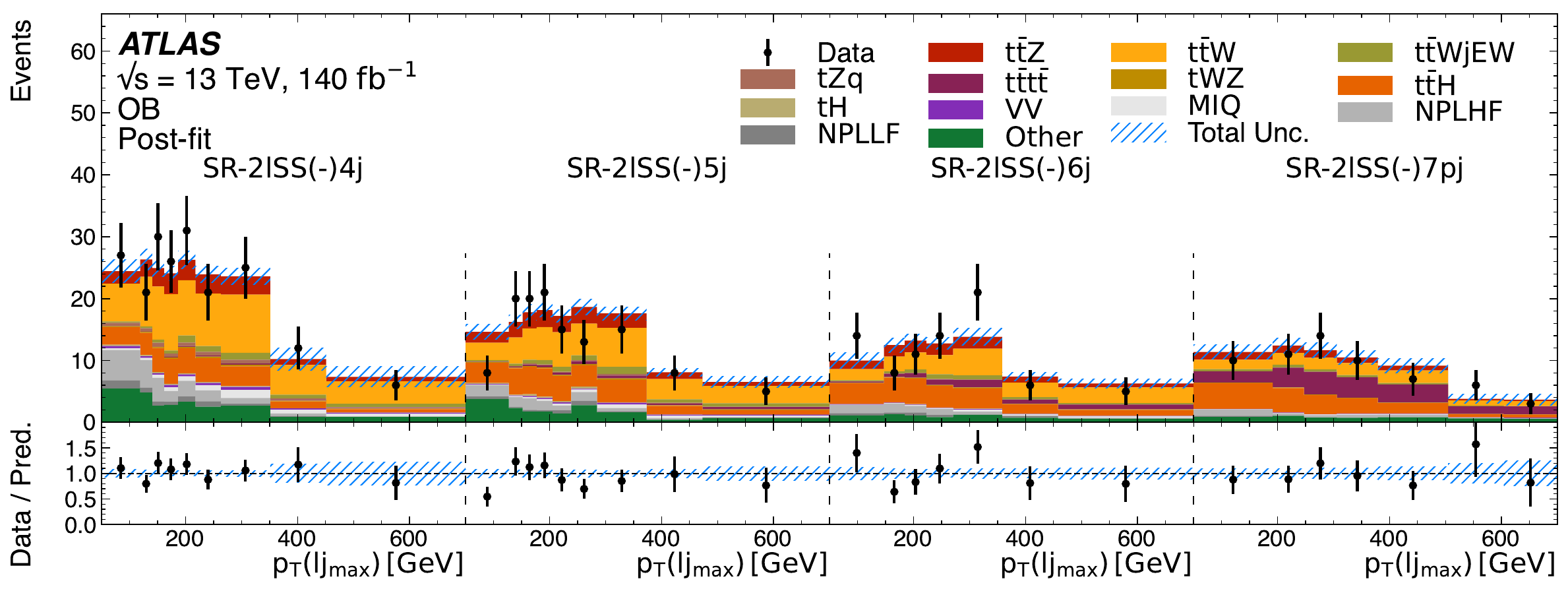}}\\
\subfloat[]{\includegraphics[width=0.75\linewidth]{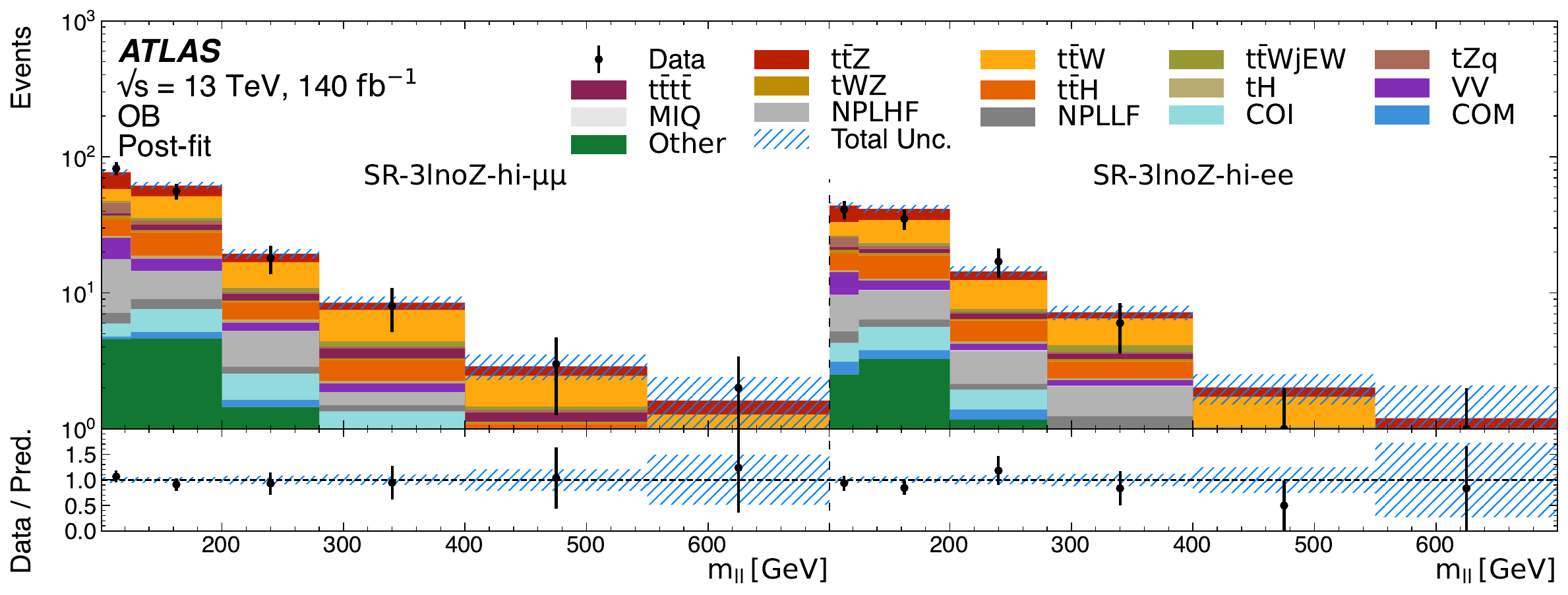}}\\
\subfloat[]{\includegraphics[width=0.75\linewidth]{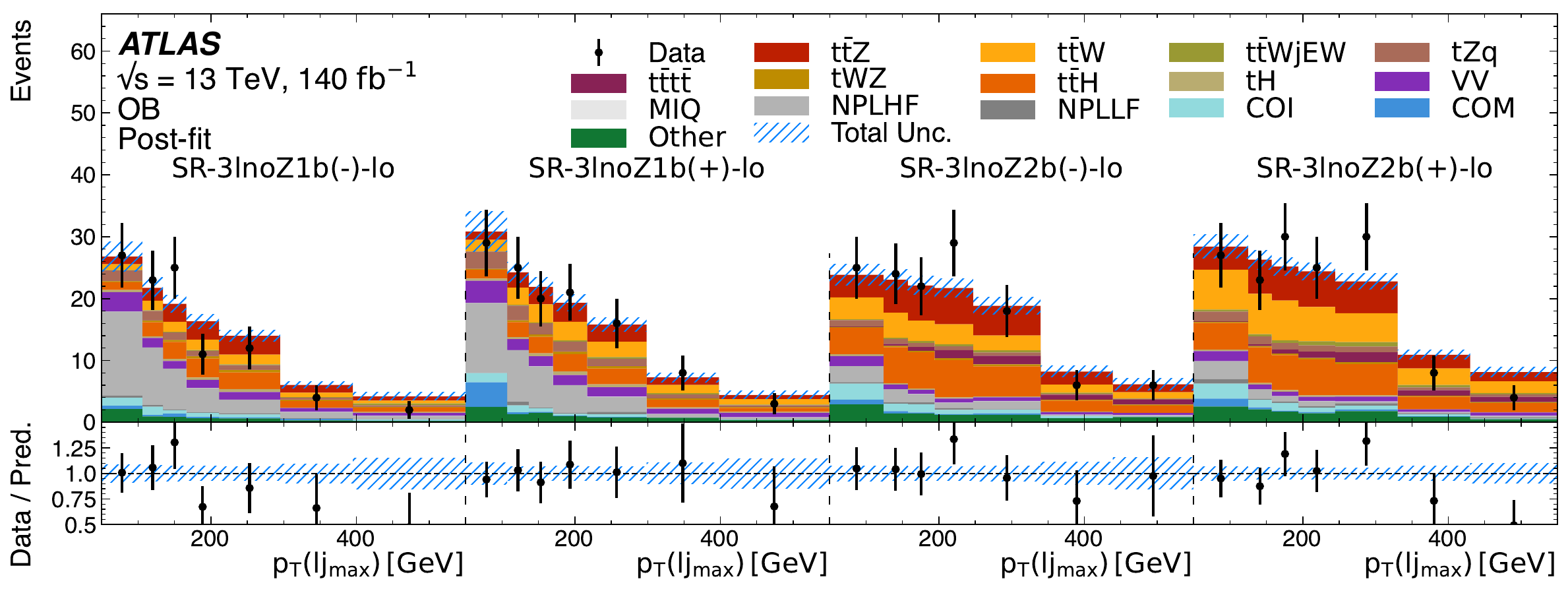}}\\
\caption{Comparison of the observed event yields and predictions in the
object-based (a) \OBtwolSSp, (b) \OBtwolSS(-), (c) high-mass \threelnoZ
regions, and (d) low-mass \OBthreelblo and \OBthreelbblo regions
after the fit to data. The EFT WCs and background normalisation
factors and fit nuisance parameters are set to their best-fit
values. \UncertaintyBand \OtherCategoryDescription}
\label{fig:Results:OB_Giga_2LSS_3LnoZ}
\end{figure}

\begin{figure}[!t]
\centering
\subfloat[]{\includegraphics[width=0.75\linewidth]{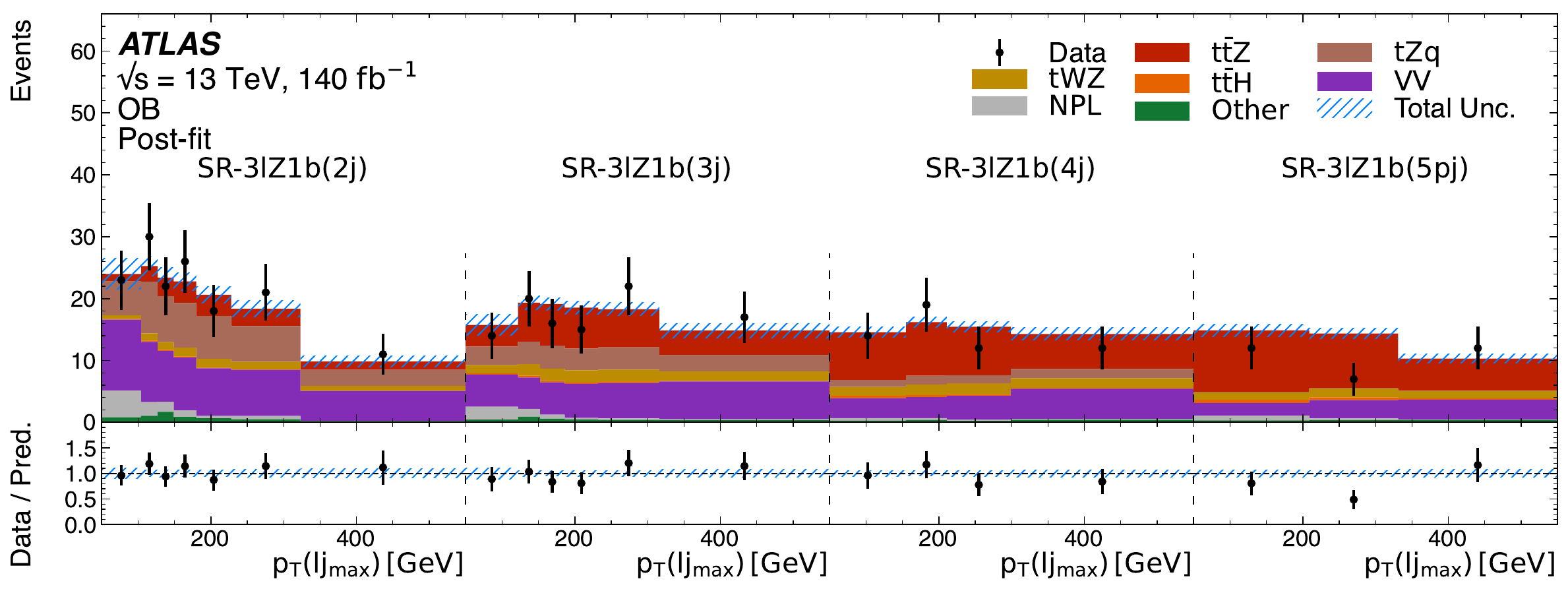}}\\
\subfloat[]{\includegraphics[width=0.75\linewidth]{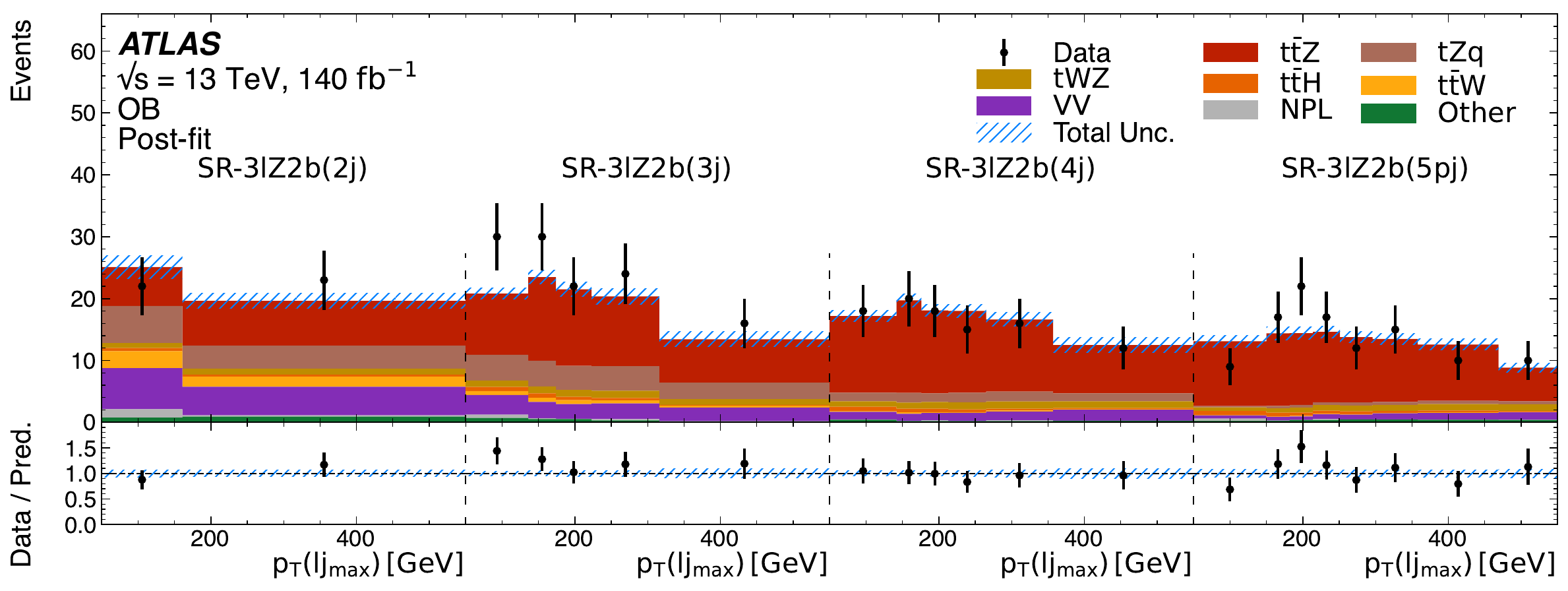}}\\
\subfloat[]{\includegraphics[width=0.75\linewidth]{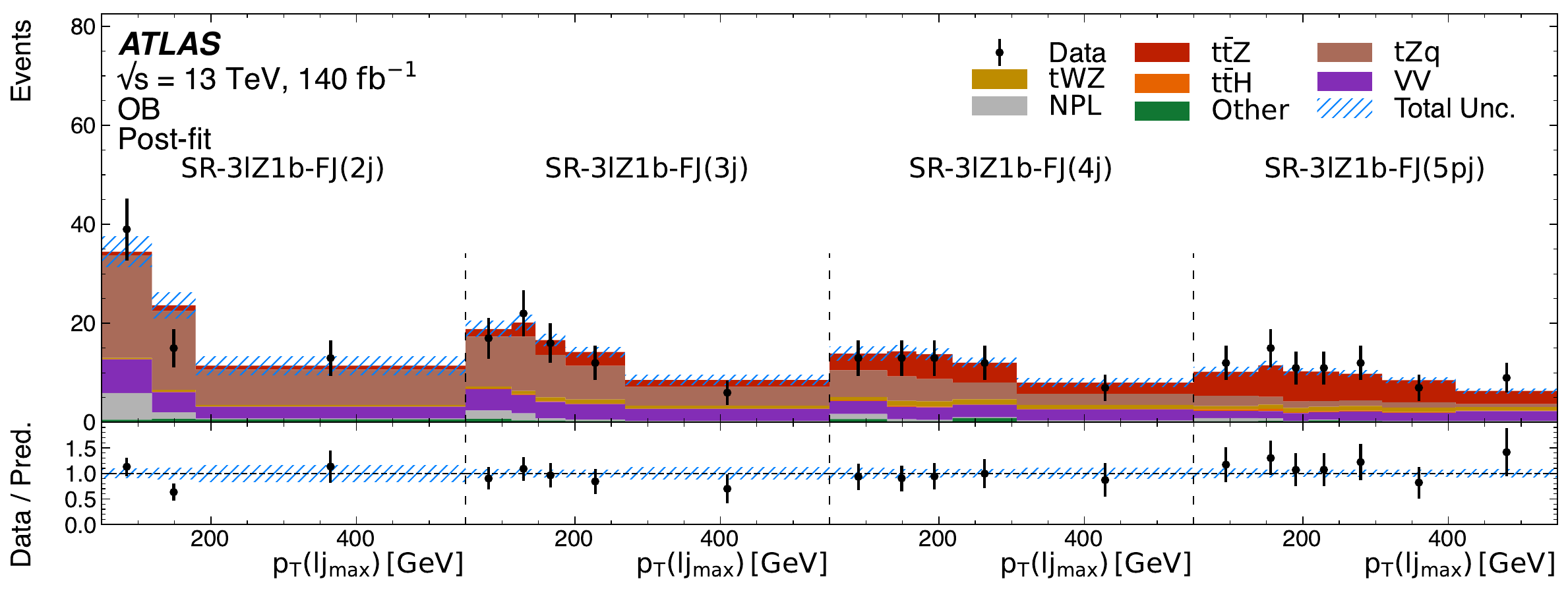}}\\
\subfloat[]{\includegraphics[width=0.75\linewidth]{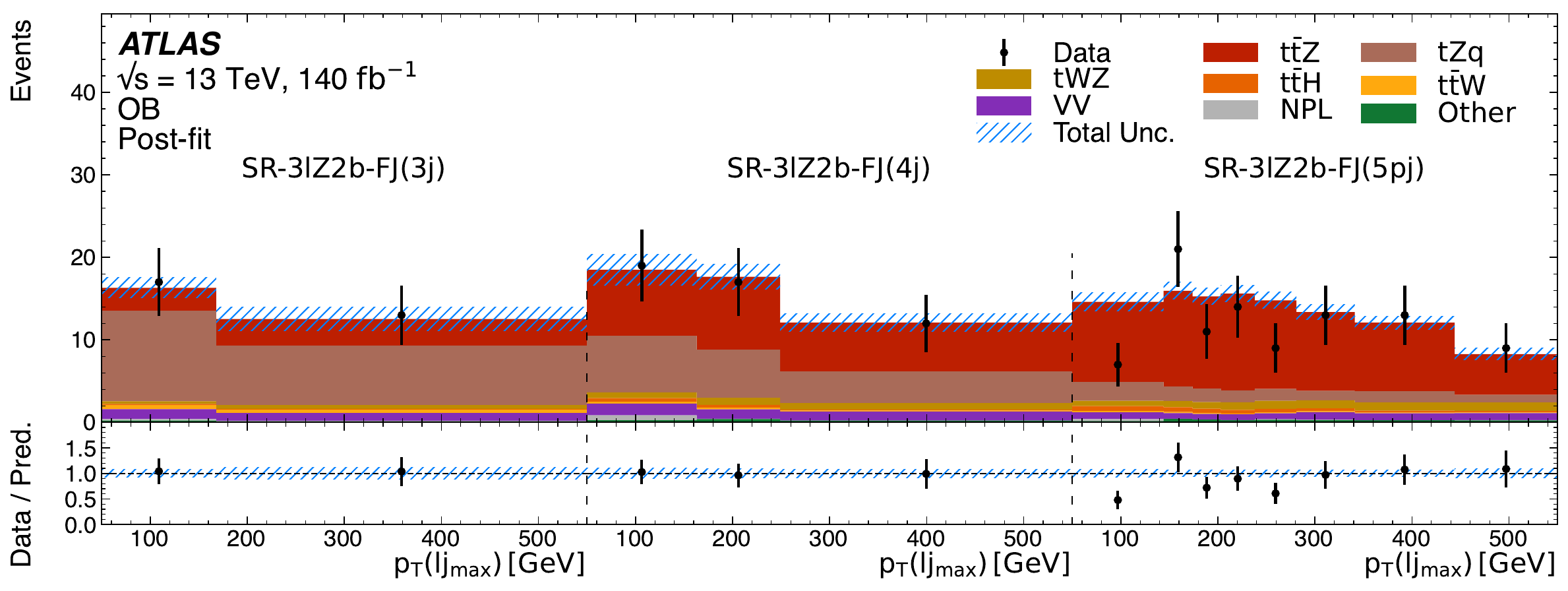}}\\
\caption{Comparison of the observed event yields and predictions in the
object-based (a) \OBthreelZb, (b) \OBthreelZbb, (c) \OBthreelZbforward and (d) \OBthreelZbbforward regions after the fit to data.
The EFT WCs, background normalisation factors and fit nuisance parameters are set to their best-fit values.
\UncertaintyBand \OtherCategoryDescription
}
\label{fig:Results:OB_Giga_3LZ}
\end{figure}

\begin{figure}[!t]
\centering
\subfloat[]{\includegraphics[width=0.75\linewidth]{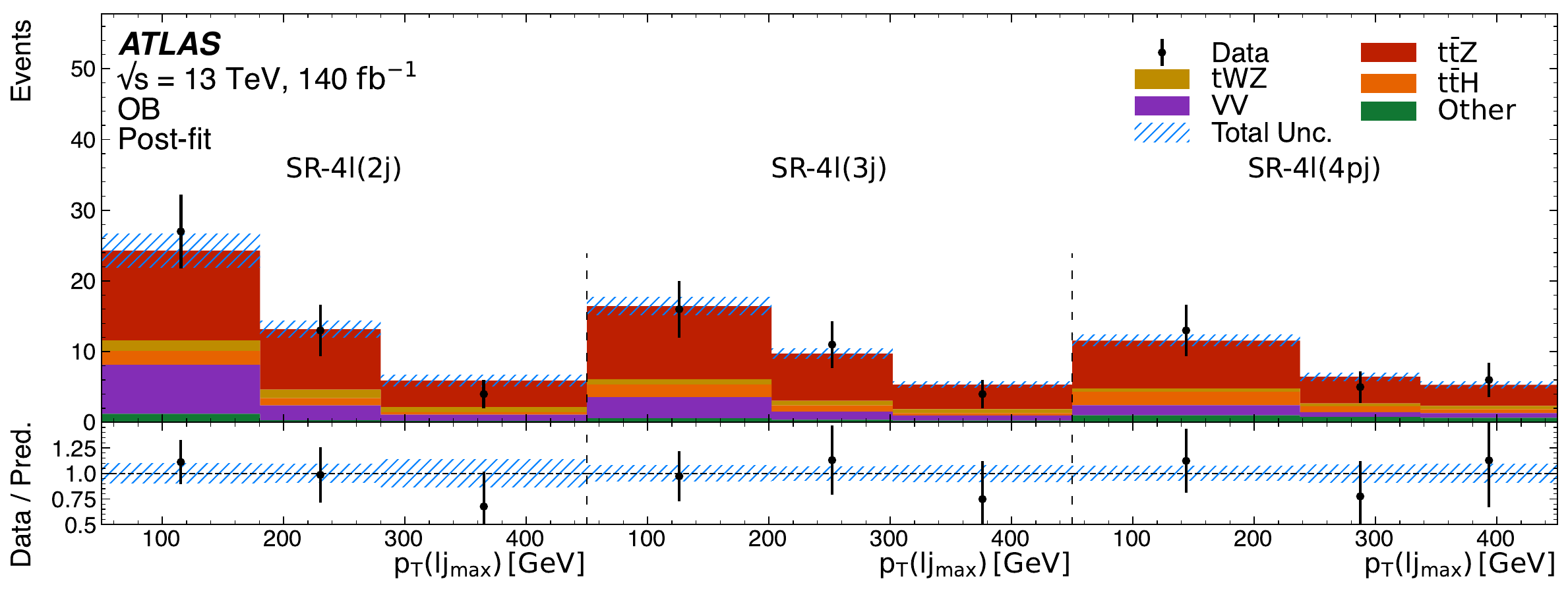}}\\
\subfloat[]{\includegraphics[width=0.75\linewidth]{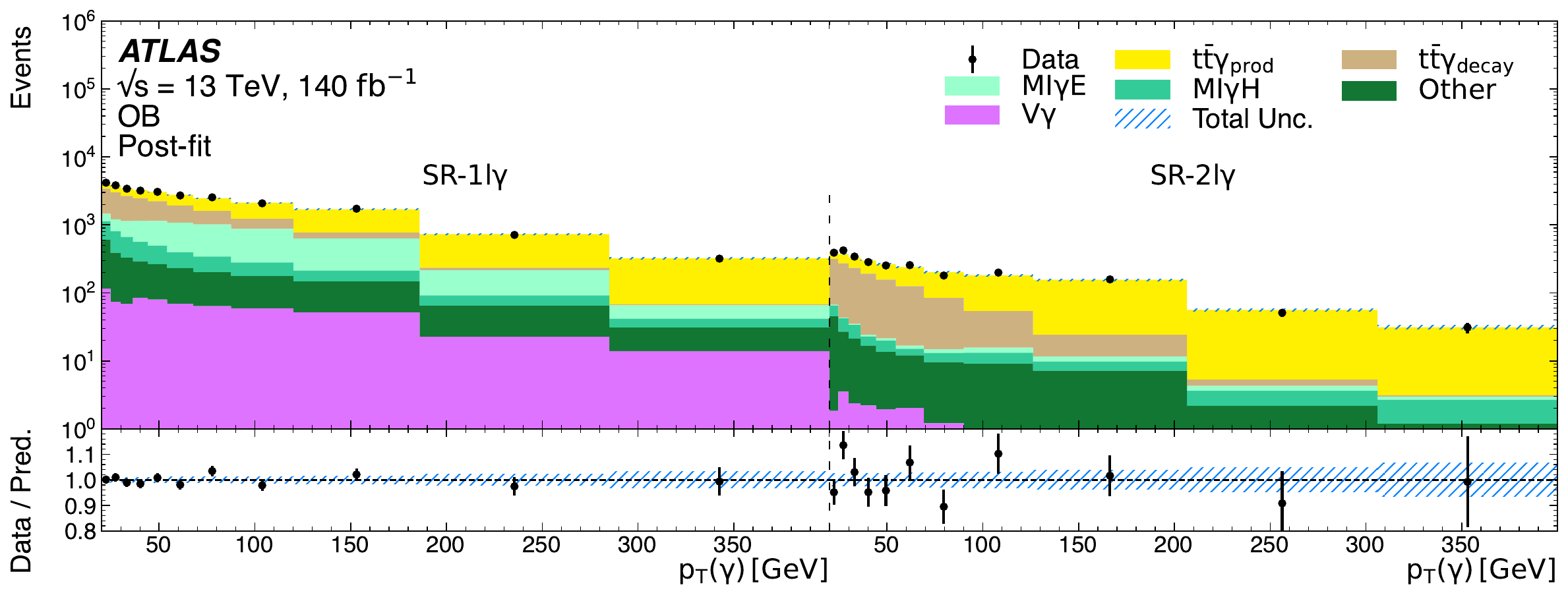}}\\
\subfloat[]{\includegraphics[width=0.75\linewidth]{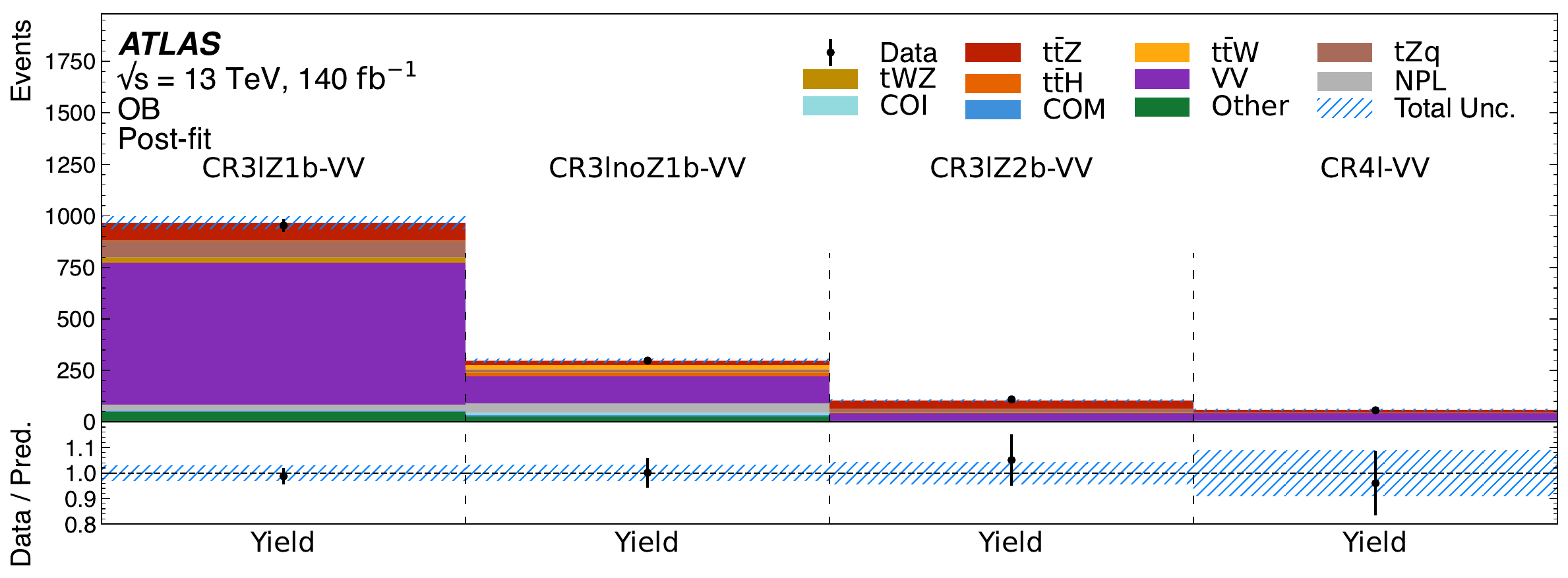}}
\caption{Comparison of the observed event yields and predictions in the
object-based (a) \OBfourl signal region (b) \OBonely and  \OBtwoly signal regions and (c) $VV$ control regions.
The EFT WCs, background normalisation factors and fit nuisance parameters are set to their best-fit values.
\UncertaintyBand \OtherCategoryDescription
}
\label{fig:Results:OB_Giga_tty_4L_CR}
\end{figure}

Similarly to the OB approach, there are few significant shifts or
constraints in the nuisance parameters in the PB approach fit. Those
that are observed are the same as previously discussed for the OB
approach. The background normalisation factors differ slightly with respect to the
OB approach results, but there are no significant differences:
$\lambda_{ZZ}=1.07^{+0.10}_{-0.09}$, $\lambda_{WZ}=1.25^{+0.20}_{-0.18}$, $\lambda_{\textrm{\NPL{LFe}}}=0.97^{+0.18}_{-0.16}$, $\lambda_{\textrm{\NPL{HF$\mu$}}}=0.96^{+0.08}_{-0.07}$, $\lambda_{\textrm{\NPL{HFe}}}=0.82^{+0.19}_{-0.19}$, $\lambda_{\textrm{\NPC{I}}}=1.03^{+0.09}_{-0.08}$ and $\lambda_{\textrm{\NPC{M}}}=0.89^{+0.09}_{-0.09}$.

\begin{figure}[!t]
\centering
\subfloat[]{\includegraphics[width=0.75\linewidth]{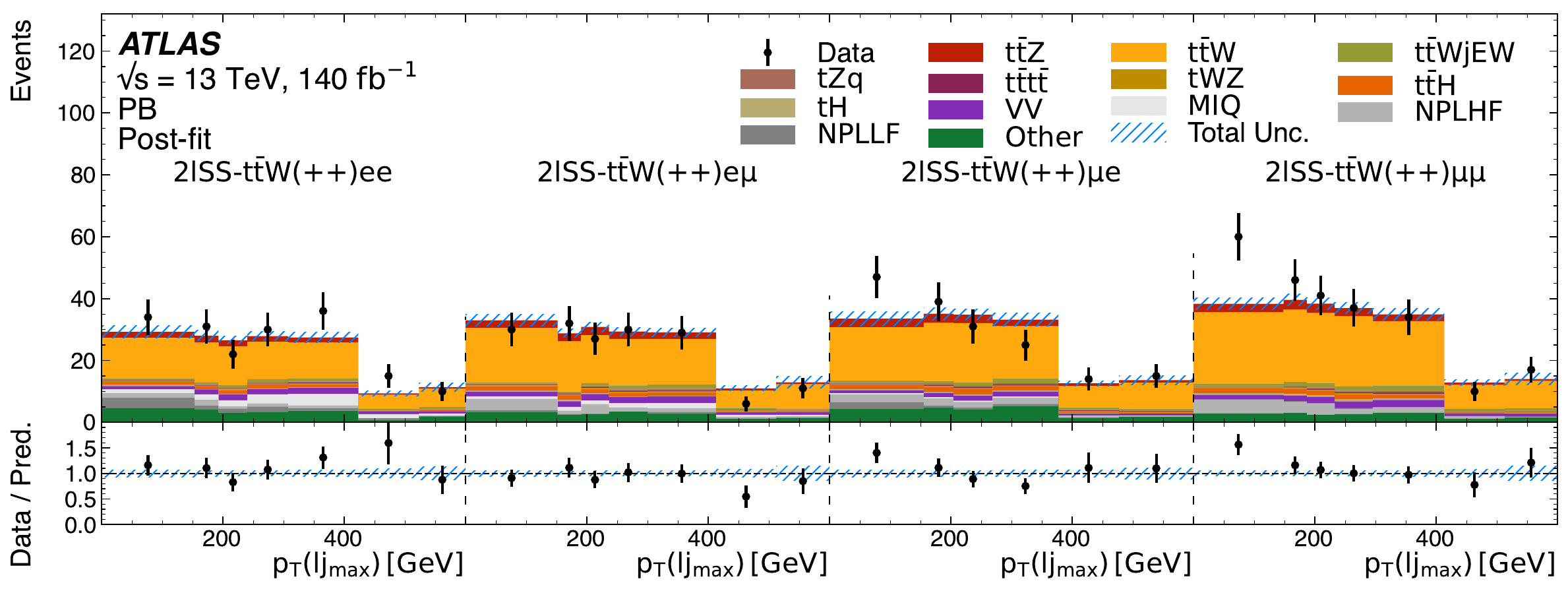}}\\
\subfloat[]{\includegraphics[width=0.75\linewidth]{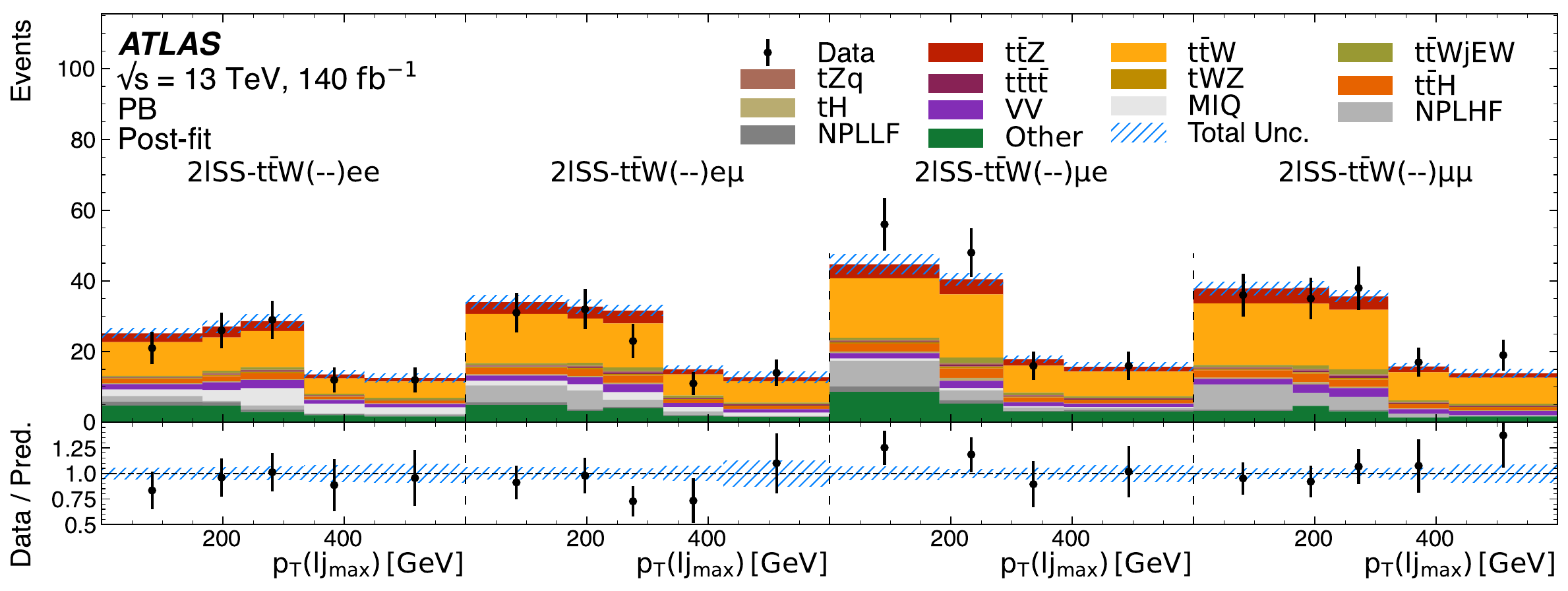}}\\
\subfloat[]{\includegraphics[width=0.75\linewidth]{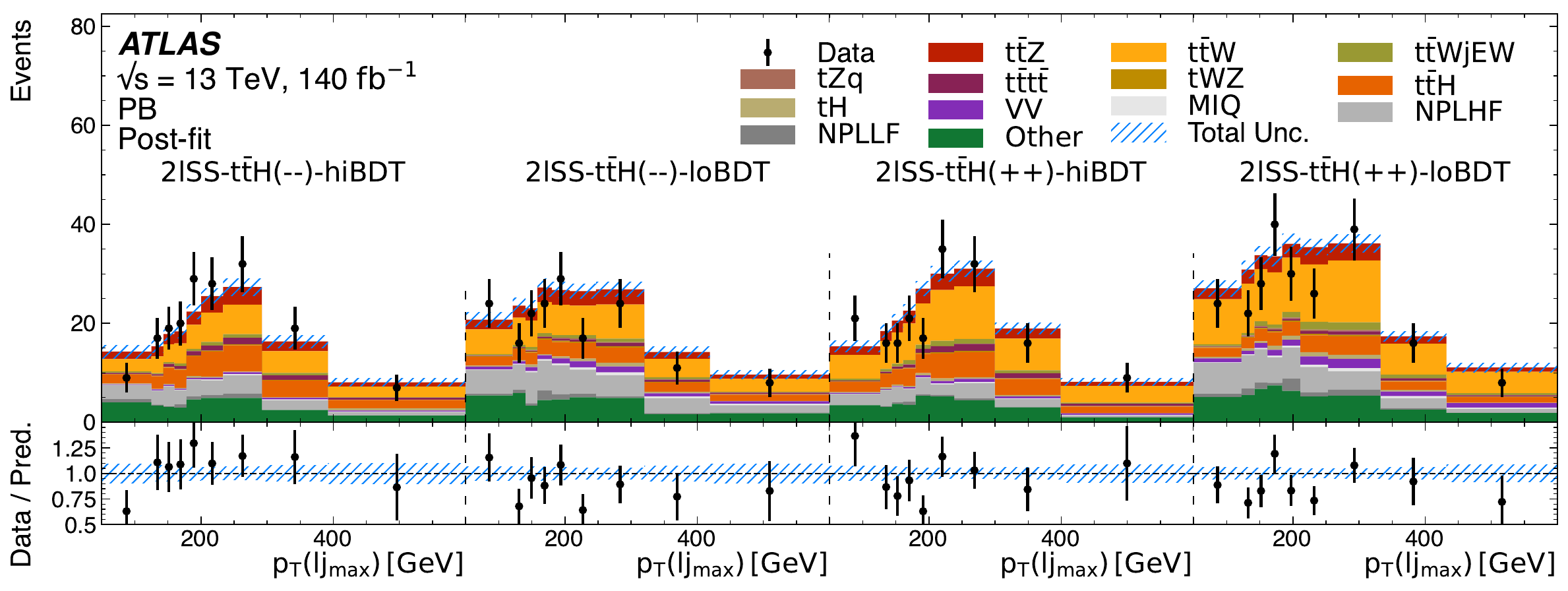}}\\
\subfloat[]{\includegraphics[width=0.75\linewidth]{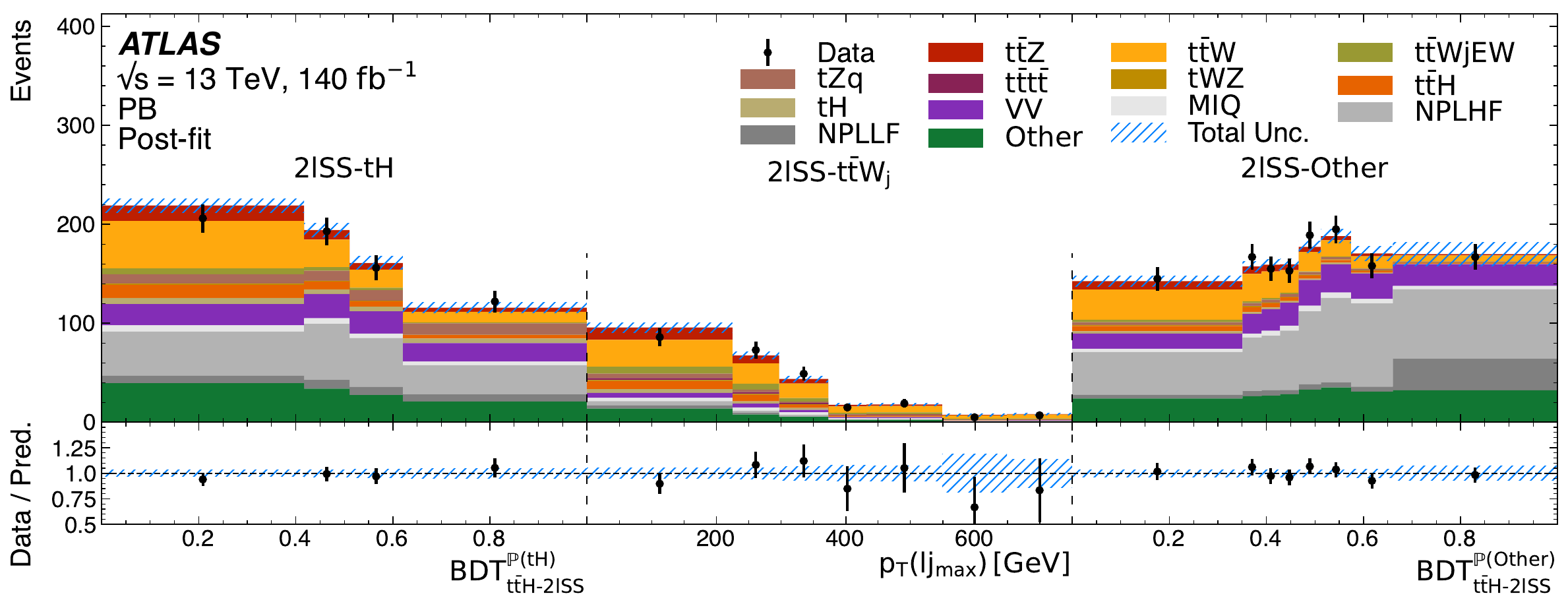}}\\
\caption{Comparison of the observed event yields and predictions in the
process-based (a) \PBtwolSSttWpp, (b) \PBtwolSSttWmm, (c) \PBtwolSSttH and (d) \PBtwolSStH and \PBtwolSSOther regions.
The EFT WCs, background normalisation factors and fit nuisance parameters are set to their best-fit values.
\UncertaintyBand \OtherCategoryDescription
\label{fig:Results:PB_Giga_2LSS}}
\end{figure}

\begin{figure}[!t]
\centering
\subfloat[]{\includegraphics[width=0.75\linewidth]{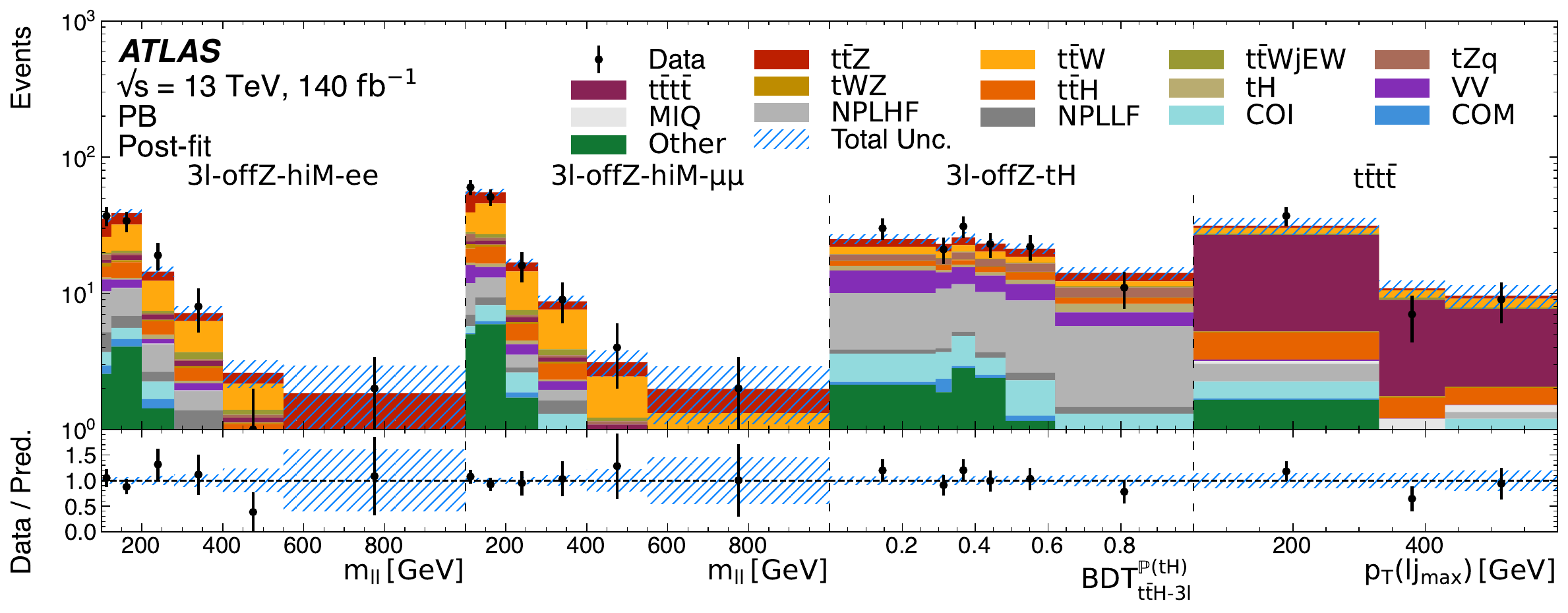}}\\
\subfloat[]{\includegraphics[width=0.75\linewidth]{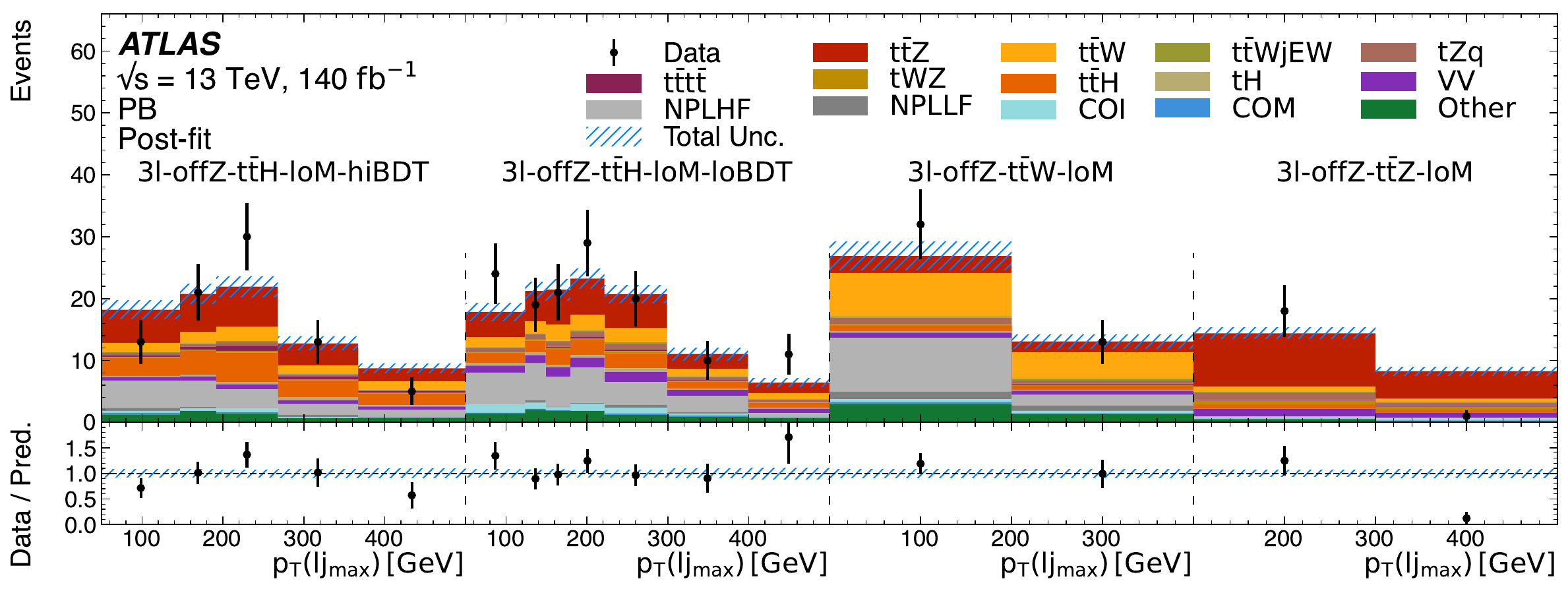}}\\
\subfloat[]{\includegraphics[width=0.75\linewidth]{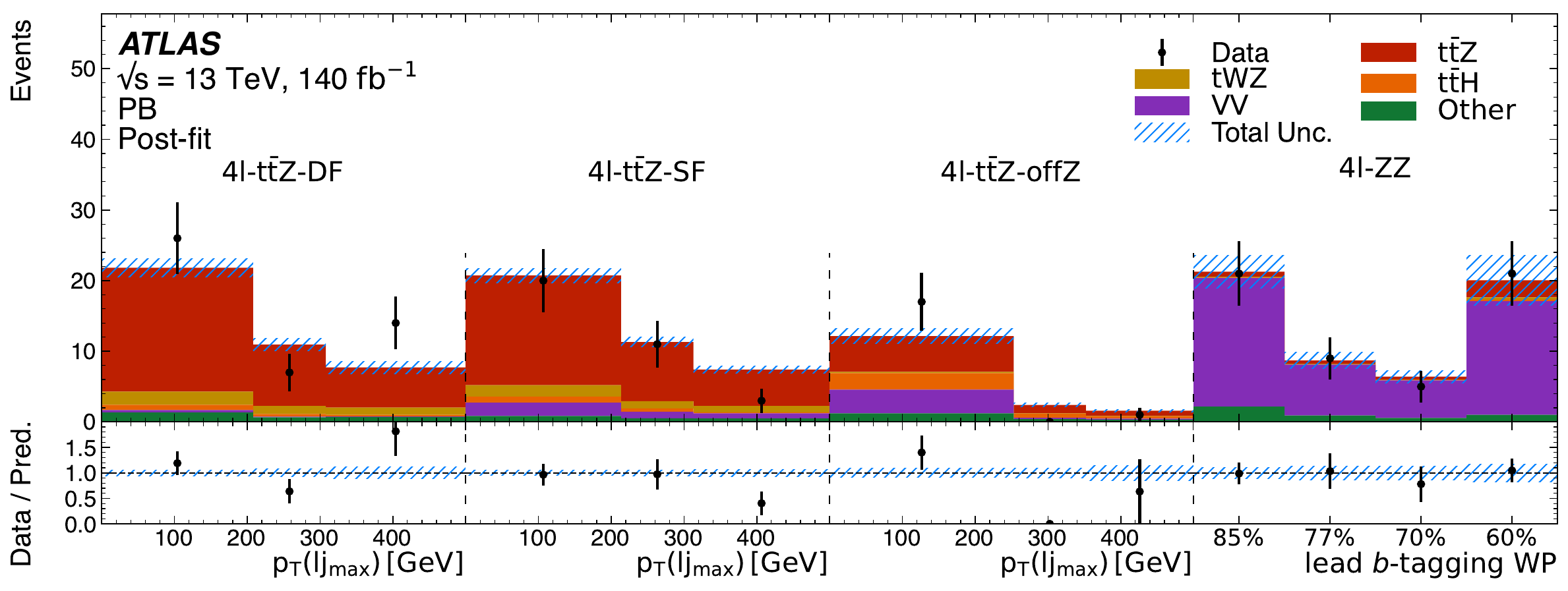}}\\
\subfloat[]{\includegraphics[width=0.75\linewidth]{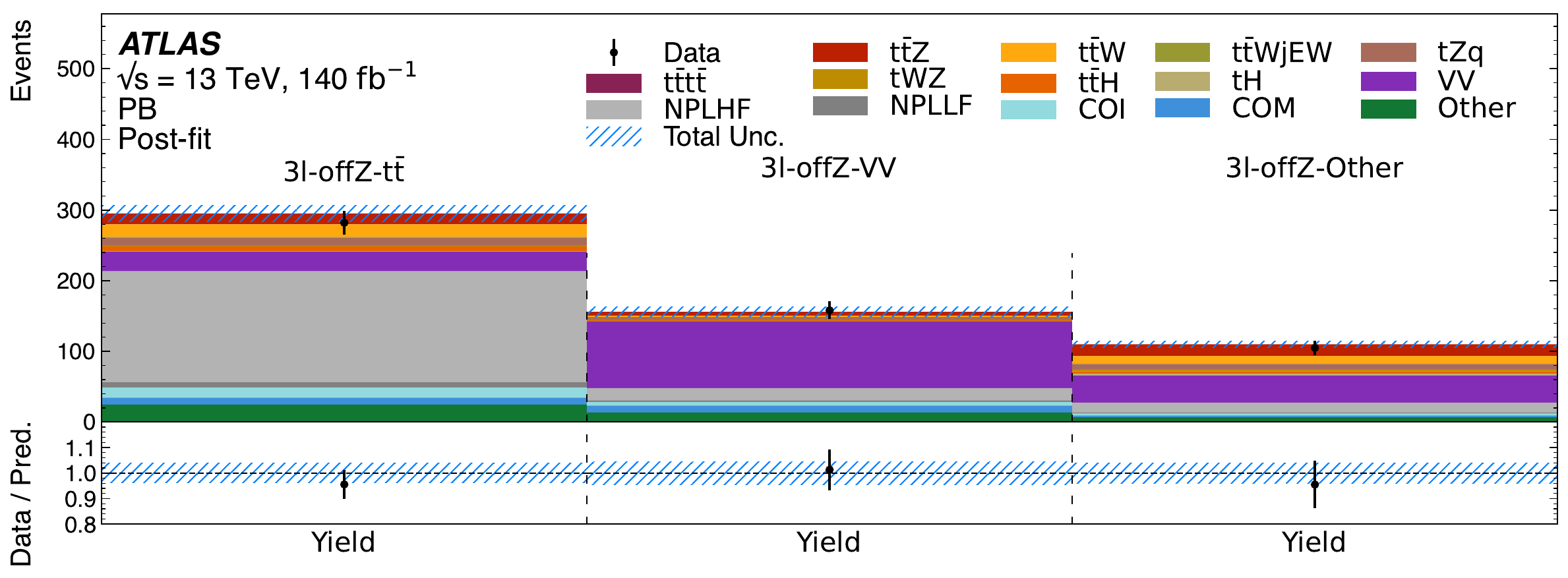}}\\
\caption{Comparison of the observed event yields and predictions in the
process-based (a) high-mass \PBthreelOffZnd, \PBthreelOffZ$tH$ and \tttt regions (b) low-mass \PBthreelOffZnd regions, (c) \PBfourl regions and (d)~\PBthreelOffZ{}$t\bar{t}$, \PBthreelOffZ{}$VV$ and \PBthreelOffZ{}Other regions.
The EFT WCs and background normalisation factors and fit nuisance parameters are set to their best-fit values.
\UncertaintyBand \OtherCategoryDescription
\label{fig:Results:PB_Giga_3Loff_4L}}
\end{figure}

\begin{figure}[!t]
\centering
\subfloat[]{\includegraphics[width=0.75\linewidth]{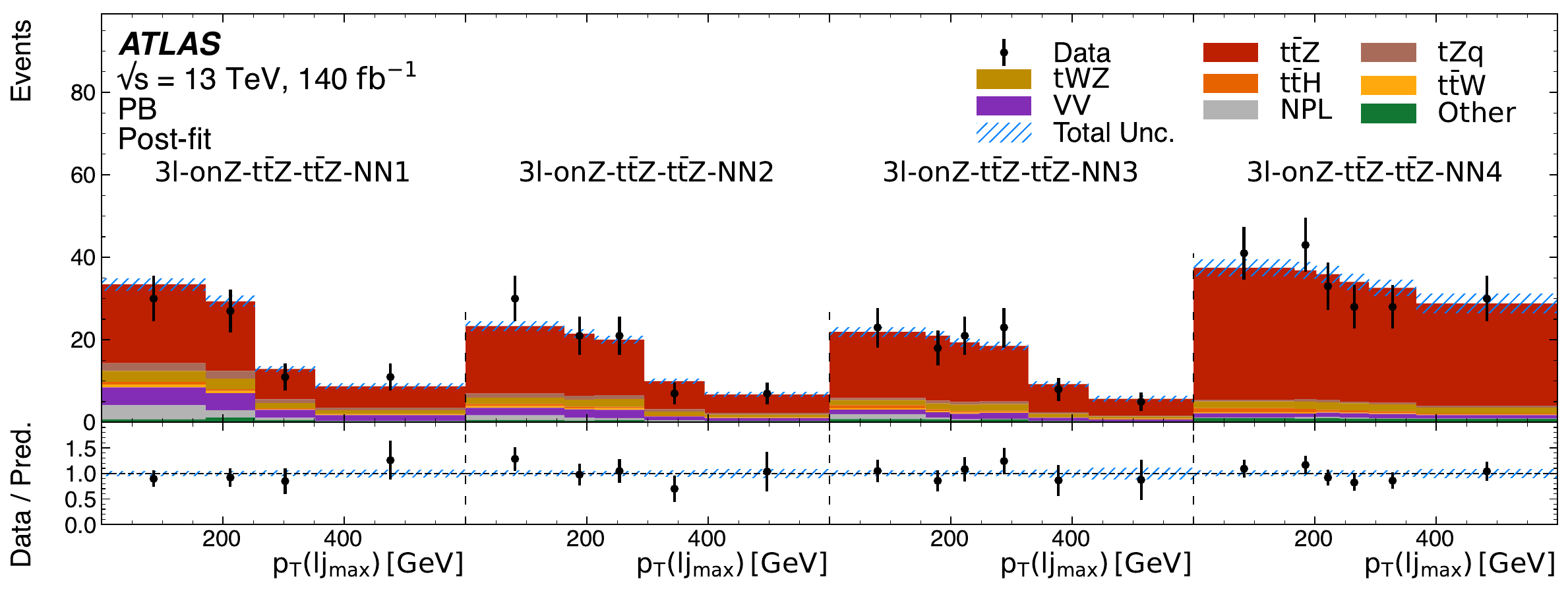}}\\
\subfloat[]{\includegraphics[width=0.75\linewidth]{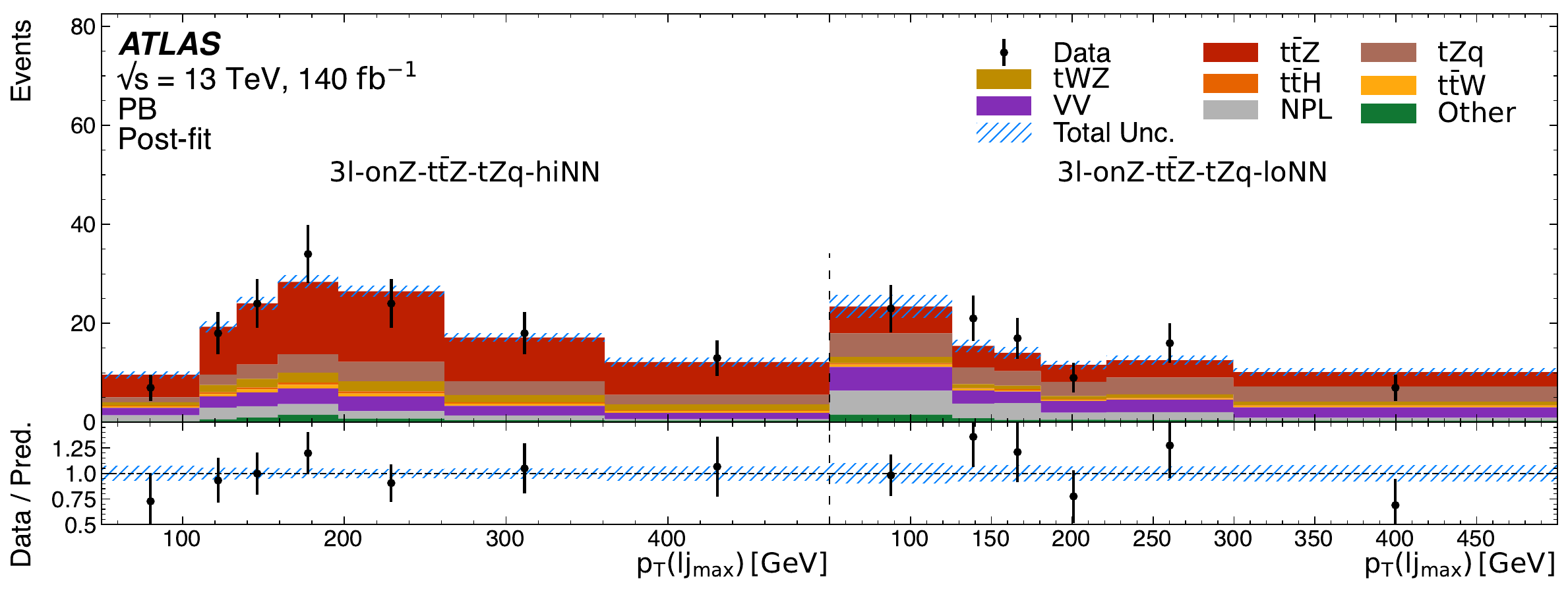}}\\
\subfloat[]{\includegraphics[width=0.75\linewidth]{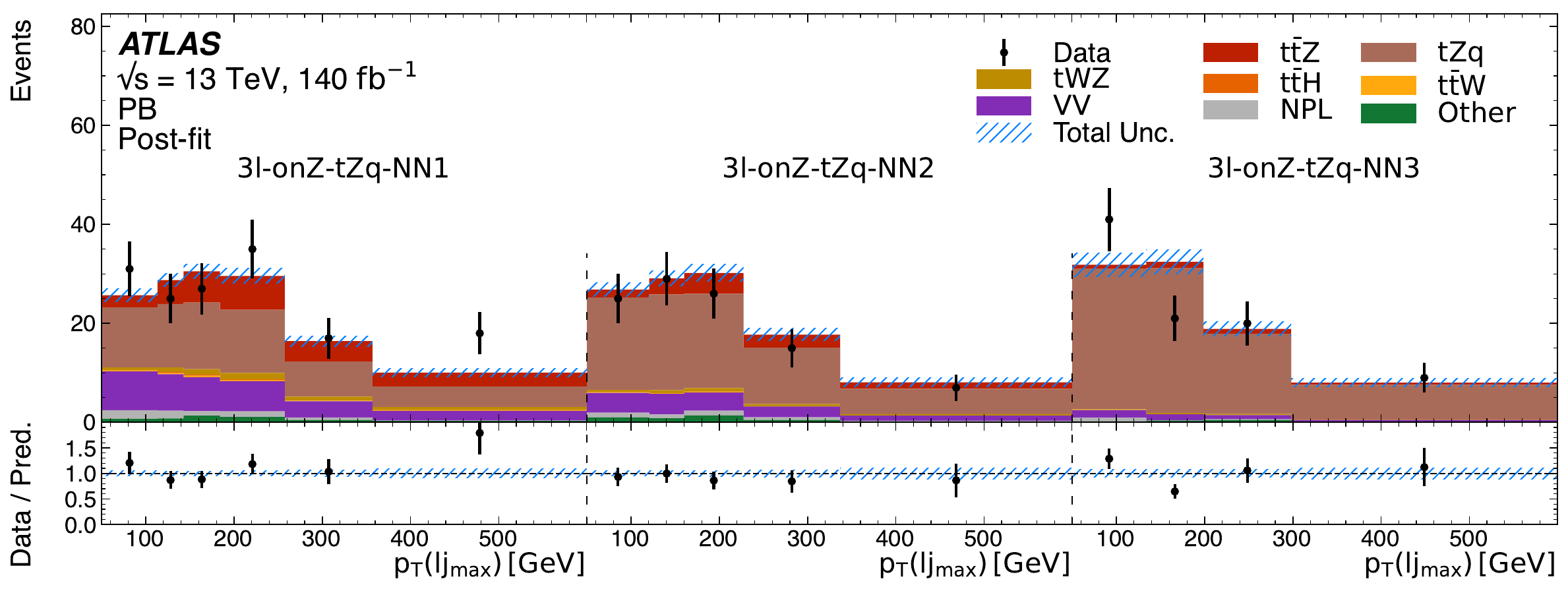}}\\
\subfloat[]{\includegraphics[width=0.75\linewidth]{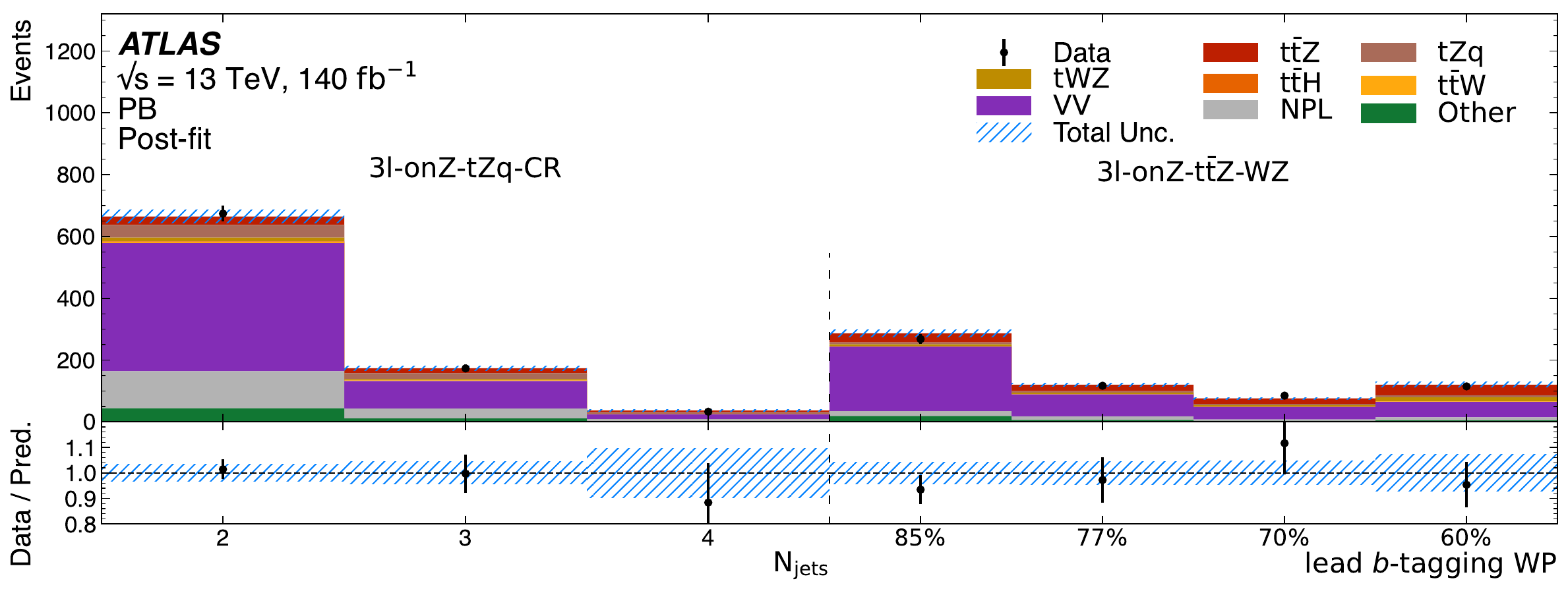}}\\
\caption{Comparison of the observed event yields and predictions in the
process-based (a) \PBthreelOnZttZttZ, (b) \PBthreelOnZ\ttZ-\tZq, (c) \PBthreelOnZtZqSR and (d) other \PBthreel regions.
The EFT WCs and background normalisation factors and fit nuisance parameters are set to their best-fit values.
\UncertaintyBand \OtherCategoryDescription
\label{fig:Results:PB_Giga_3LonZ}}
\end{figure}

\begin{figure}[!t]
\centering
\subfloat[]{\includegraphics[width=0.75\linewidth]{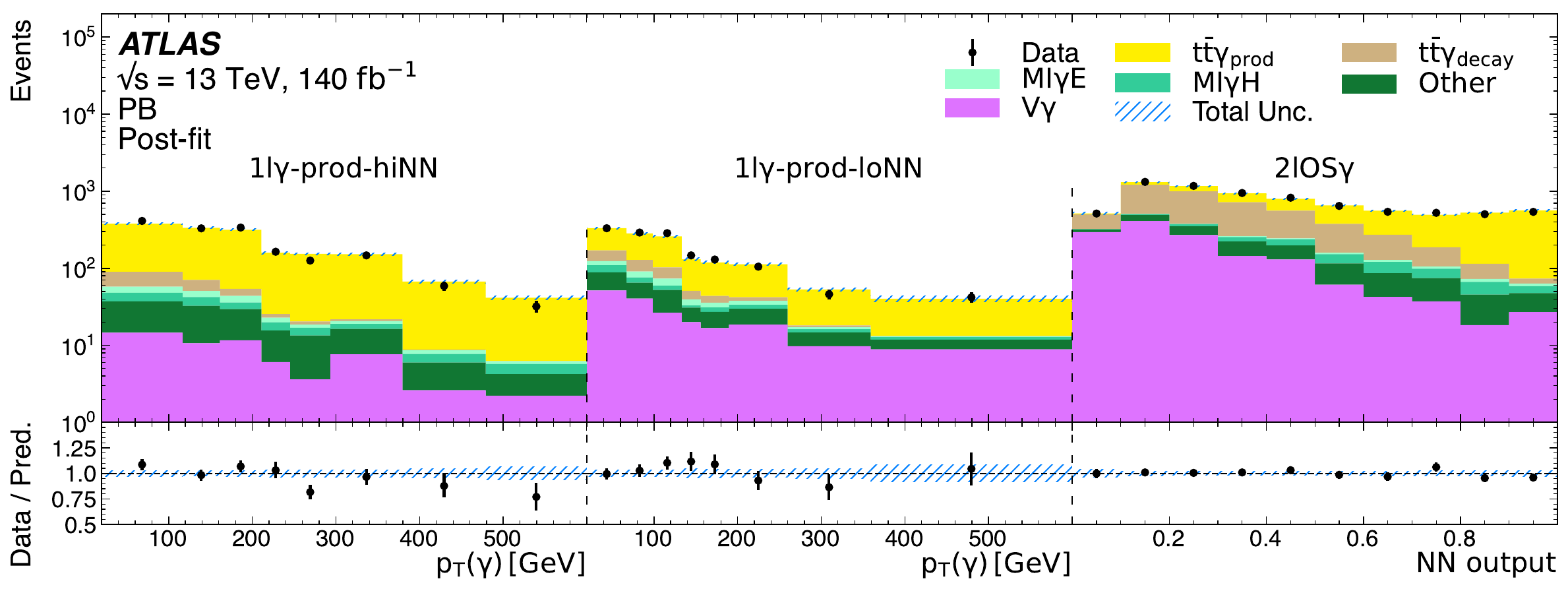}}\\
\subfloat[]{\includegraphics[width=0.75\linewidth]{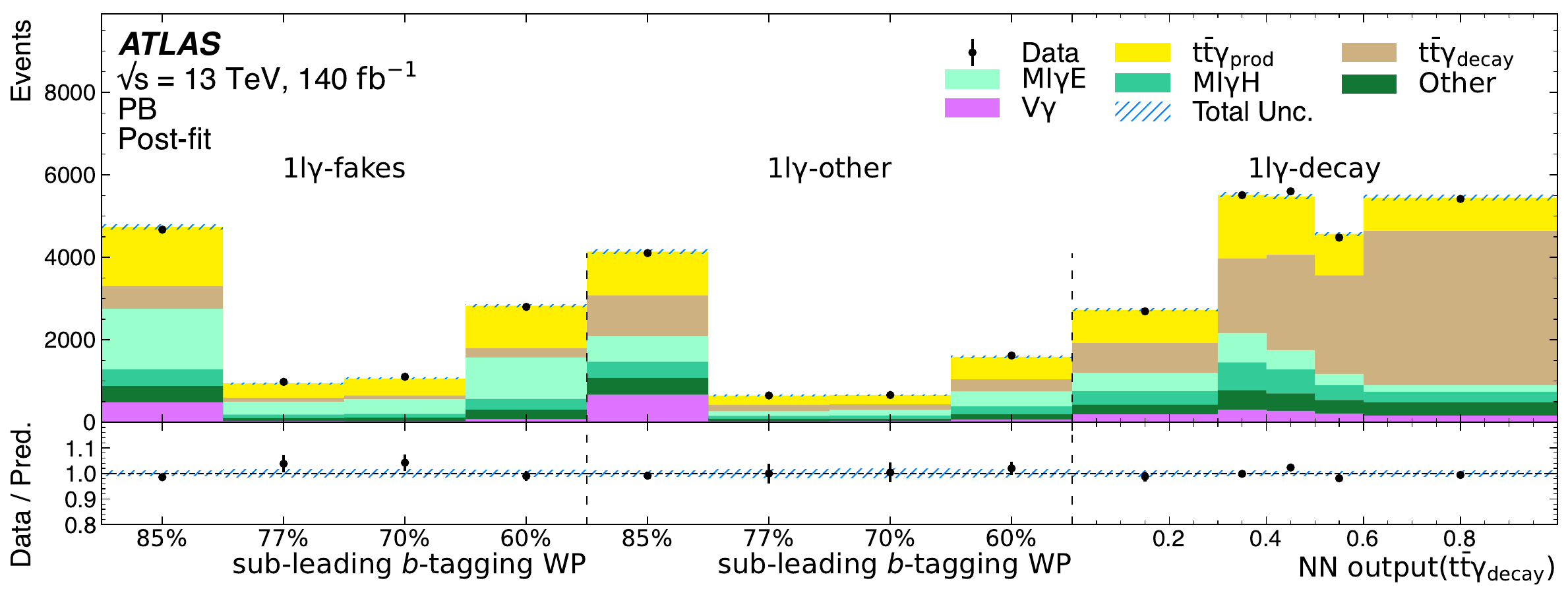}}\\
\caption{Comparison of the observed event yields and predictions in the
process-based (a) \PBonelGammaProd and \PBtwolOppSignGamma SRs and (b) \PBonelGammaFakes, \PBonelGammaOther and \PBonelGammaDecay regions.
The EFT WCs and background normalisation factors and fit nuisance parameters are set to their best-fit values.
\UncertaintyBand \OtherCategoryDescription
\label{fig:Results:PB_Giga_gamma}}
\end{figure}

\FloatBarrier

\subsection{Constraints on Wilson coefficients}
\label{sec:EFTres}

In this section the results of the fits of the WCs to data are
presented. The results for both approaches generally show good
compatibility with the SM. In some cases there are notable
mild tensions that are discussed in more detail in the following.
In Figure~\ref{fig:OBPBEFTLimits_Comp_All} and
Tables~\ref{tab:EFTSummary_OB}-\ref{tab:EFTSummary_PB} results are
shown where all WCs are allowed to float in the fit simultaneously
(Multi-Operator), and for one fitted at a time (Single-Operator).

For the top-lepton operators, all fitted WCs are found to be compatible
with the SM prediction within their 68\% confidence intervals (CIs) for both
the OB and PB approaches, in both the
single-operator and multi-operator fits. The OB approach generally
provides slightly stronger constraints for these operators, owing to
the larger event yields retained by its more inclusive event
categorisation in the relevant regions.

For the four-quark operators, the profiled 68\% CIs
for all WCs include zero in the multi-operator fit.  In contrast, the
single-operator PB fit exhibits mild deviations from the SM for the
two-heavy-two-light four-quark operators \ctj1, \ctj8, \cQj11,
\cQj18, \cQj31 and \cQj38, which derive much of their sensitivity from
the quadratic dimension-six contribution to the \ttW process. This
behaviour is consistent with the enhanced \ttW normalisation factor
observed in the SM normalisation factor fit and the excess of data
relative to the pre-fit SM prediction in the \ttW-enriched regions
that can be seen more generally in Appendix~\ref{app:prefit}. For one
particular region, \PBtwolSSttW{}(++)-$\mu e$, this is highlighted
explicitly in Figure~\ref{fig:Results:PreFitwEFT}(a) by showing the
data and the expected yields before the fit to data with the effect of
relevant EFT contributions overlaid where the size of the EFT effect
is determined by the best-fit values of relevant WCs.

For the top-Higgs operators, all WCs are found to be
consistent with the SM within their 95\% CIs in the multi-operator fit, with the exception of
\ctHIm in the PB approach.
In the PB fit, the preferred positive value of \ctHRe
is driven by its negative linear contribution to the \ttH production
rate, consistent with the low \ttH normalisation factor obtained in
the corresponding SM normalisation factor fit. The observed correlation between \ctHRe and
\ctHIm can be understood from their simultaneous impact on several
Higgs-associated top-quark processes. While the sensitivity of the
\tHq process is dominated by the quadratic dependence on \ctHRe, a
smaller quadratic contribution from \ctHIm is also
present. Consequently, a positive value of \ctHRe is compatible with
the pattern observed in the dedicated measurements, where the fitted
\tHq normalisation is above the SM expectation~\cite{HIGP-2024-03} while the fitted \ttH
normalisation is below the SM expectation. A similar behaviour is observed for
the \tttt process, where both \ctHRe and \ctHIm contribute
predominantly through quadratic terms of comparable magnitude. In this
case, a non-zero value of \ctHIm partially compensates for the
reduction in the combined \ttH+\tttt event yield associated with a
positive \ctHRe, providing a better description of the simultaneously
low fitted \ttH and high fitted \tttt normalisation. This is visualised in Figure~\ref{fig:Results:PreFitwEFT}(b) again showing the
data and the expected yields before the fit to data with the effect of
relevant EFT contributions overlaid. These important
correlations between processes and operators illustrate how a
simultaneous fit is vital to exploit complementary information across multiple
\topx processes that might not be accessible in independent
measurements.

For the other top-boson operators,
the individual WCs are generally compatible with zero in the multi-operator fits.
In the single-operator fits, the PB approach exhibits mild deviations
from the SM for a subset of operators. These preferences are
consistent with the trends observed in the corresponding dedicated
measurements, reflecting the reduced freedom of the single-operator
fit to accommodate fluctuations through correlations with other
WCs. In particular, the PB multi-operator fit prefers negative
values of \cHt and \cHQ-, consistent with the behaviour observed in the dedicated
\tZq measurement. Figure~\ref{fig:Results:PreFitwEFT}(c) gives some explanation for this preference, especially for the case of \cHQ-.

The obtained 95\% CIs for each WC can be compared to
that of Ref.~\cite{CMS-TOP-22-006}. It can be seen that the PB
approach gives markedly improved sensitivity to \cHt and both OB and
PB approaches give better sensitivity to \cHQ-, this is likely due to
the better isolation of \tZq. Similarly, several four-heavy-quark
operator WC intervals are stronger in the PB approach compared to that
of Ref.~\cite{CMS-TOP-22-006}, likely due to the improved separation
of \tttt. There are some places where the sensitivity obtained in this
result is worse or comparable to that of Ref.~\cite{CMS-TOP-22-006},
the top-lepton operator sensitivity for the PB approach is slightly
worse and the OB is very similar, while the sensitivity
to \ctHRe appears stronger in Ref.~\cite{CMS-TOP-22-006} but a direct comparison is
challenging due to the inclusion of imaginary WC components in this
result.

\begin{figure}[!t]
\centering
\subfloat[]{\includegraphics[width=0.33\linewidth]{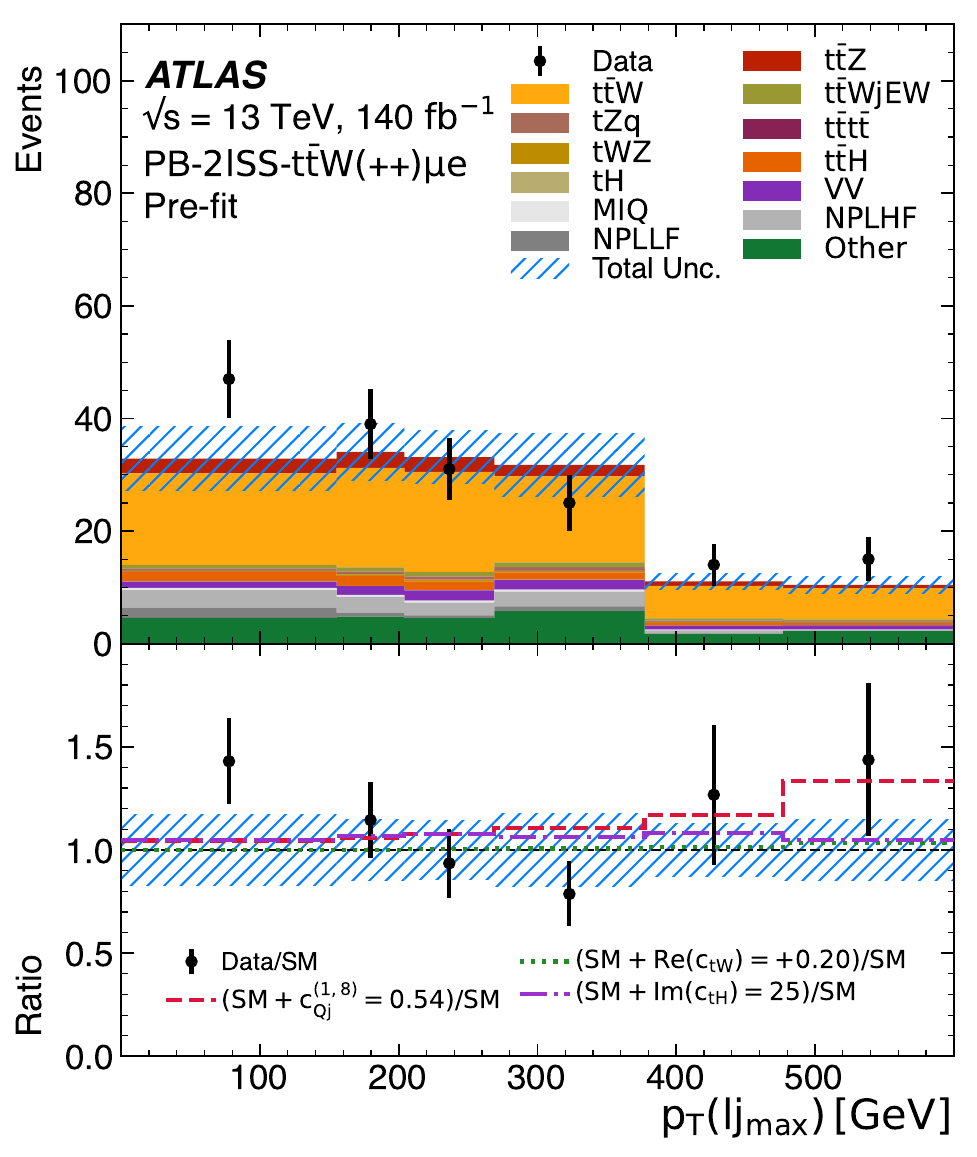}}
\subfloat[]{\includegraphics[width=0.33\linewidth]{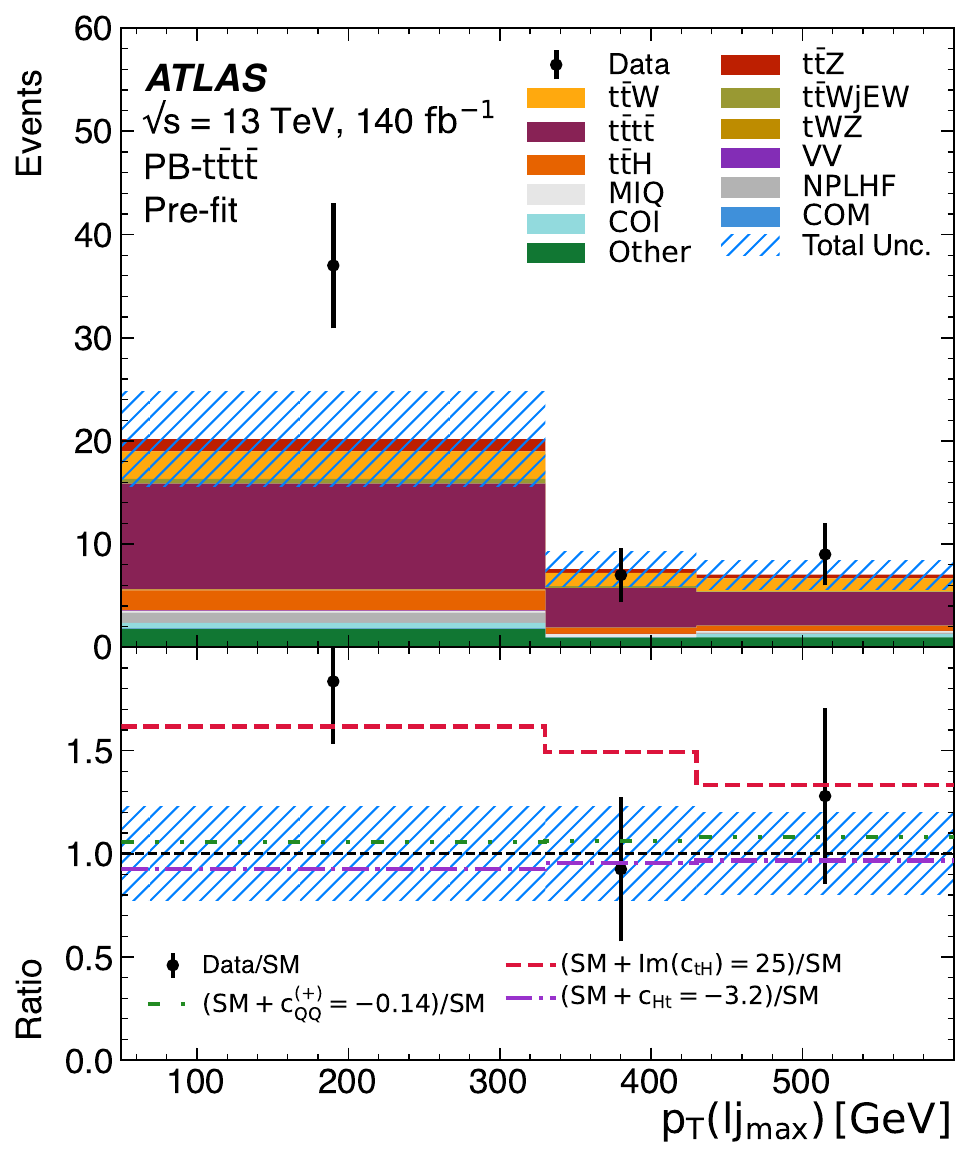}}
\subfloat[]{\includegraphics[width=0.33\linewidth]{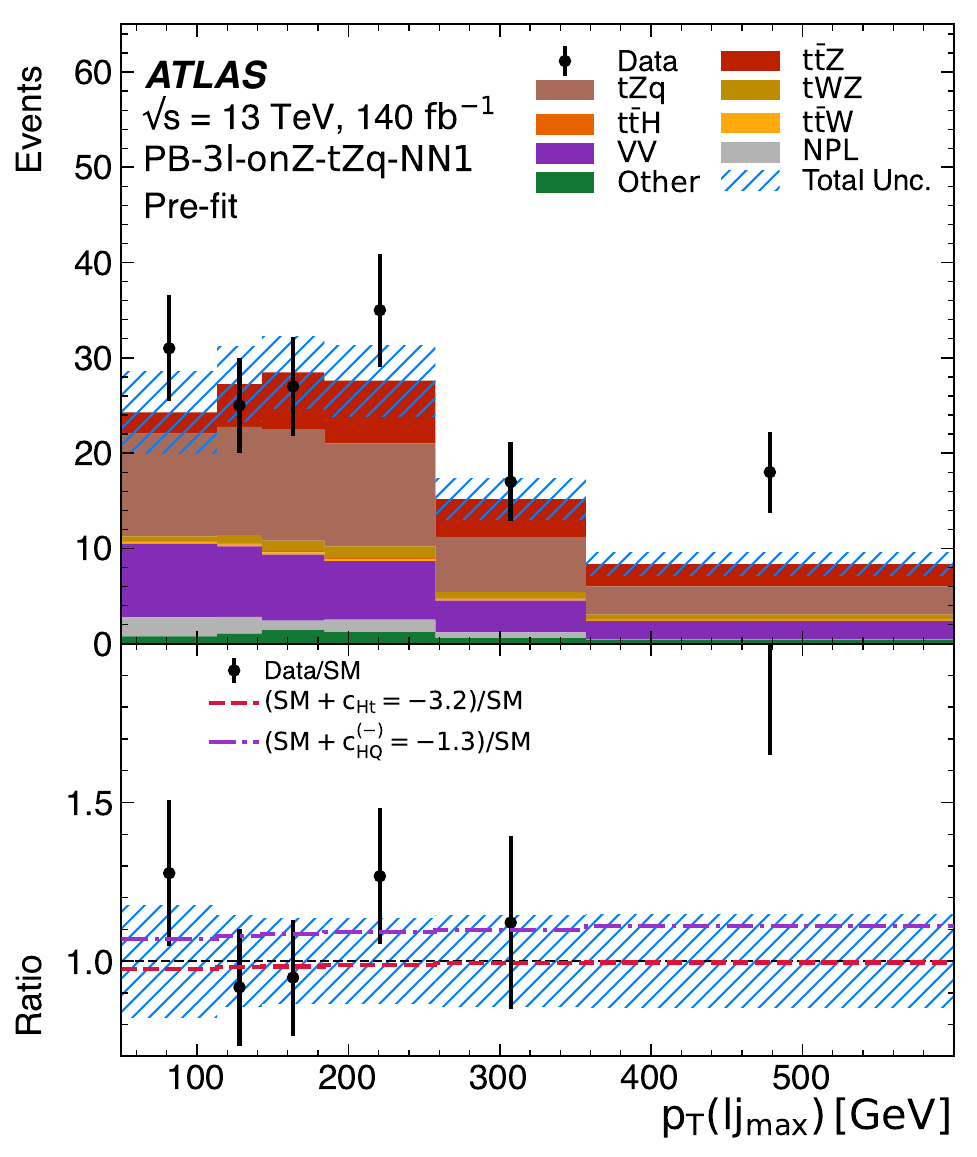}}
\caption{Comparison of event yields before the fit to data with the EFT variations overlaid where the WC values, in units of TeV{}$^{-2}$, have been set to the best-fit value after the fit to data. These are shown for three of the PB signal regions: (a)~\PBtwolSSttW{}(++)-$\mu e$, (b)~\PBtttt and (c)~\PBthreelOnZtZq{}-NN1.
\label{fig:Results:PreFitwEFT}}
\end{figure}

\subsection{Comparison of the object-based and process-based approaches}

A key aim of this measurement is a direct comparison of the
object-based and process-based approaches and the ability to compare
results between the two.
There are 245 total (209 SR) bins in the OB analysis and 281 total (242 SR) bins in the PB analysis.
Figure~\ref{fig:OBPBEFTLimits_Comp_All} shows a comparison of the
observed CIs on the set of Wilson Coefficients
between the PB and OB approaches.
In the multi-operator fits, the expected sensitivity, shown in
Appendix~\ref{app:expected}, is better in the PB approach for most
WCs.

The expected sensitivity to \cHQ-, \cHt, and \ctHRe is
improved in the PB fit compared to the OB approach.  These operators
have normalisation effects on \ttZ (\cHQ- and \cHt) and \ttH (\ctHRe),
and so the improved separation of these processes in the PB fit
increases the sensitivity to these operators.  In addition, the effect
of \cHQ- is different on \ttZ and \tZq, so the improved separation of
these processes helps to improve the limits on \cHQ-. A slightly larger systematic uncertainty for \cHt is
observed in the PB approach compared to the OB approach due to the tighter phase space requirements, but the statistical uncertainty is significantly larger in the OB approach due to the worse separation of relevant processes.  Similarly \ctHIm has a significantly larger statistical
uncertainty in the OB fit due to worse separation between processes.
The observed results from the fit to data show a more complex
situation due to the preference for non-zero WC values in some
places due to the mild tensions between the data and prediction,
especially the PB approach for example in \ttW, \ttH, \tHq and \tttt
enriched regions.
The limits on \ctBRe and \ctBIm, which gain sensitivity
mostly from \ttyprod, and on \ctWRe and \ctWIm, which gain
sensitivity mostly from \ttW, are tighter in the PB approach than the OB approach.

For the top-lepton operators, the expected limits are similar in the
OB and PB approaches. This is expected since the regions targeting
these effects are similar in the OB and PB approaches, there is no
dedicated classifier trained to enrich these regions, and the high-mass
regions in the OB (\OBthreelhiee and \OBthreelhimumu) and
PB (\PBthreelOffZ{hiMass-ee} and \PBthreelOffZ{hiMass-$\mu\mu$})
approaches have a similar purity of \ttZ
events. The OB approach is a little more sensitive, primarily due to
the mildly increased yields in this approach due to a slightly
looser event selection. The increased sensitivity in the OB approach
compared to the PB approach is enhanced in the observed results due to the
larger contribution from \ttW, which acts as a background, in the PB approach after the fit to
data.

Additionally, the expected sensitivity to the four-heavy-quark
operators is improved in the PB approach compared to the OB
approach. The PB approach is able to define a region with a high
percentage of \tttt events, which also contains enough events that it
can then be binned in \ljzero. This means it is able to better
separate out the effects of all three four-heavy-quark operators at
once, since their impact, while similar, grows differently with
energy.  For the two-heavy-two-light operators,
the PB and OB approaches give more similar results in the
multi-operator fit, in the expected results the PB approach is
marginally better for all WCs. In the observed results, the picture is more complicated. In the case of \cQj18 a larger systematic
and therefore total uncertainty is seen in the PB fit. However, this
is notably not the case for the expected limits and results
from the previously discussed tensions in the data for \ttW enriched
regions which are more significant for the PB region definitions than
for the OB.

\begin{figure}[htb]
\centering
\begin{minipage}[t]{0.48\textwidth}
\centering
\captionsetup[subfigure]{skip=-8pt}

\vspace{-6.2cm}
\begin{subfigure}{\linewidth}
\centering
\includegraphics[trim=0 0 0 10pt, clip, width=\textwidth]{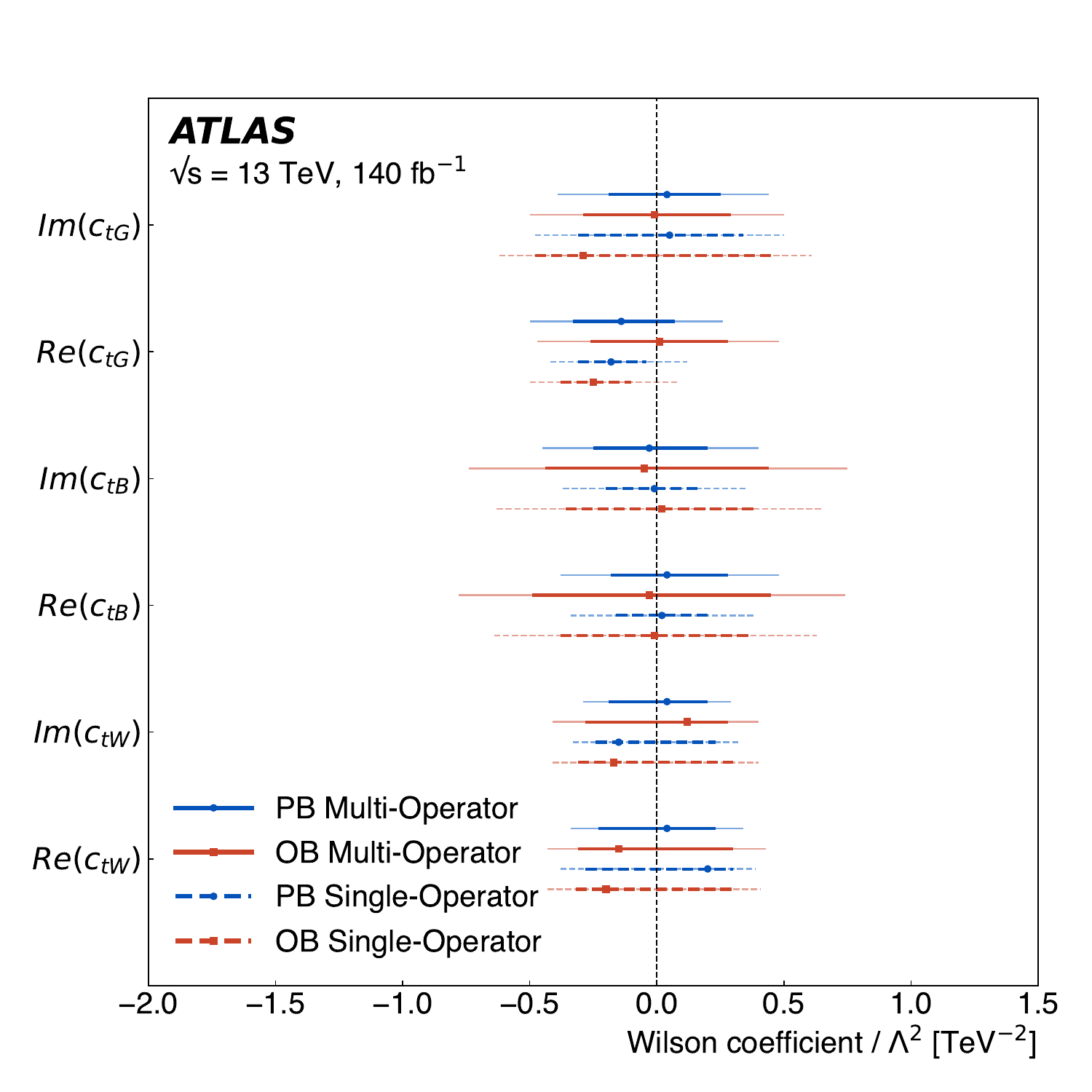}
\caption{}
\label{fig:image1}
\end{subfigure}

\vspace{-0.25em}

\begin{subfigure}{\linewidth}
\centering

\includegraphics[trim=0 0 0 10pt, clip, width=\textwidth]{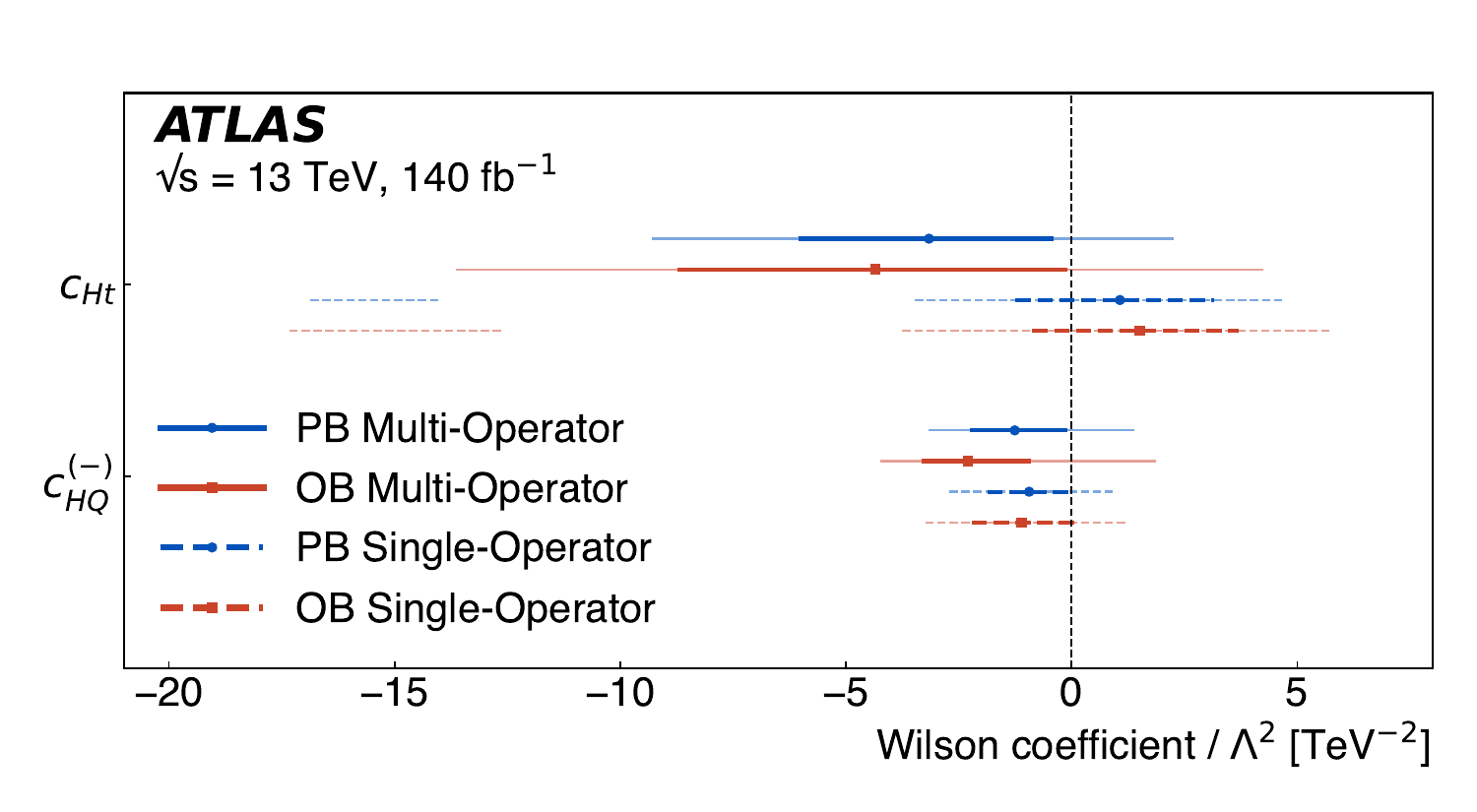}
\caption{}
\label{fig:image2}
\end{subfigure}

\vspace{-0.25em}

\begin{subfigure}{\linewidth}
\centering

\includegraphics[trim=0 0 0 10pt, clip, width=\textwidth]{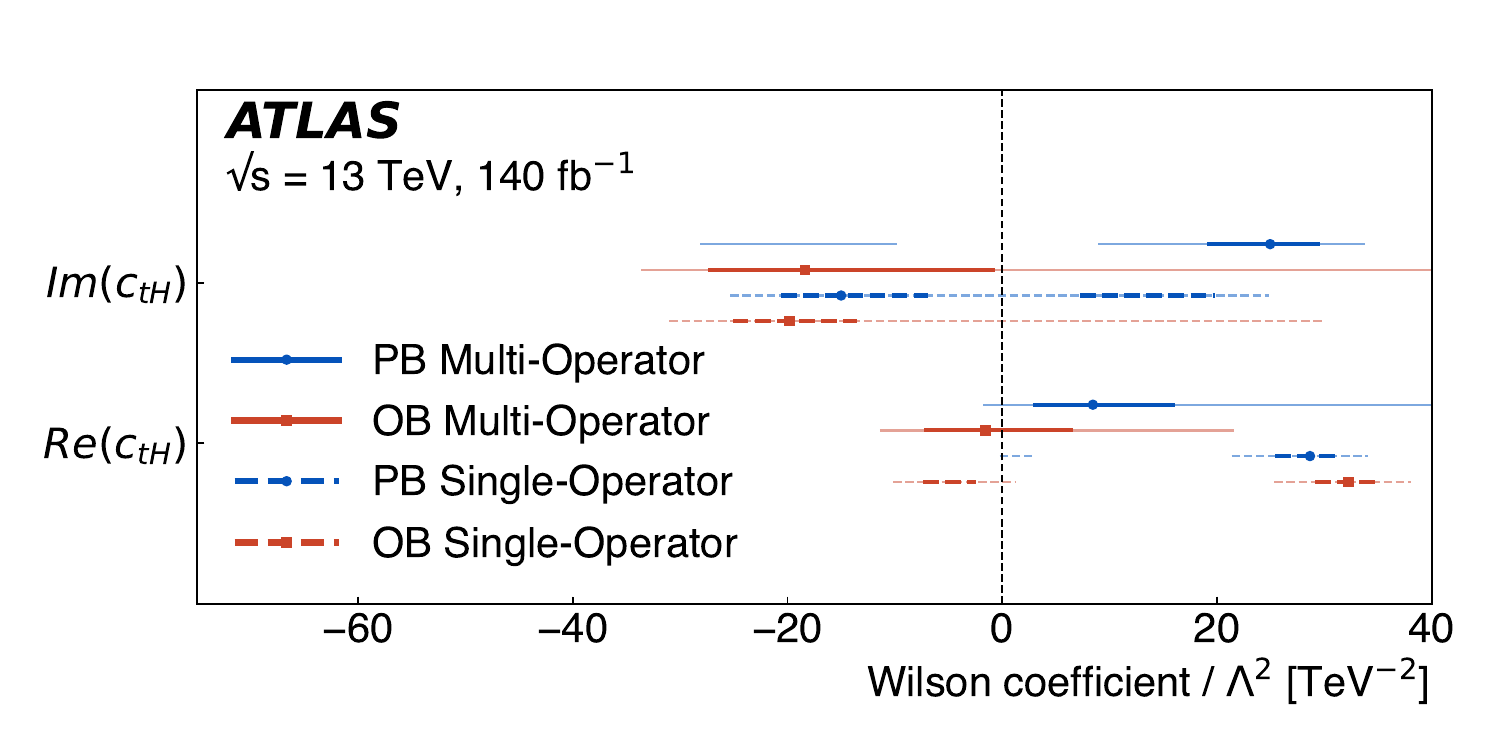}
\caption{}
\label{fig:image3}
\end{subfigure}

\vspace{-0.25em}

\begin{subfigure}{\linewidth}
\centering
\includegraphics[trim=0 0 0 10pt, clip, width=\textwidth]{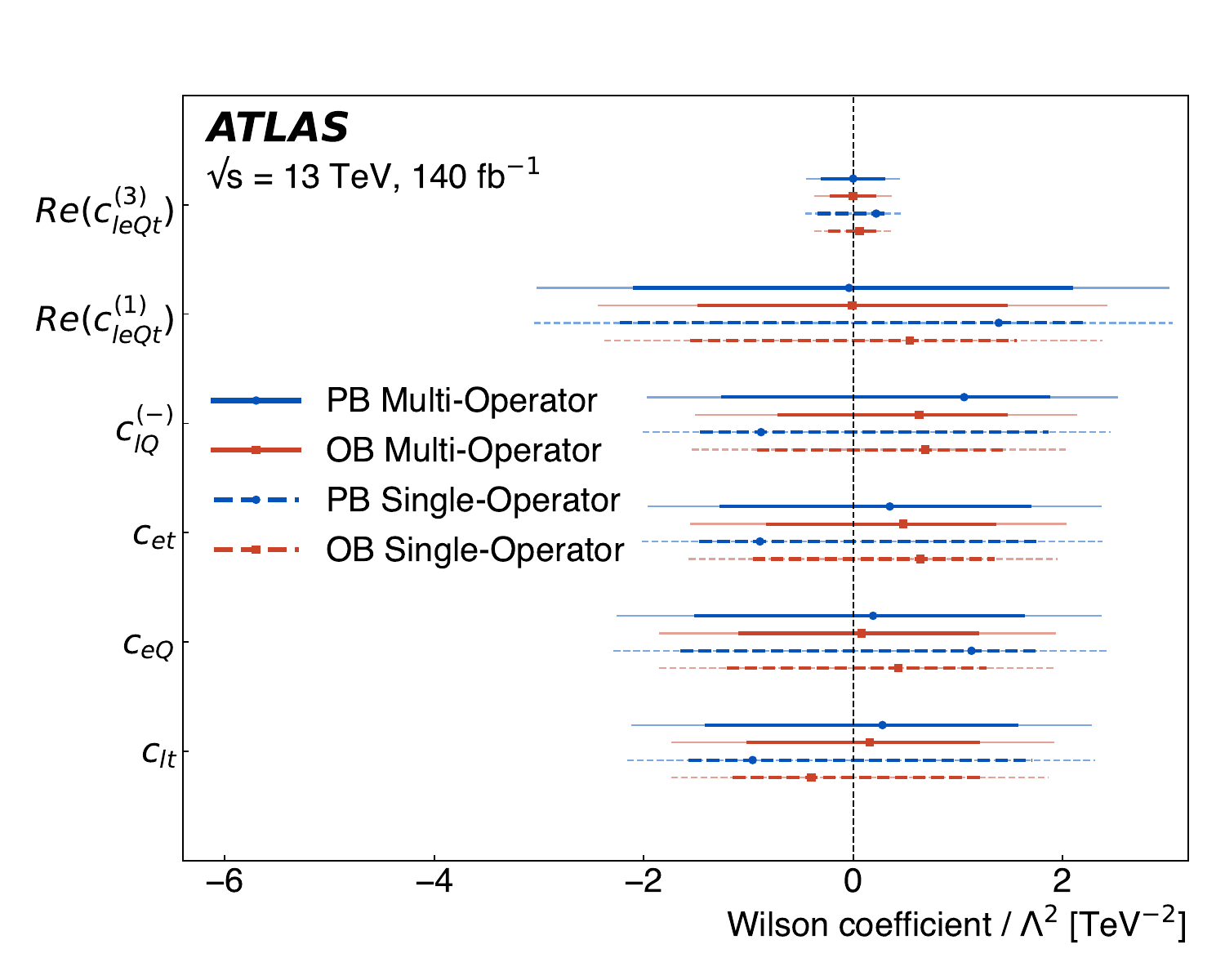}
\caption{}
\label{fig:image4}
\end{subfigure}

\end{minipage}
\hfill
\begin{minipage}[t]{0.48\textwidth}
\centering
\captionsetup[subfigure]{skip=-8pt}

\begin{subfigure}{\linewidth}
\centering
\includegraphics[trim=0 0 0 10pt, clip, width=\textwidth]{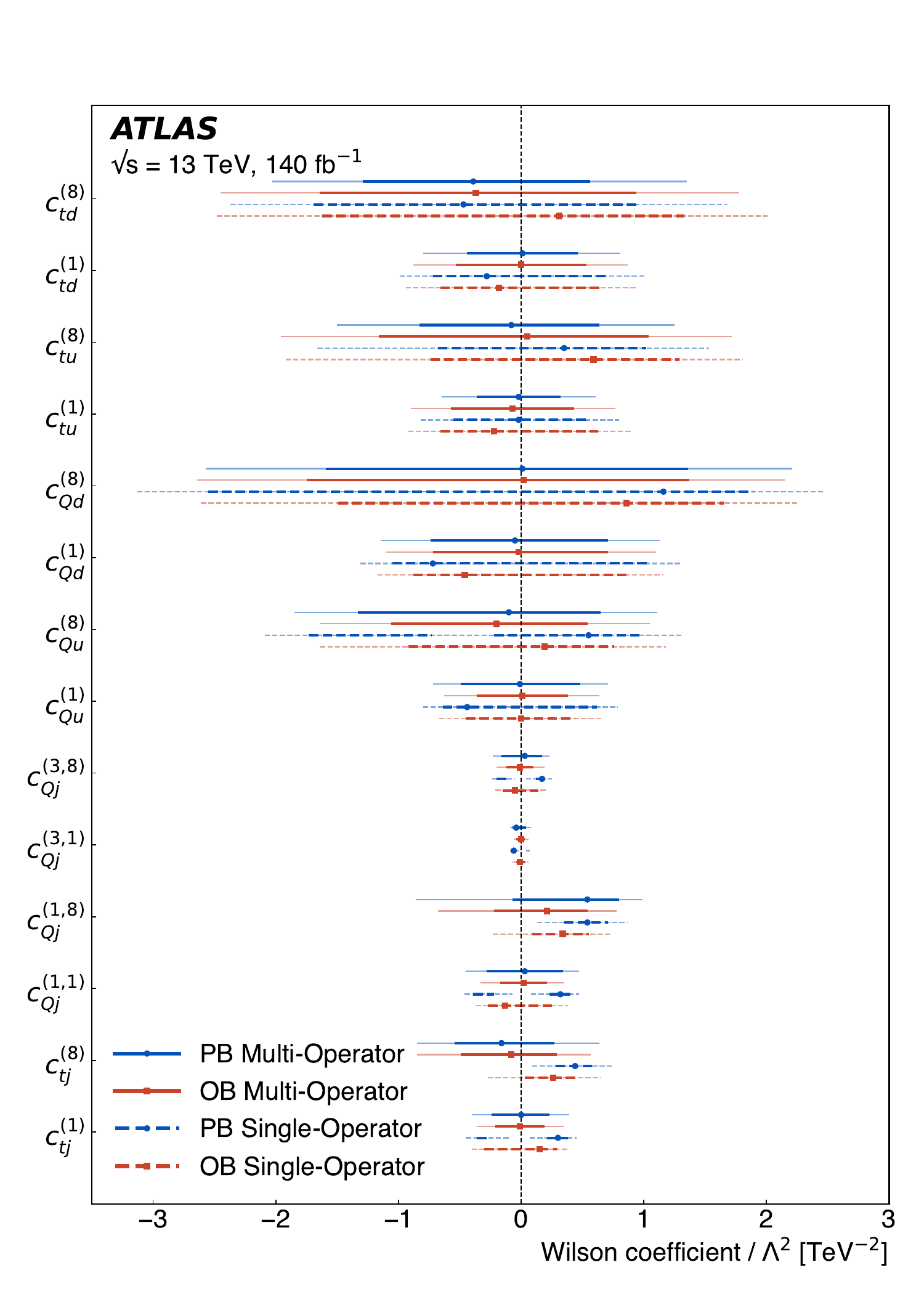}
\caption{}
\label{fig:image5}
\end{subfigure}

\begin{subfigure}{\linewidth}
\centering
\includegraphics[trim=0 0 0 10pt, clip, width=\textwidth]{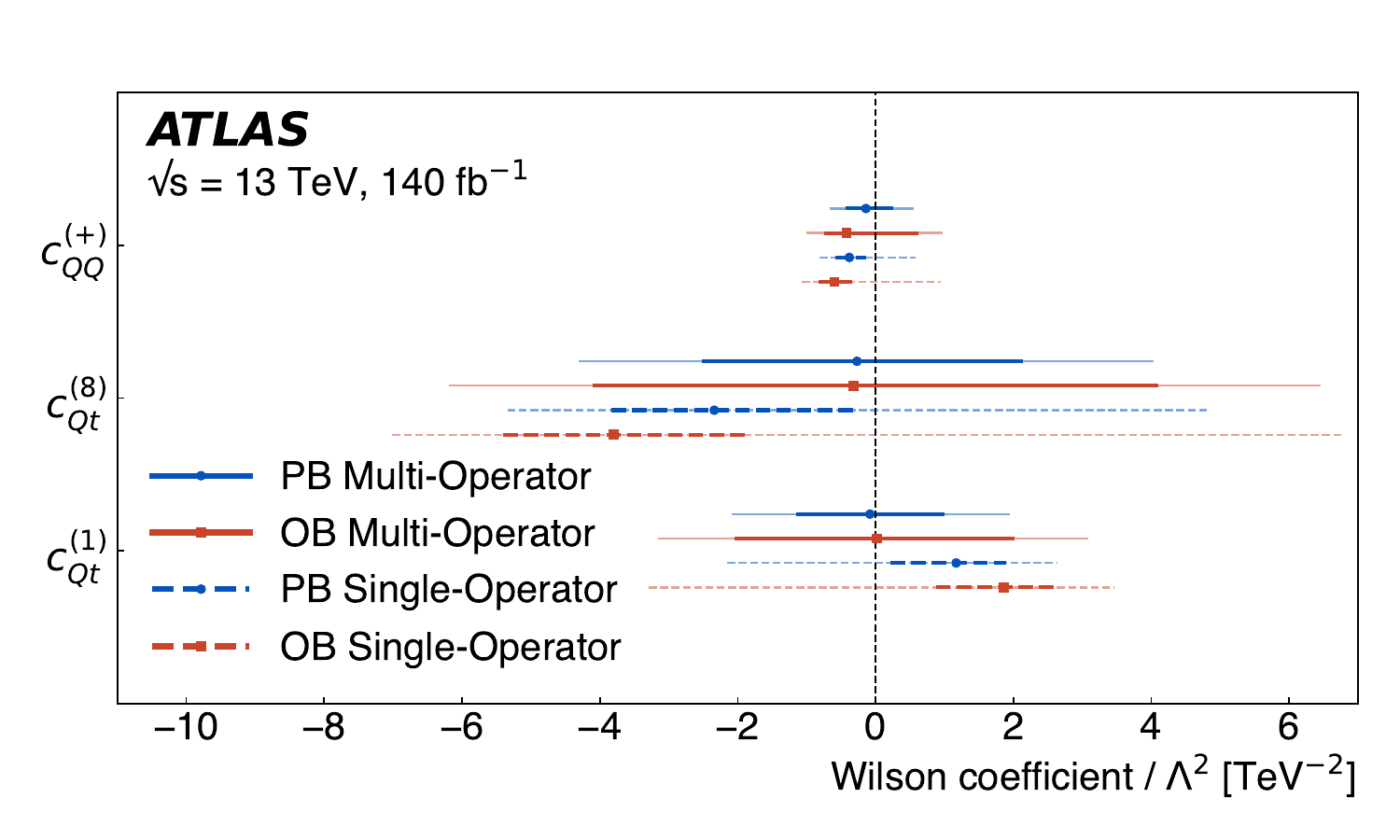}
\caption{}
\label{fig:image6}
\end{subfigure}

\end{minipage}
\caption{Comparison of the measured CIs on the WCs for the OB and PB approaches in the top-boson category for (a) \ctG, \ctB and \ctW WCs, for (b) \cHt and \cHQ- WCs, and (c) \ctHRe and \ctHIm WCs. The (d) top-lepton category and the four-quark category for (e) the two-heavy-two-light WCs and (f) the four-heavy-quark WCs. Results are for the multi-operator and single-operator fits. The thicker lines correspond to 68\% CIs while the thinner lines correspond to 95\% intervals. Lines extending to the figure boundary have no limit in that direction.}
\label{fig:OBPBEFTLimits_Comp_All}
\end{figure}

A summary of the 68\% and 95\% CIs is shown in
Tables~\ref{tab:EFTSummary_OB} and~\ref{tab:EFTSummary_PB} for the OB
and PB approaches, respectively.

%

% The next lines are included from the .//Sections/EFTLimitsTable_PDG_rounded.tex input file
\begin{table}[htbp]
\centering
\scriptsize
\resizebox{\textwidth}{!}{
{\renewcommand{\arraystretch}{1.5}
\begin{tabular}{lcccccc}
\toprule
& \multicolumn{6}{c}{$\mathcal{O}(\Lambda^{-4})$ $[\TeV^{-2}]$}\\
Coefficient & Multi best fit & Multi 68\% CI & Multi 95\% CI & Single best fit & Single 68\% CI & Single 95\% CI \\
\midrule
\cHQ- & $-2.3$ & $[-3.3,\,-0.91]$ & $[-4.2,\,1.9]$ & $-1.1$ & $[-2.2,\,0.050]$ & $[-3.2,\,1.3]$ \\
\cHt & $-4.4$ & $[-8.7,\,-0.10]$ & $[-14,\,4.2]$ & $1.5$ & $[-0.87,\,3.7]$ & $[-17,\,-13] \cup [-3.8,\,5.7]$ \\
\ctWRe & $-0.15$ & $[-0.31,\,0.30]$ & $[-0.43,\,0.43]$ & $-0.20$ & $[-0.32,\,0.31]$ & $[-0.43,\,0.41]$ \\
\ctWIm & $0.12$ & $[-0.28,\,0.28]$ & $[-0.41,\,0.40]$ & $-0.17$ & $[-0.31,\,0.30]$ & $[-0.41,\,0.40]$ \\
\ctBRe & $-0.03$ & $[-0.49,\,0.45]$ & $[-0.78,\,0.74]$ & $-0.01$ & $[-0.38,\,0.37]$ & $[-0.64,\,0.63]$ \\
\ctBIm & $-0.05$ & $[-0.44,\,0.44]$ & $[-0.74,\,0.75]$ & $0.02$ & $[-0.36,\,0.39]$ & $[-0.63,\,0.65]$ \\
\ctGRe & $0.01$ & $[-0.26,\,0.28]$ & $[-0.47,\,0.48]$ & $-0.25$ & $[-0.38,\,-0.10]$ & $[-0.50,\,0.09]$ \\
\ctGIm & $-0.01$ & $[-0.29,\,0.29]$ & $[-0.50,\,0.50]$ & $-0.29$ & $[-0.48,\,0.46]$ & $[-0.62,\,0.61]$ \\
\ctHRe & $-1.6$ & $[-7.3,\,6.5]$ & $[-11,\,22]$ & $32$ & $[-7.4,\,-2.5] \cup [29,\,35]$ & $[-10,\,1.2] \cup [25,\,38]$ \\
\ctHIm & $-18$ & $[-27,\,-0.68]$ & $[-34,\,\text{No upper limit}]$ & $-20$ & $[-25,\,-13]$ & $[-31,\,30]$ \\
\ctl & $0.16$ & $[-1.0,\,1.2]$ & $[-1.7,\,1.9]$ & $-0.40$ & $[-1.2,\,1.2]$ & $[-1.7,\,1.9]$ \\
\cQe & $0.08$ & $[-1.1,\,1.2]$ & $[-1.9,\,1.9]$ & $0.43$ & $[-1.2,\,1.3]$ & $[-1.9,\,1.9]$ \\
\cte & $0.48$ & $[-0.83,\,1.4]$ & $[-1.6,\,2.0]$ & $0.64$ & $[-0.96,\,1.4]$ & $[-1.6,\,2.0]$ \\
\cQl- & $0.63$ & $[-0.72,\,1.5]$ & $[-1.5,\,2.1]$ & $0.69$ & $[-0.92,\,1.4]$ & $[-1.5,\,2.0]$ \\
\cleQtRe1 & $-0.01$ & $[-1.5,\,1.5]$ & $[-2.4,\,2.4]$ & $0.54$ & $[-1.6,\,1.6]$ & $[-2.4,\,2.4]$ \\
\cleQtRe3 & $0.00$ & $[-0.22,\,0.22]$ & $[-0.37,\,0.37]$ & $0.06$ & $[-0.24,\,0.24]$ & $[-0.37,\,0.37]$ \\
\ctj1 & $-0.01$ & $[-0.21,\,0.19]$ & $[-0.36,\,0.35]$ & $0.15$ & $[-0.30,\,0.29]$ & $[-0.40,\,0.39]$ \\
\ctj8 & $-0.08$ & $[-0.49,\,0.29]$ & $[-0.85,\,0.57]$ & $0.26$ & $[0.03,\,0.46]$ & $[-0.27,\,0.65]$ \\
\cQj11 & $0.02$ & $[-0.17,\,0.21]$ & $[-0.33,\,0.35]$ & $-0.13$ & $[-0.27,\,0.27]$ & $[-0.37,\,0.38]$ \\
\cQj18 & $0.21$ & $[-0.22,\,0.54]$ & $[-0.68,\,0.78]$ & $0.34$ & $[0.09,\,0.55]$ & $[-0.23,\,0.74]$ \\
\cQj31 & $0.00$ & $[-0.04,\,0.03]$ & $[-0.06,\,0.06]$ & $-0.01$ & $[-0.04,\,0.04]$ & $[-0.07,\,0.06]$ \\
\cQj38 & $-0.01$ & $[-0.12,\,0.10]$ & $[-0.20,\,0.19]$ & $-0.05$ & $[-0.15,\,0.14]$ & $[-0.21,\,0.20]$ \\
\cQu1 & $0.01$ & $[-0.36,\,0.38]$ & $[-0.63,\,0.64]$ & $0.00$ & $[-0.45,\,0.44]$ & $[-0.67,\,0.66]$ \\
\cQu8 & $-0.20$ & $[-1.1,\,0.54]$ & $[-1.6,\,1.1]$ & $0.19$ & $[-0.92,\,0.76]$ & $[-1.6,\,1.2]$ \\
\cQd1 & $-0.02$ & $[-0.72,\,0.71]$ & $[-1.1,\,1.1]$ & $-0.46$ & $[-0.88,\,0.86]$ & $[-1.2,\,1.2]$ \\
\cQd8 & $0.02$ & $[-1.8,\,1.4]$ & $[-2.6,\,2.2]$ & $0.86$ & $[-1.5,\,1.7]$ & $[-2.6,\,2.3]$ \\
\ctu1 & $-0.07$ & $[-0.57,\,0.43]$ & $[-0.90,\,0.77]$ & $-0.22$ & $[-0.66,\,0.63]$ & $[-0.92,\,0.90]$ \\
\ctu8 & $0.05$ & $[-1.2,\,1.0]$ & $[-2.0,\,1.7]$ & $0.59$ & $[-0.74,\,1.3]$ & $[-1.9,\,1.8]$ \\
\ctd1 & $0.00$ & $[-0.53,\,0.53]$ & $[-0.88,\,0.87]$ & $-0.18$ & $[-0.66,\,0.66]$ & $[-0.94,\,0.95]$ \\
\ctd8 & $-0.37$ & $[-1.6,\,0.94]$ & $[-2.5,\,1.8]$ & $0.31$ & $[-1.6,\,1.3]$ & $[-2.5,\,2.0]$ \\
\cQt1 & $0.02$ & $[-2.0,\,2.0]$ & $[-3.2,\,3.1]$ & $1.9$ & $[0.88,\,2.7]$ & $[-3.3,\,3.5]$ \\
\cQt8 & $-0.32$ & $[-4.1,\,4.1]$ & $[-6.2,\,6.5]$ & $-3.8$ & $[-5.4,\,-1.8]$ & $[-7.0,\,6.8]$ \\
\cQQ+ & $-0.42$ & $[-0.74,\,0.62]$ & $[-1.0,\,0.97]$ & $-0.60$ & $[-0.83,\,-0.34]$ & $[-1.1,\,0.95]$ \\
\bottomrule
\end{tabular}
}
}
\caption{Best-fit values and 68\% and 95\% confidence intervals for all EFT coefficients in the OB fit, comparing multi-operator and single-operator fit scenarios considering $\mathcal{O}(\Lambda^{-4})$ effects. Disconnected confidence intervals are separated by $\cup$.}
\label{tab:EFTSummary_OB}
\end{table}

\begin{table}[htbp]
\centering
\scriptsize
\resizebox{\textwidth}{!}{
{\renewcommand{\arraystretch}{1.5}
\begin{tabular}{lcccccc}
\toprule
& \multicolumn{6}{c}{$\mathcal{O}(\Lambda^{-4})$ $[\TeV^{-2}]$}\\
Coefficient & Multi best fit & Multi 68\% CI & Multi 95\% CI & Single best fit & Single 68\% CI & Single 95\% CI \\
\midrule
\cHQ- & $-1.3$ & $[-2.3,\,-0.10]$ & $[-3.2,\,1.4]$ & $-0.94$ & $[-1.9,\,-0.01]$ & $[-2.7,\,0.92]$ \\
\cHt & $-3.2$ & $[-6.1,\,-0.41]$ & $[-9.3,\,2.3]$ & $1.1$ & $[-1.2,\,3.2]$ & $[-17,\,-14] \cup [-3.5,\,4.7]$ \\
\ctWRe & $0.04$ & $[-0.23,\,0.23]$ & $[-0.34,\,0.34]$ & $0.20$ & $[-0.28,\,0.30]$ & $[-0.38,\,0.39]$ \\
\ctWIm & $0.04$ & $[-0.19,\,0.20]$ & $[-0.29,\,0.29]$ & $-0.15$ & $[-0.24,\,0.23]$ & $[-0.33,\,0.32]$ \\
\ctBRe & $0.04$ & $[-0.18,\,0.28]$ & $[-0.38,\,0.48]$ & $0.02$ & $[-0.16,\,0.20]$ & $[-0.34,\,0.38]$ \\
\ctBIm & $-0.03$ & $[-0.25,\,0.20]$ & $[-0.45,\,0.40]$ & $-0.01$ & $[-0.20,\,0.17]$ & $[-0.37,\,0.35]$ \\
\ctGRe & $-0.14$ & $[-0.33,\,0.07]$ & $[-0.50,\,0.26]$ & $-0.18$ & $[-0.31,\,-0.04]$ & $[-0.42,\,0.12]$ \\
\ctGIm & $0.04$ & $[-0.19,\,0.25]$ & $[-0.39,\,0.44]$ & $0.05$ & $[-0.31,\,0.34]$ & $[-0.48,\,0.50]$ \\
\ctHRe & $8.4$ & $[2.8,\,16]$ & $[-1.8,\,\text{No upper limit}]$ & $29$ & $[25,\,32]$ & $[-0.24,\,3.0] \cup [21,\,34]$ \\
\ctHIm & $25$ & $[19,\,30]$ & $[-28,\,-9.8] \cup [9.0,\,34]$ & $-15$ & $[-21,\,-6.9] \cup [7.3,\,20]$ & $[-25,\,25]$ \\
\ctl & $0.28$ & $[-1.4,\,1.6]$ & $[-2.1,\,2.3]$ & $-0.96$ & $[-1.6,\,1.7]$ & $[-2.2,\,2.3]$ \\
\cQe & $0.19$ & $[-1.5,\,1.6]$ & $[-2.3,\,2.4]$ & $1.1$ & $[-1.7,\,1.8]$ & $[-2.3,\,2.4]$ \\
\cte & $0.35$ & $[-1.3,\,1.7]$ & $[-2.0,\,2.4]$ & $-0.89$ & $[-1.5,\,1.8]$ & $[-2.0,\,2.4]$ \\
\cQl- & $1.1$ & $[-1.3,\,1.9]$ & $[-2.0,\,2.5]$ & $-0.88$ & $[-1.5,\,1.9]$ & $[-2.0,\,2.5]$ \\
\cleQtRe1 & $-0.04$ & $[-2.1,\,2.1]$ & $[-3.0,\,3.0]$ & $1.4$ & $[-2.2,\,2.2]$ & $[-3.1,\,3.1]$ \\
\cleQtRe3 & $0.00$ & $[-0.31,\,0.31]$ & $[-0.45,\,0.45]$ & $0.22$ & $[-0.34,\,0.34]$ & $[-0.46,\,0.46]$ \\
\ctj1 & $0.00$ & $[-0.24,\,0.23]$ & $[-0.40,\,0.39]$ & $0.30$ & $[-0.36,\,-0.26] \cup [0.21,\,0.38]$ & $[-0.45,\,-0.10] \cup [0.07,\,0.45]$ \\
\ctj8 & $-0.16$ & $[-0.54,\,0.27]$ & $[-0.85,\,0.64]$ & $0.44$ & $[0.28,\,0.60]$ & $[0.09,\,0.74]$ \\
\cQj11 & $0.03$ & $[-0.28,\,0.34]$ & $[-0.45,\,0.47]$ & $0.32$ & $[-0.39,\,-0.22] \cup [0.23,\,0.40]$ & $[-0.46,\,-0.07] \cup [0.08,\,0.47]$ \\
\cQj18 & $0.54$ & $[-0.07,\,0.80]$ & $[-0.86,\,0.99]$ & $0.54$ & $[0.35,\,0.71]$ & $[0.13,\,0.87]$ \\
\cQj31 & $-0.04$ & $[-0.07,\,0.04]$ & $[-0.09,\,0.08]$ & $-0.06$ & $[-0.08,\,-0.05]$ & $[-0.09,\,-0.02] \cup [0.04,\,0.07]$ \\
\cQj38 & $0.03$ & $[-0.16,\,0.17]$ & $[-0.23,\,0.23]$ & $0.17$ & $[-0.20,\,-0.12] \cup [0.12,\,0.21]$ & $[-0.24,\,-0.05] \cup [0.04,\,0.25]$ \\
\cQu1 & $-0.01$ & $[-0.49,\,0.48]$ & $[-0.72,\,0.71]$ & $-0.44$ & $[-0.64,\,0.62]$ & $[-0.80,\,0.79]$ \\
\cQu8 & $-0.10$ & $[-1.3,\,0.65]$ & $[-1.9,\,1.1]$ & $0.55$ & $[-1.7,\,-0.73] \cup [-0.22,\,0.97]$ & $[-2.1,\,1.3]$ \\
\cQd1 & $-0.05$ & $[-0.74,\,0.71]$ & $[-1.1,\,1.1]$ & $-0.72$ & $[-1.1,\,1.0]$ & $[-1.3,\,1.3]$ \\
\cQd8 & $0.01$ & $[-1.6,\,1.4]$ & $[-2.6,\,2.2]$ & $1.2$ & $[-2.6,\,1.9]$ & $[-3.1,\,2.5]$ \\
\ctu1 & $-0.02$ & $[-0.36,\,0.32]$ & $[-0.65,\,0.61]$ & $-0.02$ & $[-0.55,\,0.53]$ & $[-0.82,\,0.80]$ \\
\ctu8 & $-0.08$ & $[-0.83,\,0.64]$ & $[-1.5,\,1.3]$ & $0.35$ & $[-0.68,\,1.0]$ & $[-1.7,\,1.5]$ \\
\ctd1 & $0.01$ & $[-0.44,\,0.46]$ & $[-0.80,\,0.81]$ & $-0.28$ & $[-0.72,\,0.72]$ & $[-0.99,\,1.0]$ \\
\ctd8 & $-0.39$ & $[-1.3,\,0.56]$ & $[-2.0,\,1.4]$ & $-0.47$ & $[-1.7,\,0.95]$ & $[-2.4,\,1.7]$ \\
\cQt1 & $-0.08$ & $[-1.2,\,1.0]$ & $[-2.1,\,2.0]$ & $1.2$ & $[0.22,\,1.9]$ & $[-2.2,\,2.6]$ \\
\cQt8 & $-0.27$ & $[-2.5,\,2.1]$ & $[-4.3,\,4.0]$ & $-2.3$ & $[-3.8,\,-0.30]$ & $[-5.3,\,4.8]$ \\
\cQQ+ & $-0.14$ & $[-0.43,\,0.26]$ & $[-0.67,\,0.56]$ & $-0.38$ & $[-0.59,\,-0.14]$ & $[-0.81,\,0.58]$ \\
\bottomrule
\end{tabular}
}
}
\caption{Best-fit values and 68\% and 95\% confidence intervals for all EFT coefficients in the PB fit, comparing multi-operator and single-operator fit scenarios considering $\mathcal{O}(\Lambda^{-4})$ effects. Disconnected confidence intervals are separated by $\cup$.}
\label{tab:EFTSummary_PB}
\end{table}

% End of text imported from the .//Sections/EFTLimitsTable_PDG_rounded.tex input file

Figure~\ref{fig:EFTsummaryPlot} provides a general overview for the OB
and PB multi-operator \EFT fits, the pulls, uncertainties, and probed
new physics scale for each operator (assuming $c=1$). The expected sensitivity
contributions from the PB fit are also shown. These are calculated in
the same way as in Ref.~\cite{STDM-2023-32} where the relative contribution from a process is determined from the change in size of the uncertainty on a given WC when removing the EFT dependence on that process. The 95\% CIs on the WCs are also shown. In the case that only one interval is present
the symmetrised value, half of the upper minus lower WC value at the
95\% $-\Delta$NLL intersection, is displayed. In the case that there are
two intervals, the one that contains the best fit value is chosen to be displayed.

Figure~\ref{fig:EFTRadialPlot} shows an
alternative representation of the 95\% CIs on the WCs
for the OB and PB approaches for both multi-operator and
single-operator results.

\begin{figure}[ht]
\centering
\includegraphics[width=0.9\textwidth]{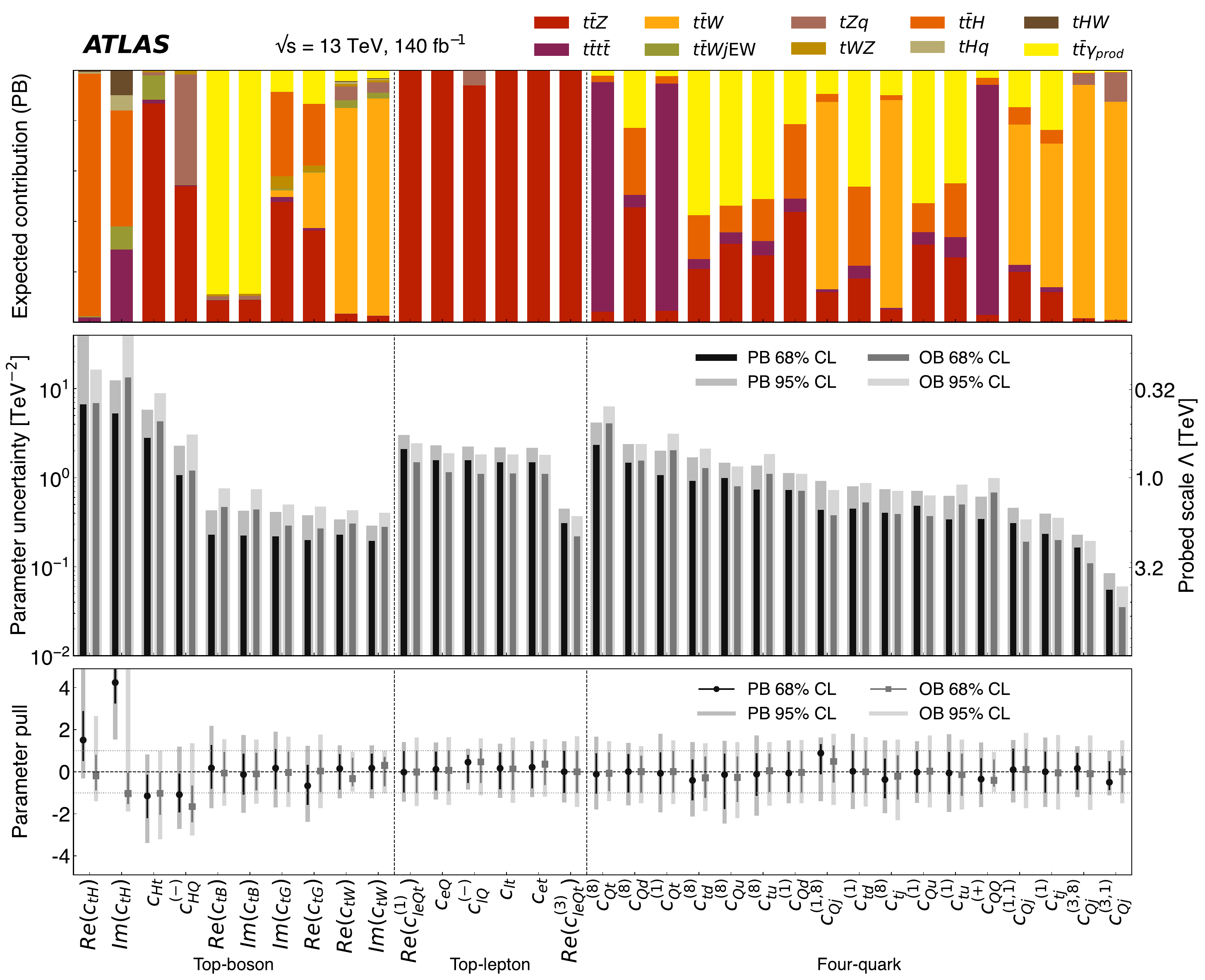}
\caption{Summary plots showing the pulls (bottom panel) and, uncertainties and probed new physics scale (middle panel) for each operator in the multi-operator PB and OB \EFT fits to data. The top-panel shows the sensitivity contributions from the various \topx processes for each operator from the PB multi-operator fits. The new physics scale assumes Wilson coefficient values of 1.}
\label{fig:EFTsummaryPlot}
\end{figure}

\begin{figure}[ht]
\centering
\subfloat[]{\includegraphics[width=0.49\textwidth]{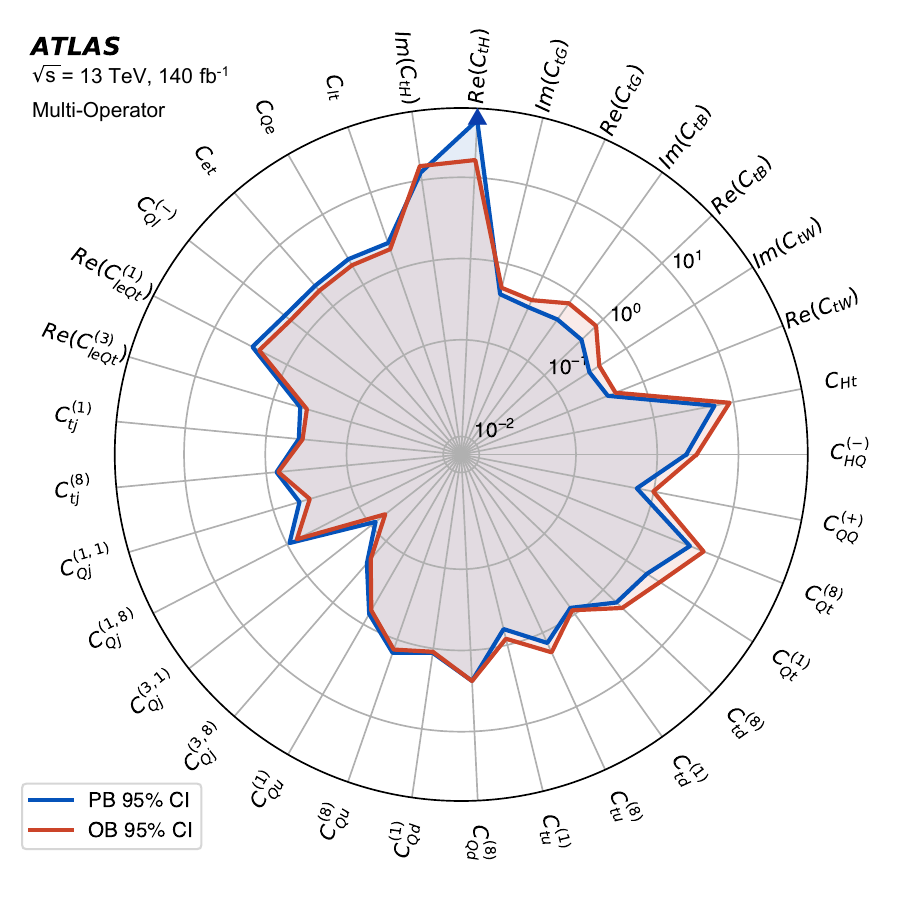}}
\subfloat[]{\includegraphics[width=0.49\textwidth]{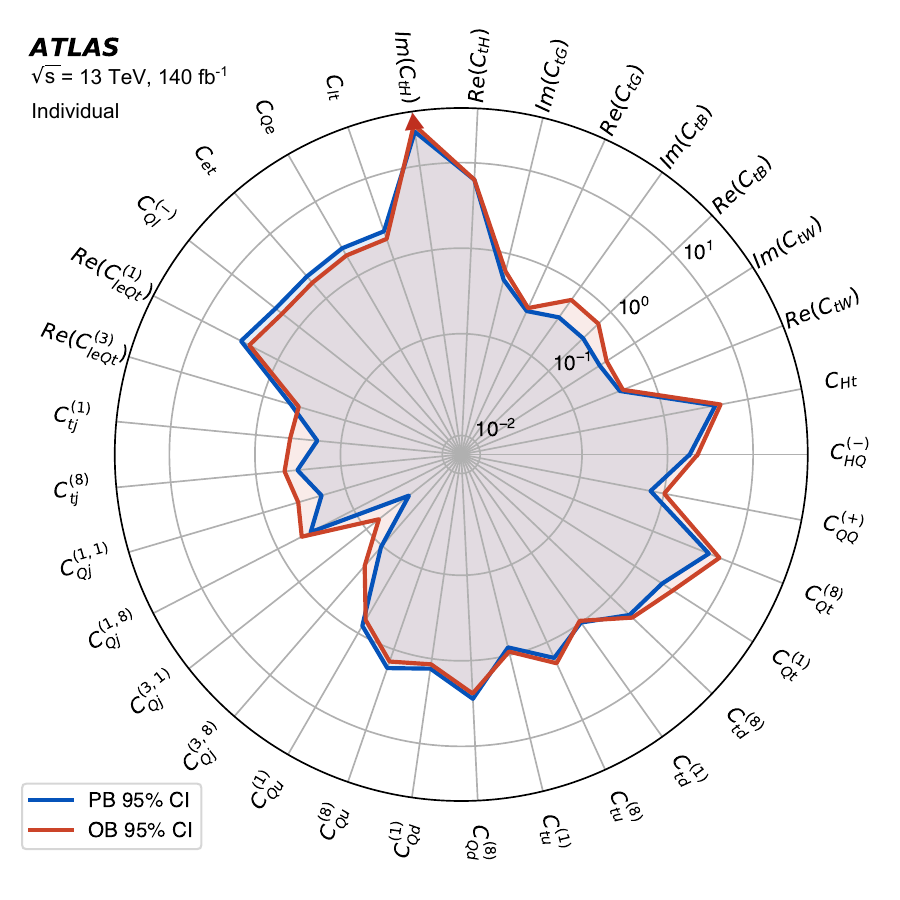}}
\caption{Radial representation of the 95\% CIs (given in $\TeV^{-2}$) on
the WCs for the OB and PB approaches for (a) multi-operator and
(b) single-operator results. A smaller area corresponds to a
smaller allowed WC value range and therefore a larger constraint. Values with a triangle symbol represent cases where the 95\%  confidence interval is unbounded on one side.}
\label{fig:EFTRadialPlot}
\end{figure}

\FloatBarrier

\subsection{Investigation of 2D correlations}

It is challenging to provide a meaningful visualisation of the complex
interplay between WCs in the 33-dimensional multi-operator
fit. However, it is possible to study correlations between pairs of
operators to get a better idea of their interplay and blind directions.
Two-dimensional confidence regions are shown in
Figure~\ref{fig:EFT2D_OBPB} illustrating the correlations between
selected pairs of WCs that highlight the relative
sensitivities of the OB and PB approaches. In general, the PB approach
yields smaller confidence regions, reflecting the enhanced
discrimination provided by the process-specific event
categorisation.
Figure~\ref{fig:EFT2D_OBPB}(a) shows the confidence regions in the
\cHQ-{}--\cHt plane. A complex correlation pattern is observed between
the two operators which is similar for the OB and PB approaches. The PB contours are shifted with respect to the
OB contours indicating slightly different preferred values.
The \ctHRe{}--\ctHIm plane, shown in Figure~\ref{fig:EFT2D_OBPB}(b), has
a shape reflecting the quadratic dependence of the relevant production
cross-sections on these operators. The PB fit with
its better separation between \ttH, \tHq and \tttt against other processes has tighter
constraints.
The 2D contour implies some sensitivity to the sign of \ctHIm, especially in the PB approach. The origin of this could be due to subtle kinematic effects in \ttH or an artefact of the EFT parameterisation extrapolation. In either case, the sensitivity is, as expected, very limited; when considering the linear-only contribution alone the fit is not able to constrain \ctHIm.
Figure~\ref{fig:EFT2D_OBPB}(c) presents the \ctBRe{}--\ctBIm confidence
regions. The contours are approximately elliptical and centred on the
SM point, indicating only weak correlations between the real and
imaginary components. The PB approach provides slightly tighter
constraints than the OB analysis, while both remain fully compatible
with the SM expectation.
The confidence regions in the \ctWIm{}--\ctBIm plane, shown in
Figure~\ref{fig:EFT2D_OBPB}(d), exhibit a mild anti-correlation
between the two operators. The PB contours are noticeably smaller than
those obtained with the OB analysis, demonstrating the improved
sensitivity obtained from the enhanced separation of \ttyprod
and \ttW from other processes.
For the four-quark operators, shown in Figure~\ref{fig:EFT2D_OBPB}(e),
the confidence regions show little correlation between \cQt1 and \cQt8, with the PB
approach providing substantially tighter constraints than the OB
analysis due to the better separation of \tttt.
Figure~\ref{fig:EFT2D_OBPB}(f) shows the confidence regions in the
\cQt1{}--\ctHRe plane. This distribution exhibits one of the largest
differences between the two analysis strategies, consistent with the better separation of \ttH and \tttt processes in the PB approach.

\begin{figure}[t]
\centering
\subfloat[]{\includegraphics[width=0.49\textwidth]{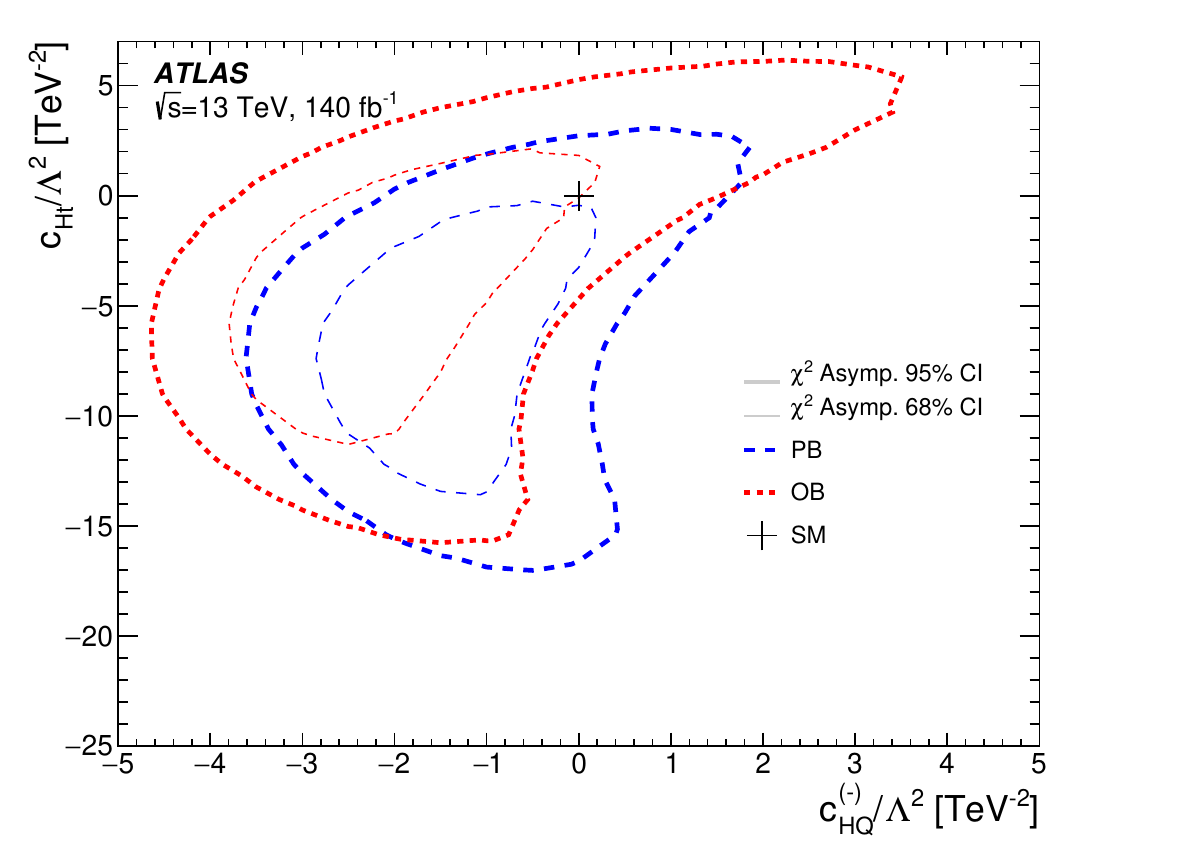}}
\subfloat[]{\includegraphics[width=0.49\textwidth]{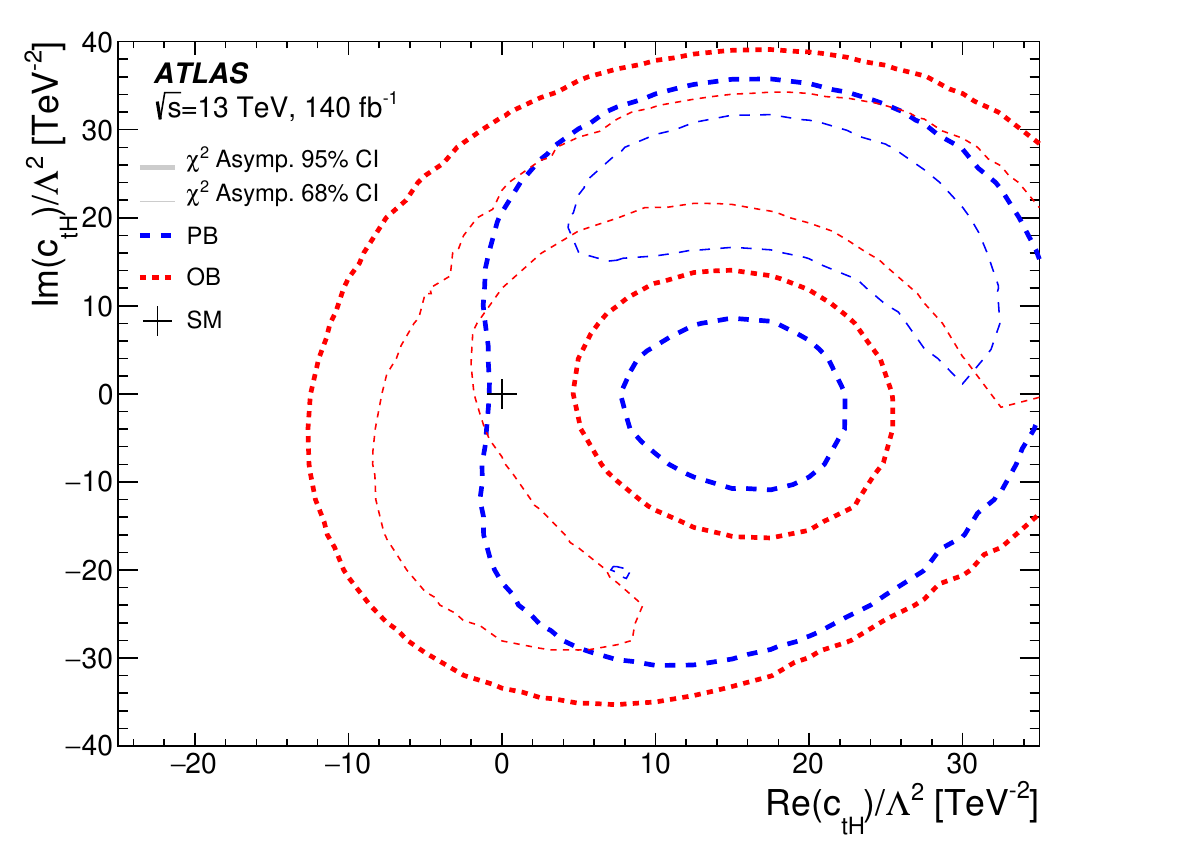}}\\
\subfloat[]{\includegraphics[width=0.49\textwidth]{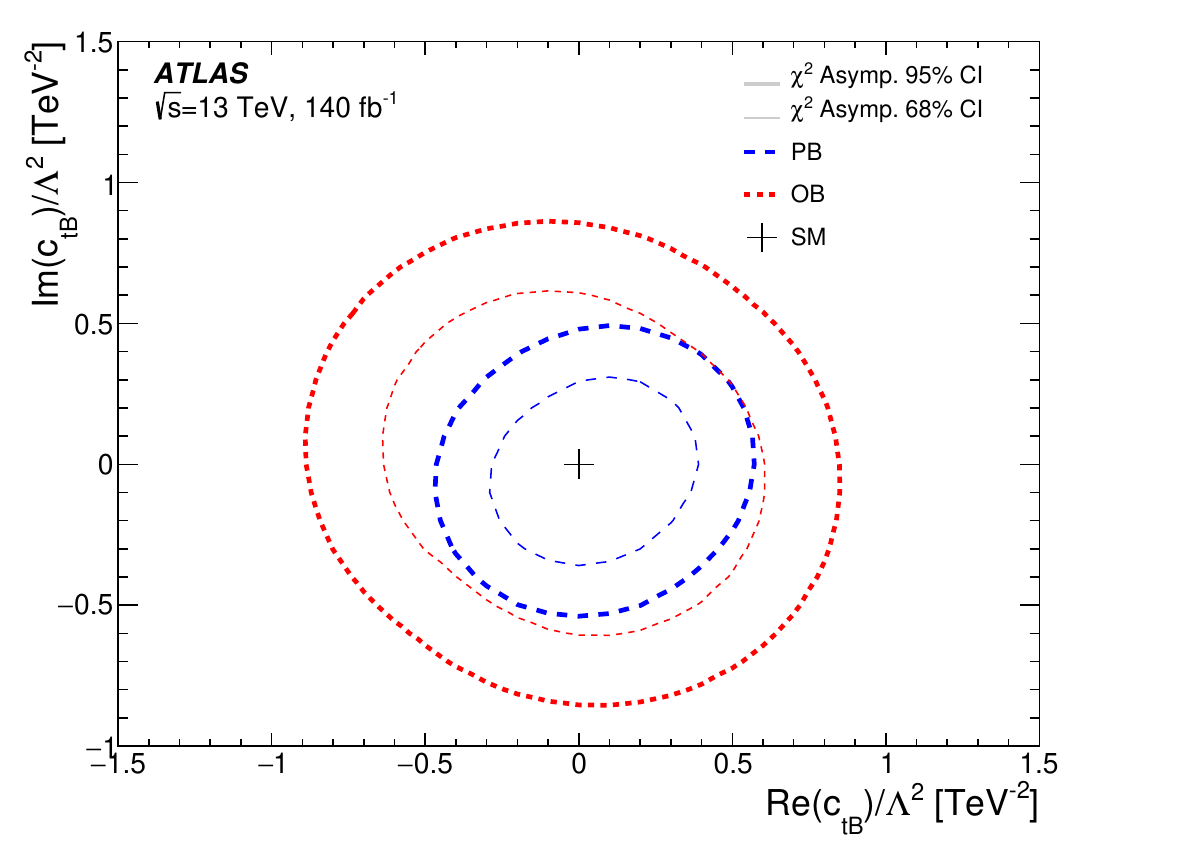}}
\subfloat[]{\includegraphics[width=0.49\textwidth]{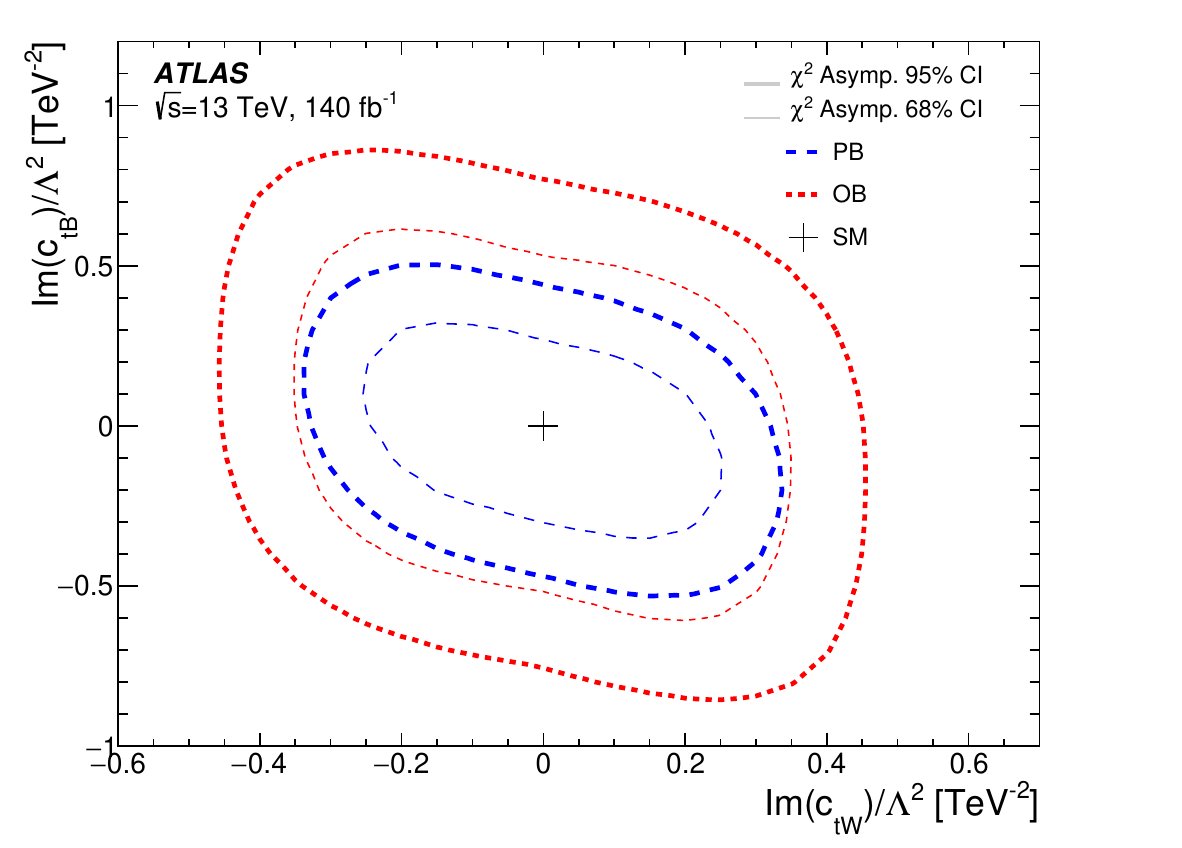}}\\
\subfloat[]{\includegraphics[width=0.49\textwidth]{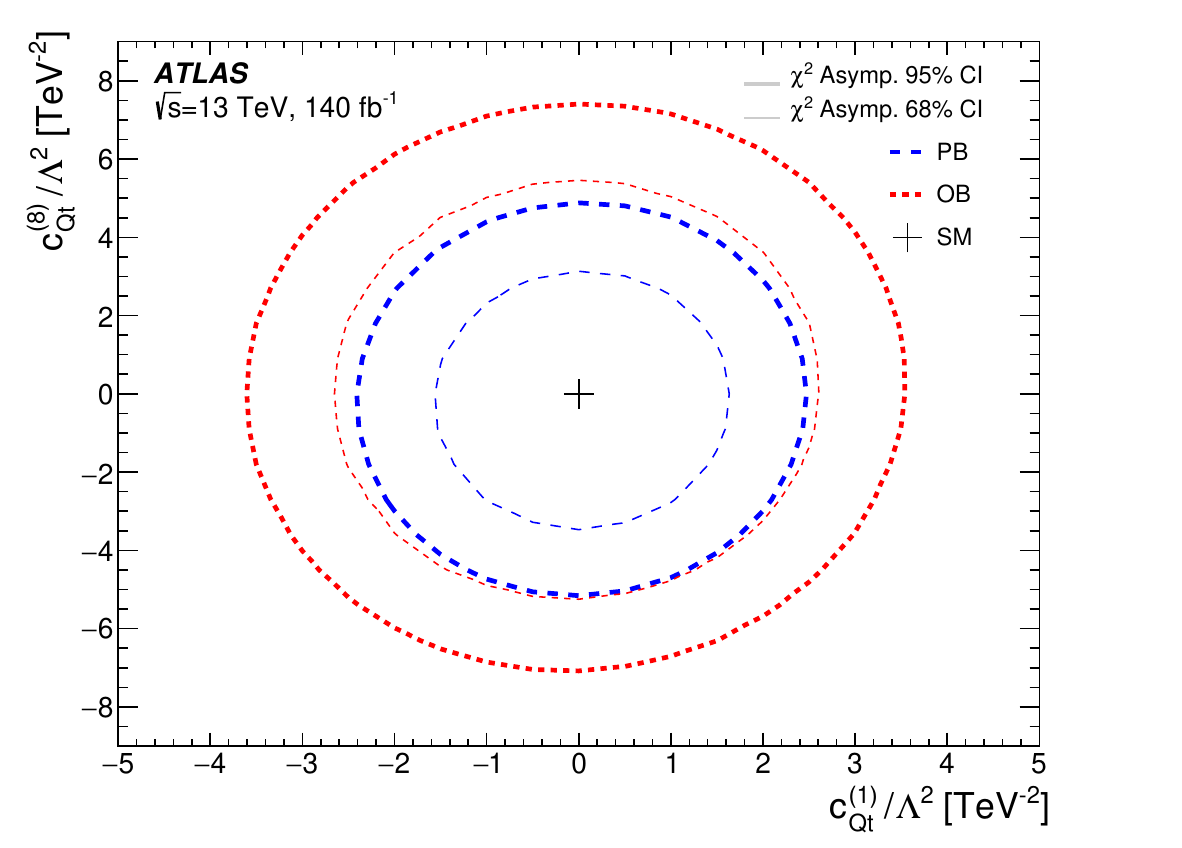}}
\subfloat[]{\includegraphics[width=0.49\textwidth]{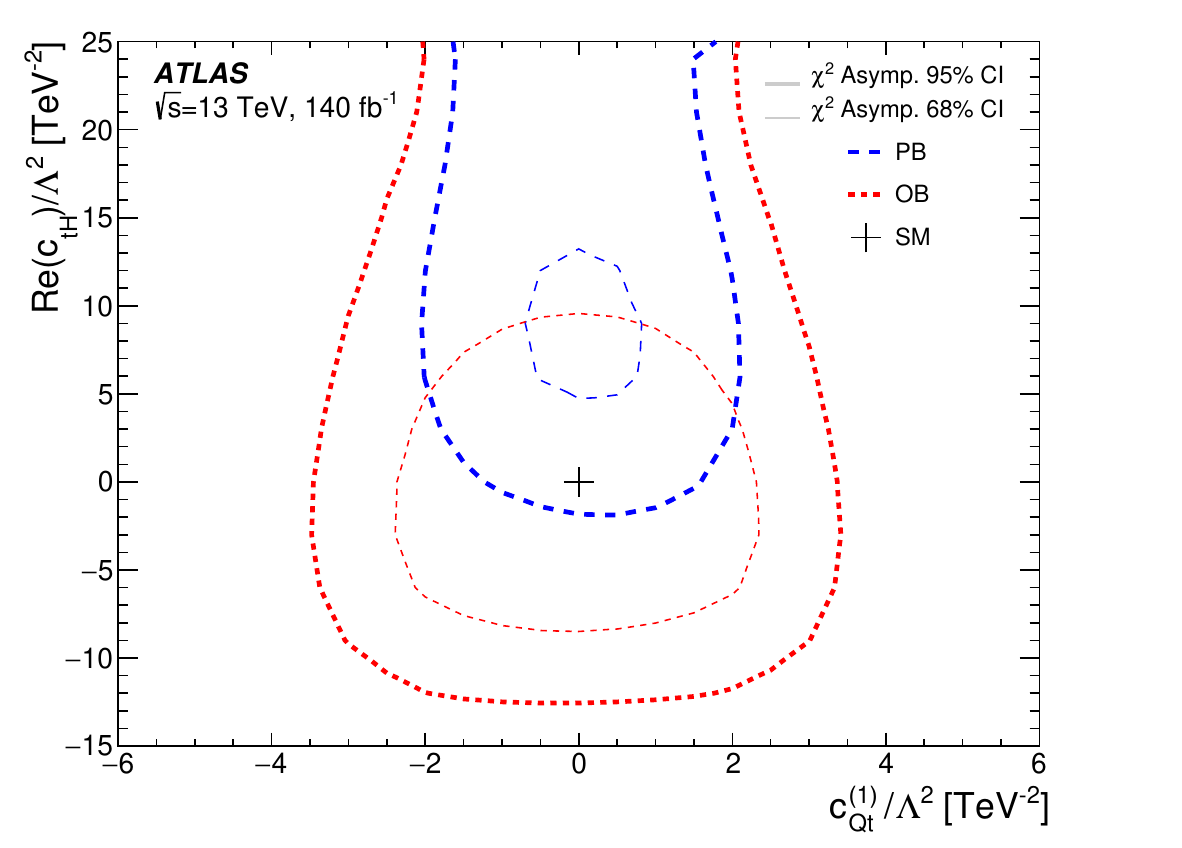}}
\caption{Two-dimensional likelihood contours of the 68\% (thin line) and 95\% (thick line) CIs are shown. The comparison of OB (dotted) and PB (dashed) results is included for the following pairs of operators: (a) \cHQ-{}--\cHt, (b) \ctHRe{}--\ctHIm, (c) \ctBRe--\ctBIm, (d) \ctWIm--\ctBIm, (e) \cQt1--\cQt8 and (f) \cQt1--\ctHRe.}
\label{fig:EFT2D_OBPB}
\end{figure}

It is also possible to use the two-dimensional likelihood scans to
better understand the contributions from individual processes to the
results. This is done in the context of the PB approach as shown in
Figure~\ref{fig:EFT2D_PBByPRoc}.
The contributions of different processes are demonstrated by observing
the effect of removing the operator dependence for a given process and rerunning the fits
to see how the CIs are affected.
Figure~\ref{fig:EFT2D_PBByPRoc}(a) shows the \cHQ-{}--\cHt plane for the
nominal PB fit as well as for two additional scenarios where dimension-six operator
effects are removed first on \tZq and second on \ttWEW. In both cases
it can be seen that there is a significant negative impact on the
sensitivity. This is understandable due to the fact that although \ttZ
has a higher cross-section than these other two processes, it has
little power to separate between the two EFT operators. However, \tZq
and \ttWEW are sensitive to interactions involving left-handed top
quarks which are important for \cHQ- whereas \ttZ is not.
Figure~\ref{fig:EFT2D_PBByPRoc}(b) shows the \ctHRe{}--\ctHIm plane, in
this case for the nominal PB fit as well as fits with the dimension-six operator effect
removed for \tHq and \tttt. The results show the importance of these
processes to the overall sensitivity to both \ctHRe and \ctHIm. As
previously described, there is a complex interplay between the
contributions of each process to both operators.  Only the inclusion
of \ttH, \tHq and \tttt all together gives sufficient sensitivity to
observe the tension with the SM at the 95\% confidence level for
\ctHIm. The preferred positive values are again consistent with the
simultaneous description of the low fitted $\ttH$ normalisation
together with the enhanced $\tttt$ and $tHq$ normalisations.
Figure~\ref{fig:EFT2D_PBByPRoc}(c) illustrates the power of including
\tty in such a multi-process EFT interpretation. As already observed
in \atty, this process is able to add significant additional
sensitivity to the \ctBRe and \ctBIm WCs compared to \ttZ
alone. Figure~\ref{fig:EFT2D_PBByPRoc}(d) demonstrates the importance
of including \ttW to allow separation between \ctWRe and \ctBRe which
have a significant degeneracy when considering only \ttZ.

\begin{figure}[t]
\centering
\subfloat[]{\includegraphics[width=0.49\textwidth]{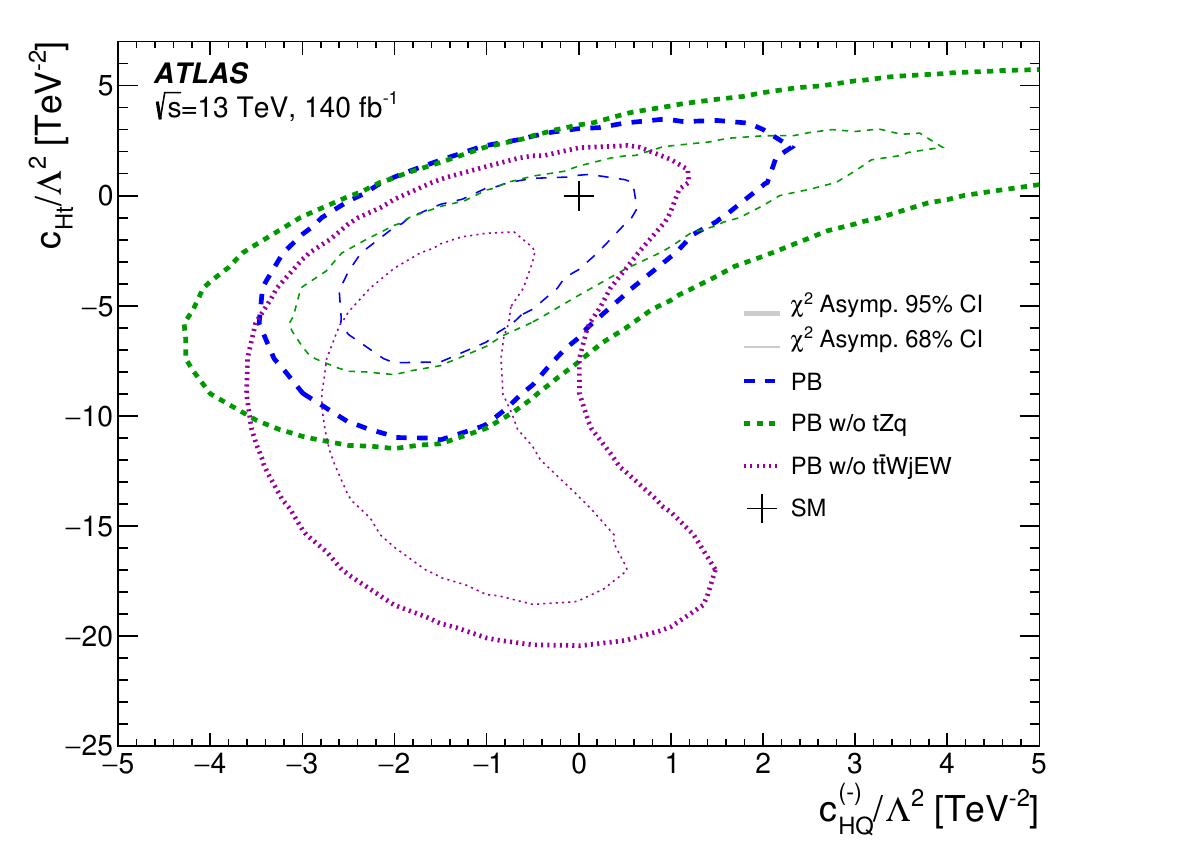}}
\subfloat[]{\includegraphics[width=0.49\textwidth]{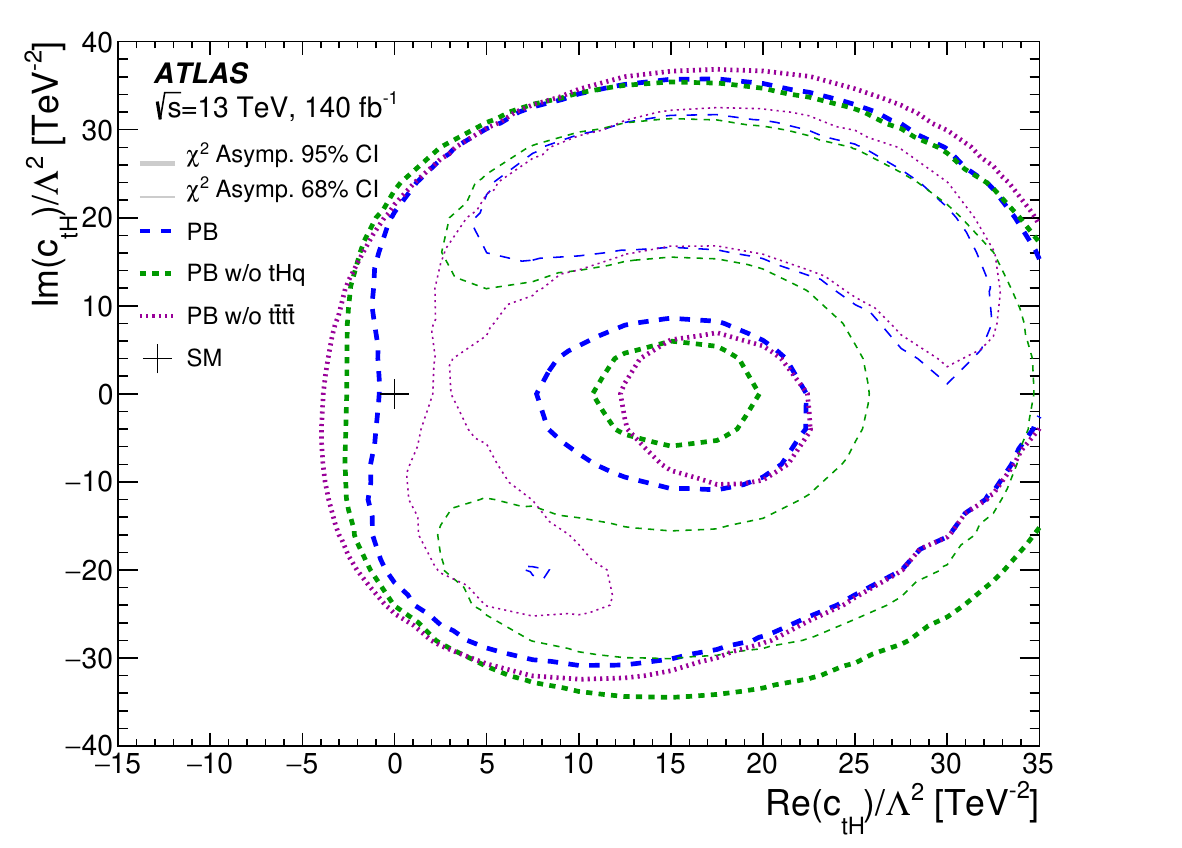}}\\
\subfloat[]{\includegraphics[width=0.49\textwidth]{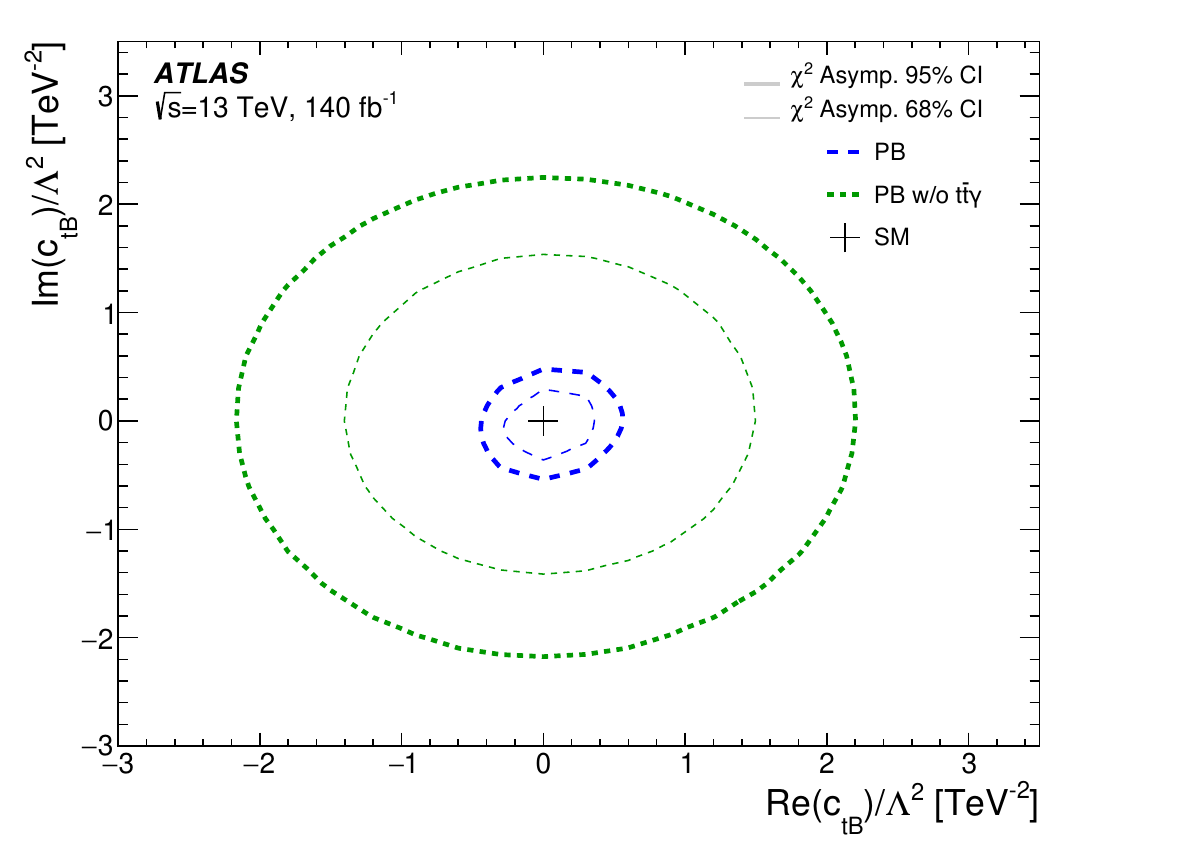}}
\subfloat[]{\includegraphics[width=0.49\textwidth]{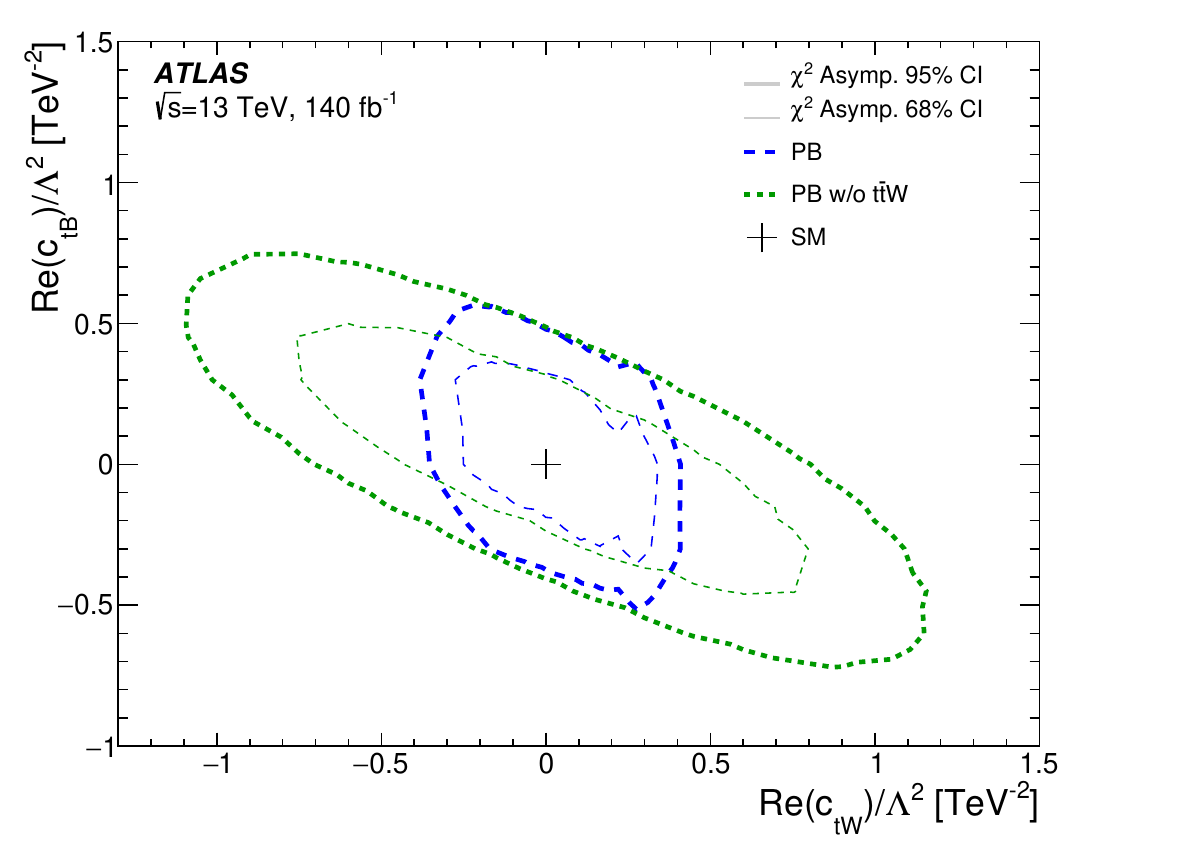}}
\caption{Two-dimensional likelihood contours of the 68\% (thin line) and 95\% (thick line) CIs are shown. The comparison of the full PB result (long dash) to results with the EFT operator dependence removed for different processes (short dash/dotted) is included for the following pairs of operators: (a) \cHQ-{}--\cHt, (b) \ctHRe{}--\ctHIm, (c) \ctBRe--\ctBIm, (d) \ctWRe--\ctBRe.}
\label{fig:EFT2D_PBByPRoc}
\end{figure}

\subsection{Coverage properties of intervals based on Wilks' theorem}

The WC best-fit values and CIs can be extracted from
likelihood scans of the WC values. Examples of such scans are shown in
Figure~\ref{fig:Results:LLH} for \cHQ-, \ctHRe, \ctl and
\cQt1. These WCs were chosen as they demonstrate a range of
interaction types: top-boson operators,
top-lepton operators and four-quark operators; as well as dependencies on the dimension-six operator contributions: linear dominated,
quadratic dominated and mixed linear and quadratic contributions.

\begin{figure}[!t]
\centering
\subfloat[]{\includegraphics[width=0.48\linewidth]{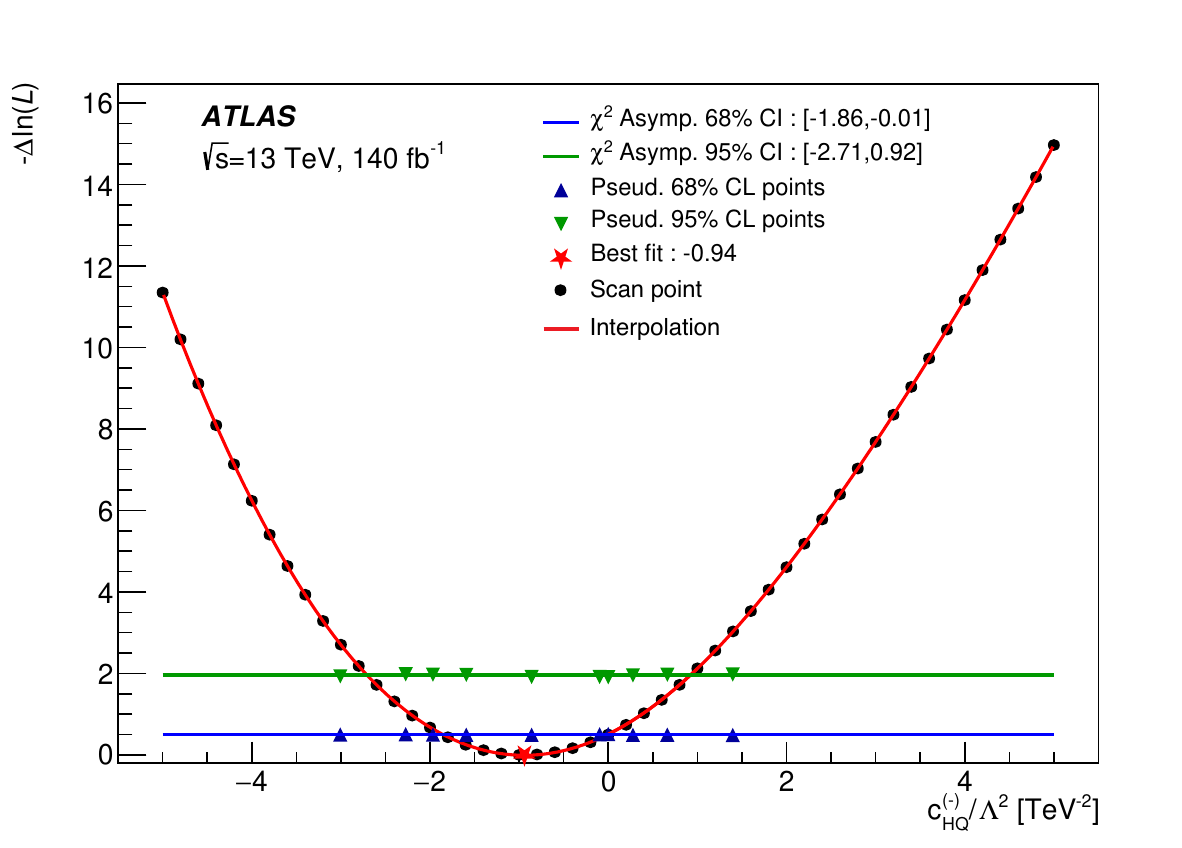}}
\subfloat[]{\includegraphics[width=0.48\linewidth]{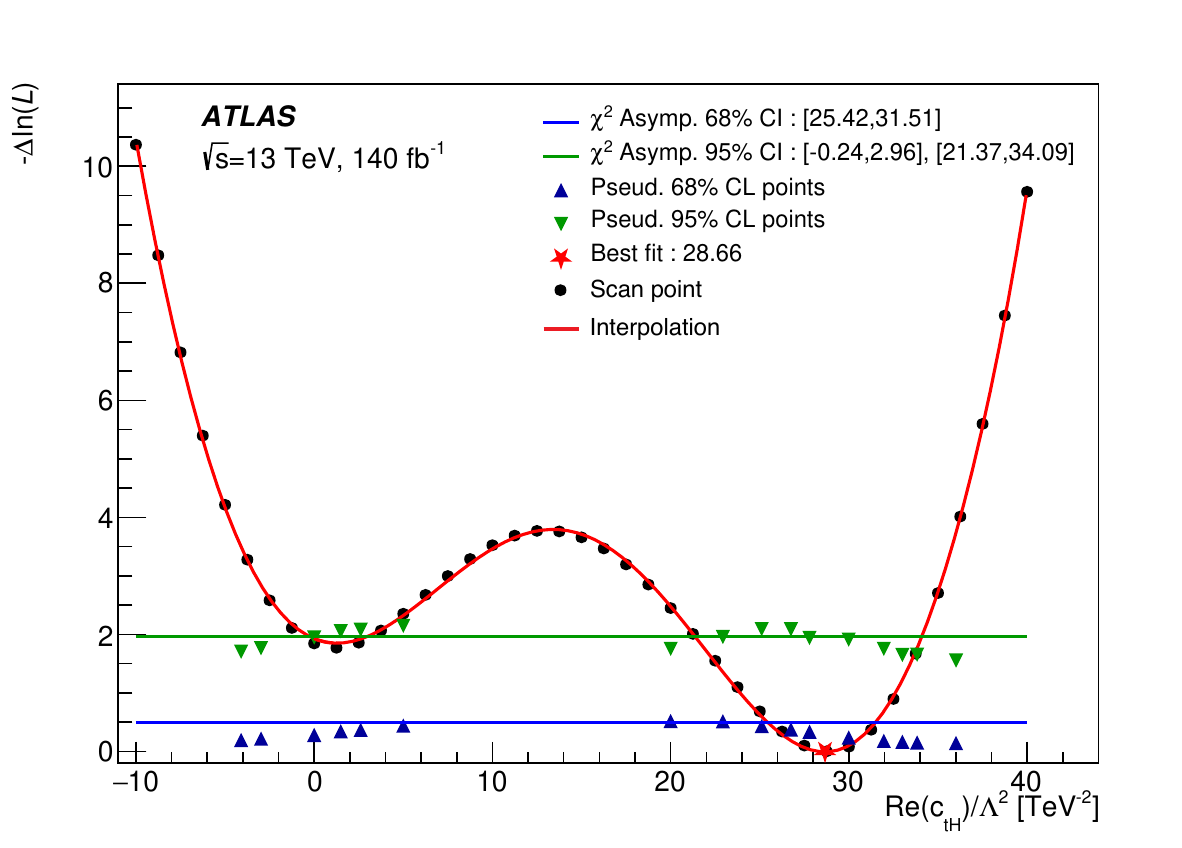}}\\
\subfloat[]{\includegraphics[width=0.48\linewidth]{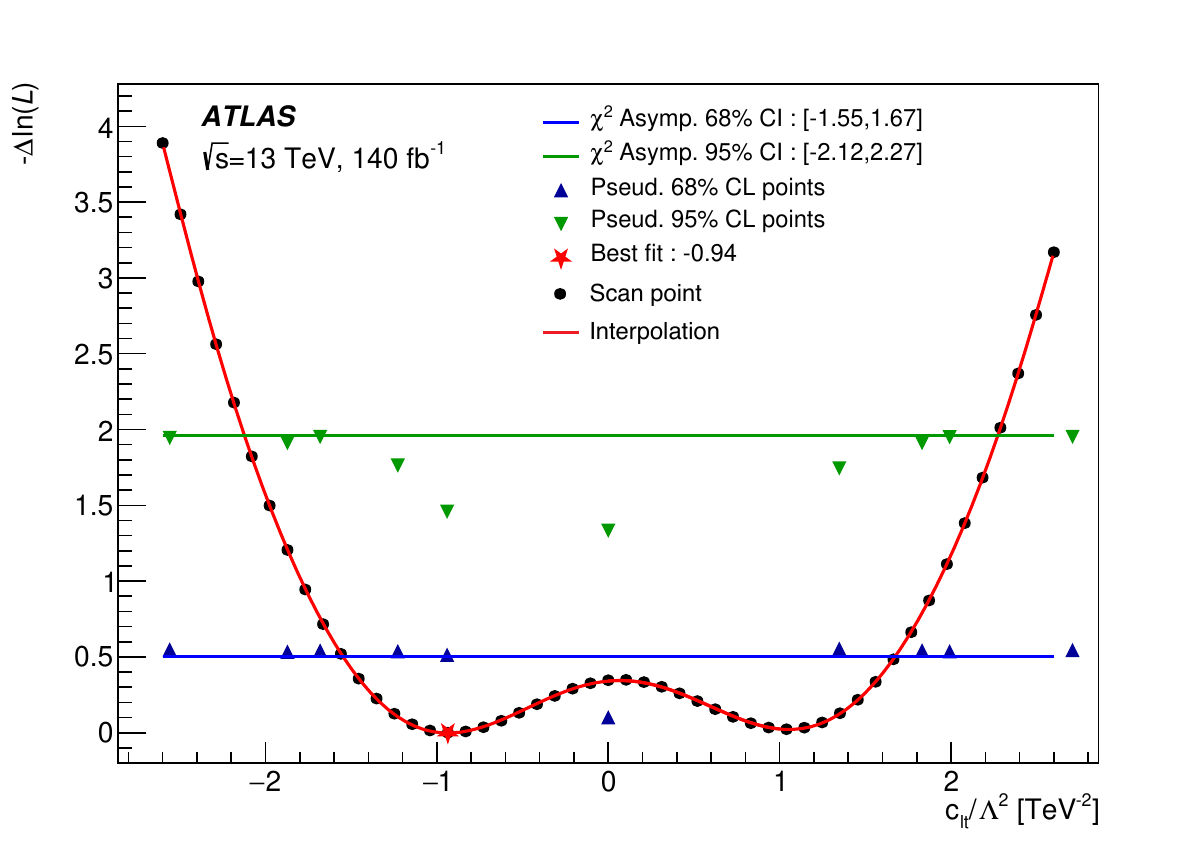}}
\subfloat[]{\includegraphics[width=0.48\linewidth]{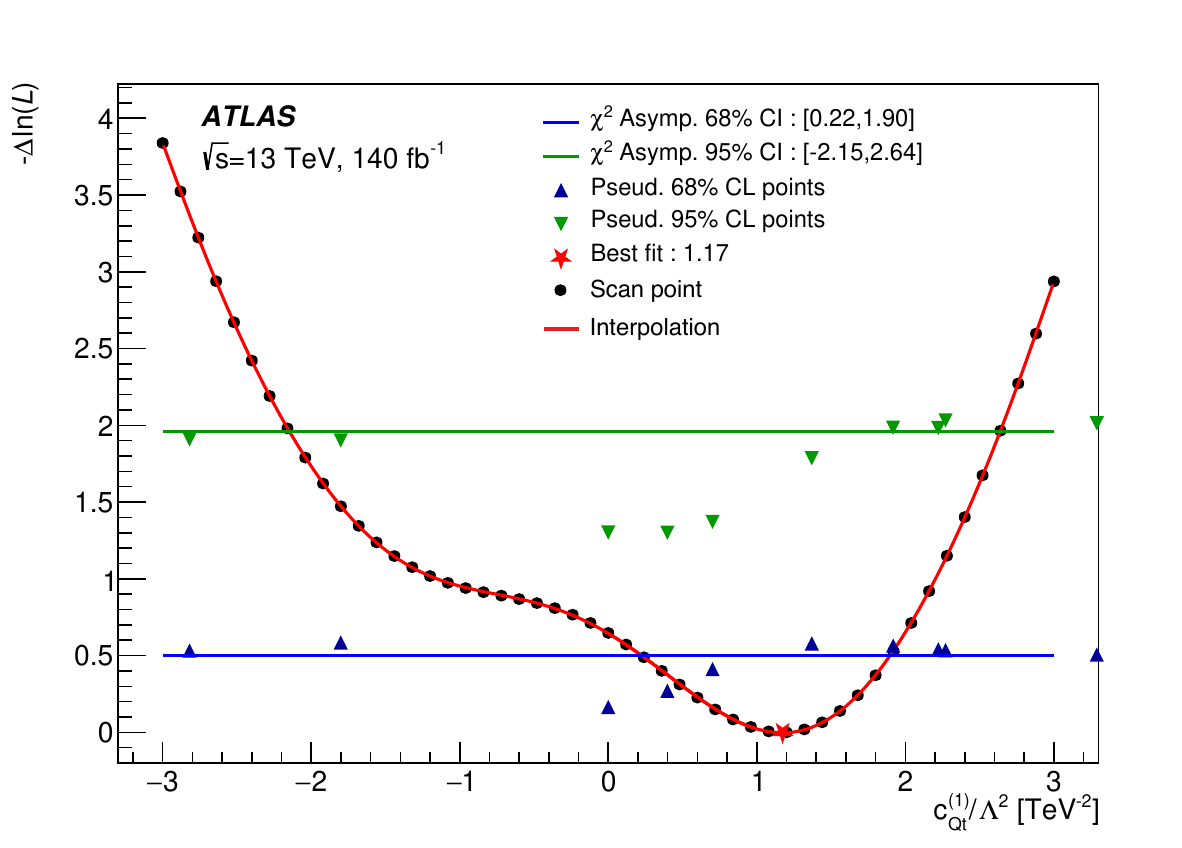}}
\caption{One-dimensional scans of the fitted negative log-likelihood value as a function of the WC value for (a) \cHQ-, (b) \ctHRe, (c) \ctl{} and (d) \cQt1. The upper (lower) horizontal lines represent the expected 95\% (68\%) $\Delta$NLL values for a $\chi^2$-distributed test statistic (Asymp.), while the downward (upward) pointing triangles show the true 95\% (68\%) coverage values as determined using pseudodata (Pseud.). The circular markers are the scan points and the line going through them is an interpolation created by a  polynomial fit to the scan points.}
\label{fig:Results:LLH}
\end{figure}

In the case of parameterisations where the quadratic dimension-six contribution
$\mathcal{O}(\Lambda^{-4})$ terms are dominant a breakdown of a key
axiom of Wilks' theorem~\cite{Wilks:1938dza} can occur, and the
PLR may not follow the expected
$\chi^2$-distribution~\cite{Bernlochner:2022oiw}. This can lead to
coverage in the test statistic different to what is expected in the
asymptotic case. The true coverage can be determined with toy
pseudodata, adjusting where to read off the CIs accordingly.  Scans of the fitted negative log-likelihood value
as a function of the WC value are shown in
Figure~\ref{fig:Results:LLH}.  The line interpolating between the scan points shows the negative
log-likelihood as a function of the WC, the upward (downward) pointing
triangles show the values of the 68\% (95\%) coverage calculated from
the fits to pseudodata. These can be compared to the coverage value in the
asymptotic assumption of a PLR,
$\Delta\textrm{NLL}=0.5$ and
$\Delta\textrm{NLL}=1.92$. Figure~\ref{fig:Results:LLH}(c) provides a
representative example, an upward fluctuation can lead to a best fit
away from WC $=0$, while a downward fluctuation cannot be reproduced in
the same way and the best fit can remain at WC $=0$, leading to large
$\Delta$NLL values being less likely than predicted by the asymptotic
assumption.  Due to the complexity of the 33-dimensional fit it is
unfeasible to run pseudodata for the multi-operator results, but a
general appreciation for the size of the coverage corrections can be
obtained from Figure~\ref{fig:Results:LLH}. A general observation is
that for the cases tested the asymptotic intervals are typically close
to correct and more often than not somewhat conservative.

\FloatBarrier

% End of text imported from the .//Sections/Results.tex input file

%

%

%
\section{Conclusion}
\label{sec:conclusion}

% The next lines are included from the .//Sections/Conclusion.tex input file

A measurement with an EFT interpretation of multiple associated top-quark
processes has been presented using the full Run 2 ATLAS dataset of
proton--proton collisions at $\sqrt{s}=13$ TeV, corresponding to an
integrated luminosity of 140 fb$^{-1}$. The analysis combines
information from the \tty, \ttZ, \ttWQCD, \ttWEW, \ttH, \tttt, \tZq, \tWZ, \tHq, and \tHW
processes within a single profile-likelihood fit, providing a coherent
treatment of processes that share common final states and are
simultaneously sensitive to many of the same EFT operators.
This measurement represents the first ATLAS simultaneous EFT
interpretation across such a broad set of associated \topx processes.
It also expands on previous similar work through the
addition of complex Wilson coefficients.
No significant deviation from the SM expectation is observed in
the simultaneous EFT interpretation.

A central objective of this study is the comparison of the two
complementary strategies for defining the signal regions, the
object-based and process-based approaches. The first employs an
object-based categorisation using detector-level object
multiplicities, while the second combines process-based regions
optimised in previous dedicated measurements.  While the object-based
approach provides a generic and process-agnostic framework suitable
for broad combinations, the process-based approach can exploit the
discriminating power developed in dedicated analyses to enhance the
separation between individual top+X processes and improve the
sensitivity to a number of EFT operators.  The comparison demonstrates
the relative strengths of the two methodologies and provides important
guidance for the design of future global EFT analyses as increasingly
precise measurements become available. The results show that for a
large number of operators the process-based approach is the stronger
of the two, while simultaneously exposing some of the complications
and subtleties with employing such an approach.

To directly compare approaches, the expected and observed results must
be considered. The expected results show the process-based approach to
be superior over the majority of Wilson coefficients.  The largest
gains, up to a 40\% reduction in the 95\% confidence interval, are
seen for the four-heavy-quark operators and for those top-boson
operators whose effect is primarily a normalisation change
on \ttZ, \ttH and \tZq, which reflects the improved separation of
those processes in the process-based approach. For the top-lepton
operators the object-based approach is stronger by 15-20\%. After the
fits to data, the strength of the process-based approach is diluted to
the extent that it is better for 16 of the 33 WCs.  The change in
relative sensitivity is primarily driven by the agreement between data
and the predictions being different in the process-based and
object-based approaches.  The process-based approach sees more
tensions and therefore a mild preference for non-zero WC values
resulting in worse 95\% confidence level intervals.

The methodology developed here provides a template for future
global combinations using the significantly larger Run 3 and
High-Luminosity LHC datasets, where simultaneous interpretations
across an even wider range of top-quark processes will play an
increasingly important role in probing the electroweak interactions of
the top quark and searching for indirect evidence of physics beyond
the Standard Model.

% End of text imported from the .//Sections/Conclusion.tex input file

%

\FloatBarrier %

\section*{Acknowledgements}
%
%

% The next lines are included from the .//acknowledgements/Acknowledgements.tex input file
%
%

%
%

We thank CERN for the very successful operation of the LHC and its injectors, as well as the support staff at
CERN and at our institutions worldwide without whom ATLAS could not be operated efficiently.

The crucial computing support from all WLCG partners is acknowledged gratefully, in particular from CERN, the ATLAS Tier-1 facilities at TRIUMF/SFU (Canada), NDGF (Denmark, Norway, Sweden), CC-IN2P3 (France), KIT/GridKA (Germany), INFN-CNAF (Italy), NL-T1 (Netherlands), PIC (Spain), RAL (UK) and BNL (USA), the Tier-2 facilities worldwide and large non-WLCG resource providers. Major contributors of computing resources are listed in Ref.~\cite{ATL-SOFT-PUB-2026-001}.

We gratefully acknowledge the support of ANPCyT, Argentina; YerPhI, Armenia; ARC, Australia; BMWFW and FWF, Austria; ANAS, Azerbaijan; CNPq and FAPESP, Brazil; NSERC, NRC and CFI, Canada; CERN; ANID, Chile; CAS, MOST, NSFC and FRFCU, China; Minciencias, Colombia; MEYS CR, Czech Republic; DNRF and DNSRC, Denmark; IN2P3-CNRS and CEA-DRF/IRFU, France; SRNSFG, Georgia; BMFTR, HGF and MPG, Germany; GSRI, Greece; RGC and Hong Kong SAR, China; ICHEP and Academy of Sciences and Humanities, Israel; INFN, Italy; MEXT and JSPS, Japan; CNRST, Morocco; NWO, Netherlands; RCN, Norway; MNiSW, Poland; FCT, Portugal; MNE/IFA, Romania; MSTDI, Serbia; MSSR, Slovakia; ARIS and MVZI, Slovenia; DSI/NRF, South Africa; MICIU/AEI, Spain; SRC and Wallenberg Foundation, Sweden; SERI, SNSF and Cantons of Bern and Geneva, Switzerland; NSTC, Taipei; TENMAK, T\"urkiye; STFC/UKRI, United Kingdom; DOE and NSF, United States of America.

Individual groups and members have received support from BCKDF, CANARIE, CRC and DRAC, Canada; CERN-CZ and PRIMUS, Czech Republic; COST, ERC, ERDF, Horizon 2020 and Marie Sk{\l}odowska-Curie Actions, European Union; Investissements d'Avenir Labex, Investissements d'Avenir Idex and ANR, France; DFG and AvH Foundation, Germany; Herakleitos, Thales and Aristeia programmes co-financed by EU-ESF and the Greek NSRF, Greece; BSF-NSF and MINERVA, Israel; NCN and NAWA, Poland; La Caixa Banking Foundation, CERCA and AGAUR programs from Generalitat de Catalunya and PROMETEO and GenT Programmes Generalitat Valenciana, Spain; G\"{o}ran Gustafssons Stiftelse, Sweden; The Royal Society and Leverhulme Trust, United Kingdom; Eric and Wendy Schmidt Fund for Strategic Innovation, United States of America.

In addition, individual members wish to acknowledge support from Chile: Agencia Nacional de Investigaci\'on y Desarrollo (FONDECYT 1230812, FONDECYT 1230987, FONDECYT 1240721, FONDECYT 1240864, FONDECYT 3240661); China: Fundamental Research Funds for the Central Universities (010-63263105), Chinese Ministry of Science and Technology (MOST-2023YFA1605700, MOST-2023YFA1609300), National Natural Science Foundation of China (NSFC 12275265, NSFC-W2543005); Czech Republic: Czech Science Foundation (GACR - 24-11373S), Ministry of Education Youth and Sports (ERC-CZ-LL2327, FORTE CZ.02.01.01/00/22\_008/0004632); EU: H2020 European Research Council (ERC - 101002463); European Union: European Research Council (BARD No. 101116429, ERC - 101219398, ERC 101089007), European Regional Development Fund (HE COFUND GA No.101081355, ERDF), Marie Sklodowska-Curie Actions (GAP-101168829); France: Agence Nationale de la Recherche (ANR-22-EDIR-0002, ANR-24-CE31-0504-01); China: Research Grants Council (GRF); Italy: Istituto Nazionale di Fisica Nucleare (LHC-MIUR - 28003/2025), Ministero dell'Università e della Ricerca (NextGenEU I53D23000820006 M4C2.1.1, SOE2024\_0000023); Japan: Japan Society for the Promotion of Science (JSPS KAKENHI  JP25H0063, JSPS KAKENHI JP22H04944, JSPS KAKENHI JP24K23939, JSPS KAKENHI JP24KK0251, JSPS KAKENHI JP25H00650, JSPS KAKENHI JP25H01291, JSPS KAKENHI JP25K01011, JSPS KAKENHI JP25K01023, JSPS KAKENHI JP25KK0047); Poland: Polish National Science Centre (NCN OPUS 2023/51/B/ST2/02507, NCN OPUS nr 2022/47/B/ST2/03059, UMO-2019/34/E/ST2/00393, UMO-2021/42/E/ST2/00350, UMO-2022/47/O/ST2/00148, UMO-2023/49/B/ST2/04085, UMO-2023/51/B/ST2/00920, UMO-2024/53/N/ST2/00869); Spain: Agència de Gestió d'Ajuts Universitaris i de Recerca. (AGAUR - 2023 BP 00141), Ministry of Science and Innovation (RYC2019-028510-I, RYC2020-030254-I, RYC2021-031273-I, RYC2022-038164-I), Ministerio de Ciencia, Innovación y Universidades/Agencia Estatal de Investigaci\'on (PID2022-142604OB-C22); Sweden: Carl Trygger Foundation (Carl Trygger Foundation CTS 22:2312), Swedish Research Council (Swedish Research Council 2023-04654, VR 2021-03651, VR 2022-03845, VR 2022-04683, VR 2023-03403, VR 2024-05451, VR 2025-05940), Knut and Alice Wallenberg Foundation (KAW 2023.0366); United Kingdom: The Binks Trust, Royal Society (NIF-R1-231091); United States of America: U.S. Department of Energy (ECA DE-AC02-76SF00515), John Templeton Foundation (John Templeton Foundation 63206), Neubauer Family Foundation.

%
%

% End of text imported from the .//acknowledgements/Acknowledgements.tex input file

%
\clearpage
\appendix
\part*{Appendix}
\addcontentsline{toc}{part}{Appendix}

\FloatBarrier
\section{Details of MC simulations}

% The next lines are included from the .//Sections/Appendices/MC.tex input file
\label{app:MC}

The following appendix provides further details of the MC generator
configurations used for the simulation of signal and background
processes in the measurement.

Unless stated otherwise all simulated samples in the measurement
showered using \PYTHIA apply the \tune{A14} set of tuned
parameters~\cite{ATL-PHYS-PUB-2014-021} with the \nnpdflo PDF
set. These samples also have the decays of bottom and charm hadrons
simulated using the \evtgen~1.7.0 package~\cite{Lange:2001uf}. The
top-quark mass is set to 172.5~\GeV for all processes involving
top-quarks.

\subsection{Nominal MC generator configurations}

\subsubsection*{\tty}
Two samples to model the \tty process are simulated with the \MGNLO~\cite{Alwall:2014hca} generator. The first MC sample of \tty events with the photon produced from the top quark or from initial-state radiation is simulated as a $2\to 3$ process at NLO accuracy in QCD with the on-shell top quarks in the final state being decayed at LO using \MADSPIN~\cite{Frixione:2007zp,Artoisenet:2012st} to preserve spin correlations. The interference effects between initial-state photon radiation (from the incoming partons) and the final-state photon radiation (from the off-shell top quarks) were taken into account in the simulation. This sample is referred to as the `\tty production sample’, \ttyprod.
The second \tty sample, where the photon arises from any of the decay products of the top quarks or from one of the on-shell top quarks, is simulated with the same version of \MGNLO but at LO precision as a $2 \to 2$ process followed by the decay of the top quarks, also simulated with \MGNLO at LO precision. This sample is referred to in the following as the `\tty decay sample’, \ttydecay. Both samples are generated by using the \NNPDF[3.0nlo]~\cite{Ball:2014uwa} PDF set. The renormalisation and factorisation scales are set to $0.5\times \sum_i \sqrt{m^2_i+p^2_{\text{T},i}}$, where $m_i$ and $p_{\text{T},i}$ are the masses and transverse momenta of the particles generated from the ME calculation. In both samples, photons are required to have $\pT > 15~\GeV$ and to be isolated according to a smooth-cone hadronic isolation criterion %
~\cite{Frixione:1998jh}, which avoids infrared divergences.
The \tty production sample is normalised to the NLO cross-section given by the MC simulation, while the normalisation of the \tty decay sample is corrected by a NLO/LO inclusive $K$-factor of 1.5 obtained in Ref.~\cite{TOPQ-2021-29}.

\subsubsection*{\tWy and \tqy}
The \tWy events were generated at LO accuracy with the \MGNLO[2.6.2] generator in the five-flavour scheme (5FS). To simulate the full process, two complementary samples were generated: one as a $2\rightarrow3$ process assuming a stable top quark and the other as a $2 \to 2$ process, where the photon is radiated from any other final-state charged particle. The decay of the top quarks is also simulated with \MGNLO. To avoid infrared divergences, the photon was required to have $\pT > 15~\GeV$ and $|\eta| < 5.0$ and to be separated by $\Delta R>0.2$ from any parton. Both samples make use of the \NNPDF[2.3lo] PDF set~\cite{Ball:2012cx} and are normalised to the cross-section provided by the MC simulation.

The production of SM \tqy events was modelled using the \MGNLO[2.9.9] generator that provides matrix elements at NLO in QCD with the \NNPDF[2.3lo] PDF set.  The events are interfaced
with \PYTHIA[8.307]~\cite{Sjostrand:2014zea} for the parton shower and its hadronisation.

\subsubsection*{\ttZ}
The \ttZ process (defined as $\ttll$, including the $\gamma^{*}$ contribution)
is modelled using the \MGNLO[2.8.1]~\cite{Alwall:2014hca} (MG5\_aMC) generator
that provides MEs at NLO in the
strong coupling constant \alphas with
the \NNPDF[3.0nlo]~\cite{Ball:2014uwa} PDF
set.  The functional form of the renormalisation and factorisation
scales (\muR, \muF) is set to the default scale
$0.5 \times \sum_i \sqrt{m^2_i+p^2_{\textrm{T},i}}$, where the sum runs over
all the particles generated from the matrix element calculation.  Top
quarks are decayed at LO
using \MADSPIN~\cite{Frixione:2007zp,Artoisenet:2012st} to preserve
all spin correlations.  The events are interfaced
with \PYTHIA[8.244]%
for the parton shower and
its hadronisation. This process is normalised to the
(on-shell) \ttZ cross-section calculated at NLO QCD and NLO EW accuracy using \MGNLO and reported in
Ref.~\cite{deFlorian:2016spz}.
The $\gamma^{*}$ contribution and the $Z/\gamma^{*}$ interference are taken into account with an off-shell (down to $m(\ell^{+}\ell^{-})=5~\GeV$) correction calculated in Ref.~\cite{ATL-PHYS-PUB-2016-005}.
The predicted value at 13~\TeV is $839^{+83}_{-97}$~fb, where the uncertainties arise from variations of renormalisation and factorisation scales, PDF uncertainties, and \alphas variations.

\subsubsection*{\tZq and \tWZ}
The \tZq sample is simulated using the \MGNLO[3.5.1]~\cite{Alwall:2014hca}
generator at NLO with the \NNPDF[3.0nlo] PDF. The events are interfaced with
\PYTHIA[8.309]. Off-resonant events away from
the Z mass peak are included, down to 5 GeV.
The top quark is decayed at LO using
\MADSPIN~\cite{Frixione:2007zp,Artoisenet:2012st} to preserve spin
correlations.
The four-flavour scheme is used, where all the quark masses are set to
zero, except for the top and bottom quarks.
Following the discussion in Ref.~\cite{tZq-Pagani}, the
scales are set to $H_\mathrm{T}/6$.
The \tZq sample is normalised via $k$-factors to a reference cross-section from
dedicated samples for both top and anti-top with all settings consistent
to the nominal sample except for the use of the five-flavour scheme, where quark masses are set to zero except for the top quark.
The $k$-factor for \tZqt is 1.29, while for \tZqtbar it is 1.32.

The production of single top quarks in association with a \Wboson and
a \Zboson boson (\tWZ) is modelled using \MGNLO[2.2.2] at NLO, with
the \NNPDF[2.3lo] PDF set %
interfaced with \PYTHIA[8.212].

\subsubsection*{\ttWQCD}
The \ttW QCD sample is generated with \SHERPA[2.2.10]~\cite{Bothmann:2019yzt,Gleisberg:2008ta} at \NLO accuracy
in QCD with $\mu_R$ = $\mu_F$ = $H_\textrm{T}/2$. The ME was
calculated for up to one additional parton at \NLO and up to two
partons at \LO using
\textsc{Comix}~\cite{Gleisberg:2008fv}
and \OPENLOOPS~\cite{Cascioli:2011va,Denner:2016kdg,Buccioni:2019sur},
and merged with the \Sherpa parton shower~\cite{Schumann:2007mg} using
the \MEPSatNLO
prescription~\cite{Hoeche:2012yf,Hoeche:2011fd,Hoeche:2012yf,Catani:2001cc,Hoeche:2009rj}
with a merging scale of 30 \GeV. The sample is normalised to the
next-to-next-to-leading-order (\NNLO) cross section of
$710.7^{+37.4}_{-48.1}$~fb quoted in
Ref.~\cite{PhysRevLett.131.231901}.

\subsubsection*{\ttWEW}
The \ttWEW process is generated
with \SHERPA[2.2.10]~\cite{Bothmann:2019yzt,Gleisberg:2008ta} at LO in
QCD for the ME and showered with the default \SHERPA[2.2.10] settings
and \NNPDF[3.0nnlo]\cite{NNPDF:2017mvq}. The renormalisation and
factorisation scales, $\mu_\mathrm{R}$ and $\mu_\mathrm{F}$, are set
to $H_\textrm{T}/2 = \frac{1}{2}\sum_i\sqrt{m_i^2 + p_{\text{T},i}^2}$
where $i$ runs over all outgoing partons from the ME calculation.  The
cross section from the generator is $\sigma_{\ttWEW} =
42.1^{+1.7}_{-1.3}$~fb. This is scaled (by a factor of 1.13) to match
the theoretical cross section predicted at \NLO in QCD from
Ref.~\cite{Frederix:2021agh}, $\sigma_{\ttWEW} =
47.7^{+1.9}_{-1.4}$~fb.

\subsubsection*{\ttH}
The $\ttH$ process is simulated using \POWHEGBOX[v2]~\cite{Nason:2009ai,Alioli:2010xd,Nason:2004rx,Frixione:2007vw}.
The \hdamp\ parameter, which controls the \pt\ of the first additional emission beyond the Born configuration and therefore regulates the high-\pt\ emission,  was set to $3 \times (2 m_t + m_H) / 4=352.5$~\GeV.
The functional form of the renormalisation (\muR) and factorisation (\muF) scales was set to $\sqrt[3]{m_\text{T}(t)\cdot m_\text{T}(\bar{t}) \cdot m_\text{T}(H)}$, where
$m_\text{T}=\sqrt{m^2+p_\text{T}^2}$ is the transverse mass of a generated particle, $m$ is its mass, and $p_\text{T}$ is its transverse momentum.
For this sample, the \NNPDF[3.0nlo] PDF set~\cite{Ball:2014uwa} was used.
The Monte Carlo prediction was normalised to cross-sections calculated at NLO in QCD and EW couplings~\cite{deFlorian:2016spz,Beenakker:2002nc,Dawson:2003zu,Yu:2014cka,Frixione:2014qaa}.

\subsubsection*{\tH}
The \tHq\ (\tHW) samples were produced with \MGNLO[2.6] (2.8)~\cite{Alwall:2014hca} in the four-flavour (five-flavour) scheme with the \NNPDF[2.3lo] PDF.
The same flavour scheme was used in the ME calculation and the PDF.
The top-quark and $W$ boson decays were handled by \MADSPIN~\cite{Artoisenet:2012st} to account for spin correlations in the decay products.
The overlap of the \tHW\ process with \ttH\ at NLO was removed by using a diagram removal technique~\cite{Frixione:2008yi,Demartin:2016axk}.

\subsubsection*{\tttt}
The production of SM \tttt events was modelled using the \MGNLO[v2.6.2] generator that provides matrix elements at NLO in QCD with the \NNPDF[3.1nlo] PDF set.
The functional form of the renormalisation and factorisation scales was set to \(0.25\times \sum_i \sqrt{m^2_i+p^2_{\text{T},i}}\),
where the sum runs over all the particles generated from the matrix element calculation, following Ref.~\cite{Frederix:2017wme}.
Top quarks were decayed at LO using \MADSPIN to preserve all spin correlations.
The production of \tttt\  events is normalised to a cross-section of 13.37~fb computed at NLO in QCD including EWK corrections~\cite{Frederix:2017wme}.

\subsubsection*{Diboson}
Diboson ($VV$) processes with four charged leptons, three charged
leptons and a neutrino, or two charged leptons and two neutrinos (all
with or without extra partons) are simulated with \SHERPA[2.2.2] with
the PDF set \NNPDF[3.0nnlo]. The processes are generated with up to one
extra parton at \NLO and up to three extra partons at \LO using the
same formalism as in \ttW.

\subsubsection*{\ttbar}
The MC production
of \ttbar events is modelled using the \POWHEGBOX[2]
generator~\cite{Alioli:2010xd} at \NLO with the \NNPDF[3.0nlo] PDF set
and the damping factor $h_\text{damp}$ set to 1.5 times the top quark
mass. The events are interfaced with \PYTHIA[8.2].  Single top-quark
production in the $s$- and $t$-channels, and $tW$ production, is
simulated similarly.

\subsubsection*{$V$+jets}
The production of \Wboson and \Zboson bosons with multiple jets
($V+$jets) is simulated with the \SHERPA[2.2.1] generator using
NLO-accurate MEs for up to two jets and LO-accurate MEs for up to four
jets, calculated with the Comix~\cite{Gleisberg:2008fv} and \OPENLOOPS
libraries. They are matched with the \SHERPA parton shower using
the \MEPSatNLO prescription.  The \NNPDF[3.0nnlo] set of PDFs is used
and the samples are normalised to \NNLO predictions.

\subsubsection*{Triboson}
Processes with three heavy gauge bosons ($WWW$, $WWZ$, $WZZ$ and
$ZZZ$) with up to six leptons in the final state are simulated
with \SHERPA[2.2.2] and the \NNPDF[3.0nlo] PDF set.  Final states with
no additional partons are calculated at NLO, whereas final states with
one, two or three additional partons are calculated at LO.

\subsection{Systematic uncertainties}

The uncertainties in \ttyprod and \ttydecay are applied separately. The uncertainty in the parton shower and hadronisation model is estimated by comparing the \tty production nominal sample, produced with \MGNLO + \PYTHIA[8], with showering by \HERWIG[7.2.1].
The uncertainty in the modelling of initial-state radiation is estimated using a \tty production sample produced with \MGNLO + \PYTHIA[8], with the \PYTHIA \tune{A14} \tune{Var3c} eigentune for low and high radiation scenarios implemented.
For the \ttZ process uncertainties associated with the parton showering algorithm and the underlying event model are evaluated by comparing the nominal samples generated with \MGNLO interfaced to \PYTHIA[8], to equivalent samples interfaced to \HERWIG[7] instead.
Uncertainties related to the modelling of initial-state radiation are obtained by varying the \tune{Var3c} parameter of the \PYTHIA \tune{A14} tune in dedicated alternative samples.

For \ttW all the systematic uncertainties are applied separately on the QCD and the EW samples. Uncertainties considered include theory uncertainties in the generator for QCD production (defined by the \MGNLO interfaced to \PYTHIA[8] MC sample with FxFx merging scheme~\cite{Frederix:2012ps}, serving as a combined ME+PS uncertainty), a theory uncertainty in the generator for EW production (defined by the LO \MGNLO sample which models subleading EWK effects, again serving as a combined ME+PS uncertainty; uncorrelated from the QCD ME+PS uncertainty), a shape-only uncertainty in the relative yields of the EW sample within the QCD+EW $t\bar{t}W$ templates (varying the EW fraction between the values of the Sherpa and the MG5$\_$aMC@NLO prediction). A separate PS uncertainty based on the \POWHEG interfaced to \PYTHIA[8] and \POWHEG interfaced to \HERWIG[7] MC samples is included. The nominal PDF set for Sherpa 2.2.10 is \texttt{NNPDF30\_nnlo\_as\_0118\_hessian}. An uncertainty in the choice of $\alpha_s = 0.118$ in the PDF is estimated using the \NNPDF[3.0nlo] sets with $\alpha_s = 0.117$ and $\alpha_s = 0.119$, defining a two-sided uncertainty. An uncertainty in the PDF set itself is defined using \texttt{CT14nnlo} and \texttt{MMHT2014nnlo68cl} PDF sets defining a two-sided uncertainty.

Uncertainties in the \tZq processes include a parton shower uncertainty obtained by comparing an alternative sample of \tZq events generated with \MGNLO interfaced to \HERWIG[7] to the nominal setup (\MGNLO interfaced to \PYTHIA[8]).
As for \ttZ, variations of the \tune{Var3c} parameter of the \PYTHIA \tune{A14} tune, as well as of the factorisation and renormalisation scales, are considered.

For the \tWZ process, a flat \SI{30}{\percent} normalisation uncertainty is added based on the latest experimental sensitivity of Ref.~\cite{CMS-TOP-24-009-fix}. No parton shower uncertainty is considered, but instead the difference between samples generated with the DR1 and DR2 diagram removal schemes~\cite{Demartin:2016axk} is treated as a modelling uncertainty.
Furthermore, both the shape and normalisation (acceptance) components of this systematic uncertainty are considered: in the absence of higher-order theoretical calculations for the \tWZ process, the comparison of the cross sections obtained in the five-flavour scheme in the DR1 and DR2 setups leads to a $10\text{-}15\%$ normalisation uncertainty in the SRs.

The uncertainty in \ttH modelling due to the initial-state-radiation (ISR) is estimated using weights in the matrix element calculation and in the parton shower. Uncertainties in the matching between the ME and PS are estimated by comparing the nominal \POWHEG interfaced to \PYTHIA[8] MC simulation with an alternative \POWHEG interfaced to \PYTHIA[8] sample where the definition of the hardness of the Powheg emission, calculated via the parameter $\pt^{\textrm{hard}}$, is changed following Ref.~\cite{ATL-PHYS-PUB-2023-029}. Uncertainties in the PS algorithm (parton shower, hadronisation and underlying event modelling) are assessed by comparing the nominal sample to  \POWHEG interfaced to \HERWIG[7]. The uncertainties in the amount of initial and final-state QCD radiation (ISR and FSR) predicted by the PS are estimated by varying the scale in $\alpha_S^{ISR}$ according to the values given by \tune{Var3c} in the \PYTHIA[8] \tune{A14} tune and by varying the scale in $\alpha_S^{FSR}$ by a factor two up and down.

An uncertainty related to the generator choice for the \tttt process, including ME, PS and hadronisation,
is considered by comparing the nominal \tttt sample with a sample generated using \Sherpa[2.2.11].
An uncertainty related to the parton shower generator choice is considered by comparing the nominal \tttt sample
with the sample produced with \HERWIG[7.0.4] as the parton shower generator.

The production of SM \tWy and \tqy events are assigned a 50\% normalisation uncertainty.

Uncertainties on the $WZ \rightarrow \ell\ell\ell\nu$ ($WZ$+jets) and $ZZ \rightarrow \ell\ell\ell\ell$ ($ZZ$+jets) backgrounds related to the CKKW matching scale, QSF resumation scale and alternative recoil scheme are considered. Uncertainties related to the choice of PDF are derived by comparing the nominal value to those from the \tune{CT14} and \tune{MMHT14} PDF sets. Variations of \alphas in the nominal PDF are also included. In addition, a normalisation uncertainty of $40\%$ is applied to the \VV$+c$ and of $50\%$ to the \VV$+b$ components of the \WZ and \ZZ backgrounds separately. All of the above uncertainties are applied as separate nuisance parameters for each of the light-, $c$- and $b$-flavour components of the $WZ+$jets and $ZZ+$jets background processes.

It is observed in the signal regions targeting photons that there is a significant background from prompt production of $V\gamma+$jets. Based on the
measurement of $V\gamma+$jets cross sections in Ref.~\cite{STDM-2020-14}, these events are assigned a $20\%$ normalisation uncertainty.

For other minor background processes, such as $VV$ with less than two charged leptons in the decay, $HV$, $VVV$, $\ttWW$ or multi-top-quark (\ttt) production, a common conservative overall normalisation uncertainty of $50\%$ is applied. These background components contribute to about $2.5\%$ of the total event yields in the signal regions.

The MC template for \NPL{}s and \NPC{}s that cannot be normalised in data as part of the template method is assigned a $50\%$ normalisation uncertainty.
%

% End of text imported from the .//Sections/Appendices/MC.tex input file

\FloatBarrier
\section{Object definitions}
\label{app:objectdefinitions}

% The next lines are included from the .//Sections/Appendices/ObjectDefinitions.tex input file

The following section outlines how the reconstructed objects, common
to both OB and PB approaches, are defined.

Interaction vertices from the $pp$ collisions are reconstructed from at least two tracks with \pt larger than $500~\mev$ that are consistent with
originating from the beam collision region in the $x$--$y$ plane. If more than one primary vertex candidate is found in the event, the candidate for which
the associated tracks form the largest sum of squared \pt is selected as the hard-scatter primary vertex~\cite{IDTR-2021-01}.

\subsection{Electrons}
Electron candidates are reconstructed from clusters of energy deposits in the electromagnetic calorimeter
that are matched to a track in the ID. They are required to satisfy $\pT>7$~\GeV, $|\eta|<2.47$ and
a likelihood-based electron identification~\cite{EGAM-2018-01} is required.
For pre-selected electrons, the \texttt{LooseAndBLayerLH} identification is used, while for the signal electrons, \texttt{TightLH} is required.
Electrons in the LAr transition region ($1.37 < \abseta < 1.52$) are rejected and requirements on the transverse ($d_{0}$) and longitudinal ($z_{0}$) impact parameter are applied to mitigate the contribution from electrons from reducible backgrounds and pile-up. To suppress electrons with incorrect charge assignment, a boosted decision tree~(BDT) discriminant based on calorimeter and tracking quantities is applied, referred as \nplqbdt, and finally a procedure to resolve ambiguities between overlapping reconstructed objects is applied.
The electron energy scale is calibrated using data~\cite{EGAM-2018-01,EGAM-2021-02}. The associated scale factors~(SFs) for electron reconstruction, identification, and trigger efficiency are applied in MC, to correct for the efficiency differences between data and simulation~\cite{EGAM-2021-01}.

\subsection{Muons}
Muon candidates are reconstructed from MS tracks matched to ID tracks
in the pseudorapidity range of $|\eta|<2.5$ and are required to have
$\pt > 7$~\GeV.  Both the pre-selected and signal muons are required to
satisfy \texttt{Medium} identification requirements defined in
Refs.~\cite{MUON-2022-01,MUON-2018-03}. Selections on
the longitudinal and transverse impact parameters are applied as for the electrons.
The associated SFs for identification and trigger efficiencies are applied as multiplicative factors to
the MC event weight, to correct for the efficiency differences between
data and MC.

\subsection{Lepton isolation}
The selection criteria above largely suppress the contribution from non-prompt leptons
arising from hadron decays that contain bottom- or charm-quarks (referred to as
``heavy-flavour (HF) non-prompt leptons''), misidentified jets or photon conversions.
However, as these measurements are particularly sensitive to such backgrounds, signal
electrons have a further selection requirement from a non-prompt lepton BDT discriminant
(\nplbdt)~\cite{HIGG-2017-02}, based on isolation and lifetime information about a track-based
jet that matches the selected light lepton. Two versions of the discriminant exist; \plv which
has two working points (WP) `Loose' and `Tight', and \pliv has `Tight', `VeryTight', and
`Tight-not-VeryTight' (the latter of which is not considered in this analysis). The \plv{Tight}
WP is defined with a prompt electron (muon) efficiency of 65\% (80\%) for a candidate
lepton $\pt=20$~\GeV, increasing to 90\% (95\%) at $\pt=45$~\GeV. The rejection factors for
non-prompt electrons (muons) from heavy-flavour decays range from 60 (30) at low $\pt$ to 30
(20) at high $\pt$. The rejection factor for electrons from photon conversions is 5 at low $\pt$
and 1.5 at high $\pt$. The \pliv{Tight} WP allows prompt muons (barrel/endcap electrons) satisfying
the calorimeter and track-based isolation criteria to be selected with an efficiency that is about
60\% (60\%/70\%) for $\pt \sim 20$~\GeV and reaches a plateau of 95\% (95\%/90\%) for $\pt\sim 40
(40/65)$~\GeV. The prompt-lepton efficiency of the \pliv{VeryTight} WP for muons (barrel/endcap electrons)
that satisfy the calorimeter- and track-based isolation criteria is about 55\% (55\%/60\%) for
$\pt \sim 20$~\GeV and reaches a plateau of 90\% (85\%/83\%) for $\pt \sim 40 (40/65)$~\GeV.
The corresponding rejection factor for muons (electrons) from the decay of $b$-hadrons ranges
from 33 to 50 (20 to 50) for the \pliv{Tight} WP, and from 50 to 100 (33 to 66) for
the \pliv{VeryTight} WP, depending on $\pt$ and $\eta$, after resolving ambiguities
between overlapping reconstructed objects.
The associated SFs for isolation are applied as multiplicative factors to
the MC event weight, to correct for the efficiency differences between
data and the MC simulation.

\subsection{Jets}
Jets are reconstructed using the \antikt jet algorithm~\cite{Cacciari:2008gp} as implemented in the \textsc{FastJet} package~\cite{Fastjet} with the distance parameter $R$ set to $0.4$. Topological clusters are used~\cite{ATL-PHYS-PUB-2015-036} along with ID tracks as input to the particle flow reconstruction algorithm~\cite{PERF-2015-09, }. The algorithm removes calorimeter energy deposits due to charged hadrons and uses more precise measurements of their momenta from tracks instead. Jets are calibrated~\cite{JETM-2018-05} with the \texttt{EMPFlow} scheme applying the jet area pile-up corrections.
The jets are kept only if they have $\pt > 25$~\GeV\ and are inside a pseudorapidity range of $\abseta < 4.5$.
The \tool{JetVertexTagger}~(JVT)~\cite{ATLAS-CONF-2014-018} is employed to mitigate pile-up effects, and specifically rejects any selected jets after the overlap removal procedure with $\pt < 60$~\GeV, $\abseta < 2.4$ and for which $\JVT < 0.5$ (\texttt{Tight} working point).
In order to reject forward jets originating from additional proton--proton interactions,
a forward jet vertex tagger (fJVT) requirement is applied where jets after the overlap removal procedure with $\pT < 60$~\GeV, $2.5 < \abseta < 4.5$, $\mathrm{fJVT}>0.4$ and $|\mathrm{timing}|>10$~ns are rejected, corresponding to the fJVT \texttt{Tight} working point~\cite{PERF-2016-06}.
Jets containing a $b$-hadron, referred to as $b$-jets, are identified with the \texttt{DL1r} $b$-tagging algorithm~\cite{FTAG-2018-01} within the pseudorapidity range of $\abseta < 2.5$.
A WP corresponding to an $85\%$ $b$-tagging efficiency is used in most regions for pre-selections.
Exclusive bins in the $b$-tagging discriminant corresponding to different $b$-tagged jet identification efficiencies ($60$, $70$, $77$ and $85\%$, are also used.
To account for residual differences in efficiencies and mistag rates between data and simulation, separate correction factors for JVT, fJVT and $b$-tagging are applied.

\subsection{Photons}
Photons are reconstructed from energy deposits in the central region
of the electromagnetic calorimeters. If the cluster is considered not
matched to any reconstructed tracks in the ID system, an unconverted
photon candidate is reconstructed. If the cluster is matched with
reconstructed tracks that are consistent with originating from a
photon conversion and if in addition a conversion vertex is
reconstructed, a converted  photon candidate is
reconstructed. Photons are required to be isolated~\cite{EGAM-2018-01} fulfilling
the criteria $\ET^{\textrm{iso}}\big|_{\Delta R=0.4}<0.022\cdot\ET(\gamma)+2.45$~\GeV
and $\pT^{\textrm{iso}}\big|_{\Delta R=0.2}<0.05\cdot\ET(\gamma)$, where
$\ET^{\textrm{iso}}\big|_{\Delta R=0.4}$ is the calorimeter isolation within
a core of radius $\Delta R=0.4$ in the direction of the photon candidate,
$\pT^{\textrm{iso}}\big|_{\Delta R=0.2}$ is the track isolation within
$\Delta R=0.2$ and $\ET(\gamma)$ is the transverse energy of a photon.
The photons are also required to satisfy a \texttt{tight} identification
criteria~\cite{EGAM-2018-01}.
Photon energies are calibrated with the method described
in~\cite{EGAM-2021-02}. Only photons with calibrated transverse energy \ET >
\qty{20}{\GeV} and \abseta< 2.37, excluding the LAr crack region (1.37
< \abseta < 1.52), are considered.

\subsection{Missing transverse momentum}
The missing transverse momentum $\vec{p}_{\mathrm{T}}^{\mathrm{miss}}$
(with magnitude $\met$) is defined as the negative vector sum of the
$\pT$ of all selected and calibrated objects in the event, including a
term to account for the momenta of soft particles that are not
associated with any of the selected objects~\cite{JETM-2020-03}.  This
soft term is calculated from inner-detector tracks matched to the
primary vertex, which makes it more resilient to contamination
from \pileup interactions.

\subsection{Overlap removal}
After the identification, overlaps between objects are resolved to avoid double-counting of physics objects.
Pre-selected leptons, photons and jets are used for this procedure, following a standard
sequence of operations that are performed to solve these ambiguities,
referred to as the overlap removal (OR) procedure~\cite{TOPQ-2018-24}.
%

% End of text imported from the .//Sections/Appendices/ObjectDefinitions.tex input file

\FloatBarrier
\section{Analysis regions}
\label{app:regions}

% The next lines are included from the .//Sections/Appendices/OBRegions.tex input file
%

The following subsections provide tables with the details of event
selection criteria applied to define the various region definitions in
the OB and PB approaches. Most variables used are defined in the main
body of the text. In the case of
events with more than one same-flavour di-lepton pair, the pair giving the lowest invariant
mass, \mllmin, is also used for some selections.

\subsection{OB approach region definitions}
\label{sec:app:OBregions}

Tables~\ref{tab:OBPhotonRegions}-\ref{tab:OBLeptonRegions_forward}
provide details of the OB approach signal region definitions. They are
grouped into categories based on the requirements of the presence of a
photon, and the selection or veto of lepton pairs consistent with an
on-shell $Z$ boson. Table~\ref{tab:DibosonCRs_reduced} provides the details of the diboson control region definitions.

\begin{table}
\caption{Summary of event selection requirements for the OB approach photon regions, \OBonely and \OBtwolOSy.}
\label{tab:OBPhotonRegions}
\begin{center}{\renewcommand{\arraystretch}{1.2}
\begin{tabular}{ c|ccp{3cm} }
\toprule
Variable & \OBonely& \OBtwolOSy\\
\midrule
\nphotons & \multicolumn{2}{c}{1} \\
$\pt^{\gamma}$ & \multicolumn{2}{c}{$\geq 20$~\GeV}\\
\nleps & 1 & 2\\
\ptlep{1} & \multicolumn{2}{c}{$\geq 27$~\GeV} \\
\ptlep{2} & --- & $\geq 15$~\GeV\\
$m_{\ell\ell\gamma}$ & --- & $\geq 15$~\GeV \\
\mllSFmz & --- & $\geq  10$~\GeV\\
$|m_{\ell\ell\gamma}-m_{Z}|$ & --- & $\geq 10$~\GeV \\
\modsumq & --- & $0$ \\
\met & --- & $\geq 40$~\GeV \\
\ncjets & $\geq 4$ & $\geq 2$ \\
\nbjets$@ 85\%$ & \multicolumn{2}{c}{$\geq 2$} \\
\nbjets$@ 60\%$ & $\geq 1$ & ---\\
\bottomrule
\end{tabular}
}
\end{center}
\end{table}

\begin{table}[h]
\caption{Summary of event selection requirements for the OB approach 2- and 3-lepton regions that veto events consistent with an on-shell $Z$ boson: \OBtwolSS, \OBthreelblo and \OBthreelbblo, \OBthreelhiee and \OBthreelhimumu.}
\label{tab:OBLeptonRegions_reduced}
\begin{center}{\renewcommand{\arraystretch}{1.0}
\begin{tabular}{ c|ccccc }
\toprule
Variable & \OBtwolSS & \OBthreelblo & \OBthreelbblo & \OBthreelhiee & \OBthreelhimumu \\
\midrule
\nleps & 2 & 3 & 3 & 3 & 3 \\
\ptlep{1} & \multicolumn{5}{c}{$\geq 27$~\GeV} \\
\ptlep{2} & \multicolumn{5}{c}{$\geq 15$~\GeV} \\
\ptlep{3} & --- & \multicolumn{4}{|c}{$\geq 15$~\GeV} \\
\mllmin & --- & \multicolumn{2}{|c}{$\geq 12$~\GeV, $\leq 81$~\GeV} & \multicolumn{2}{|c}{$\geq 101$~\GeV}  \\
\mllSF & $\geq 12$~\GeV & \multicolumn{4}{|c}{---} \\
\mllSFmz & \multicolumn{3}{c}{$> 10$~\GeV} \\
\modsumq & 2 & \multicolumn{4}{|c}{1} \\
\ntjets & $\geq 4$ & \multicolumn{4}{|c}{$\geq 2$} \\
\nbjets$@ 85\%$ & $\geq 2$ & 1 & $\geq 2$ & $\geq 1$ & $\geq 1$ \\
\nbjets$@ 60\%$ & $\geq 1$ & 1 & --- & --- & --- \\
\bottomrule
\end{tabular}
}
\end{center}
\end{table}

\begin{table}[h]
\caption{Summary of event selection requirements for the OB approach
3- and 4-lepton regions that select events consistent with an
on-shell $Z$ boson: \OBthreelZb, \OBthreelZbforward, \OBthreelZbb
, \OBthreelZbbforward and \OBfourl. Some of the regions apply
dedicated requirements on forward jets.}
\label{tab:OBLeptonRegions_forward}
\begin{center}{\renewcommand{\arraystretch}{1.0}
\begin{tabular}{ c|ccccc }
\toprule
Variable
& \OBthreelZb
& \OBthreelZbforward
& \OBthreelZbb
& \OBthreelZbbforward
& \OBfourl \\
\midrule
\nleps & \multicolumn{4}{c|}{3} & 4 \\
\ptlep{1} & \multicolumn{5}{c}{$\geq 27$~\GeV} \\
\ptlep{2} & \multicolumn{4}{c|}{$\geq 15$~\GeV} & --- \\
\ptlep{3} & \multicolumn{4}{c|}{$\geq 15$~\GeV} & --- \\
\mllmin & \multicolumn{5}{c}{$\geq 12$~\GeV} \\
\mllSFmz
& \multicolumn{4}{c|}{$\leq 10$~\GeV}
& --- \\
\nbjets$@ 85\%$
& \multicolumn{2}{c|}{1} & \multicolumn{2}{c|}{$\geq 2$} & $\geq 1$ \\
\nbjets$@ 60\%$
& \multicolumn{2}{c|}{1} & \multicolumn{2}{c|}{$\geq 1$} & $\geq 1$ \\
\ntjets & \multicolumn{5}{c}{$\geq 2$} \\
\nfjets & 0 & $\geq 1$ & 0 & $\geq 1$ & --- \\
\bottomrule
\end{tabular}
}
\end{center}
\end{table}

\begin{table}[h]
\caption{Summary of requirements for object-based diboson control regions for three and four leptons.}
\label{tab:DibosonCRs_reduced}
\begin{adjustbox}{width=\columnwidth,center}
{\renewcommand{\arraystretch}{1.2}
\begin{tabular}{ c|cccc }
\toprule
Variable & \OBDibosonthreelZb & \OBDibosonthreelZbb & \OBDibosonthreelb & \OBDibosonfourl \\
\midrule
\nleps & \multicolumn{3}{c|}{3} & 4\\

\ptlep{1} & \multicolumn{4}{c}{$\geq 27$~\GeV} \\

\ptlep{2} & \multicolumn{3}{c|}{$\geq 15$~\GeV} & ---\\

\ptlep{3} & \multicolumn{3}{c|}{$\geq 15$~\GeV} & --- \\

\mllmin & \multicolumn{4}{c}{$\geq 12$~\GeV} \\

\mllSFmz & \multicolumn{2}{c|}{$\leq 10$~\GeV} & $> 10$~\GeV & --- \\

\modsumq & \multicolumn{2}{c|}{---} & 1 & ---\\

\ntjets & \multicolumn{4}{c}{$\geq 2$} \\

\nbjets$@ 85\%$ & $1$ & $\geq 2$ & $1$ & $\geq 1$ \\

\nbjets$@ 60\%$ & \multicolumn{4}{c}{0} \\
\bottomrule
\end{tabular}
}
\end{adjustbox}
\end{table}

% End of text imported from the .//Sections/Appendices/OBRegions.tex input file

\FloatBarrier

% The next lines are included from the .//Sections/Appendices/PBRegions.tex input file

\subsection{PB approach region definitions}

Tables~\ref{tab:PBPhotonRegionSelection}-\ref{tab:PB4lRegionSelection}
provide details of the PB approach region definitions. They are
grouped into categories based on the requirements of the presence of a
photon, regions targeting the \tttt process, two-lepton selections,
three-lepton selections with either the selection or veto of lepton
pairs consistent with an on-shell $Z$ boson, and four-lepton
selections.

\begin{table}[h]
\caption{Summary of selection requirements for the \PBonelGamma regions.}
\label{tab:PBPhotonRegionSelection}
\begin{adjustbox}{width=\columnwidth,center}{\renewcommand{\arraystretch}{1.5}
\begin{tabular}{ c|cccccc }
\toprule
Variable & \PBonelGammaDecay & \PBonelGammaProd & \PBonelGammaFakes & \PBonelGammaOther \\
\midrule
Preselection & \multicolumn{4}{c}{\PreselOnelY} \\

\nbjets  & \multicolumn{4}{c}{$\geq1 @ 70\%$}\\
$|m_{\ell\gamma}-m_{Z}|$ & \multicolumn{4}{c}{$\geq 10$~\GeV} \\
& \ttYdecayNN$>0.25$ & \multicolumn{2}{|c|}{\ttYdecayNN$<0.15$} &  (\ttYdecayNN$>0.15$ \&\& \ttYdecayNN$<0.25$)~~|| \\

NN cuts & --- & \multicolumn{1}{|c}{\ttYfakesNN$<0.2$} & \multicolumn{1}{|c|}{\ttYfakesNN$>0.2$} &  (\ttYotherNN $>0.4$ \&\& \ttYdecayNN $>0.25$)~~|| \\

&  \ttYotherNN$<0.4$ & \multicolumn{2}{|c|}{\ttYotherNN$<0.5$} & (\ttYotherNN  $>0.5$) \\
\bottomrule
\end{tabular}
}
\end{adjustbox}
\end{table}

\begin{table}
\caption{Summary of event selection requirements for the PB approach \PBtwolOppSignGamma photon region.}
\label{tab:OBPhotonRegions2}
\begin{center}{\renewcommand{\arraystretch}{1.2}
\begin{tabular}{ c|c }
\toprule
Variable & \PBtwolOppSignGamma\\
\midrule
Preselection & \PreselTwolOSY \\
\mllSFmz & $\geq  5$~\GeV\\
\met & $\geq 30$~\GeV \\
\bottomrule
\end{tabular}
}
\end{center}
\end{table}

\begin{table}[h]
\caption{Selection requirements for the \PBtttt selection which includes events from both \PreselTwolSS and \PreselThreel preselections.}
\label{tab:PB_tttt_only}
\begin{adjustbox}{width=0.6\columnwidth,center}{\renewcommand{\arraystretch}{1.5}
\begin{tabular}{ c|c }
\toprule
Variable & \PBtttt \\
\midrule
Preselection &  \PreselTwolSS ||  \PreselThreel  \\

\ncjets & $\geq6$ \\

\nbjets & $\geq2 @ 77\%$ \\

$H_{\mathrm{T}}$ & $\geq500$~\GeV \\
{\ensuremath{GNN_{\tttt}}} & $>0.6$ \\
Region Exclusions & None \\
\bottomrule
\end{tabular}
}
\end{adjustbox}
\end{table}

\begin{table}[h]
\caption{Summary of selection requirements for the \PBtwolSS regions. The exclusion of \PBtttt in the final row refers to the \PBtttt selection in Table~\ref{tab:PB_tttt_only}. }
\label{tab:PB2lSSRegionSelection}
\begin{adjustbox}{width=\columnwidth,center}{\renewcommand{\arraystretch}{1.5}
\begin{tabular}{ c|ccccc }
\toprule
Variable & \PBtwolSSttWj & \PBtwolSSttH & \PBtwolSSttW & \PBtwolSStH & \PBtwolSSOther\\
\midrule
Preselection & \multicolumn{5}{c}{\PreselTwolSS}\\

\njets & \ntjets$\geq4$ & \multicolumn{4}{|c}{\ncjets$\geq3$}\\

\nbjets  & $\geq1 @ 77\%$ & \multicolumn{4}{|c}{$\geq1 @ 85\%$}\\

\ptlep{1}\;/\;\ptlep{2} & $\geq80$~\GeV\;/\;$\geq40$~\GeV & \multicolumn{4}{|c}{---}\\

\ptjet{1} & $\geq60$~\GeV & \multicolumn{4}{|c}{---}\\

\mllSF & $\geq125$~\GeV & \multicolumn{4}{|c}{---}\\

\htjets & $\geq250$~\GeV & \multicolumn{4}{|c}{---}\\

\detaj & $>2.0$ & \multicolumn{4}{|c}{---}\\

\ttHtwolBDT{X} & --- & \ttHtwolBDT{\ttH} & \ttHtwolBDT{\ttW} & \ttHtwolBDT{\tH} & \ttHtwolBDT{\text{Other}} \\

\multirow{2}{*}{Region Exclusions} &
\multirow{2}{*}{\PBtttt} &
\multicolumn{4}{|c}{\PBtttt} \\
& & \multicolumn{4}{|c}{\PBtwolSSttWj} \\
\bottomrule
\end{tabular}
}
\end{adjustbox}
\end{table}

\begin{table}[h]
\caption{Summary of selection requirements for the \PBthreelOnZ regions.}
\label{tab:PB3lonZRegionSelection}
\begin{adjustbox}{width=\columnwidth,center}{\renewcommand{\arraystretch}{1.5}
\begin{tabular}{ c|ccccc }
\toprule
Variable & \PBthreelOnZtZqSR & \PBthreelOnZtZqCR & \PBthreelOnZ{\ttZ\text{-}\ttZ} & \PBthreelOnZttZtZq & \PBthreelOnZttZWZ\\
\midrule
Preselection & \multicolumn{5}{c}{\PreselThreel}\\

\mllmz & \multicolumn{5}{c}{<10~\GeV}\\

\njets & \multicolumn{2}{c|}{$2\leq$ \ntjets$\leq5$,} & \multicolumn{3}{c}{$3\leq$ \ncjets}\\

\nbjets & \multicolumn{2}{c|}{$1-2 @ 85\%\;(\mathrm{excluding}\,2j2b)$} & \multicolumn{3}{c}{$\geq1 @ 85\%$} \\

\tZqNN & $>0.7$ & $\leq0.7$ & --- & --- & ---\\

\ttZNN{\tZq} & --- & --- & $<0.43$ & $\geq0.43$ & ---\\

\ttZNN{\WZ} & --- & --- & $<0.27$ & $<0.27$ & $\geq0.27$\\

\multirow{3}{*}{Region Exclusions} & \multirow{3}{*}{---} & \PBthreelOnZttZttZ & \multicolumn{3}{|c}{\multirow{3}{*}{\PBthreelOnZtZqSR}}\\
&     & \PBthreelOnZttZtZq & \multicolumn{3}{|c}{}\\
&     & \PBthreelOnZttZWZ  & \multicolumn{3}{|c}{}\\
\bottomrule
\end{tabular}
}
\end{adjustbox}
\end{table}

\begin{sidewaystable}[h]
\caption{Summary of selection requirements for the \PBthreelOffZ regions. Requirements marked with an asterisk should be interpreted as a logical ``or'' of the corresponding selections in that column.
The exclusion of \PBtttt in the final row refers to the \PBtttt selection in Table~\ref{tab:PB_tttt_only}. }
\label{tab:PB3lnoZRegionSelection}
\begin{adjustbox}{width=1.0\columnwidth,center}{\renewcommand{\arraystretch}{1.5}
\begin{tabular}{ c|cccccccccc }
\toprule
Variable & \PBthreelOffZee & \PBthreelOffZmumu & \PBthreelOffZttH & \PBthreelOffZtH & \PBthreelOffZttW & \PBthreelOffZttZ & \PBthreelOffZVV & \PBthreelOffZtt & \PBthreelOffZOther\\
\midrule
Preselection & \multicolumn{9}{c}{\PreselThreel}\\

\mllmz & \multicolumn{9}{c}{$\geq10$~\GeV}\\

\mll & \multicolumn{2}{c|}{$\geq 101$~\GeV} & \multicolumn{4}{c|}{$\leq 81$~\GeV} & --- & --- & ---  \\

Lepton Flavour & ee & \(\mu\mu\)  & --- & --- & --- & --- & --- & --- & ---  \\

\ncjets & \multicolumn{9}{c}{$\geq2$}\\

\nbjets & \multicolumn{9}{c}{$\geq1 @ 85\%$}\\

\ttHthreelBDT{\ttH} & $>0.2$* & $>0.2$* & $>0.2$ & \multicolumn{6}{|c}{$\leq0.2$}\\

\ttHthreelBDT{\tH}  & --- & --- & --- & $>0.25$ & \multicolumn{5}{|c}{$\leq0.25$}\\

\ttHthreelBDT{\ttW}  & $>0.3$* & $>0.3$* & --- & --- & $>0.3$ & \multicolumn{4}{|c}{$\leq0.3$}\\

\ttHthreelBDT{\ttZ}  & $>0.45$* & $>0.45$* & --- & --- & --- & $>0.45$ &\multicolumn{3}{|c}{$\leq0.45$}\\

\ttHthreelBDT{\VV} & --- & --- & --- & --- &  --- &  --- &  $>0.65$ & \multicolumn{2}{|c}{$\leq0.65$}\\

\ttHthreelBDT{\ttbar} & --- & --- & --- & ---  & --- &  --- & --- & $>0.25$ & $\leq0.25$\\

Region Exclusions & \multicolumn{9}{c}{\PBtttt}\\
\bottomrule
\end{tabular}
}
\end{adjustbox}
\end{sidewaystable}

\begin{table}[h]
\begin{center}
\caption{Summary of selection requirements for the \PBfourl regions.}
\label{tab:PB4lRegionSelection}
{\renewcommand{\arraystretch}{1.2}
\begin{tabular}{ c|cccc }
\toprule
Variable & \PBfourlSF & \PBfourlDF & \PBfourlZZ & \PBfourloffZ \\
\midrule
Preselection & \multicolumn{4}{c}{\PreselFourl} \\

$\ell\ell^{\text{non-Z}}$ & $e^{+}e^{-}$ or $\mu^{+}\mu^{-}$ & $e^{\pm}\mu^{\mp}$ & $e^{+}e^{-}$ or $\mu^{+}\mu^{-}$ & --- \\

\ttZNN{4lttZ} & $\geq 0.4$ & --- & $\leq 0.4$ & --- \\

\mllmz & \multicolumn{3}{c|}{<10~\GeV} & $\geq10$~\GeV \\
\bottomrule
\end{tabular}
}
\end{center}
\end{table}

% End of text imported from the .//Sections/Appendices/PBRegions.tex input file

\FloatBarrier

% The next lines are included from the .//Sections/Appendices/NNCuts.tex input file

Table~\ref{tab:SMRegionMLSplits} provides the details of the ML
classifiers, their type, and the cuts applied on their output values used
to define the PB approach regions.

\begin{table}[h]
\caption{Details of the ML classifiers, the classifier output used and the corresponding cut on that output score for the PB signal regions. Where multiply cuts are specified the classifier output is split into more than two regions.}
\label{tab:SMRegionMLSplits}
\begin{adjustbox}{width=0.8\columnwidth,center}{\renewcommand{\arraystretch}{1.5}
\begin{tabular}{ cccc }
\toprule
Region             & Target Process & Classifier output & Optimised Score Cut(s) \\
\midrule
\PBtttt      & \tttt    & {\ensuremath{GNN_{\tttt}}}              & 0.60      \\
\PBtwolSSttHplus   & \ttH     & {\ensuremath{BDT_{\ttH-2\ell\mathrm{SS}}^{\mathbb{P}({\ttH})}}\xspace}                    & 0.46   \\
\PBtwolSSttHminus  & \ttH     & {\ensuremath{BDT_{\ttH-2\ell\mathrm{SS}}^{\mathbb{P}({\ttH})}}\xspace}                    & 0.46  \\
\PBthreelOnZtZqSR  & \tZq     & {\ensuremath{NN_{\tZq}}}               & 0.85,0.94     \\
\PBthreelOnZttZttZ & \ttZ     & \ttZNN{\ttZ}                          & 0.6, 0.7, 0.8     \\
\PBthreelOnZttZtZq & \tZq     & \ttZNN{\ttZ}                          & 0.22     \\
\PBthreelOffZttH   & \ttH     & \ttHthreelBDT{\ttH}                    & 0.34     \\
\PBonelGammaProd   & \ttyprod & {\ensuremath{NN_{\ttyprod}}}           & 0.65    \\
\bottomrule
\end{tabular}
}
\end{adjustbox}
\end{table}
\FloatBarrier

% End of text imported from the .//Sections/Appendices/NNCuts.tex input file

%
%
\FloatBarrier

% The next lines are included from the .//Sections/Appendices/BackgroundEstimation.tex input file

\subsection{Common control region definitions}
\label{app:backgroundest}

Tables~\ref{app:tab:CommonControlRegionsHad}-\ref{app:tab:CommonControlRegionsConv}
provide details of the common control region definitions.

\begin{table}[h]
\footnotesize
\begin{center}
\caption{Summary of requirements for control regions targeting non-prompt lepton backgrounds from hadron decays.}
\label{app:tab:CommonControlRegionsHad}
{\renewcommand{\arraystretch}{1.2}
\begin{tabular}{ c|ccc }
\toprule
Variable & \FakeCRtwoSSHF & \FakeCRthreelLFe & \FakeCRthreelHF \\
\midrule
\nleps & 2 & \multicolumn{2}{|c}{3} \\

\nlooseleps & \multicolumn{3}{c}{1} \\

\ptlep{1} & \multicolumn{3}{c}{$\geq 27$~\GeV} \\

\ptlep{2} & \multicolumn{3}{c}{$\geq 15$~\GeV}\\

\ptlep{3} & --- & \multicolumn{2}{|c}{$\geq 10$~\GeV}\\

\mllSF & $\geq 12$~\GeV & --- & ---\\

\mllmz & --- & $>10$~\GeV & --- \\

\mllmin & --- & \multicolumn{2}{|c}{$\geq 10$~\GeV}\\

\modsumq & $\equiv 2$ & \multicolumn{2}{|c}{$\equiv 1$}\\

\ntjets & $\geq 3 $ & $\geq 2$ & {---}\\

\nbjets $@ 85\%$ & $\geq 1$ & 0 & $\geq 1$\\
\bottomrule
\end{tabular}
}
\end{center}
\end{table}

\begin{table}[h]
\footnotesize
\begin{center}
\caption{Summary of requirements for control regions targeting non-prompt lepton backgrounds from conversion.}
\label{app:tab:CommonControlRegionsConv}
{\renewcommand{\arraystretch}{1.2}
\begin{tabular}{ c|cccc }
\toprule
Variable & \FakeCRthreelIntConv & \FakeCRthreelExtConv & \FakeCRthreelPromptConv & \FakeCRfourlZeroB \\
\midrule
\nleps & \multicolumn{3}{c|}{2$\mu$, 1$e$} & 4\\

\ptlep{1} & \multicolumn{4}{c}{$\geq 27$~\GeV} \\

\ptlep{2} & \multicolumn{3}{c|}{$\geq 15$~\GeV} & $\geq 7$~\GeV \\

\ptlep{3} & \multicolumn{3}{c|}{$\geq 10$~\GeV} & $\geq 7$~\GeV \\

\ptlep{4} & \multicolumn{3}{c|}{---} & $\geq 7$~\GeV \\

\nossf & \multicolumn{3}{c|}{1} & $\geq 1$ \\

\mll & \multicolumn{3}{c|}{$\geq 20$~\GeV} & --- \\

\mllmz & \multicolumn{3}{c|}{$\geq 10$~\GeV} & --- \\

\mlllmz & \multicolumn{3}{c|}{$\leq 10$~\GeV} & --- \\

\mllmin & \multicolumn{3}{c|}{---} & $\geq 10$~\GeV \\

\nbjets $@ 85\%$ & \multicolumn{3}{c|}{$0$} & $0$\\

\met &  \multicolumn{3}{c|}{$\geq 50$~\GeV} & --- \\

Conversion Vertex & \multicolumn{2}{c|}{\cmark} & \xmark  & ---\\

Conversion \mll & \multicolumn{2}{c|}{$<100$~\MeV} & --- & ---\\

Conversion radius & $<20$ mm & $>20$ mm & --- & --- \\
\bottomrule
\end{tabular}
}
\end{center}
\end{table}

% End of text imported from the .//Sections/Appendices/BackgroundEstimation.tex input file

%
%
%
%

\FloatBarrier
\section{Pre-fit distributions}
\label{app:prefit}

% The next lines are included from the .//Sections/Appendices/PreFitPlots.tex input file

This section provides distributions from all analysis signal regions
before the fit to data. The purpose is to show the agreement between
data and the predictions to first give an indication of the quality of
the MC modelling and size of systematic uncertainties, and second to
have a reference for where possible tensions in the fits to data may arise.

Figures~\ref{fig:App:OB_Giga_2LSS_3LnoZ}-\ref{fig:App:OB_Giga_tty_4L_CR}
show the OB approach regions before the fit to data.

\begin{figure}[ht]
\centering
\subfloat[]{\includegraphics[width=0.75\linewidth]{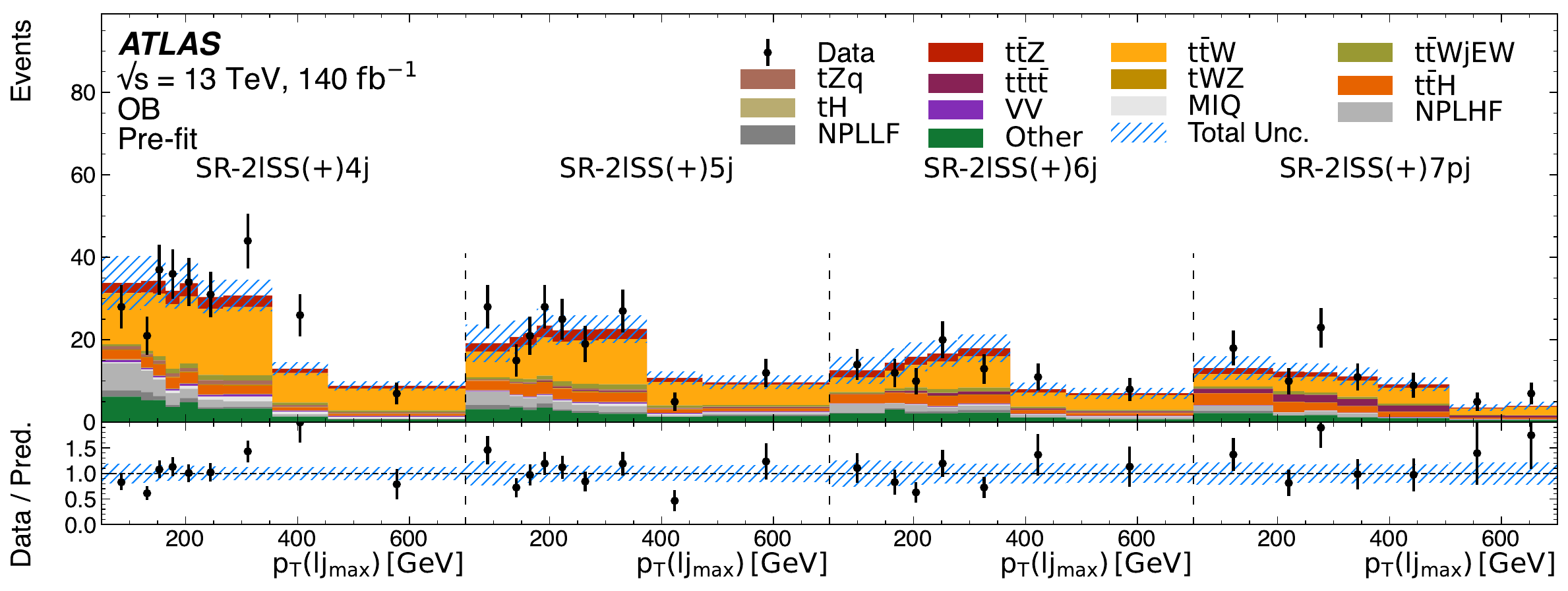}}\\
\subfloat[]{\includegraphics[width=0.75\linewidth]{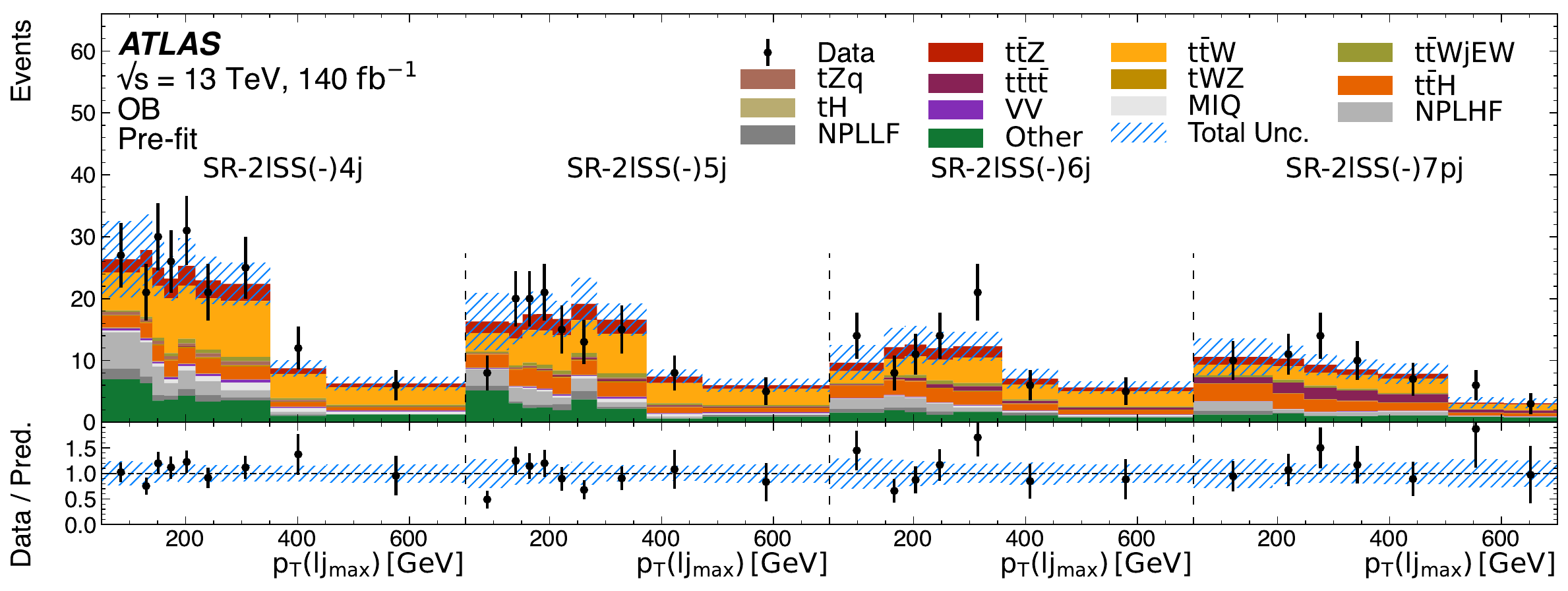}}\\
\subfloat[]{\includegraphics[width=0.75\linewidth]{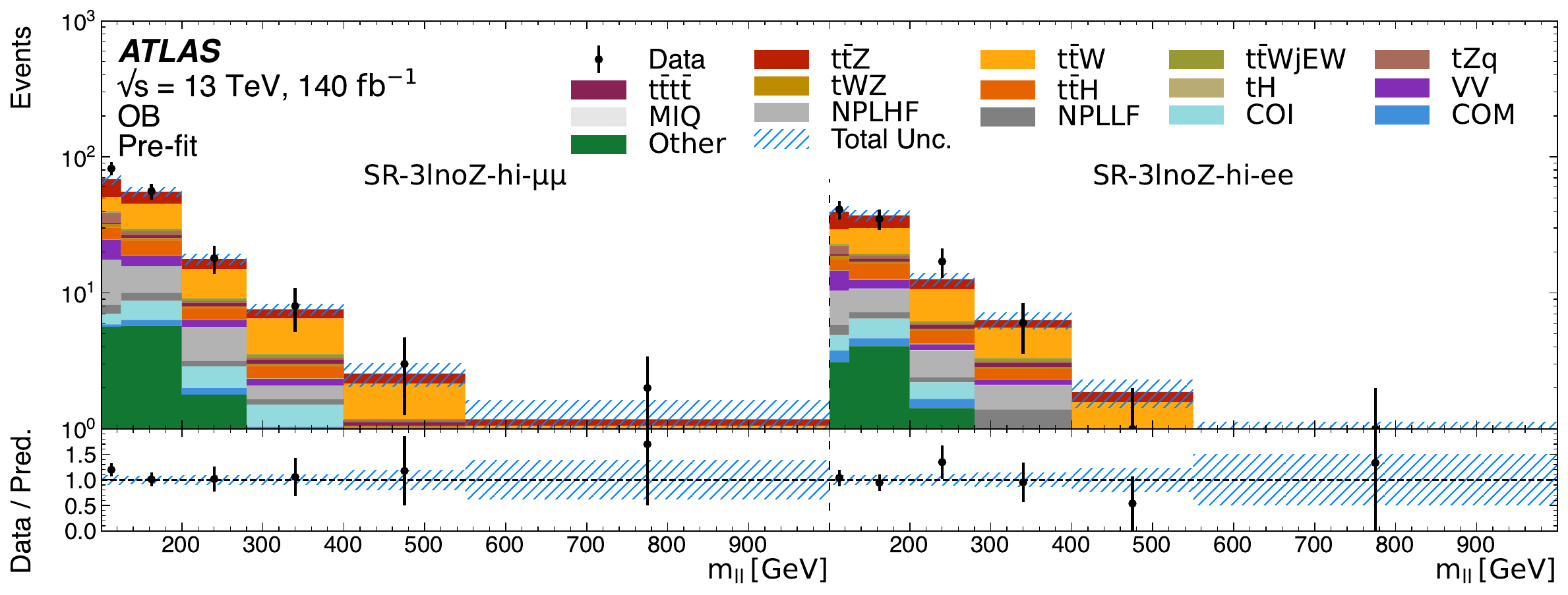}}\\
\subfloat[]{\includegraphics[width=0.75\linewidth]{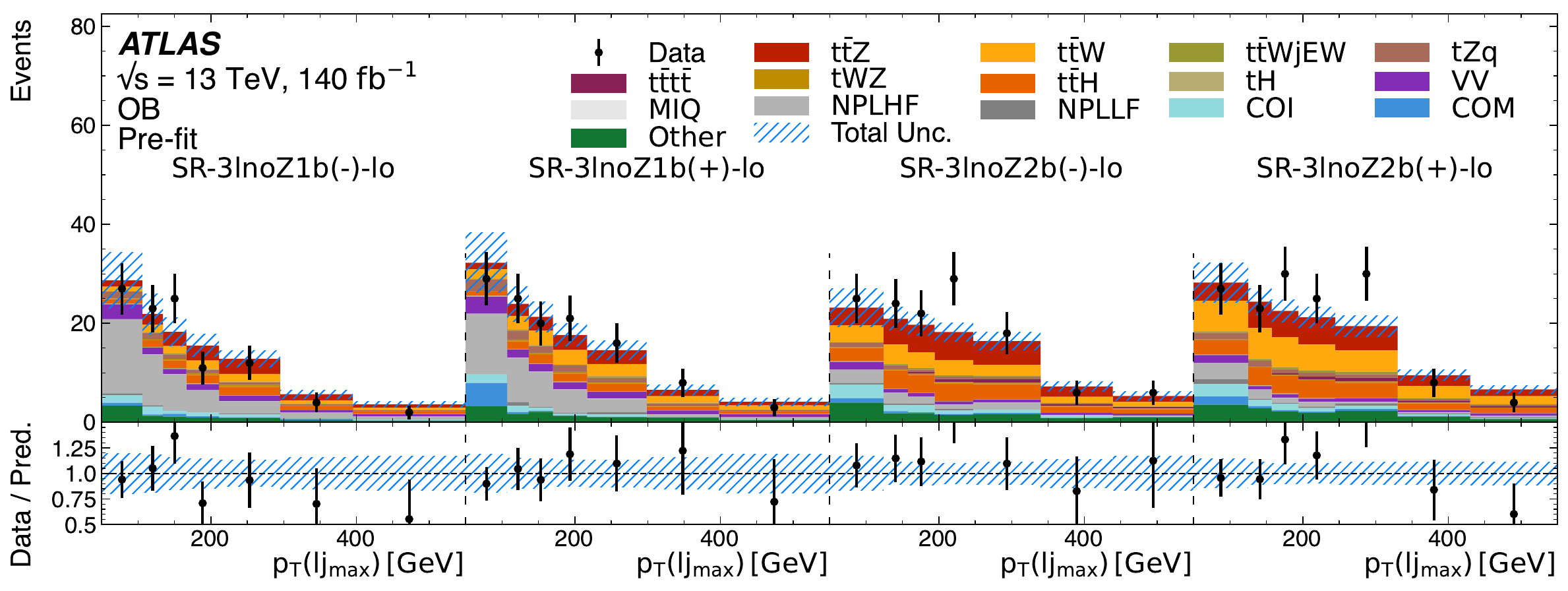}}\\
\caption{Distributions of the event yields observed in data compared to the expected yields before the fit to data for the
object-based (a) \OBtwolSSp, (b) \OBtwolSS-, (c) high-mass \threelnoZ , and (d) low-mass \OBthreelblo and \OBthreelbblo regions.
The EFT WCs are set to the SM (all equal to zero) and background normalisation factors and fit nuisance parameters are set to their pre-fit predictions.
\UncertaintyBand \OtherCategoryDescription
}
\label{fig:App:OB_Giga_2LSS_3LnoZ}
\end{figure}

\begin{figure}[!t]
\centering
\subfloat[]{\includegraphics[width=0.75\linewidth]{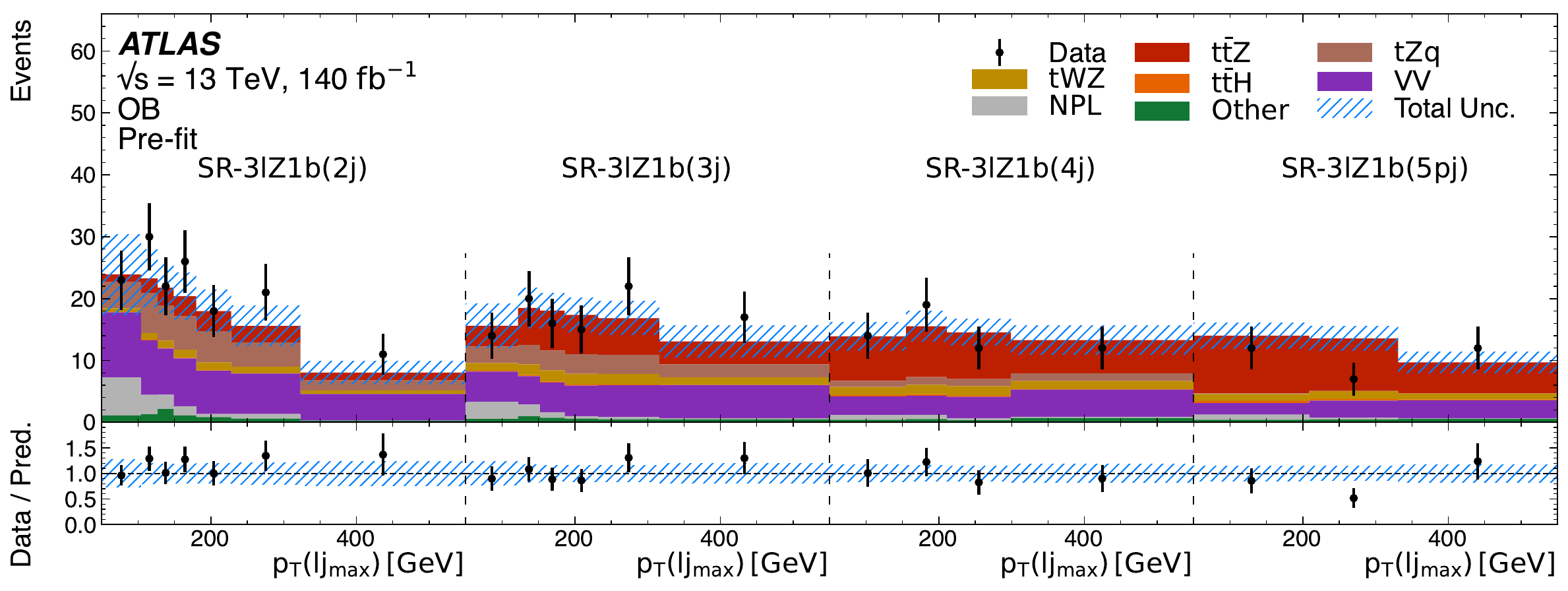}}\\
\subfloat[]{\includegraphics[width=0.75\linewidth]{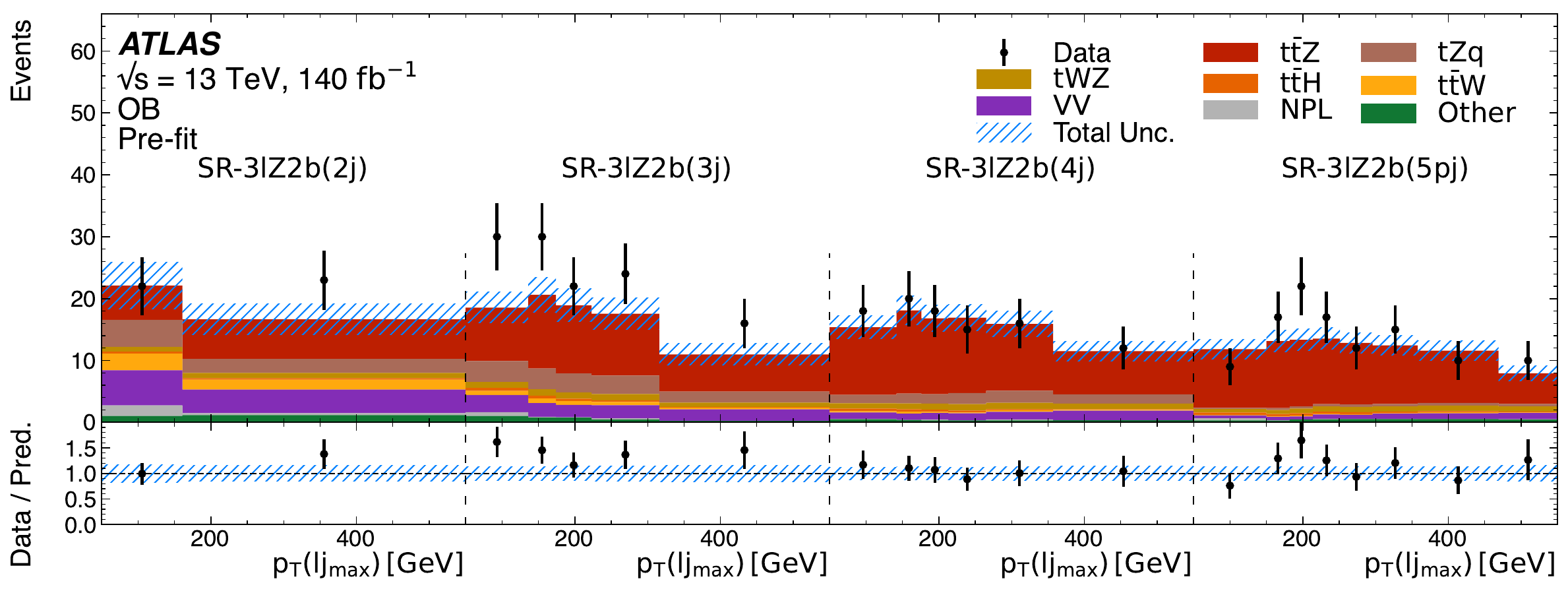}}\\
\subfloat[]{\includegraphics[width=0.75\linewidth]{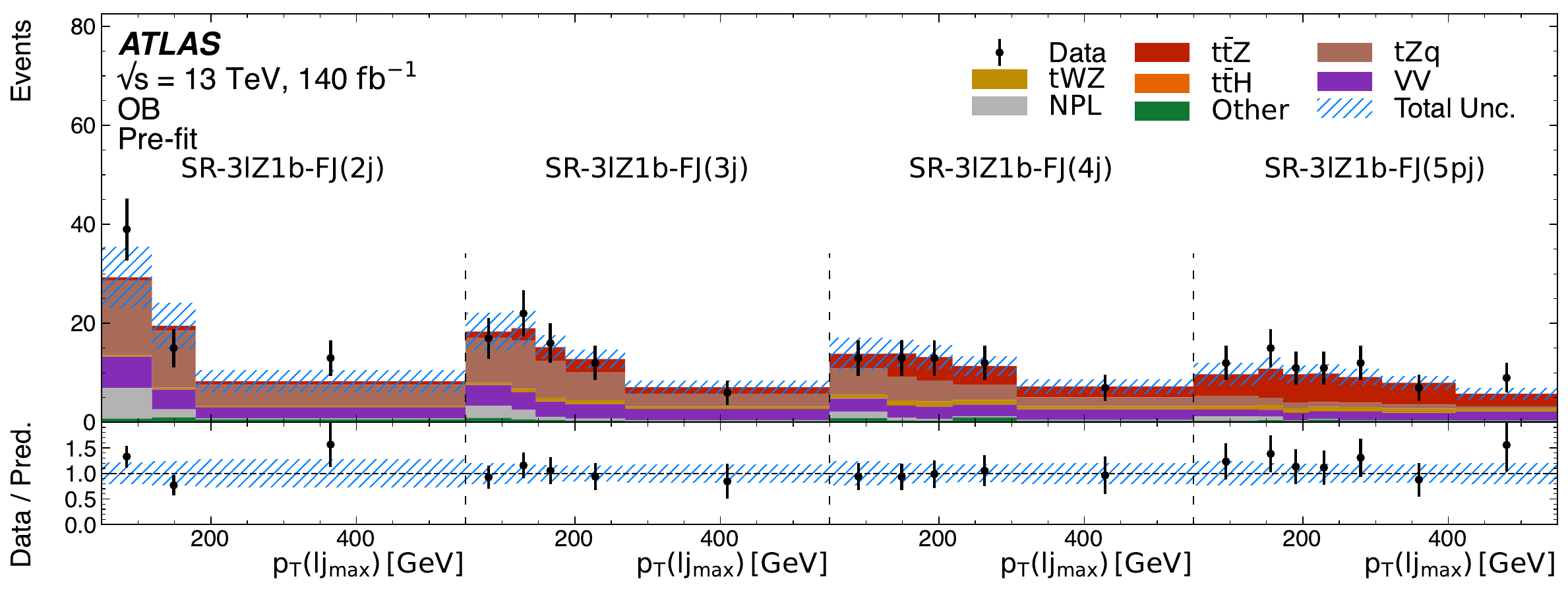}}\\
\subfloat[]{\includegraphics[width=0.75\linewidth]{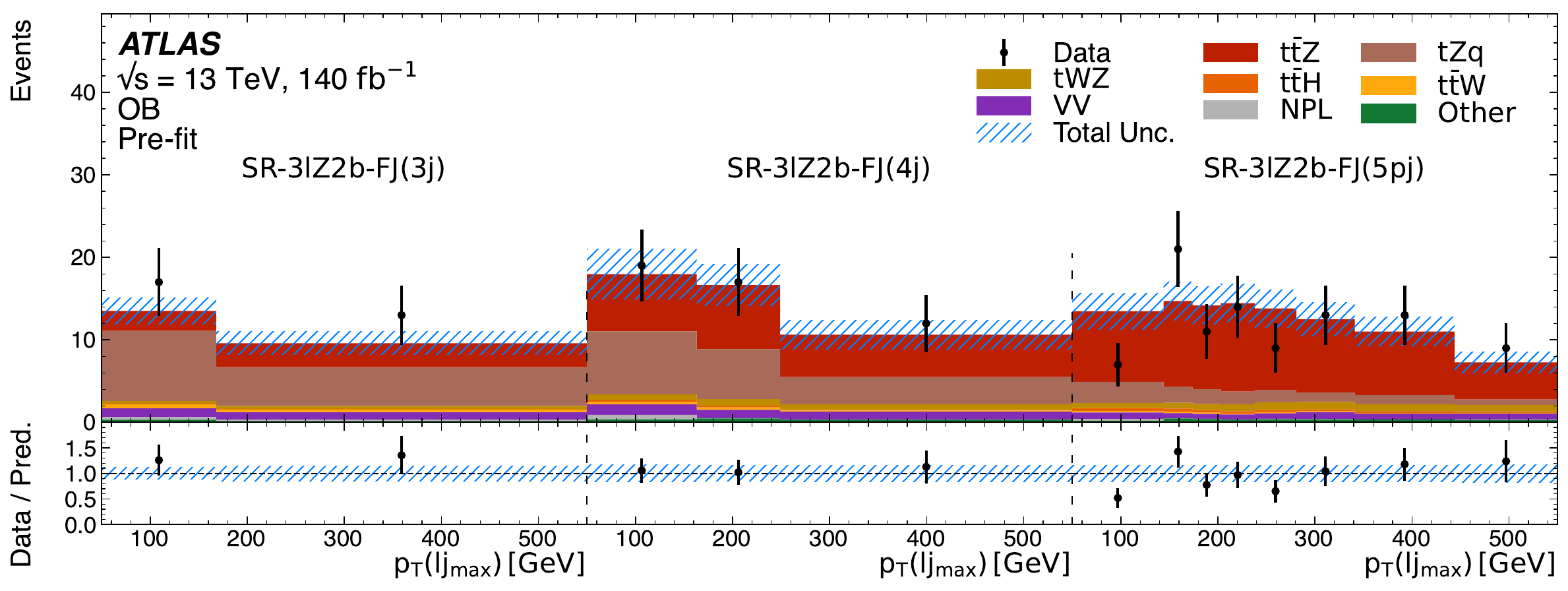}}\\
\caption{Distributions of the event yields observed in data compared to the expected yields before the fit to data for the
object-based (a) \OBthreelZb, (b) \OBthreelZbb, (c) \OBthreelZbforward and (d) \OBthreelZbbforward regions.
The EFT WCs are set to the SM (all equal to zero) and background normalisation factors and fit nuisance parameters are set to their pre-fit predictions.
\UncertaintyBand \OtherCategoryDescription
}
\label{fig:App:OB_Giga_3LZ}
\end{figure}

\begin{figure}[!t]
\centering
\subfloat[]{\includegraphics[width=0.75\linewidth]{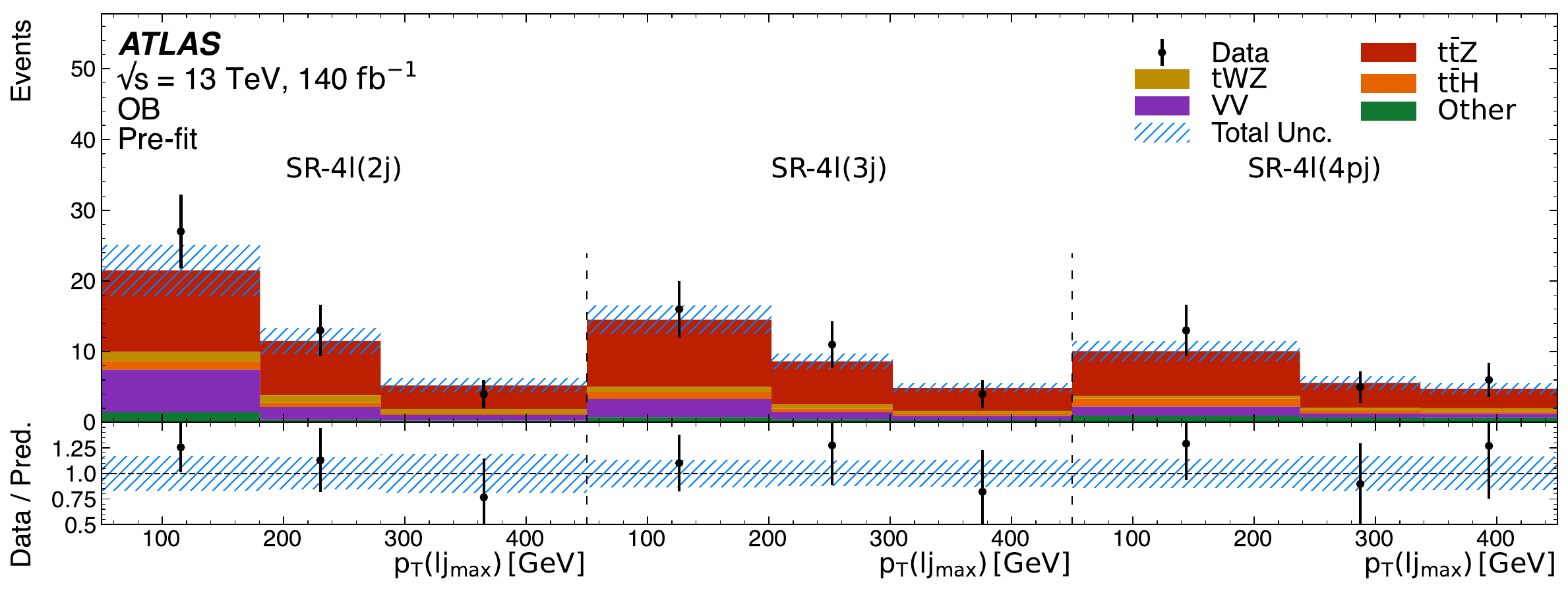}}\\
\subfloat[]{\includegraphics[width=0.75\linewidth]{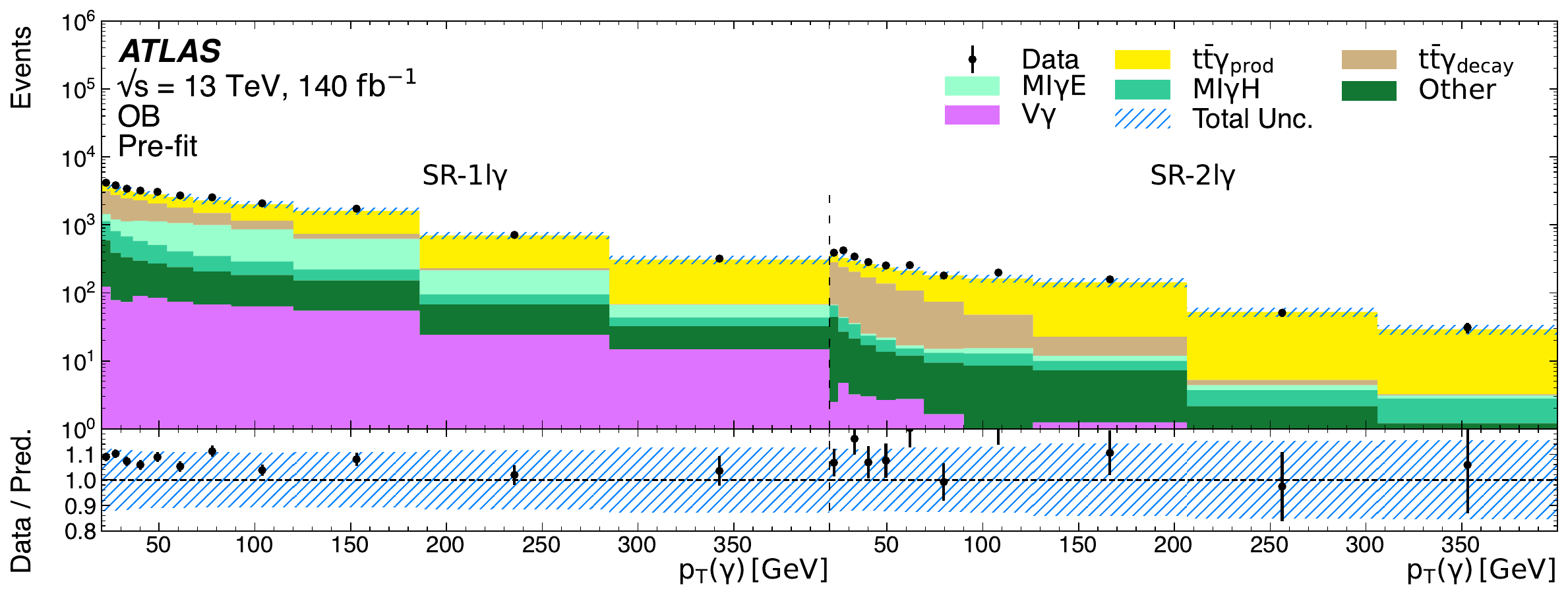}}\\
\subfloat[]{\includegraphics[width=0.75\linewidth]{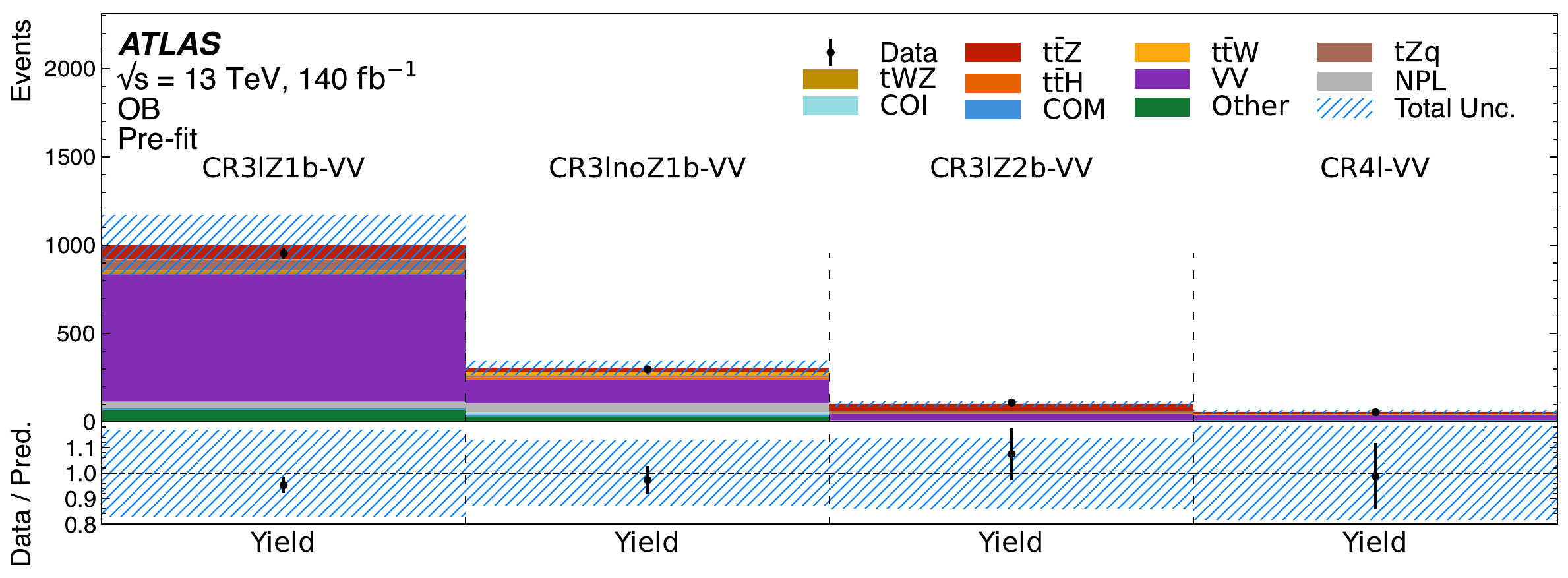}}
\caption{Distributions of the event yields observed in data compared to the expected yields before the fit to data for the
object-based (a) \OBfourl signal region, (b) \OBonely and  \OBtwoly signal regions and (c) $VV$ control regions.
The EFT WCs are set to the SM (all equal to zero) and background normalisation factors and fit nuisance parameters are set to their pre-fit predictions.
\UncertaintyBand \OtherCategoryDescription
}
\label{fig:App:OB_Giga_tty_4L_CR}
\end{figure}

Figures~\ref{fig:App:PreFit_PB_Giga_2LSS}-\ref{fig:App:Prefit_PB_Giga_gamma}
show the PB approach regions before the fit to data.

\begin{figure}[!t]
\centering
\subfloat[]{\includegraphics[width=0.75\linewidth]{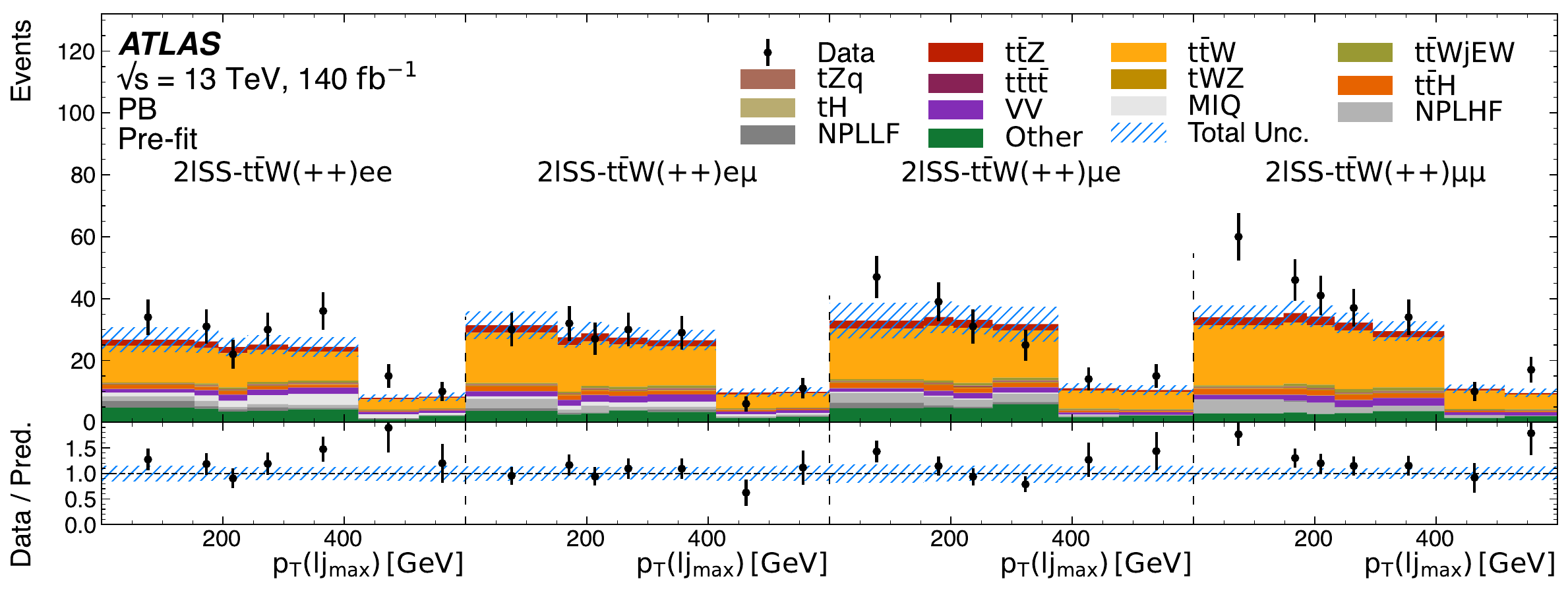}}\\
\subfloat[]{\includegraphics[width=0.75\linewidth]{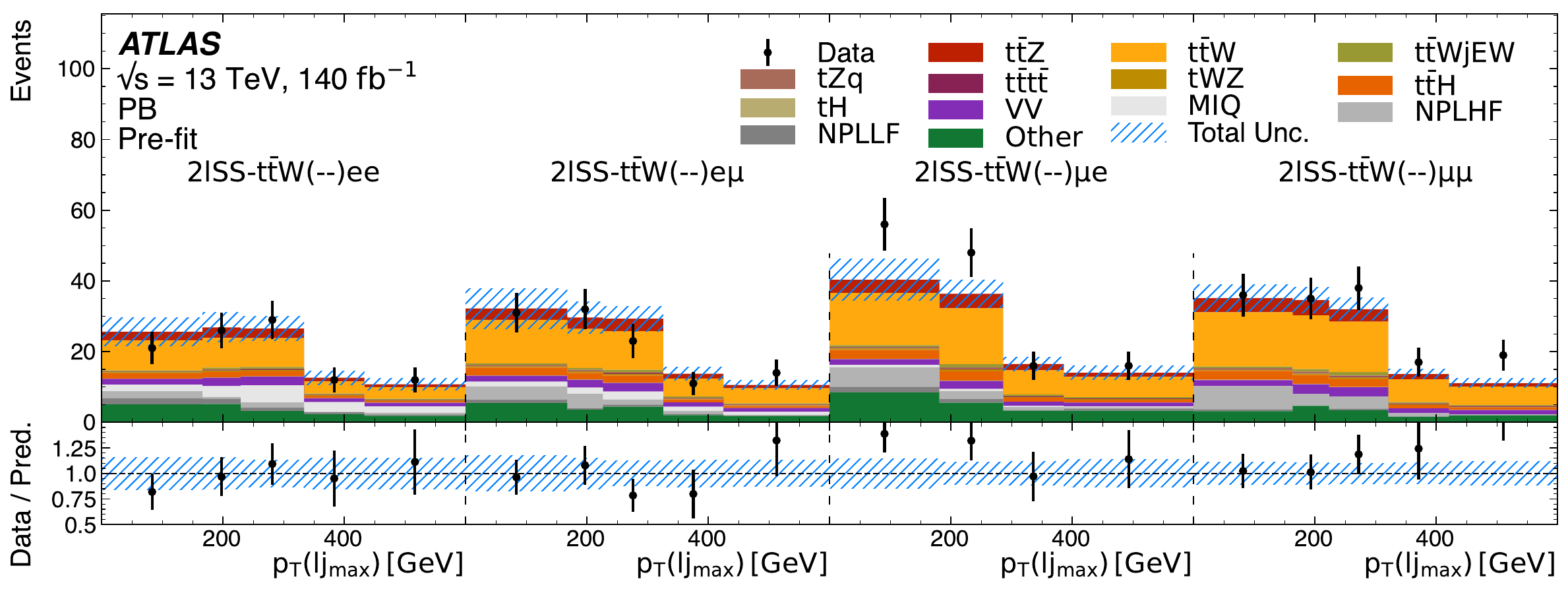}}\\
\subfloat[]{\includegraphics[width=0.75\linewidth]{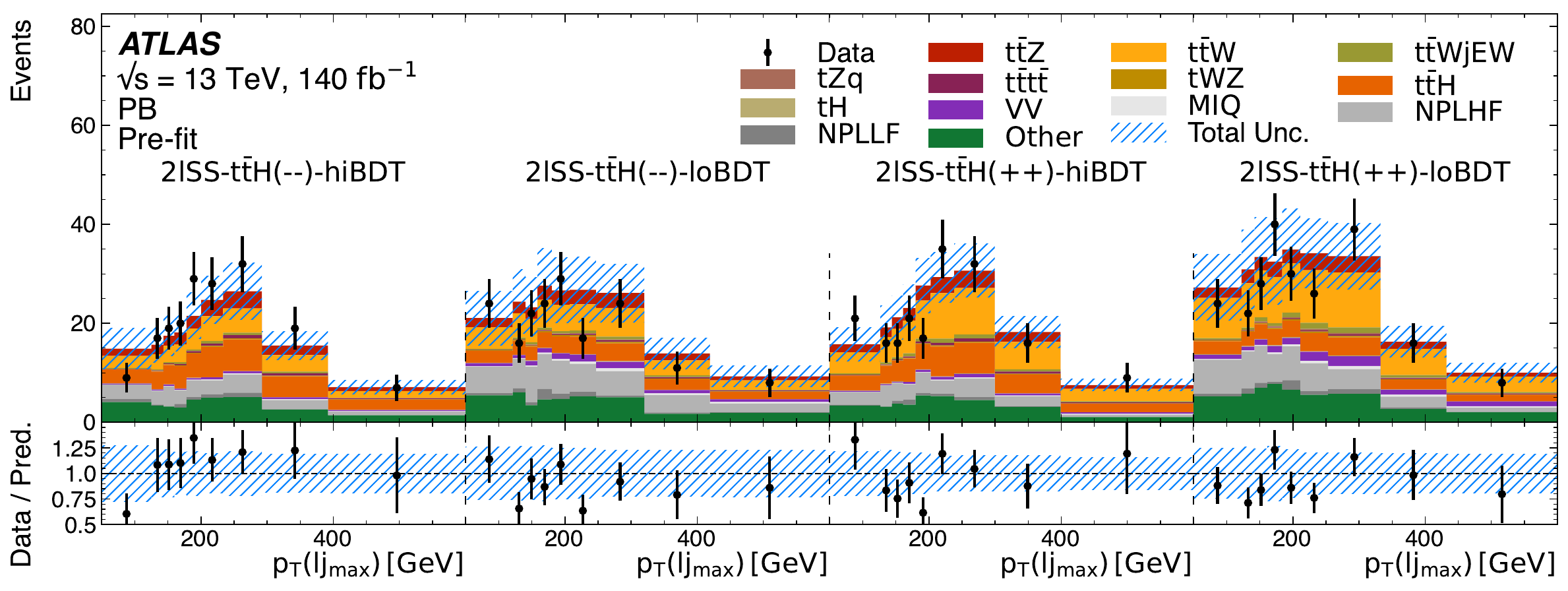}}\\
\subfloat[]{\includegraphics[width=0.75\linewidth]{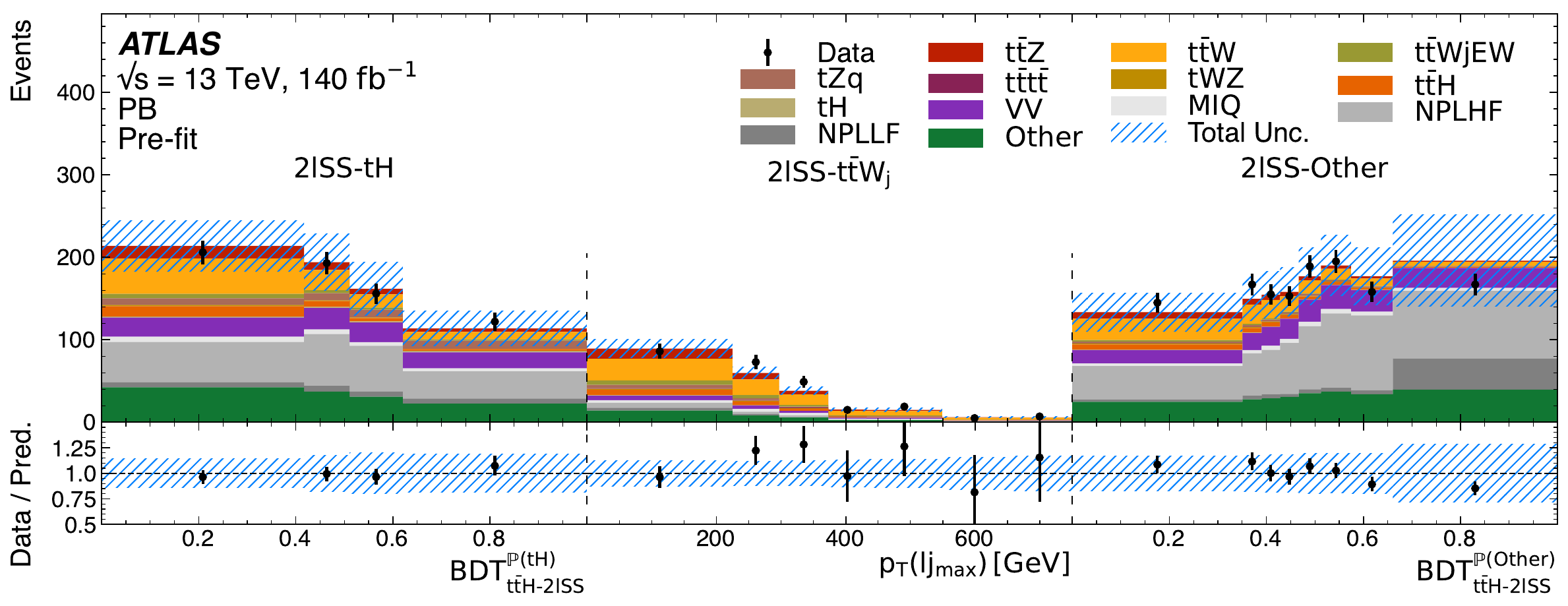}}\\
\caption{Distributions of the event yields observed in data compared to the expected yields before the fit to data for the
process-based (a) \PBtwolSSttWpp, (b) \PBtwolSSttWmm, (c) \PBtwolSSttH and (d) \PBtwolSStH and \PBtwolSSOther regions.
The EFT WCs are set to the SM (all equal to zero) and background normalisation factors and fit nuisance parameters are set to their pre-fit predictions.
\UncertaintyBand \OtherCategoryDescription
}
\label{fig:App:PreFit_PB_Giga_2LSS}
\end{figure}

\begin{figure}[!t]
\centering
\subfloat[]{\includegraphics[width=0.75\linewidth]{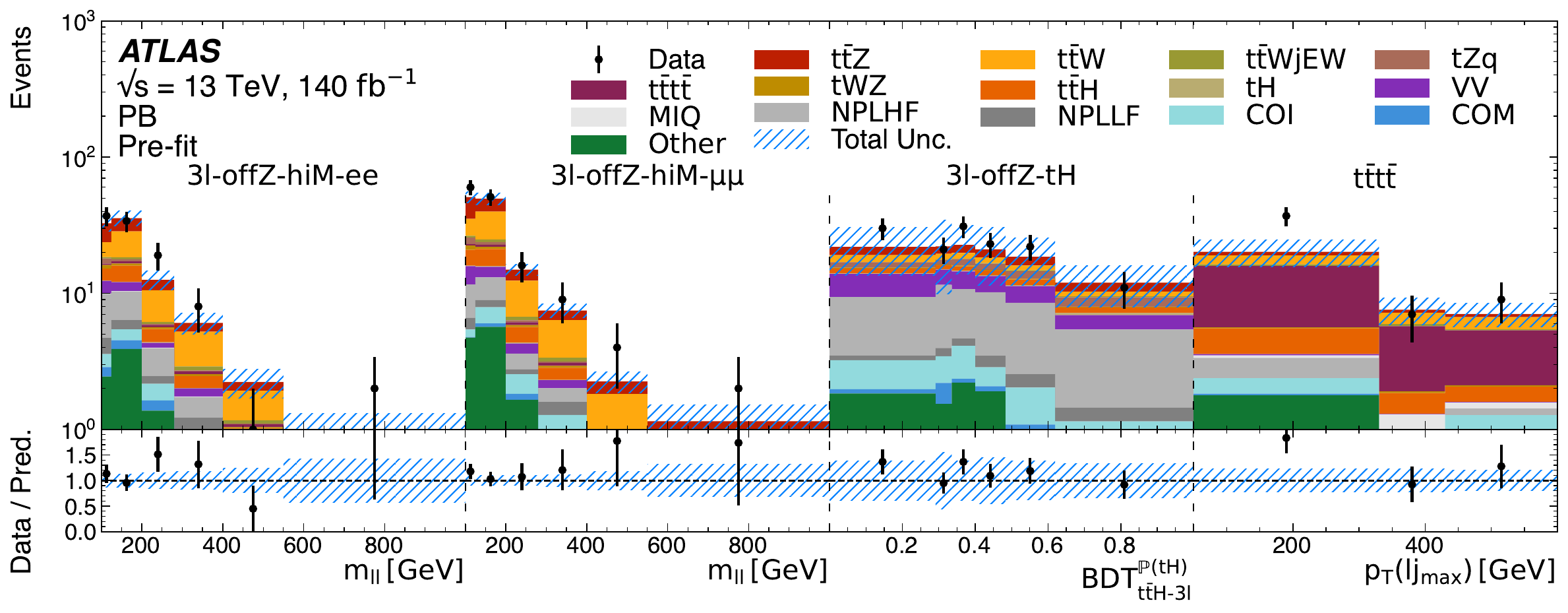}}\\
\subfloat[]{\includegraphics[width=0.75\linewidth]{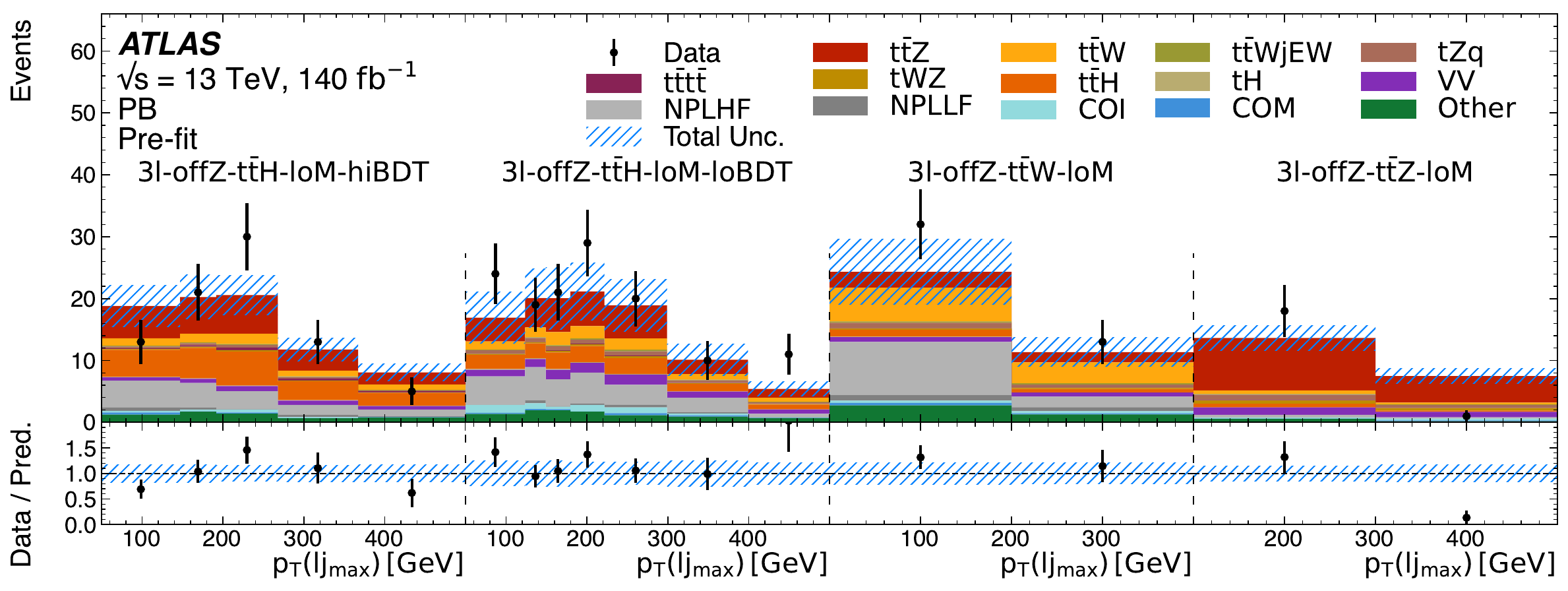}}\\
\subfloat[]{\includegraphics[width=0.75\linewidth]{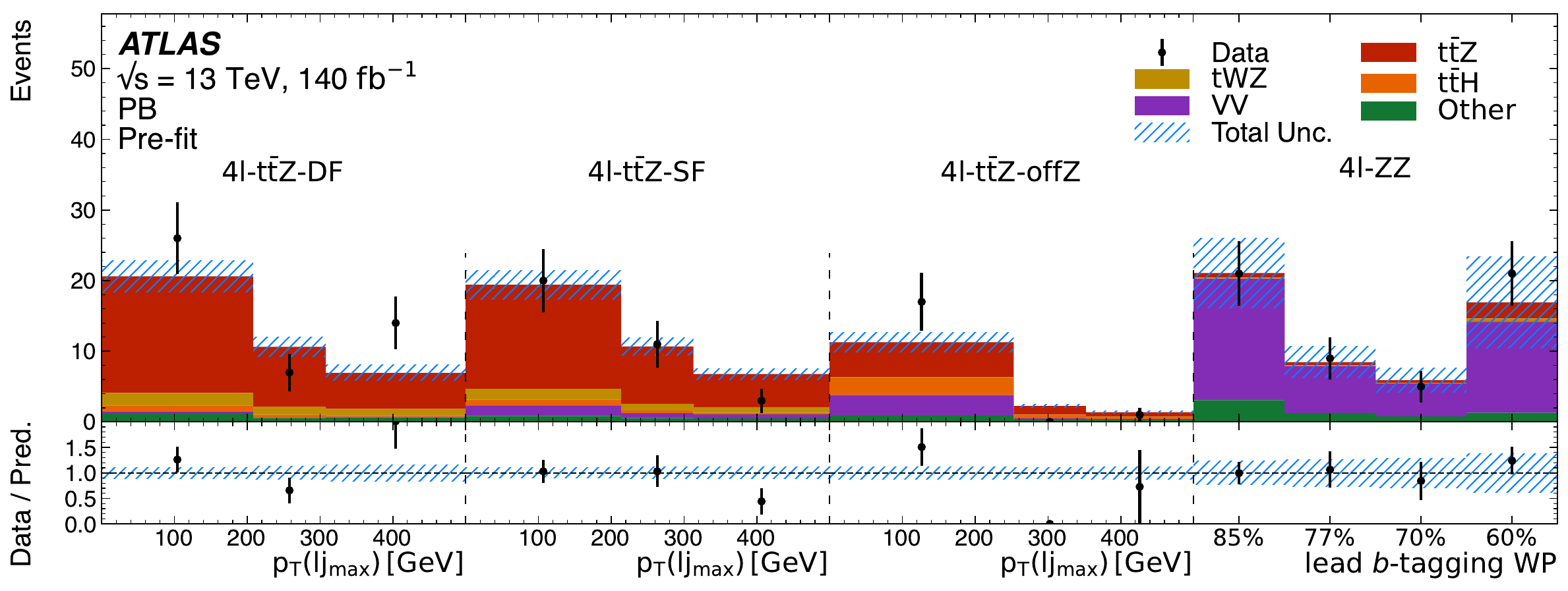}}\\
\subfloat[]{\includegraphics[width=0.75\linewidth]{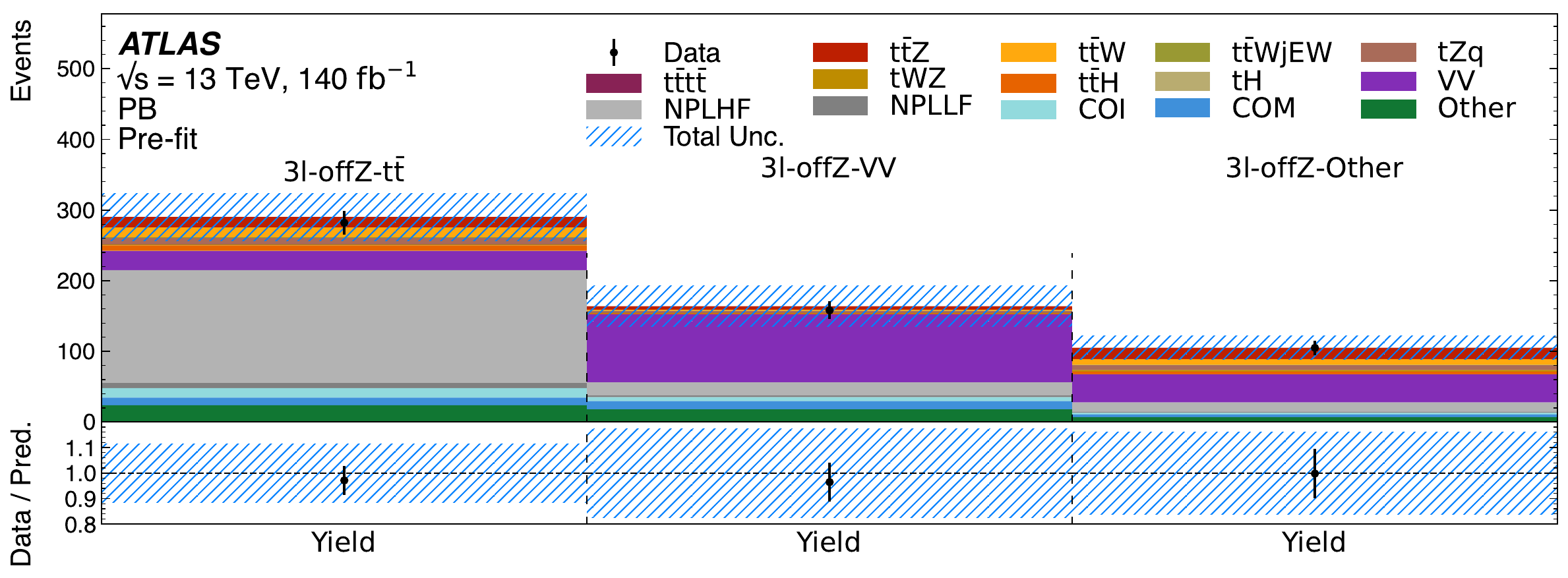}}\\
\caption{Distributions of the event yields observed in data compared to the expected yields before the fit to data for the
process-based (a) high-mass \PBthreelOffZnd, \PBthreelOffZ$tH$ and \tttt regions, (b) low-mass \PBthreelOffZnd regions, (c) \PBfourlnd regions and (d)~\PBthreelOffZ{}$t\bar{t}$, \PBthreelOffZ{}$VV$ and \PBthreelOffZ{}Other regions.
The EFT WCs are set to the SM (all equal to zero) and background normalisation factors and fit nuisance parameters are set to their pre-fit predictions.
\UncertaintyBand \OtherCategoryDescription
}
\label{fig:App:PreFit_PB_Giga_3Loff_4L}
\end{figure}

\begin{figure}[!t]
\centering
\subfloat[]{\includegraphics[width=0.75\linewidth]{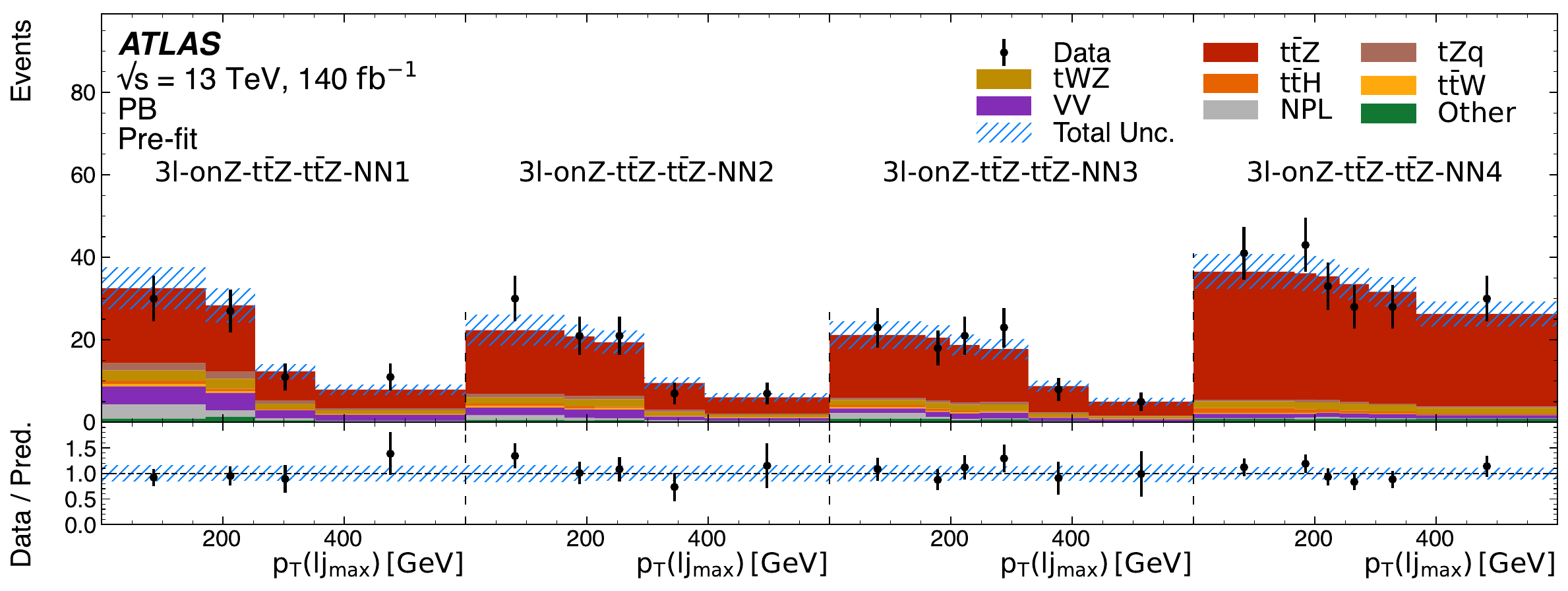}}\\
\subfloat[]{\includegraphics[width=0.75\linewidth]{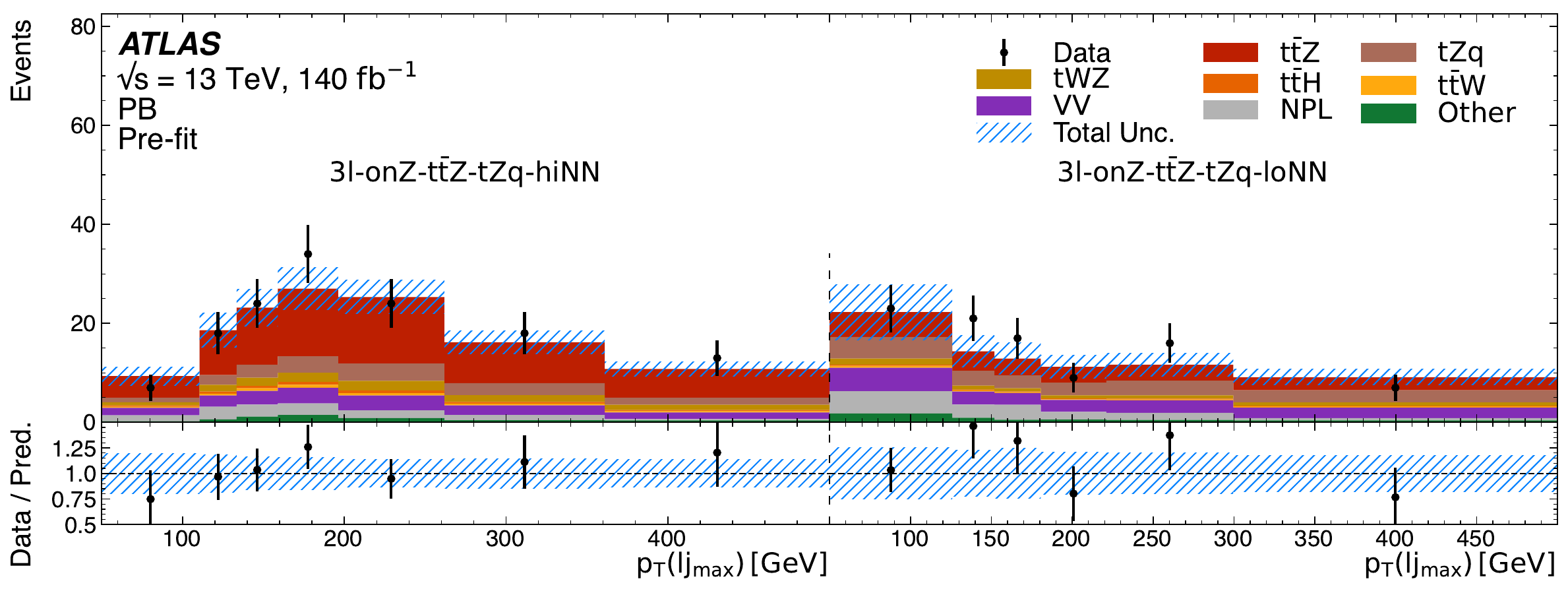}}\\
\subfloat[]{\includegraphics[width=0.75\linewidth]{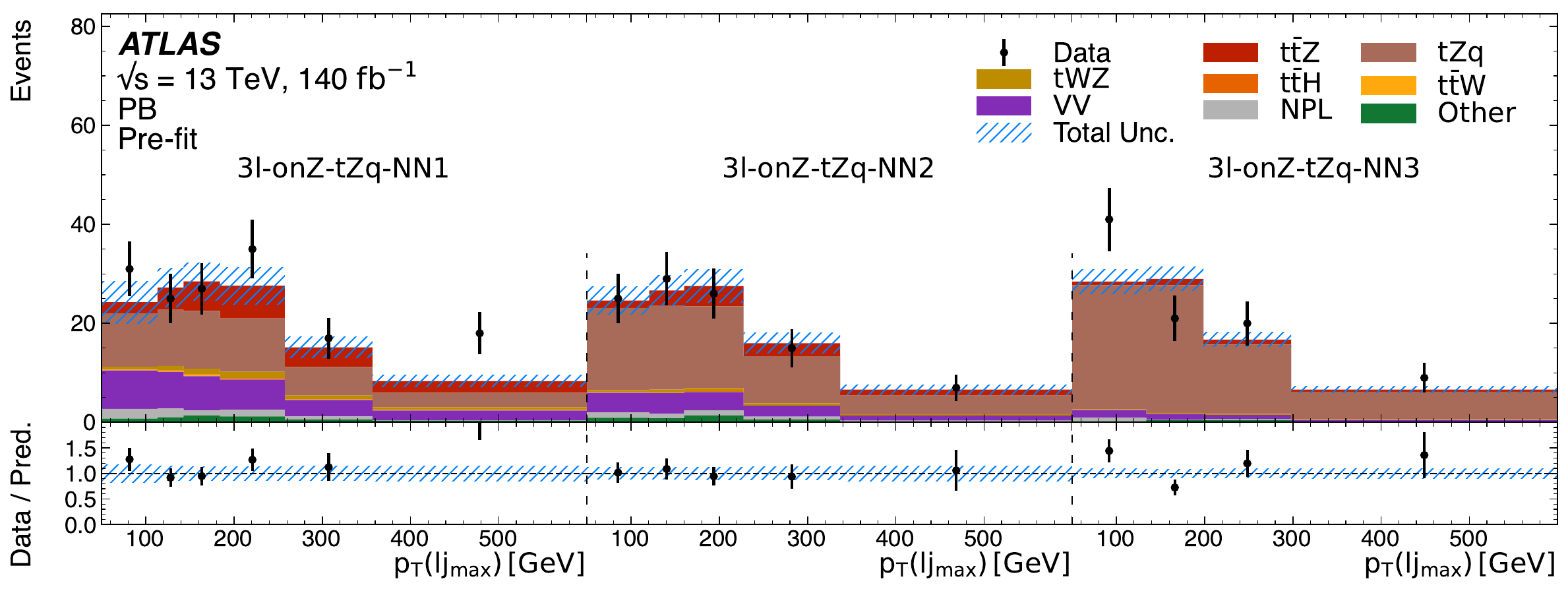}}\\
\subfloat[]{\includegraphics[width=0.75\linewidth]{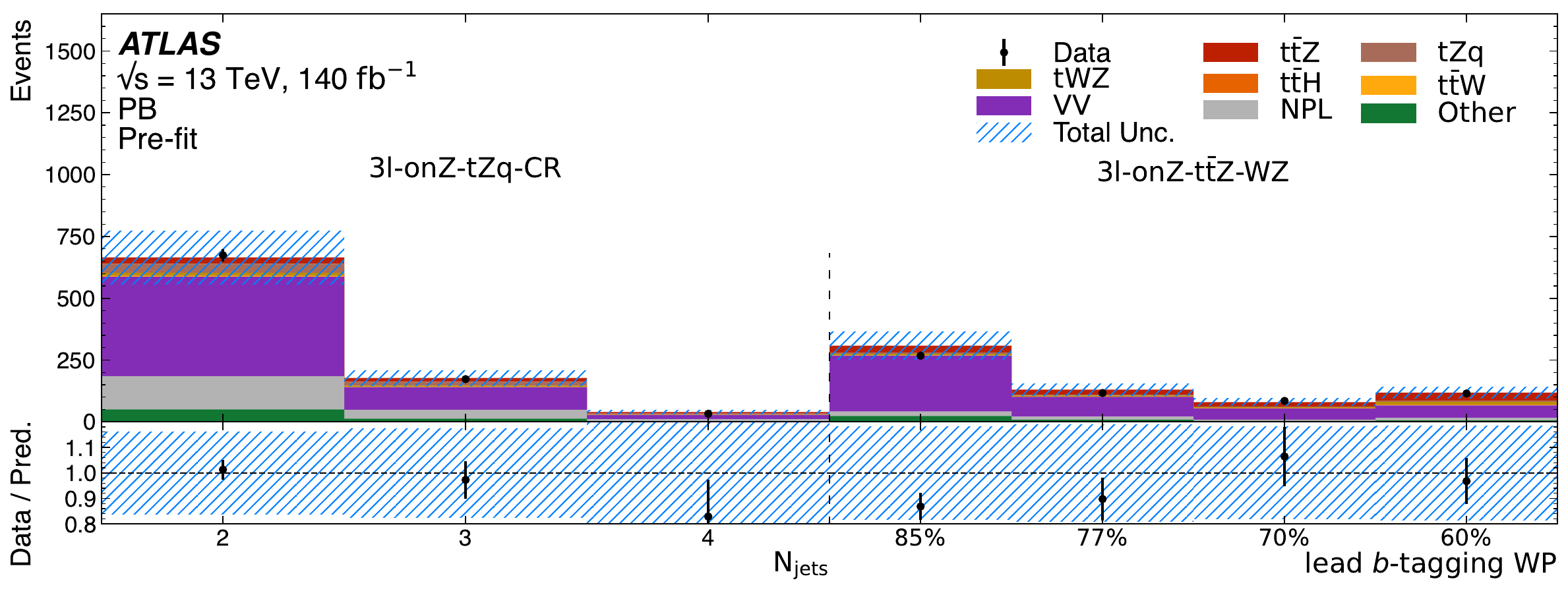}}\\
\caption{Distributions of the event yields observed in data compared to the expected yields before the fit to data for the
process-based (a) \PBthreelOnZ\ttZ, (b) \PBthreelOnZ\ttZ-\tZq, (c) \PBthreelOnZtZqSR and (d) other \PBthreel regions.
The EFT WCs are set to the SM (all equal to zero) and background normalisation factors and fit nuisance parameters are set to their pre-fit predictions.
\UncertaintyBand \OtherCategoryDescription
}
\label{fig:App:PreFit_PB_Giga_3LonZ}
\end{figure}

\begin{figure}[!t]
\centering
\subfloat[]{\includegraphics[width=0.75\linewidth]{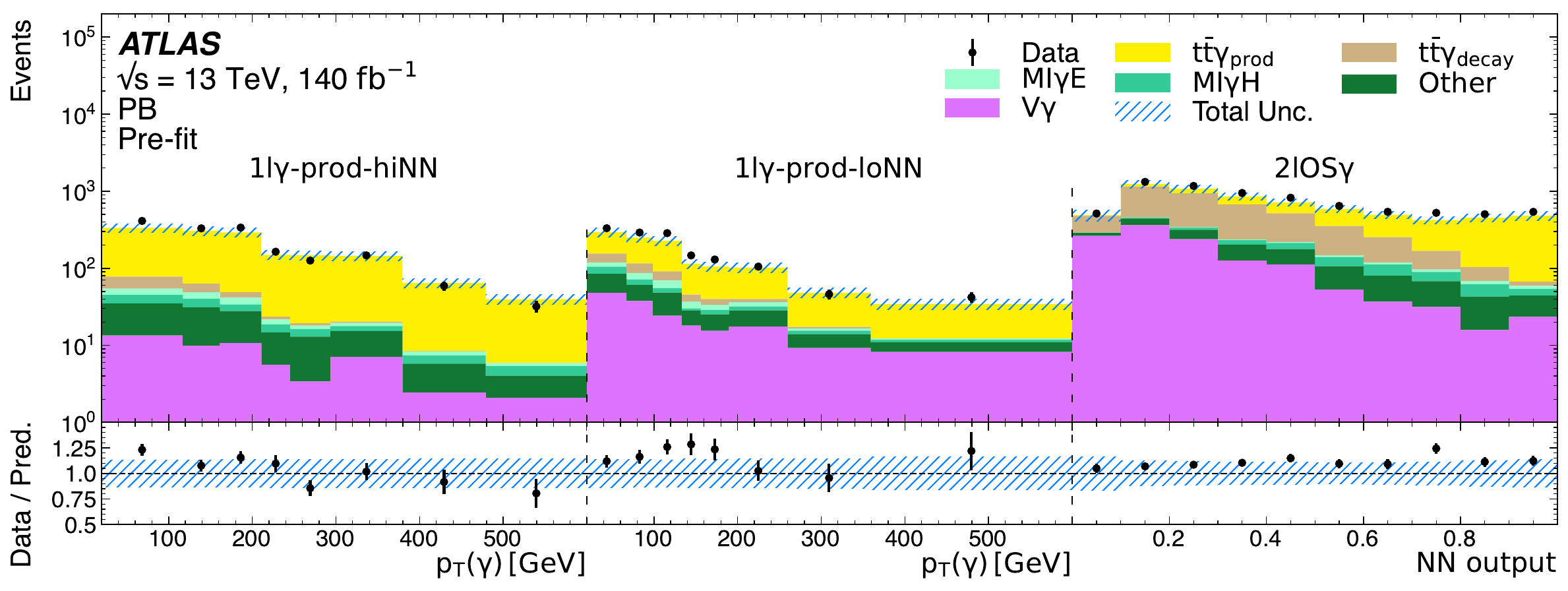}}\\
\subfloat[]{\includegraphics[width=0.75\linewidth]{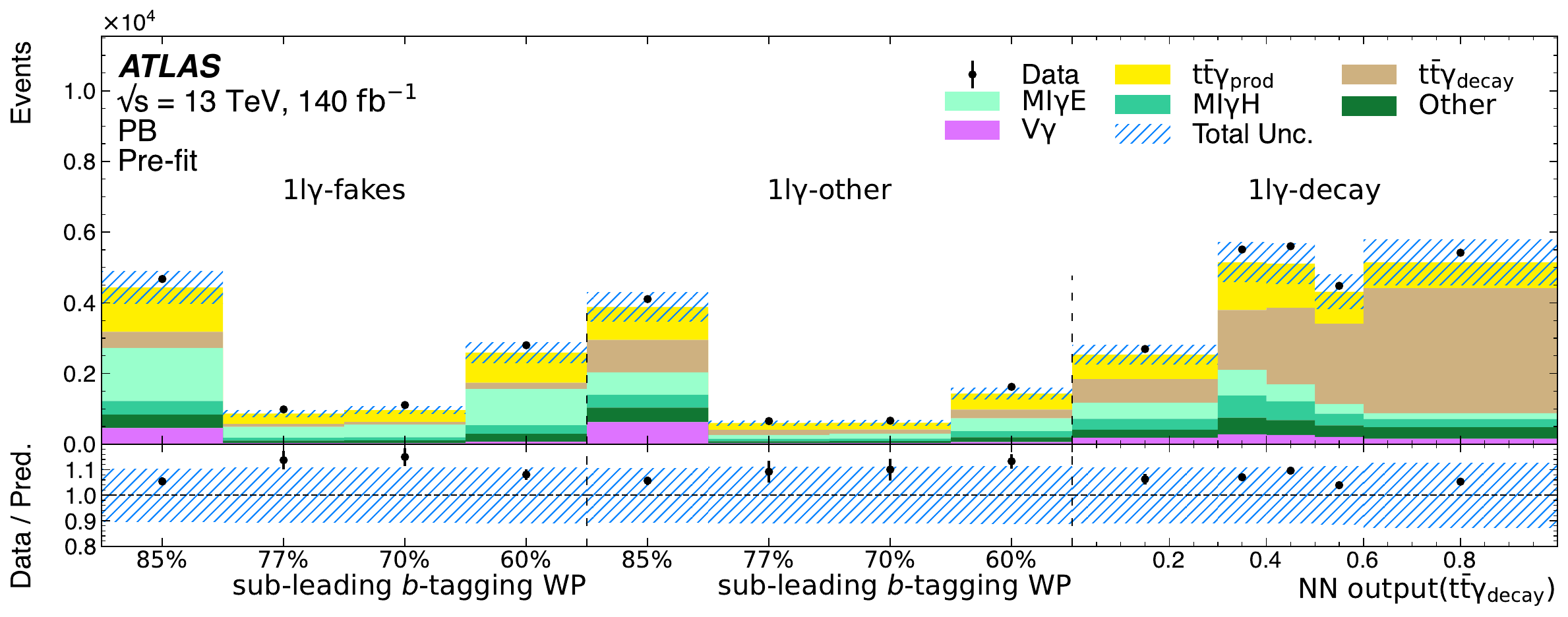}}\\
\caption{Distributions of the event yields observed in data compared to the expected yields before the fit to data for the
process-based (a) \PBonelGammaProd and \PBtwolOppSignGamma SRs and (b) \PBonelGammaFakes, \PBonelGammaOther and \PBonelGammaDecay regions.
The EFT WCs are set to the SM (all equal to zero) and background normalisation factors and fit nuisance parameters are set to their pre-fit predictions.
\UncertaintyBand \OtherCategoryDescription
\label{fig:App:Prefit_PB_Giga_gamma}
}
\end{figure}

\FloatBarrier

% End of text imported from the .//Sections/Appendices/PreFitPlots.tex input file

%
%
%
%
%
%
\FloatBarrier
\section{Expected results for WC sensitivity}
\label{app:expected}

% The next lines are included from the .//Sections/Appendices/HA_EFT_Results.tex input file

When comparing the OB and PB approaches, there is important
information contained in the expected results which can be somewhat
clearer to interpret than when looking at the fit to data. In this
section the expected sensitivity to the WCs is shown using a
realistic-Asimov fit. The realistic-Asimov fit sets normalisation
factors and nuisance parameter values to those observed in a fit data
in the CRs alone. These values are propagated to the full fit regions
including the SRs to obtain a more realistic estimate than the
full-Asimov configuration that considers what is already known from
the data in the control regions.
Figure~\ref{fig:app:OBPBEFTLimits_Comp_All_expected} shows these
results, grouped by operator type. It can be seen that the PB approach
has the best expected sensitivity to the majority of operators.

\begin{figure}[htb]
\centering
\begin{minipage}[t]{0.48\textwidth}
\centering

\vspace{-6cm}
\begin{subfigure}{\linewidth}
\centering
\includegraphics[width=\textwidth]{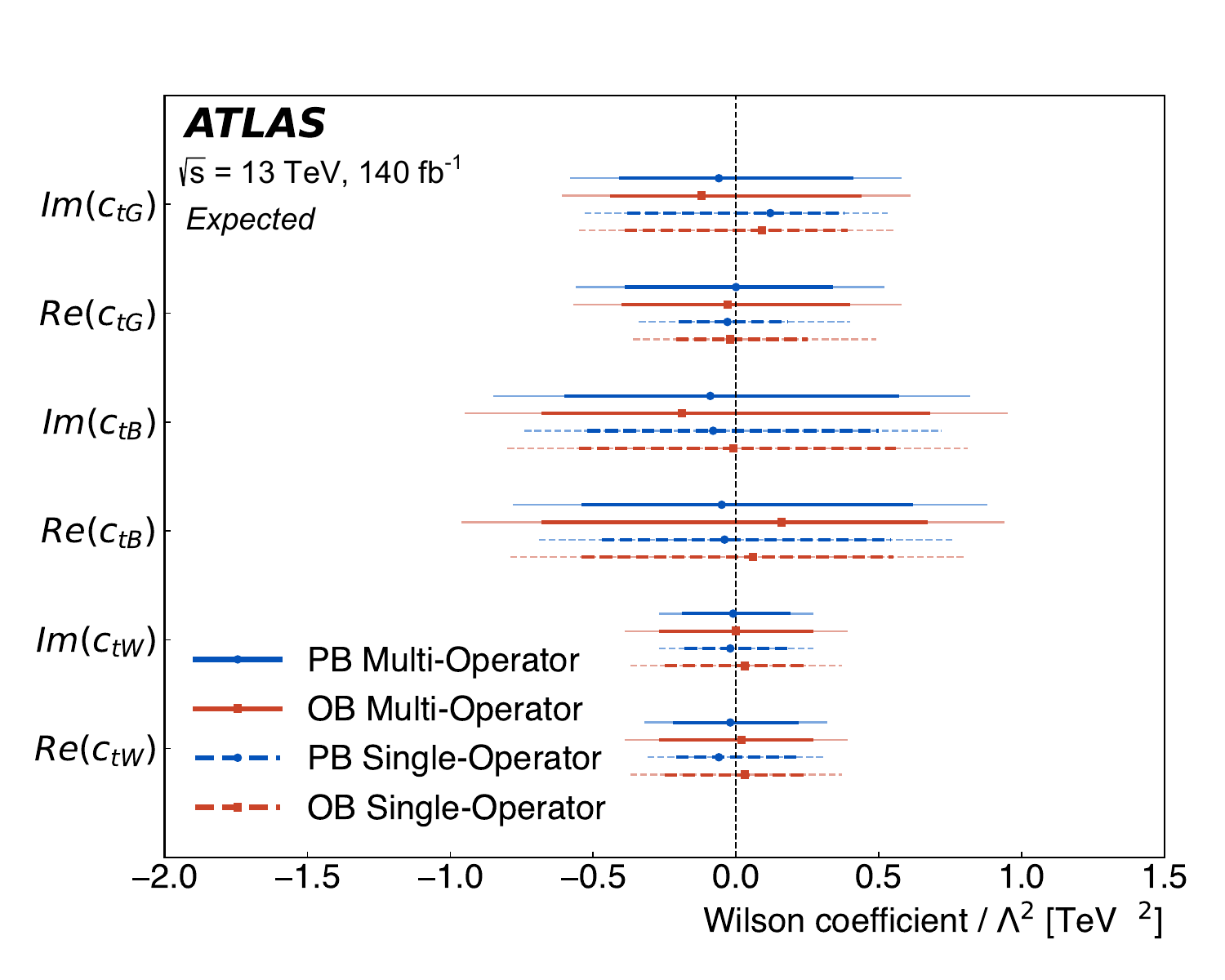}
\caption{}
\label{fig:HAimage1}
\end{subfigure}

\begin{subfigure}{\linewidth}
\centering

\includegraphics[width=\textwidth]{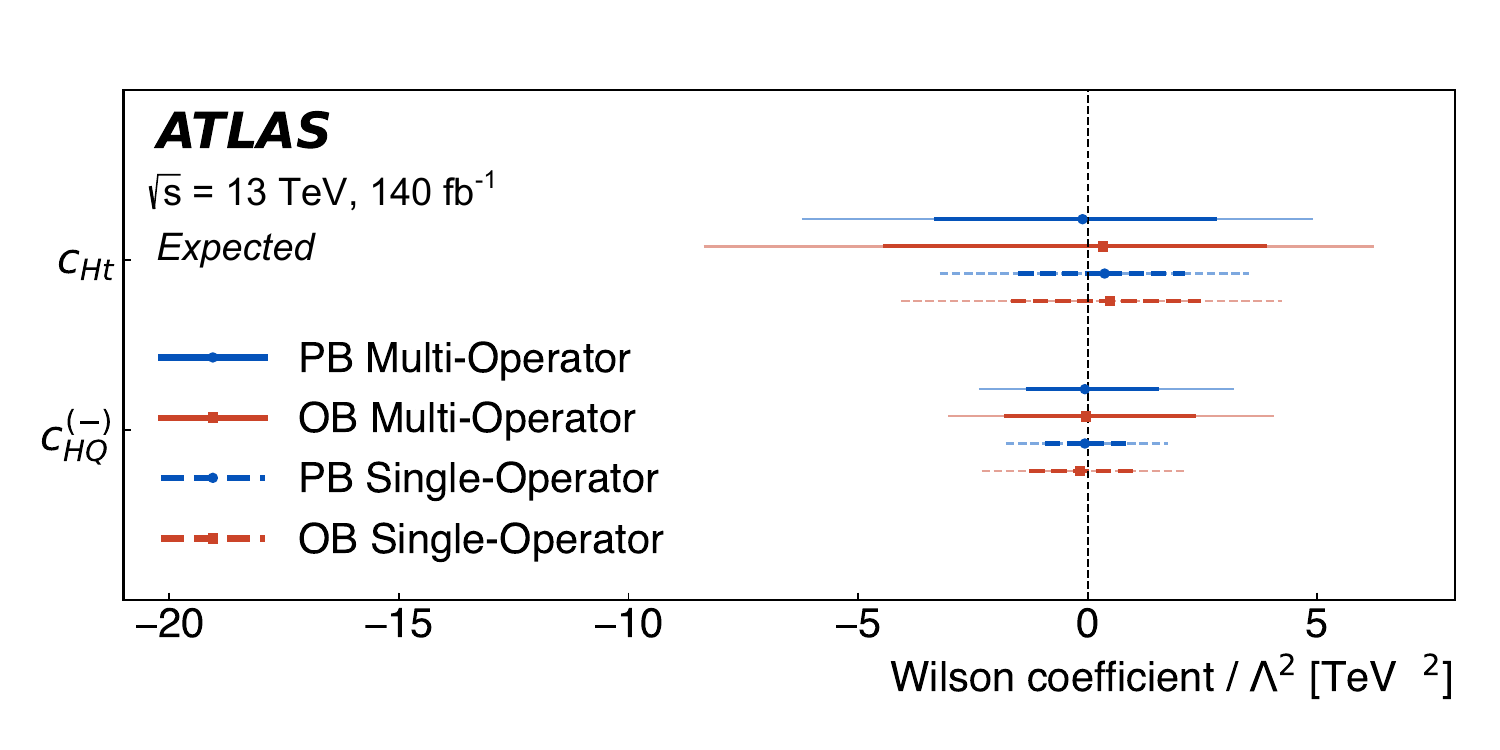}
\caption{}
\label{fig:HAimage2}
\end{subfigure}

\begin{subfigure}{\linewidth}
\centering

\includegraphics[width=\textwidth]{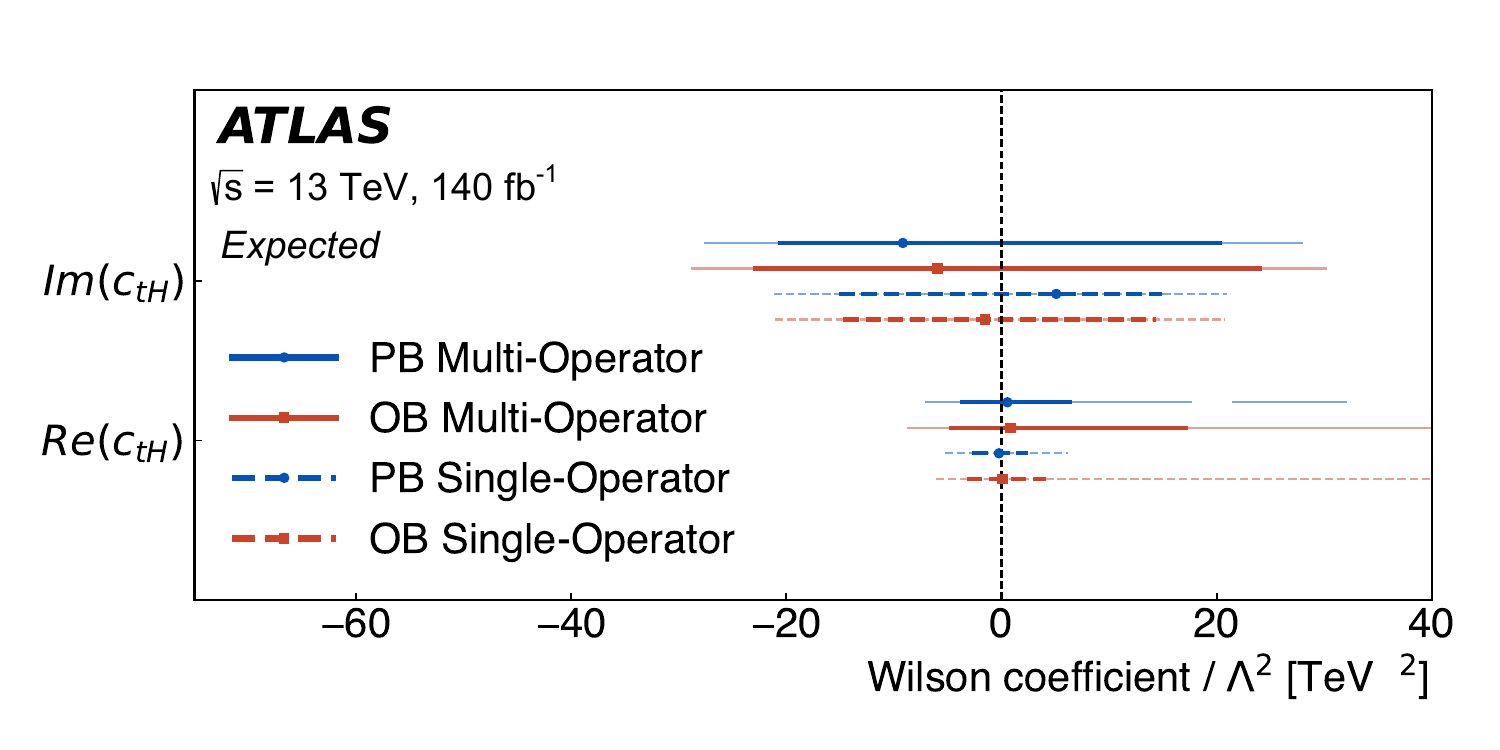}
\caption{}
\label{fig:HAimage3}
\end{subfigure}

\begin{subfigure}{\linewidth}
\centering
\includegraphics[width=\textwidth]{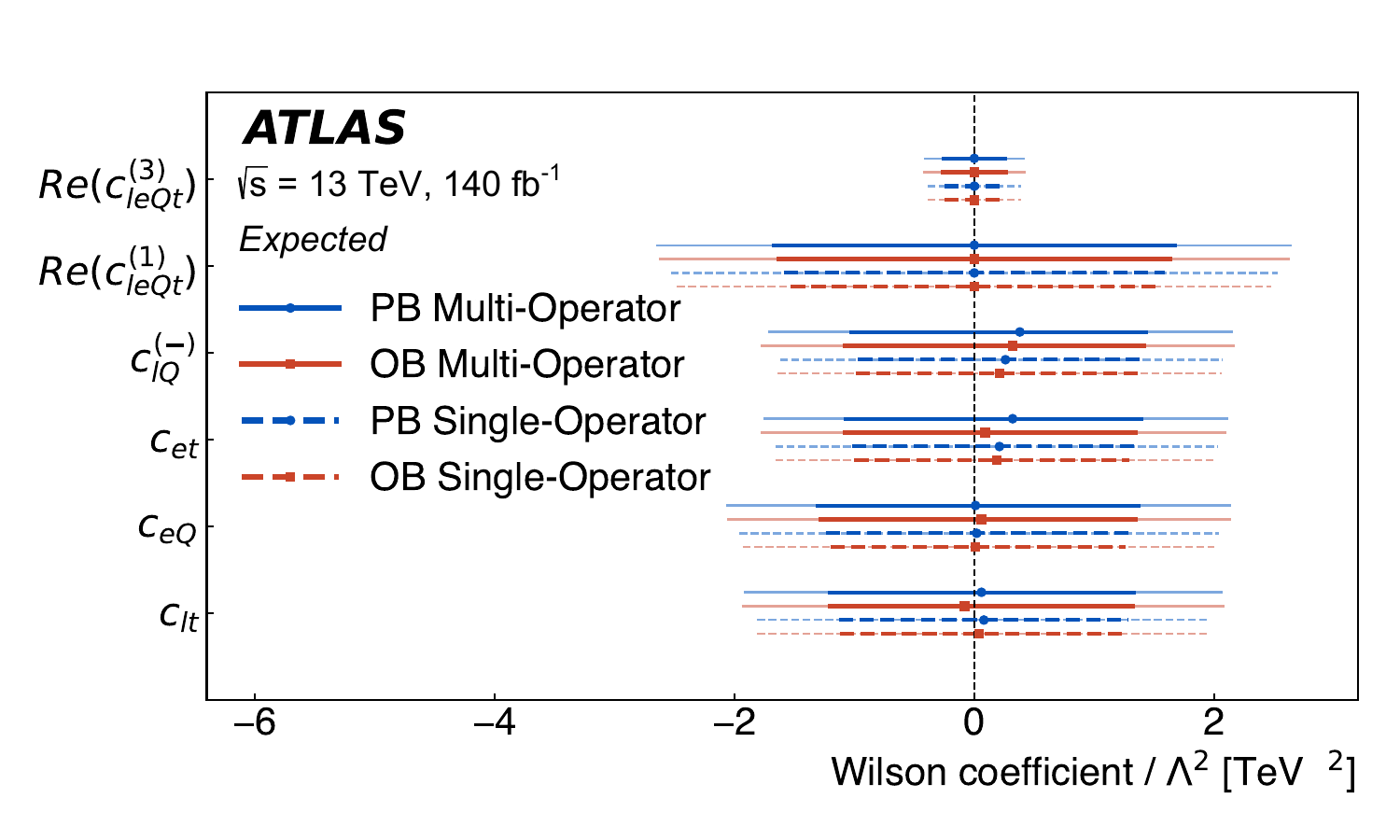}
\caption{}
\label{fig:HAimage4}
\end{subfigure}

\end{minipage}
\hfill
\begin{minipage}[t]{0.48\textwidth}
\centering

\begin{subfigure}{\linewidth}
\centering
\includegraphics[width=\textwidth]{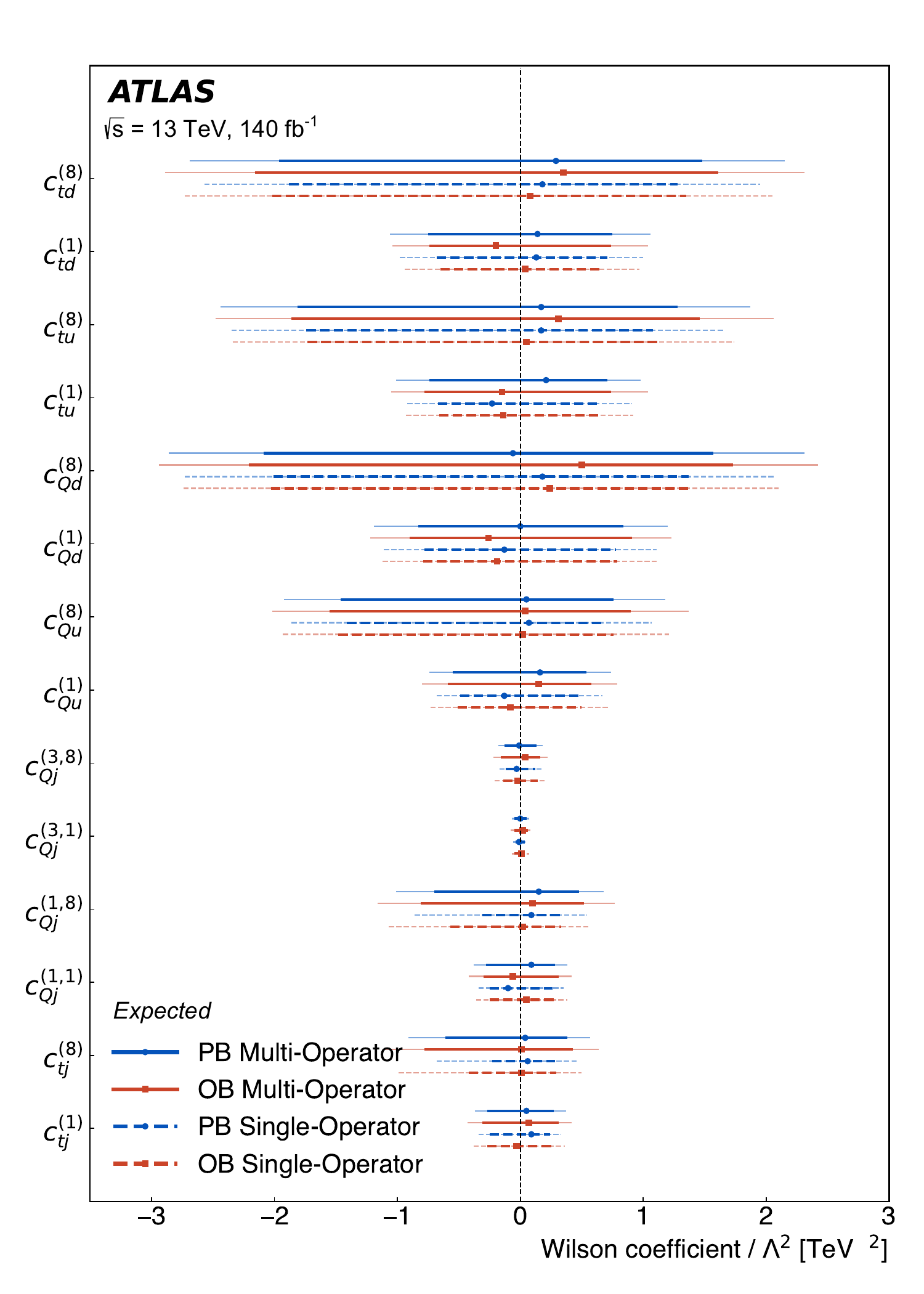}
\caption{}
\label{fig:HAimage5}
\end{subfigure}

\vspace{0.4em}

\begin{subfigure}{\linewidth}
\centering
\includegraphics[width=\textwidth]{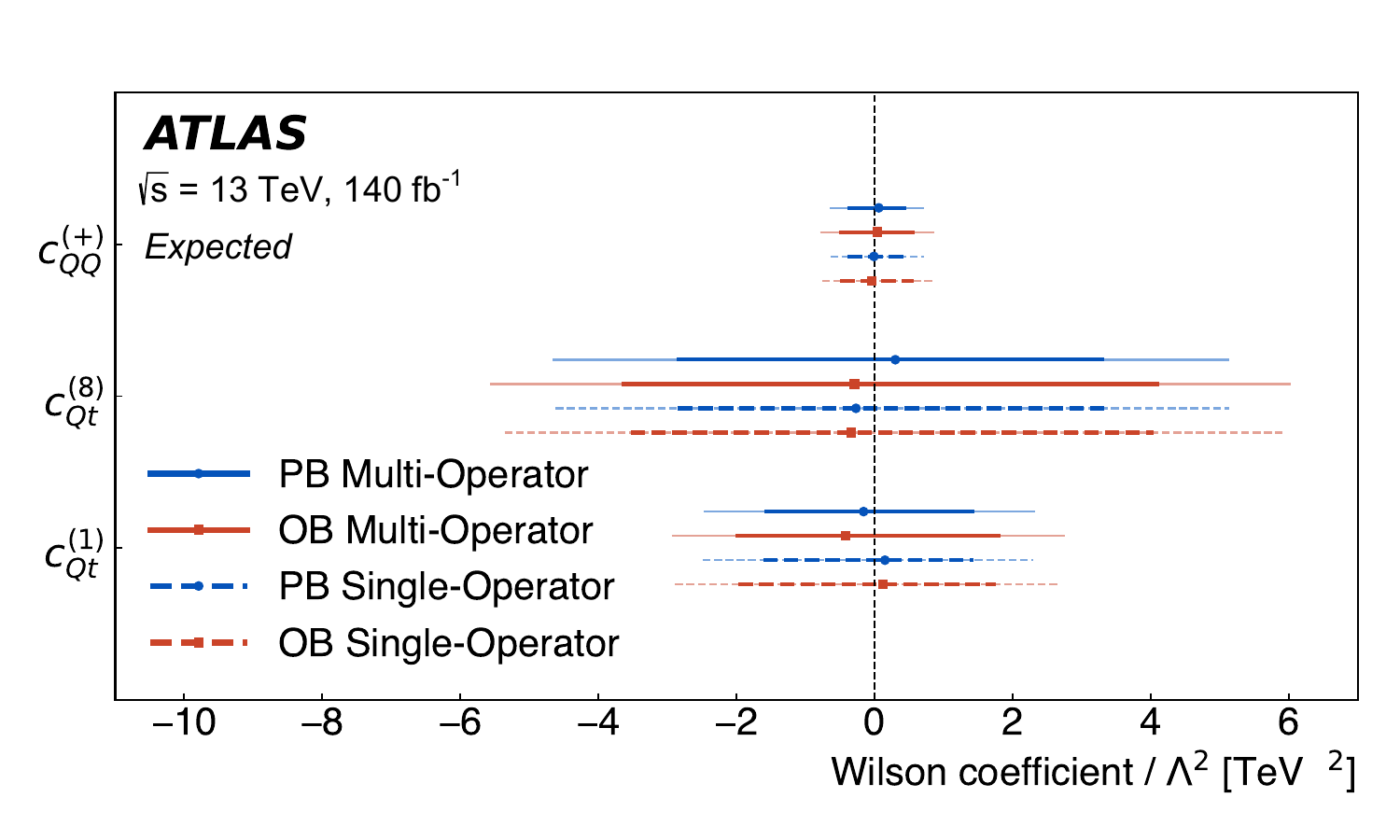}
\caption{}
\label{fig:HAimage6}
\end{subfigure}

\end{minipage}
\caption{Comparison of the expected (based on a realistic-Asimov fit model using observed data in the CRs and the corresponding SR background expectations) confidence intervals on the WCs for the object-based (OB) and process-based (PB) approaches in the top-boson category for (a) \ctG, \ctB and \ctW WCs, for (b) \cHt and \cHQ- WCs, and (c) \ctHRe and \ctHIm WCs. The top-lepton category (d) and the four-quark category for (e) the two-heavy-two-light WCs (f) the four-heavy-quark WCs. Results are shown where all WCs are allowed to float in the fit simultaneously (Multi-Operator), and for one fitted at a time (Single-Operator). The thicker lines correspond to 68\% CIs while the thinner lines correspond to 95\% intervals.}
\label{fig:app:OBPBEFTLimits_Comp_All_expected}
\end{figure}

\FloatBarrier

% End of text imported from the .//Sections/Appendices/HA_EFT_Results.tex input file

%
%
%
%

\FloatBarrier

\printbibliography
\clearpage
% ATLAS Collaboration author list
% Reference date of TOPQ-2023-34 is 2026-08-19
% Author list last updated on date 11-NOV-26
% Data extracted on 11-Nov-2026 for paper reference TOPQ-2023-34
% at 8:47am
 
\begin{flushleft}
\hypersetup{urlcolor=black}
{\Large The ATLAS Collaboration}

\bigskip

\AtlasOrcid[0000-0002-6665-4934]{G.~Aad}$^\textrm{\scriptsize 102}$,
\AtlasOrcid[0000-0001-7616-1554]{E.~Aakvaag}$^\textrm{\scriptsize 17}$,
\AtlasOrcid[0000-0002-5888-2734]{B.~Abbott}$^\textrm{\scriptsize 121}$,
\AtlasOrcid[0000-0002-0287-5869]{S.~Abdelhameed}$^\textrm{\scriptsize 83b}$,
\AtlasOrcid[0000-0002-1002-1652]{K.~Abeling}$^\textrm{\scriptsize 54}$,
\AtlasOrcid[0000-0001-5763-2760]{N.J.~Abicht}$^\textrm{\scriptsize 48}$,
\AtlasOrcid[0000-0002-8496-9294]{S.H.~Abidi}$^\textrm{\scriptsize 30}$,
\AtlasOrcid[0009-0003-6578-220X]{M.~Aboelela}$^\textrm{\scriptsize 44}$,
\AtlasOrcid[0000-0002-9987-2292]{A.~Aboulhorma}$^\textrm{\scriptsize 36e}$,
\AtlasOrcid[0000-0001-5329-6640]{H.~Abramowicz}$^\textrm{\scriptsize 154}$
% ...
\bigskip
\\

% ...
$^{17}$Department for Physics and Technology, University of Bergen, Bergen; Norway.\\
% ...
$^{30}$Physics Department, Brookhaven National Laboratory, Upton NY; United States of America.\\
% ...
$^{36}$$^{(a)}$Facult\'e des Sciences Ain Chock, Universit\'e Hassan II de Casablanca;$^{(b)}$Facult\'{e} des Sciences, Universit\'{e} Ibn-Tofail, K\'{e}nitra;$^{(c)}$Facult\'e des Sciences Semlalia, Universit\'e Cadi Ayyad, LPHEA-Marrakech;$^{(d)}$LPMR, Facult\'e des Sciences, Universit\'e Mohamed Premier, Oujda;$^{(e)}$Facult\'e des sciences, Universit\'e Mohammed V, Rabat;$^{(f)}$Institute of Applied Physics, Mohammed VI Polytechnic University, Ben Guerir; Morocco.\\
% ...
$^{44}$Physics Department, Southern Methodist University, Dallas TX; United States of America.\\
% ...
$^{48}$Fakult\"{a}t Physik, Technische Universit{\"a}t Dortmund, Dortmund; Germany.\\
% ...
$^{54}$II. Physikalisches Institut, Georg-August-Universit\"{a}t G\"ottingen, G\"ottingen; Germany.\\
% ...
$^{83}$$^{(a)}$Khalifa University of Science and Technology, Abu Dhabi;$^{(b)}$New York University Abu Dhabi, Abu Dhabi;$^{(c)}$United Arab Emirates University, Al Ain;$^{(d)}$University of Sharjah, Sharjah; United Arab Emirates.\\
% ...
$^{102}$CPPM, Aix-Marseille Universit\'e, CNRS/IN2P3, Marseille; France.\\
% ...
$^{121}$Homer L. Dodge Department of Physics and Astronomy, University of Oklahoma, Norman OK; United States of America.\\
% ...
$^{154}$Raymond and Beverly Sackler School of Physics and Astronomy, Tel Aviv University, Tel Aviv; Israel.\\
% ...
$^{a}$ Also at Affiliated with an institute formerly covered by a cooperation agreement with CERN.\\
$^{b}$ Also at An-Najah National University, Nablus; Palestine.\\
$^{c}$ Also at Borough of Manhattan Community College, City University of New York, New York NY; United States of America.\\
$^{d}$ Also at Center for Interdisciplinary Research and Innovation (CIRI-AUTH), Thessaloniki; Greece.\\
$^{e}$ Also at Centre of Physics of the Universities of Minho and Porto (CF-UM-UP); Portugal.\\
$^{f}$ Also at CERN, Geneva; Switzerland.\\
% ...

\end{flushleft}

% Created with Glance <Atlas.Glance@cern.ch>

%

%
%

%

%
%
%
%

\end{document}